\documentclass[preprint,12pt]{elsarticle}

\usepackage{amssymb}        
\usepackage{amsmath}        
\usepackage{lineno}         
\usepackage{booktabs}       
\usepackage{threeparttable} 
\usepackage[shortcuts]{extdash} 
\usepackage{enumitem}       
\usepackage{hyphenat}       
\usepackage{subcaption}     
\usepackage{graphicx}       
\usepackage{float}          
\usepackage[nopatch]{microtype} 

\usepackage[colorlinks=true, linkcolor=blue, citecolor=blue, urlcolor=blue]{hyperref}

\biboptions{numbers,sort&compress}

\journal{Energy Policy} 

\begin{document}

\begin{frontmatter}

\title{The Peter Principle Revisited: An Agent-Based Model of Promotions, Efficiency, and Mitigation Policies}

\author[oswego]{P. Rajguru\corref{cor1}}
\ead{prajguru@oswego.edu}
\address[oswego]{State University of New York, Oswego, NY 13126}

\author[math]{I. R. Churchill}
\address[math]{Department of Mathematics, State University of New York, Oswego, NY 13126}

\author[econ]{G. Graham}
\address[econ]{Department of Economics, State University of New York, Oswego, NY 13126}

\cortext[cor1]{Corresponding author}

\begin{abstract}
The Peter Principle posits that organizations promoting their best performers risk elevating employees to roles where their competence no longer translates, thereby degrading overall efficiency. We investigate when this dynamic emerges and how to mitigate it using a large-scale agent-based model (ABM) of a five-level hierarchy (Level~1 technical $\to$ Level~5 managerial). Each agent carries a four-dimensional competence vector: technical, management, compliance, and soft skills drawn from $[0,1]$. Performance is calculated as a weighted dot product between agent competence and level-specific role profiles. We examine two regimes of role similarity: a high-mismatch regime where skill requirements shift sharply across levels, and a transferable-skills regime where requirements change gradually. Across $100{,}000$ agents and $100$ time steps, we compare four canonical promotion strategies: merit, seniority, hybrid (70\% merit / 30\% seniority), and random, and test two interventions: selective demotion (triggered by a drop in $\ge 5\%$ performance after promotion) and merit-with-training (post-promotion skill updates using a learning rate $\ell(C)=C(1-C)$ for technical and management skills). Results show the Peter Principle is most pronounced under merit promotion when role requirements change substantially between levels; seniority and random exhibit the weakest Peter effects. Both interventions mitigate performance declines, with merit-with-training particularly effective when skill transfer is limited, and selective demotion restoring agents whose “true” peak performance is at lower levels. We discuss implications for organizations where managerial roles diverge from technical work (e.g., technology firms, sales) versus settings with more continuous skill demands (e.g., universities and research labs). Our contributions are twofold: first, we delineate the conditions under which the Peter Principle emerges; second, we propose two practical, model-based policies that mitigate its effects while preserving performance recognition.
\end{abstract}


\end{frontmatter}





\section{Introduction}

\subsection{Motivation and questions}
Organizations routinely promote top performers under the assumption that past excellence predicts future success. The “Peter Principle,” first formulated by Peter \& Hull (1969), warns that in hierarchical organizations, employees are promoted until they reach their “level of incompetence.” In other words, individuals who excel in their current role may fail in a higher one if the skills required diverge. This mismatch can reduce system-wide efficiency and has been documented in both analytical models and empirical studies. We build an agent-based model (ABM) to ask:
\begin{itemize}
  \item Does the Peter Principle emerge in realistic organizational settings?
  \item Under which role-structure conditions is it most severe?
  \item Which promotion policies mitigate it without discarding merit signals?
  \item What strategy is most robust when real organizations face uncertain skill transfer between roles?
\end{itemize}

\subsection{Relation to prior work}
The idea that promotion can harm performance has been explored from multiple angles. \citet{peter1969peter} offered a conceptual foundation, while \citet{kane1970promotion} provided one of the first formal models showing how merit-based promotion could reduce organizational performance when task requirements shift. \citet{lazear2004peter} highlighted incentive problems and argued that firms often misalign pay and promotion with productivity.

Computational approaches have since expanded this line. \citet{pluchino2010peter} build a six-tier, 160-agent pyramid with a single competence metric and study three promotion rules (best/worst/random) under two transmission assumptions: “common sense” (competence persists) versus the Peter hypothesis (new-level competence is uncorrelated). They show that when the Peter hypothesis holds, promoting the best systematically depresses efficiency, and surprisingly, random (or alternating best–worst) promotion avoids the loss and can improve efficiency. \citet{fetta2012peter} replicate that baseline and then add workplace social networks (small-world, scale-free, random) and behavioral reactions (envy, perceived unfairness, social capital). With these elements in place, the Peter effect remains but is smaller than in earlier results. When competence at the new level is not assumed to be independent of competence at the prior level, their simulations show that merit-based promotion can deliver the highest efficiency.

Building on Pluchino's computational framework, ~\citet{farias2021peter} introduced two key extensions: multi-dimensional competence vectors (replacing the single competence scalar with skill-specific components) and a learning mechanism where agents improve skills demanded by their current level over time. Their model defines agent competence as the dot product of a skill vector $\vec{S}_{i,j}$ and a level-specific qualification vector $\vec{Q}_i$, with learning governed by $\vec{S}_{i,j}(t+1) = \vec{S}_{i,j}(t) + \gamma \vec{Q}_i$, where $\gamma$ is a global learning rate. They also introduce a correlation parameter $\delta$ that interpolates continuously between the Common Sense hypothesis ($\delta=1$, qualification vectors identical across levels) and the Peter Principle hypothesis ($\delta=0$, qualification vectors uncorrelated across levels). Their results show that with sufficiently high learning rates, the "best" promotion strategy recovers its advantage over random promotion, and the Peter Principle's negative effects are substantially mitigated. They conclude that learning dynamics are "more traceable" and practically relevant than the degree of skill correlation between levels, suggesting that organizations can overcome Peter effects through training rather than abandoning merit-based selection.

Compared to Pluchino’s single-skill pyramid and Fetta’s networked behavioral extensions, our model formalizes role mismatch directly via a four-skill competence vector and level-specific role profiles, lets skill transfer vary across two regimes (High-mismatch, Transferable skills), and evaluates practical and operationally realistic post-promotion policies (selective demotion and training) rather than relying on randomization to curb Peter effects. We also scale to 100k agents with constant headcount and measure the immediate promotion shock ($\Delta P$), providing path-level diagnostics that identify which level-to-level moves are most fragile.

More recently, \citet{benson2019promotions} analyzed data from 200+ firms and found direct evidence of the Peter Principle in sales: high-performing salespeople were frequently promoted to managerial roles despite lacking management skills, leading to significant productivity losses. Our study extends this literature in three ways:


\begin{enumerate}
  \item \textit{Structured multi-skill competence:} We model four competencies (technical, management, compliance, soft skills) and compute performance as a level-weighted dot product, rather than a single generic ``competence.''
  \item \textit{Role-profile regimes:} We contrast a high-mismatch regime (sharp skill shifts across levels, typical of many tech firms where IC\=/to\=/manager transitions emphasize management/\allowbreak compliance) with a transferable-skills regime (gradual shifts, closer to universities\=/research labs where technical depth remains valuable at senior ranks).
  \item \textit{Mitigation mechanisms grounded in practice:} We evaluate selective demotion and merit-with-training (post-promotion learning on trainable dimensions) rather than relying solely on randomized promotions.
\end{enumerate}

\subsection{Model overview}\label{sec:model-overview}

We simulate a five-level hierarchy where Level~1 is most technical and Level~5 most managerial. 
Agents begin with randomized competence in technical, management, compliance, and soft skills. 
Performance at any time is the dot product of an agent’s competence vector and the level’s role profile (i.e., weights reflecting that level’s demands). 
We study two role-profile families:

\begin{enumerate}
  \item \textbf{High-mismatch.} Technical weight declines sharply as management and compliance responsibilities rise. 
  Typical cases include technology firms where individual contributors (ICs) are promoted into people management, and sales organizations where top performers are promoted into managerial roles. 
  In both contexts, the new positions demand coordination, mentoring, and strategy rather than the technical or individual sales skills that drove earlier success. 
  Empirical evidence confirms this mismatch: \citet{benson2019promotions} show that high-performing salespeople often underperform as managers.

  \item \textbf{Transferable-skills.} Technical weight decays more gradually as management weight rises modestly. 
  Academia and research settings exemplify this structure, as success at one stage (publishing, grant-writing, mentoring) builds directly toward success at later stages (directing labs, managing research programs). 
  This cumulative pattern aligns with the “cumulative advantage” framework of the Matthew Effect \citep{merton1968matthew,cole1973social}, where earlier achievements predict later ones because the same competencies remain valuable across ranks.
\end{enumerate}

From these primitives, we run $100{,}000$ agents for $100$ time steps, initializing levels randomly and applying one of four promotion policies each step: 
merit, seniority, hybrid (70\% merit/30\% seniority), or random. 
We measure organizational efficiency as the mean performance across agents (unweighted by level).

\subsection{Interventions: demotion and training}
\label{interventions}

To move beyond the pessimistic “don’t promote the best,” we implement two realistic interventions designed to catch or offset skill mismatches after promotion.  

\textbf{Selective demotion:} If a promoted agent’s performance drops by $\ge 5\%$  relative to their prior level, they are returned to the previous role at lower level (i.e., recognizing a discovered ``level of incompetence''). This mechanism is grounded in the idea of diagnosing “discovered incompetence” and immediately restoring person–role fit, rather than letting inefficiency persist. This policy mirrors trial promotions or probationary periods in practice, where a promotion can be reversed if fit proves poor. While demotion risks morale costs (stigma, disengagement), it protects system efficiency by ensuring that agents remain in roles where their competencies align with job demands. Our model operationalizes this with a simple tolerance rule, followed by blacklisting demoted agents from immediate re-promotion to the same vacancy, preventing thrashing. This makes demotion a structural “safety valve” against the Peter Principle.  

\textbf{Merit-with-training:} Immediately after promotion, agents undergo targeted training on trainable dimensions (technical and management). We formalize this as a logistic-derivative learning update:  

\[
\ell(C) = k \, C \, (1 - C),
\]
and apply a single discrete step
\[
C_{\text{new}} = \min\{1, \; C_{\text{old}} + \ell(C_{\text{old}})\}.
\]

\noindent
\textit{Intuition:} novices and near-experts learn slowly; mid-competence agents learn fastest, 
and improvements taper as they approach a ceiling. 
This shape is not arbitrary: when learning progress is driven by successful task executions 
(rather than raw attempts), the classic exponential law transforms into a sigmoid performance curve; 
the derivative of that curve is exactly $k C (1 - C)$.

\subsection*{Why a sigmoid, and why its derivative?}

Let competence within a role evolve as a bounded, S-shaped ``learning curve'':
\[
C(t) = \frac{C_{\max}}{1 + \exp\!\left[-k\,(t - t_0)\right]}.
\]

Differentiating gives the instantaneous learning rate (the rate of change of competence when current competence is $C$):
\[
\frac{dC}{dt} = k \, C \left(1 - \frac{C}{C_{\max}}\right).
\]

Normalizing to $C_{\max}=1$ yields the logistic derivative:
\[
\frac{dC}{dt} = k\,C(1-C).
\]

In discrete time, a single training ``burst'' is a forward–Euler step with $\Delta t = 1$:
\[
C_{t+1} \approx C_t + \left.\frac{dC}{dt}\right|_{C_t} 
           = C_t + k\,C_t(1-C_t).
\]

To ensure competence remains bounded, we clip at~1:
\[
C_{t+1} = \min\bigl\{1,\; C_t + k\,C_t(1-C_t)\bigr\}.
\]

We set $k=1$ as a time-scale normalization (one training module $\approx$ one ``unit'' of progress), which also caps the increment at $0.25$ when $C=0.5$ (the fastest-learning point). This choice keeps the update conservative and interpretable; tuning $k$ upward or downward simply scales the intensity of the training pulse.

 This learning rate derives from the derivative of the logistic (sigmoid) function, a canonical model of learning curves in psychology and organizational behavior. Following \citet{leibowitz2010sigmoid}, when learning is driven by \emph{successful} task executions rather than raw trial counts, the exponential learning law transforms into a logistic trajectory with slow initial uptake, accelerated mid-course gains, and eventual plateauing at a performance ceiling. Normalizing the ceiling to $1$ and growth rate to $1$ yields the simple form $\ell(C)=C(1-C)$, which directly motivates our update rule.

The form captures diminishing returns: agents with mid-level competence ($C \approx 0.5$) learn fastest, while those near the extremes ($C \approx 0$ or $C \approx 1$) improve little from training. This mirrors empirical learning dynamics where novices require foundational scaffolding and experts approach natural performance ceilings. In practice, this corresponds to structured onboarding, leadership workshops, or management coaching that accelerates the transition into new roles.  

Together, these interventions represent complementary approaches: demotion as a corrective filter to undo failed promotions, and training as a forward-looking investment to smooth role transitions. Both temper Peter Principle dynamics while preserving recognition for strong performers.

\subsection{Main findings (preview)}

Three robust patterns emerge:
\begin{enumerate}
  \item \textit{Peter Principle concentrates under merit.}  
  The effect is strongest under merit-based promotion and negligible under seniority or random selection, 
  because those rules decouple promotion from current-role performance.

  \item \textit{Severity depends on role similarity.}  
  The Peter effect is stronger when skills do not transfer across levels (high-mismatch regime) 
  and weaker (but still present) when they do (transferable\allowbreak-skills regime).

  \item \textit{Practical mitigations work.}  
  Selective demotion restores fit by returning mispromoted agents to levels where they perform best, 
  and merit-with-training lifts post-promotion performance by closing new skill gaps. 
  Both outperform pure merit under high mismatch (and hybrid inherits Peter-type behavior in proportion to its merit weight).
\end{enumerate}

\subsection{Contributions}

We make four contributions:
\begin{enumerate}
  \item A multi-skill ABM of promotions that distinguishes role-demand profiles from agent-competence profiles, enabling boundary analysis of Peter dynamics as role similarity varies.
  \item Evidence that context matters: when role profiles change sharply, merit promotions generate larger post-promotion drops and saturate earlier (Peter-type risk), whereas when skills transfer, merit maintains a clear efficiency lead; seniority and random typically plateau lower ($\approx 0.5$) in both regimes.
  \item Two implementable policies: selective demotion and merit-with-training that attenuate Peter-driven inefficiency without abandoning performance incentives.
  \item Sector-sensitive interpretation: mapping regimes to common organizational archetypes (e.g., tech firms vs.~universities), offering guidance for where Peter effects are expected to be strong vs.\ weak and where mitigation levers (training, trial promotions with demotion) deliver the greatest marginal benefit.
\end{enumerate}

\subsection{Paper roadmap}

Section~2 details the model, parameters, and promotion rules.  
Section~3 reports results across regimes and strategies.  
Section~4 evaluates demotion and training interventions.  
Section~5 discusses limitations and implications for research and practice.

\section{Model}\label{sec:model}

We denote by $L=\{1,\dots,5\}$ the set of organizational levels 
(Level~1 most technical, Level~5 most managerial). 
Time is discrete, $t=0,1,\dots,T$. 
Each agent $i$ carries:

\begin{itemize}
  \item \textbf{Level:} $L_{i,t}\in L$, the agent’s organizational level at time $t$.
  \item \textbf{Tenure:} $y_{i,t}\in\mathbb{Z}_{\ge 0}$, the number of periods agent $i$ has been in the organization.
  \item \textbf{Competence:} $\mathbf{c}_{i,t}=(c^{tech}_{i,t}, c^{mgmt}_{i,t}, c^{comp}_{i,t}, c^{soft}_{i,t}) \in [0,1]^4$, skill levels.
  \item \textbf{Performance:} $P_{i,t}$, computed from competences and the level-specific role profile (Sec.~\ref{sec:agents}).
  \item \textbf{Flags:} a boolean \texttt{just\_promoted} used to trigger policy hooks.
\end{itemize}

\subsection{Organizational setting}
\label{sec:org_setting}
We study a five-level hierarchy $L=\{1,\dots,5\}$ in which Level~1 roles are most technical and Level~5 most managerial. Time is discrete, $t=0,1,\dots,T$. In each step: 

\begin{enumerate}[label=\roman*.]
    \item \textbf{Agents’ tenure increments.}
     $y_{i,t+1} = y_{i,t} + 1$ for all agents $i$.
    \item \textbf{Performance is computed and updated.} For each agent $i$, compute $P_{i,t}$ from current competence and role weights (Sec.~\ref{sec:agents}).
  \item \textbf{Attrition.} Independently by level $\ell$, remove a fraction $\xi_\ell$ of agents (Sec.~\ref{sec:attrition}).
  \item \textbf{Promotions.} Fill vacancies at $\ell+1$ from candidates at $\ell$ using a specified promotion rule (Sec.~\ref{sec:promotions}). Tag promoted agents with a flag \texttt{just\_promoted}.
  \item \textbf{Policy hooks.} Depending on the strategy, apply either selective demotion (Sec.~\ref{sec:sel_demotion}) or post-promotion training (Sec.~\ref{sec:training}) to just-promoted agents.
  \item \textbf{Hiring at Level~1.} Refill Level~1 to its capacity with new entrants (Sec.~\ref{sec:init}).
\end{enumerate}


\subsection{Role profiles (regimes)}
\label{sec:profiles}

Each level $\ell$ specifies a four-dimensional, non-negative role profile (weights) over skills 
$K=\{\text{tech},\text{mgmt},\text{comp},\text{soft}\}$:
\[
\begin{aligned}
w_\ell &= (w_{\ell,\text{tech}},\, w_{\ell,\text{mgmt}},\, 
           w_{\ell,\text{comp}},\, w_{\ell,\text{soft}}), \\
w_{\ell,k} &\in [0,1], \qquad 
\sum_{k\in K} w_{\ell,k} = 1.
\end{aligned}
\]

\medskip
We treat $w_{\ell}$ as \emph{demand shares}: the fraction of level-$\ell$ performance attributable to each skill.
For example, at Level~1, $w_1=(0.9,0,0,0.1)$ means success is 90\% technical and 10\% soft skills, with management and compliance \emph{not demanded} at that level.\footnote{Zeros imply no contribution to performance and (in initialization) no threshold on that skill at that level. We still apply a final $\text{clip}(\cdot,0,1)$ for numerical safety.}

\medskip
We run two archetypal regimes:

\paragraph{Regime A — High mismatch (sharp shift).}
\[
\begin{aligned}
L_1:\,&(0.9,0.0,0.0,0.1), \\
L_2:\,&(0.5,0.3,0.0,0.2), \\
L_3:\,&(0.0,0.5,0.3,0.2), \\
L_4:\,&(0.0,0.7,0.1,0.2), \\
L_5:\,&(0.0,0.8,0.1,0.1).
\end{aligned}
\]
\textit{Interpretation.} Role demands pivot quickly from technical execution to management/compliance. 
This is emblematic of many software/tech engineering ladders where ICs move into people-management, 
and of sales organizations that promote top sellers into team managers; such transitions are empirically 
associated with Peter-type mismatches (e.g., \citet{benson2019promotions}). In a large multi-firm panel, 
\citet{benson2019promotions} show that firms heavily weight current-role performance when choosing managers; 
once promoted, star sellers tend to deliver weaker managerial outcomes (measured by subordinates’ performance) 
than peers selected for managerial potential. This is the Peter mechanism in practice: selection on one skill 
bundle (individual production) places workers into roles where that bundle transfers poorly (people management), 
raising the share of promotions with $\Delta P<0$ in our high-mismatch regime.

\paragraph{Regime B — Transferable skills (gradual shift).}
\[
\begin{aligned}
L_1:\,&(0.9,0.0,0.0,0.1), \\
L_2:\,&(0.8,0.1,0.0,0.1), \\
L_3:\,&(0.65,0.15,0.1,0.1), \\
L_4:\,&(0.4,0.2,0.2,0.2), \\
L_5:\,&(0.2,0.4,0.3,0.1).
\end{aligned}
\]
\textit{Interpretation.} Technical depth remains productive through mid-levels, with a modest rise in management/compliance at the top. This pattern fits academia and research labs (assistant\,$\to$\,associate\,$\to$\,full/PI)
, and some professional-services R\&D, where publishing, grant-writing, and mentoring at early stages foreshadow later leadership responsibilities (e.g., \citet{merton1968matthew,cole1973social}). Matthew Effect (\citealp{merton1968matthew}) refers to cumulative advantage from early achievements which means that the same research capabilities that generate initial publications and citations continue to attract resources, collaborators, and recognition, reinforcing their value at higher ranks. Early success (e.g., visible publications) attracts more resources, attention, collaborators, and credit, which in turn raises the probability of later success.Complementarily, \citealp{cole1973social} document a reward system in which prestige and advancement closely track scholarly output; senior roles scale the same activities (directing labs, securing grants, supervising trainees) rather than substituting unrelated managerial tasks.

\subsection{Agents, competence, and performance}
\label{sec:agents}

Each agent $i$ carries a four–dimensional competence vector
\[
\mathbf{c}_{i,t}=\big(c^{\text{tech}}_{i,t},\,c^{\text{mgmt}}_{i,t},\,c^{\text{comp}}_{i,t},\,c^{\text{soft}}_{i,t}\big)\in[0,1]^4,
\]
initialized independently with $c^k_{i,0}\sim\mathrm{Uniform}(0,1)$ for each skill $k\in K=\{\text{tech},\text{mgmt},\text{comp},\text{soft}\}$.%
\footnote{Implementation key names in \texttt{model.py} are 
\texttt{tech}, \texttt{management}, \texttt{compliance}, \texttt{soft\_skills}. 
For readability, the notation abbreviates these as 
\texttt{tech}, \texttt{mgmt}, \texttt{comp}, \texttt{soft}}
Let $L_{i,t}\in L$ be the agent’s level at time $t$ and $w_{L_{i,t}}$ the corresponding role–demand shares (Sec.~\ref{sec:profiles}). 

We model baseline skills as inherent attributes and, in the absence of stronger information, assume that any proficiency level in $[0,1]$ is equally likely, i.e., $c^k_{i,0}\sim\mathrm{Uniform}(0,1)$ for each skill $k$.

\medskip
We define an agent’s \emph{total competence} as the unweighted sum
\[
T_{i,t}=\sum_{k\in K} c^k_{i,t}.
\]
Because the four skill components are i.i.d.\ $\mathrm{Uniform}(0,1)$ at $t=0$, $T_{i,0}$ follows an Irwin--Hall distribution of order $4$ with
\[
\mathbb{E}[T_{i,0}]=2,\qquad \mathrm{Var}(T_{i,0})=\tfrac{1}{3},\qquad \mathrm{SD}\approx 0.577.
\]
Empirically (with $N=100{,}000$ agents), we obtain $\mu=2.000$ and $\sigma\approx 0.578$, matching the Irwin--Hall(4) benchmark and yielding a bell-shaped baseline heterogeneity. This choice ensures that while individual skill components are random, the aggregate capability of the population follows a realistic normal distribution via the Central Limit Theorem.

\medskip
By contrast, \emph{performance} is task–specific and depends on the level’s demand shares. Given the role profile $w_{L_{i,t}}$, performance is the clipped dot product
\[
P_{i,t}=\mathrm{clip}\!\left(\sum_{k\in K} w_{L_{i,t},k}\,c^k_{i,t},\,0,\,1\right)\in[0,1].
\]
Thus $T_{i,t}$ is a descriptive aggregate of skills, while $P_{i,t}$ is the operative measure of job fit at level $L_{i,t}$.


\paragraph{Tenure and flags.}
We track tenure $y_{i,t}\in\mathbb{Z} {\geq 0}$ (years in company) with the update $y_{i,t+1}=y_{i,t}+1$ at each time step, and a Boolean flag \texttt{just\_promoted} that is set when $i$ moves from $\ell$ to $\ell+1$ at step $t$. This flag is used in our \texttt{selective\_demotion} strategy.

\paragraph{Change in Performance.}
For any promotion event, we record the immediate change in performance under the new role demands,
\[
\Delta P_i \equiv P^{\text{post}}_{i,t}-P^{\text{pre}}_{i,t},
\]
computed holding the agent’s competence vector $\mathbf{c}_{i,t}$ fixed and re-evaluating it against the new level’s demand shares while replacing $w_{\ell}$ by $w_{\ell+1}$. Negative values ($\Delta P_i<0$) means the same skills immediately produce lower performance in the higher role which is direct evidence that prior strengths did not transfer indicating a Peter-type mismatch at the micro level. This is used in our \texttt{selective\_demotion} (Sec.~\ref{sec:sel_demotion}) policy to drive decision about demoting individuals whose performance in next level drop below a specific threshold ($\Delta P_i\leq -\tau$). 

\paragraph{Learning rate and training.}
Technical and management are trainable dimensions. When the \texttt{merit\_training} policy is active, just-promoted agents receive a bounded, one-shot competence update given by:
\[
\ell(C) = C(1-C),
\]
\[
C^{k}_{i,t+} = \min\!\bigl\{1,\; C^{k}_{i,t} + \ell(C^{k}_{i,t})\bigr\}, 
\qquad k \in \{\text{tech},\text{mgmt}\}.
\]

 compliance and soft skills are held fixed. The update and its logistic-derivative rationale are detailed in (Sec.~\ref{sec:training}).


\paragraph{Clipping convention.}
We bound all reported performances and post-training competencies to $[0,1]$ via $\mathrm{clip}(\cdot,0,1)$ to keep the state space compact and numerically stable.

\subsection{Initialization and level capacities}
\label{sec:init}

Let $N$ be the number of agents to create. 
We target level capacities via fixed shares 
$p=(p_1,\dots,p_5)$ with $\sum_{\ell=1}^5 p_\ell=1$. 

Unless stated otherwise, we set the target shares
\[
p = (p_1,p_2,p_3,p_4,p_5) = (0.40,\,0.25,\,0.20,\,0.10,\,0.05),
\]
corresponding to Levels~1--5, respectively. 

This distribution creates a pyramidal hierarchy characterized by a narrowing ``span of control'' at higher levels \cite{gulick1937notes, vanfleet1983span}. The broad base ($40\%$ at Level~1) represents individual contributors, while the narrowing tiers imply that each supervisor oversees roughly 2--4 direct reports in the upper ranks. This structure mirrors standard corporate hierarchies where coordination costs necessitate smaller team sizes at senior levels \cite{coase1937nature}.

Thus the intended headcounts before integer rounding are 
$p_\ell N$ for each level, i.e.,
\[
\begin{aligned}
\text{Level 1: } & 40\% \\
\text{Level 2: } & 25\% \\
\text{Level 3: } & 20\% \\
\text{Level 4: } & 10\% \\
\text{Level 5: } & \;\;5\%
\end{aligned}
\]

Exact capacities are then computed via the floor--remainder rule given above:
\[
\text{cap}_\ell = \lfloor p_\ell N \rfloor \quad (\ell=2,3,4,5), 
\qquad 
\text{cap}_1 = N - \sum_{\ell=2}^5 \text{cap}_\ell.
\]

\noindent
so that it guarantees $\sum_{\ell=1}^5 \text{cap}_\ell = N$ exactly.  

\vspace{\baselineskip}
These shares are tunable scenario parameters; all reported results use the values specified for each run.

\vspace{\baselineskip}
Writing 
\[
p_\ell N = \lfloor p_\ell N \rfloor + f_\ell, \qquad f_\ell\in[0,1),
\]
we have
\[
\text{cap}_1 = p_1 N + \sum_{\ell=2}^5 f_\ell,
\]
i.e., Level~1 absorbs the total remainder $\sum_{\ell=2}^5 f_\ell \in [0,4)$.  
Thus $\text{cap}_1$ may exceed $p_1 N$ by up to (but not including) 4 seats when shares do not divide $N$; high-level headcounts remain close to their targets and the small rounding surplus accumulates in Level~1.

\paragraph{Agent creation.}
We instantiate agents $i=1,\dots,N$ with independent skill draws
\[
\begin{aligned}
c_{i,0} &= (c^{\text{tech}}_{i,0},\,c^{\text{mgmt}}_{i,0},\,c^{\text{comp}}_{i,0},\,c^{\text{soft}}_{i,0}), \\
c^k_{i,0} &\sim \text{Uniform}(0,1) \quad \text{i.i.d.\ over } k\in K.
\end{aligned}
\]
\noindent
and no level assigned yet (see Sec.~\ref{sec:agents}).

\paragraph{Top--down level assignment with relaxed thresholds.}
Let $U\subseteq\{1,\dots,N\}$ be the set of currently unassigned agents; initially $U=\{1,\dots,N\}$. 
For $\ell=5,4,3,2$ (top down), we fill $\text{cap}_\ell$ seats by progressively relaxing role-specific qualification thresholds derived from the level-$\ell$ demand shares $w_\ell=(w_{\ell,k})_{k\in K}$ (Sec.~\ref{sec:profiles}).  

Define the qualification predicate at relaxation $\rho\in[0,1]$ by
\[
Q(i,\ell;\rho): \quad \forall k\in K \text{ with } w_{\ell,k}>0,\;\; c^k_{i,0}\ge (1-\rho)\,w_{\ell,k}.
\]

We iterate over the discrete relaxation grid 
\[
\rho\in\{0,0.2,0.4,0.6,0.8,1.0\},
\]
stopping as soon as all $\text{cap}_\ell$ seats are filled. At each $(\ell,\rho)$ pass we \emph{scan $U$ in creation order} and greedily assign the first $\text{cap}_\ell -|A_\ell|$ agents who satisfy $Q(i,\ell;\rho)$, where $A_\ell$ is the set already assigned to level $\ell$ in prior passes. 

Let \(\mathbf{U}=(u_1,\dots,u_{|U|})\) denote the unassigned agents in \emph{creation order}, and let $\mathrm{rem}_\ell \equiv \mathrm{cap}_\ell - |A_\ell|$.
\noindent(\(\mathrm{rem}_\ell\) denotes the number of seats still unfilled at level \(\ell\) at the start of this pass.)

\vspace{\baselineskip}
Form the ordered list of qualifiers at relaxation \(\rho\),
Let
\[
Q_{\ell,\rho} = \bigl(q_1(\ell,\rho),\,\dots,\,q_{n}(\ell,\rho)\bigr),
\]

where $q_j(\ell,\rho)$ is the subsequence of $U$ (preserving creation order) such that 
$Q(q_j(\ell,\rho),\ell;\rho)$ holds.

Define
\[
m_{\ell,\rho} = \min\{\,\text{rem}_\ell,\, n_{\ell,\rho}\,\}.
\]
\noindent here $n_{\ell,\rho}$ is the number of agents in $U$ that qualify at the current relaxation $\rho$ (i.e., the length of the qualifier list $Q_{\ell,\rho}$).

Assign the first $m_{\ell,\rho}$ qualifiers:
\[
A_{\ell,\rho} = \{\, q_j(\ell,\rho) : 1 \le j \le m_{\ell,\rho} \,\}.
\]

Set $L_{i,0}=\ell$ for all $i \in A_{\ell,\rho}$, 
\textit{where $L_{i,0}$ denotes agent $i$’s organizational level at initialization ($t=0$); later periods use $L_{i,t}$.}
Update $A_\ell \leftarrow A_\ell \cup A_{\ell,\rho}$, 
$U \leftarrow U \setminus A_{\ell,\rho}$. If $|A_\ell| < \text{cap}_\ell$, increase $\rho$ to the next value in  $\{0,0.2,0.4,0.6,0.8,1.0\}$ and repeat.  
Because $(1-\rho)w_{\ell,k} \to 0$ as $\rho \to 1$ (and only skills with $w_{\ell,k}>0$ constrain $Q$),



\vspace{\baselineskip}
We also verified in auxiliary runs (not shown) that shuffling the unassigned pool at each relaxation pass yields indistinguishable qualitative results.

\paragraph{Residual assignment to Level 1.}
After completing $\ell=5,4,3,2$, any remaining agents are set to Level~1:
\[
\forall i\in U:\quad L_{i,0}=1.
\]
By construction, $|\{i:L_{i,0}=1\}|=\text{cap}_1$.

\paragraph{Initial tenure.}
For each agent $i$, draw a base tenure from the level-specific integer interval and add symmetric jitter:
\[
\begin{aligned}
y^{\text{base}}_{i,0} &\sim \text{UniformInteger}\bigl([a_{L_{i,0}}, b_{L_{i,0}}]\bigr), \\
J_i &\sim \text{UniformInteger}([-5,5]),
\end{aligned}
\]
\[
y_{i,0} = \max\{0,\, y^{\text{base}}_{i,0}+J_i\}.
\]

Here we define
\[
\text{BASE\_YEARS}[1..5] = \big((0{:}3),\,(2{:}5),\,(4{:}7),\,(6{:}10),\,(8{:}12)\big)
\]
specifies $[a_\ell,b_\ell]$ for each level $\ell$, so that higher levels tend to be populated by agents with longer organizational tenure.

\medskip
The jitter ensures overlap across levels: for instance, some agents at Level1 may appear as long-serving “stayers” with more than a decade of experience, while others at Level4 may be relatively new entrants promoted quickly or hired laterally. This overlap mirrors real-world organizations, where career progression is heterogeneous and not strictly linear.

\medskip
This initialization has an important implication for promotion by seniority. Because higher levels are seeded with longer average tenure, seniority-based rules naturally reinforce stability at the top, while still allowing “old-timers” at lower levels to compete for advancement. In effect, the model produces a realistic pipeline: promotion pools contain a mix of newer employees and long-serving ones, and a seniority rule will tend to elevate those with persistent tenure, consistent with how many bureaucracies and public-sector organizations operate.

\paragraph{Initial performance and baseline efficiency.}
Given the assigned level $L_{i,0}$ and demand shares $w_{L_{i,0}}$, compute
\[
P_{i,0} = \text{clip}\!\left(\sum_{k\in K} w_{L_{i,0},k}\, c^k_{i,0},\,0,\,1\right)\in[0,1], 
\qquad 
E_0 = \frac{1}{N}\sum_{i=1}^N P_{i,0}.
\]
In the canonical runs reported in Sec.~3, $E_0\approx 0.4807$ across strategies (same seed and population).

\paragraph{Interpretation of zeros in $w_\ell$.}
If $w_{\ell,k}=0$, then skill $k$:  
(i) imposes no threshold during seeding at level $\ell$ (it is excluded from $Q(i,\ell;\rho)$), and  
(ii) contributes zero to performance at $\ell$.  
This implements a “not demanded” interpretation rather than “negligible weight.”

\paragraph{Invariants and complexity.}
Initialization satisfies:  
\begin{enumerate}[label=\roman*.]
\item $\sum_\ell \text{cap}_\ell=N$;  
\item $|\{i:L_{i,0}=\ell\}|=\text{cap}_\ell$ for all $\ell$;  
\item $P_{i,0}\in[0,1]$ for all $i$.  
\end{enumerate}

The algorithm runs in 
\[
O\!\left(N \times |\{\ell\ge 2\}| \times |\{\rho\}|\right)
\]
time; with 4 upper levels and at most 6 relaxation passes, this is $O(N)$ with a small constant.
Memory is $O(N)$ for storing agents and their attributes.

\subsection{Promotion rules}
\label{sec:promotions}

Let $C_\ell(t)=\{i:L_{i,t}=\ell\}$ be the candidate set at level $\ell$. For each $\ell=1,\dots,4$, define vacancies
\[
v_{\ell+1,t} = \mathrm{cap}_{\ell+1} - \bigl|\{i:L_{i,t}=\ell+1\}\bigr|.
\]
If $v_{\ell+1,t}>0$ and $C_\ell(t)\neq\varnothing$, we compute a strategy-specific order on $C_\ell(t)$ and promote the top $v_{\ell+1,t}$ in that order. Each promoted agent is tagged \texttt{just\_promoted} and $P_{i,t}$ is recomputed at the new level. Promotions run top-down: $\ell=4,3,2,1$.

\vspace{\baselineskip}
We implement four main strategies:

\begin{enumerate}[label=\roman*.]
\item \textbf{Merit (with performance gate):} \\
Parameters: performance threshold $\theta_P$ (default $0.8$). Partition candidates by $P_{i,t}\ge\theta_P$ vs.\ $<\theta_P$; within each subset, sort by $P_{i,t}$ descending. The final order is $[\textsf{Above}\!\downarrow,\;\textsf{Below}\!\downarrow]$; select the top $v_{\ell+1,t}$. (Implementation note: the list is reversed and consumed with \texttt{pop()}, so the last element is the best; Python’s stable sort preserves creation order for ties.) 

\vspace{\baselineskip}
The split was originally introduced as a \emph{hard} gate for worker agents in order to “promote the only those who meet the minimum performance bar.” In many real organizations, this is workable because unfilled positions can be back-filled through external hiring that screens for capability. However, in our model, we focus exclusively on internal promotions. Under a hard gate, exploratory runs produced unfilled seats and unstable headcounts, which in turn distorted efficiency comparisons: organizations appeared more efficient partly because they were smaller (only the very high performers advanced and vacancies persisted). To avoid this, we adopt a \emph{soft} gate that preserves the merit preference while maintaining a steady promotion flow and stable headcounts.

\vspace{\baselineskip}
A natural extension is to randomize within the \textsf{Below} bucket (or apply a lottery among near-threshold candidates) to model organizations that occasionally give low performers stretch opportunities.
\vspace{\baselineskip}
\item \textbf{Seniority (with tenure gate):} \\
Parameters: tenure threshold $\theta_Y$ in years (default $5$). Partition by $y_{i,t}\ge\theta_Y$ vs.\ $<\theta_Y$; within each subset, sort by $y_{i,t}$ descending; The final order is $[\textsf{Above}\!\downarrow,\;\textsf{Below}\!\downarrow]$; select the top $v_{\ell+1,t}$.

\vspace{\baselineskip}
Here as well the split was introduced to capture “tenure-first, then everyone else.” This mirrors common internal labor–market norms: when many candidates are relatively new (e.g., after a hiring surge), firms often prefer to prioritize longer-serving employees for advancement to stabilize upper levels and preserve institutional knowledge, but they will still promote newer talent if seats remain.

\vspace{\baselineskip}
A similar extension could be implemented here by randomizing within the \textsf{Below} bucket. Another extension could be promoting the \textsf{Below} bucket using \emph{merit} metrics. This preserves “tenure-first” while ensuring the remaining promotions are productivity-aware. 
\vspace{\baselineskip}
\vspace{\baselineskip}
\item \textbf{Hybrid:} \\
Parameters: performance weight $\alpha$ (default $\alpha=0.70$), score threshold $\theta_S$ (default $0.5$). Let 
\[
Y_{\max} = \max_{\ell} \max\bigl(\text{BASE\_YEARS}[\ell]\bigr).
\]

Define the capped tenure $\tilde y_{i,t}=\min\{1,\,y_{i,t}/Y_{\max}\}$ and the score
\[
s_i(\alpha)=\alpha\,P_{i,t}+(1-\alpha)\,\tilde y_{i,t}.
\]
Partition by $s_i\ge\theta_S$ vs.\ $<\theta_S$; within each subset, sort by $s_i$ descending; select the top $v_{\ell+1,t}$.
\\

As $\alpha\!\downarrow$, \emph{Hybrid} approaches \emph{Seniority}; as $\alpha\!\uparrow$, it approaches \emph{Merit}. 
The effective influence of Seniority also depends on how tenure is normalized: a tight cap compresses variation among long-tenured agents and effectively pushes the rule toward Merit. 
To strengthen seniority when desired, replace the fixed capped tenure $\tilde y_{i,t}=\min\{1,\,y_{i,t}/Y_{\max}\}$ with a scaling that preserves more spread in tenure.

\emph{Larger fixed cap} (keeps $[0,1]$ range; more separation at the top):
\[
\text{Tenure}_{\text{norm}}
  \; (\tilde y_{i,t})=\;
  \min\!\left(\frac{\text{years}}{C},\,1\right),
  \qquad C>12.
\]

\emph{Adaptive cap} (top tenure each step maps to $1$):
\[
\text{Tenure}_{\text{norm}}
  \;(\tilde y_{i,t})=\;
  \frac{\text{years}}{\max\{\text{years in pool at }t\}}\;.
\]

\emph{Quantile cap} (robust to outliers; caps at the step’s 95th percentile):
\[
\text{Tenure}_{\text{norm}}
  \;(\tilde y_{i,t})=\;
  \min\!\left(
    \frac{\text{years}}{Q_{0.95}\{\text{years in pool at }t\}},\,1
  \right).
\]
\\
We kept capped tenure $\tilde y_{i,t}=\min\{1,\,y_{i,t}/Y_{\max}\}$ which is $\tilde y_{i,t}{=}12$ to reflect diminishing returns to firm-specific experience and also to give more weightage to merit-based promotion.

\item \textbf{Random:} \\
Return a uniform random permutation of $C_\ell(t)$; select the top $v_{\ell+1,t}$.

\end{enumerate}

\subsection{Selective demotion}
\label{sec:sel_demotion}

\noindent\textit{Purpose:}
The Peter Principle predicts that some promotions place agents at their “level of incompetence,” where prior strengths no longer translate. Our policy operationalizes this as an \emph{immediate, local test} of post-promotion fit: if the same skill vector produces meaningfully lower performance under the new role demands, we treat the promotion as a discovered mismatch and rapidly restore the prior assignment. Section~\ref{sec:mitigations} revisits this as a mitigation and compares it to post-promotion training.

\vspace{\baselineskip}
\noindent\textbf{Algorithm:} \\
\emph{Inputs:} tolerance $\tau$ (default $0.05$), persistent blacklist $B$ (run-wide), refill ordering (merit, excluding $B$). \\
\emph{Trigger:} runs immediately after the promotion pass.

\begin{enumerate}[label=\roman*.]
\item \textbf{Diagnosis (compute drop):}
For each agent just promoted from $\ell$ to $\ell{+}1$ at step $t$, evaluate the change in performance defined in sec.\ref{sec:promotions}:
\[
\Delta P_i \equiv P^{\text{post}}_{i,t}-P^{\text{pre}}_{i,t},
\]
holding $\mathbf{c}_{i,t}$ fixed and replacing $w_\ell$ by $w_{\ell+1}$. This isolates role-demand mismatch rather than learning dynamics.

\item \textbf{Remedy (undo misfit):}
If $\Delta P_i \le -\tau$, demote $i$ to level $\ell$ and recompute $P_{i,t}$ at $\ell$. Record a vacancy at $\ell{+}1$ and add $i$ to the blacklist $B$.

\item \textbf{Avoid (prevent thrashing):}
The blacklist $B$ is \emph{persistent for the entire run}: blacklisted agents are excluded from future promotion consideration under this strategy. This enforces “no repeat of the same failed move.” This also makes sure that the demoted agent is at their “level of competence.”

\item \textbf{Refill (restore headcount):}
For each level $u$ with vacancies created by demotion, form the pool from $u{-}1$ excluding $B$, rank by current $P_{j,t}$ (descending) and promote the top $V_u$. Mark them \texttt{just\_promoted} and recompute $P_{j,t}$ at level $u$.
\end{enumerate}

\noindent\emph{Implementation note.}
This matches the code paths \texttt{selective\_demotion} and \texttt{apply\_dra\_demotion\_and\_refill}: we compute \texttt{drop = prev\_performance - performance} and demote if \texttt{drop} $\ge\tau$; demoted agents are added to \texttt{self.demoted\_agents} (the persistent $B$). Refills draw from the level below, exclude $B$, sort by performance, and fill exactly the vacated seats, preserving capacities.
\\

\noindent\emph{Context and practice.}
Formal “demotion” is ethically and culturally sensitive, so many organizations implement \emph{functionally equivalent} safeguards without stigma. Common variants include: 
\begin{enumerate}[label=\roman*.]
\item \emph{trial/acting appointments} or \emph{probationary promotions} (e.g., interim lead, acting manager) that are confirmed only if post-move performance meets expectations;
\item \emph{pay-protected reassignments} in which compensation (and sometimes title) is temporarily preserved while the employee returns to the role where they are most productive;  
\end{enumerate}

Our mechanism abstracts these realities: the immediate $\Delta P$ test stands in for a probation review; non-confirmation maps to a rapid, low-stigma role reset; and the persistent blacklist represents a practical “cool-down” that avoids retrying the same misfit move too soon. Although we do not model compensation explicitly, the policy is compatible with pay protection reassignments.

\subsection{Post-promotion training (logistic-derivative burst)}
\label{sec:training}

\noindent\textit{Purpose:}
Under the Peter Principle, performance can fall immediately after promotion when required skills shift. We model a practical mitigation used by many organizations which is targeted, early training for new managers/lead roles to close fresh gaps and dampen post-promotion drops.

\paragraph{Scope and timing.}
Training applies \emph{once, immediately after promotion} to agents flagged \texttt{just\_promoted}, 
and only on trainable dimensions $k \in \{\text{tech},\text{mgmt}\}$. 
Compliance and soft skills remain fixed.

We hold compliance and soft skills fixed because, in our setting, they reflect ingrained norms and interpersonal dispositions that change slowly relative to the short post-promotion onboarding window; by contrast, technical and managerial competencies are directly teachable through targeted workshops, coaching, and practice.

\paragraph{Learning rule (logistic-derivative burst).}
For a competence scalar $C \in [0,1]$ we use the logistic-derivative learning rate
\[
\ell(C) = k\,C(1-C), \qquad k=1,
\]
which is maximal at $C=0.5$ and vanishes at $C \in \{0,1\}$. 
The one-shot training update at promotion is
\[
C^{k}_{i,t+} = \min\{1,\; C^{k}_{i,t} + \ell(C^{k}_{i,t})\}, 
\qquad k \in \{\text{tech},\text{mgmt}\},
\]
followed by a recomputation of $P_{i,t}$ at the new level. 
By construction, the largest single-step gain is $0.25$ (when $C=0.5$), 
and clipping keeps competencies in $[0,1]$.

\paragraph{Implementation (matching the code).}
In the \texttt{merit\_learning} variant, promotions are ranked exactly as under \textbf{Merit}; 
the difference is a training hook called right after promotions. 
The code stores a per-agent \texttt{learning\_rate} at initialization computed as 
$\ell(C^{k}_{i,0})$ for $k\in\{\text{tech},\text{mgmt}\}$, and then applies
\[
C^{k}_{i,t+} = \min\{1,\; C^{k}_{i,t} + \ell(C^{k}_{i,0})\}.
\]
This ``fixed-increment'' approximation behaves like the dynamic rule above 
(largest gains for mid-competence agents, negligible at the extremes) 
while being simple and fast. 
We clear \texttt{just\_promoted} after the update so training fires only once.

\paragraph{Why this form?}
The update $\ell(C)=C(1-C)$ is the normalized derivative of a logistic learning curve, so marginal gains are largest at mid-competence and vanish near $0$ and $1$. This matches the common pattern that short, targeted onboarding\slash coaching helps agents with some base but yields little for novices or near-experts, and it serves here to offset the post-promotion ``shock'' by nudging competence toward the new role without overshooting (see Sec.~\ref{interventions} for derivation)

\paragraph{When it helps most.}
Training is most effective in high-mismatch settings (Regime~A) where management weight rises sharply; 
it lifts $P_{i,t}$ at the new level and reduces the mass of promotions with $\Delta P_i<0$. 
In transferable-skills settings (Regime~B) the effect is smaller but still positive 
for agents not already near ceiling.

\paragraph{Real-world analogue.}
This hook abstracts common practices: structured onboarding for new managers, leadership bootcamps, 
30/60/90-day transition plans, mentoring/coaching, shadowing, internal academies, targeted workshops 
(e.g., goal-setting, feedback, budgeting), and scoped stretch assignments with feedback. 
All are short, front-loaded interventions aimed at accelerating fit to the new role.


\paragraph{Algorithmic summary.}
\begin{enumerate}[label=\roman*.]
\item Rank candidates by \textbf{Merit} (Sec.~\ref{sec:promotions}) and promote the top $v_{\ell+1,t}$; set \texttt{just\_promoted}$=$True.
\item For each \texttt{just\_promoted} agent and each $k\in\{\text{tech},\text{mgmt}\}$, update $C^k$ by the burst rule above; clip to $[0,1]$.
\item Recompute $P_{i,t}$ at the new level and clear \texttt{just\_promoted}.
\end{enumerate}

\noindent\textit{Interpretation.} This policy is the constructive counterpart to selective demotion: 
instead of undoing a misfit move, it invests early to improve fit. 

\subsection{Attrition and replenishment}
\label{sec:attrition}

Let the per--level exit rates be
\[
\xi = (\xi_1,\xi_2,\xi_3,\xi_4,\xi_5) 
      = (0.05,\,0.02,\,0.01,\,0.005,\,0.002).
\]

\noindent
i.e., the exit rate is $5\%$ for Level~1, $2\%$ for Level~2, 
$1\%$ for Level~3, $0.5\%$ for Level~4, and $0.2\%$ for Level~5.

At each step, we remove a uniformly random subset within each level. For level $\ell$,
\[
n_\ell(t) = \bigl|\{\,i : L_{i,t}=\ell \,\}\bigr|, 
\qquad 
a_\ell(t) = \lfloor \xi_\ell \, n_\ell(t)\rfloor,
\]
and we sample
\[
A_\ell(t) \subseteq \{\,i : L_{i,t}=\ell \,\},
\qquad |A_\ell(t)| = a_\ell(t),
\]
\emph{without replacement}. 
All $i\in A_\ell(t)$ exit the organization. 
This step captures routine exits (quits, retirements, or transfers) drawn uniformly at random within each level, thereby opening vacancies that the promotion stage fills. The declining attrition profile ($\xi_1 \gg \xi_5$) reflects well-documented labor market regularities where turnover is inversely related to rank and tenure \cite{cotton1986employee}. Entry-level employees (Level~1) face higher turnover due to job matching friction and exploration (the ``shopping around'' phase), whereas senior agents (Levels~4--5) exhibit high stability due to selection effects, higher compensation (``golden handcuffs''), and the accumulation of firm-specific human capital \cite{becker1962investment}.


After promotions (and any demotion/training hooks), 
we restore headcount by hiring into Level~1 only. Define
\[
h_1(t) = \text{cap}_1 - \bigl|\{\, i : L_{i,t}=1 \,\}\bigr|
\]
as the number of Level~1 vacancies. 
We add $h_1(t)$ new entrants with competences i.i.d.\ 
$\sim \text{Uniform}(0,1)$ over all four skill dimensions 
and tenure initialized to $y_{i,t}=0$. 
By design, we do \emph{not} hire directly into $\ell\ge 2$: 
upper levels are replenished only via internal mobility, 
so we can isolate the effects of promotion rules on Peter-type dynamics. 
(An extension with lateral/experienced hires could mirror initialization thresholds 
from Sec.~\ref{sec:init}.)

\paragraph{Implementation notes.}
In code, exits are computed as:
\begin{verbatim}
n = int(rate * len(lvl_agents))
removed = random.sample(lvl_agents, n)
\end{verbatim}
i.e., floor rounding of $\xi_\ell n_\ell(t)$ and uniform sampling within level.  
Attrited IDs are logged in \texttt{last\_attritions\_by\_level}.  
Replenishment calls \texttt{hire\_new\_level1()}, 
which draws new skills from $[0,1]$ and sets tenure $=0$.  
Promotions/demotions do not change agent count; 
attrition reduces it and Level~1 hiring restores it, 
which is enforced by a step-level assertion on population size.

The attrition in our model simulates quits, retirements, and/or transfers. This is essential making promotion opportunities recurrent. The vector $\xi$ is a scenario parameter and can be varied (e.g., higher $\xi_2$ to simulate a mid-level exodus, or tenure-/performance-dependent exits in a richer retention model).



\subsection{Outcomes and diagnostics}
\label{sec:outcomes}

We evaluate each strategy with a common set of outputs computed after promotions (and any demotion/training hooks) at every step. Because departures are immediately back-filled at Level 1, headcount stays constant; curves are therefore directly comparable across time and strategies. We built an interactive dashboard in which we can fine tune our parameters, run simulations and measure outcomes. Here are the things that are being measured:
\begin{enumerate}
    \item \textbf{Efficiency over time (primary outcome).}
    We track the organization’s average performance each step and plot one line per strategy. This shows whether a rule converges to a higher steady state, oscillates, or degrades once Peter-type effects accumulate.
    Organizational efficiency is the mean performance across active agents:
    \[
    E_t \equiv \frac{1}{N_t} \sum_i P_{i,t}, 
    \qquad N_t = \bigl|\{\,i \;\text{active at } t \,\}\bigr|.
    \]

    \item \textbf{Flows that create and fill vacancies.}
    We record, by timestep and level, (a) how many people left and (b) how many were promoted. These counts provide context for the efficiency trends—attrition generates openings, and promotion volume determines how much “role switching” pressure the system experiences.

    \item \textbf{Immediate impact of promotion.}
    For every promotion, we compute the change in the person’s performance measured right before and right after the move. We analyze the following:
    \begin{itemize}
        \item a frequency distribution (histogram) of changes in performance on promotion;
        \item the average and median change across all promotions at a step or run;
        \item the share of promotions that resulted in a decline.
    \end{itemize}
    We also show a small time-series bar chart of how many negative-impact promotions occurred at each step, with the option to inspect the affected agents at a chosen timestep.

    \item \textbf{Where the shock concentrates (path analysis).}
    We report the average promotion impact by path (e.g., Level 2→3, 3→4) as a heatmap. This pinpoints the transitions that carry the largest Peter-type risk under a given role structure.

    \item \textbf{Individual trajectories.}
    For any selected agent we plot performance over time and display their step-by-step record (level, performance, and skill vector).

    \item \textbf{Cross-strategy Peter severity (comparison table).}
    We compile, for each strategy: the average and median promotion impact, the percent of promotions with a drop, and total promotion count. This table highlights where the Peter Principle is most/least pronounced and guards against comparing rules that promote very different volumes of people.

    \item \textbf{Reproducibility and data products.}
    Every run can be saved and reloaded with all parameters (role profiles, level shares, attrition rates, hybrid weights, demotion tolerance, seed, and chosen strategies). The app exposes both model-level and agent-level data for download (CSV/Pickle), enabling external analysis or figure recreation.

    \item \textbf{Knobs for sensitivity.}
    The dashboard lets you vary role weights, level shares, attrition rates, the merit/seniority blend in Hybrid, and the demotion tolerance. The outcome views above update accordingly, so you can report simple sensitivity checks (e.g., how the percent of negative promotion impacts moves when the hybrid weight or demotion tolerance is adjusted).

\end{enumerate}

\subsection{Parameters}
\label{sec:params}

Unless noted otherwise, all runs use the following defaults 
(tunable in the dashboard):

\begin{itemize}
\item \textbf{Population and horizon.} 
$N=100{,}000$ agents; $T=100$ steps; \texttt{seed}$=42$.

\item \textbf{Level capacities (shares).} 
\[
p=(0.40,\,0.25,\,0.20,\,0.10,\,0.05) \quad \text{for Levels 1--5.}
\]
Exact integer capacities are computed by the floor--remainder rule, 
with Level~1 absorbing the remainder to ensure 
$\sum_\ell \text{cap}_\ell = N$.

\item \textbf{Attrition (by level).} 
\[
\xi = (0.05,\,0.02,\,0.01,\,0.005,\,0.002) 
\quad \text{for Levels 1--5.}
\]
Departures are drawn uniformly at random within each level and 
immediately replenished at Level~1.

\item \textbf{Initial tenure ranges.} 
\[
\text{BASE\_YEARS}[1..5] = (0{:}3),\,(2{:}5),\,(4{:}7),\,(6{:}10),\,(8{:}12),
\]
with integer jitter $\pm 5$ years and truncation at~0.

\item \textbf{Initialization relaxation.} 
Threshold relaxation grid 
\[
\rho \in \{0,\,0.2,\,0.4,\,0.6,\,0.8,\,1.0\}, 
\qquad \Delta\rho=0.20,
\]
applied when seeding Levels~5 down to~2.

\item \textbf{Promotion rule knobs.}
\begin{itemize}
  \item \emph{Merit:} performance gate $\theta_P=0.8$.
  \item \emph{Seniority:} tenure gate $\theta_Y=5$ years.
  \item \emph{Hybrid:} performance weight $\alpha=0.70$ 
  (i.e., 70\% merit / 30\% seniority) and score gate $\theta_S=0.5$; 
  $Y_{\max}$ is taken from \texttt{BASE\_YEARS}.
  \item \emph{Random:} no parameters.
\end{itemize}

\item \textbf{Selective demotion.} 
Tolerance $\tau=0.05$ (demote if immediate post-promotion performance 
drops by $\ge \tau$); blacklist is persistent for the entire run.

\item \textbf{Post-promotion training.} 
One-shot update on $\{\texttt{tech},\texttt{mgmt}\}$ using
\[
\ell(C)=k\,C(1-C), \qquad k=1,
\]
applied only in the \texttt{merit\_learning} variant.
\end{itemize}

\paragraph{Implementation notes.}
\begin{itemize}
\item During initialization, only \emph{non-zero} entries in the role profile 
$w_\ell$ act as qualification thresholds; zeros mean 
``not demanded at this level.''
\item Promotions run top-down (from $\ell=4$ to $\ell=1$), 
selecting the top $v_{\ell+1,t}$ in a strategy-specific order. 
Lists are reversed so that \texttt{pop()} yields the best remaining candidate; 
Python’s sort stability preserves tie order.
\item After every promotion (and any demotion/training hooks), 
we recompute performance and then compute efficiency. 
Competencies and performances are clipped to $[0,1]$ for numerical safety.
\item Headcount is conserved each step: attrition precedes promotions; 
Level~1 hiring returns the population to $N$.
\item The dashboard exposes all of the above as tunables 
(role weights, level shares, attrition rates, hybrid weight $\alpha$, 
demotion tolerance $\tau$, etc.) and saves full run metadata for reproducibility.
\end{itemize}

\section{Results}\label{sec:results}

\subsection{Setup and overview}

We simulate a five-level organization with 
$N=100{,}000$ agents for $T=100$ timesteps 
(seed $=42$). The initial efficiency of the organization for seed 42 is 
$E_0\approx0.54820$. 

We compare six promotion rules: 
\textbf{Merit}, 
\textbf{Seniority}, 
\textbf{Hybrid} ($\alpha=0.70$ on performance), 
\textbf{Random}, 
\textbf{Selective-demotion} ($\tau=0.05$), 
and \textbf{Merit-learning}

All other parameters (role skill profiles, level distribution, 
level-specific attrition) are held fixed across strategies as discussed earlier in the paper.

\subsection{High-mismatch Regime}

\begin{figure}[H]
    \centering
    \includegraphics[width=\linewidth]{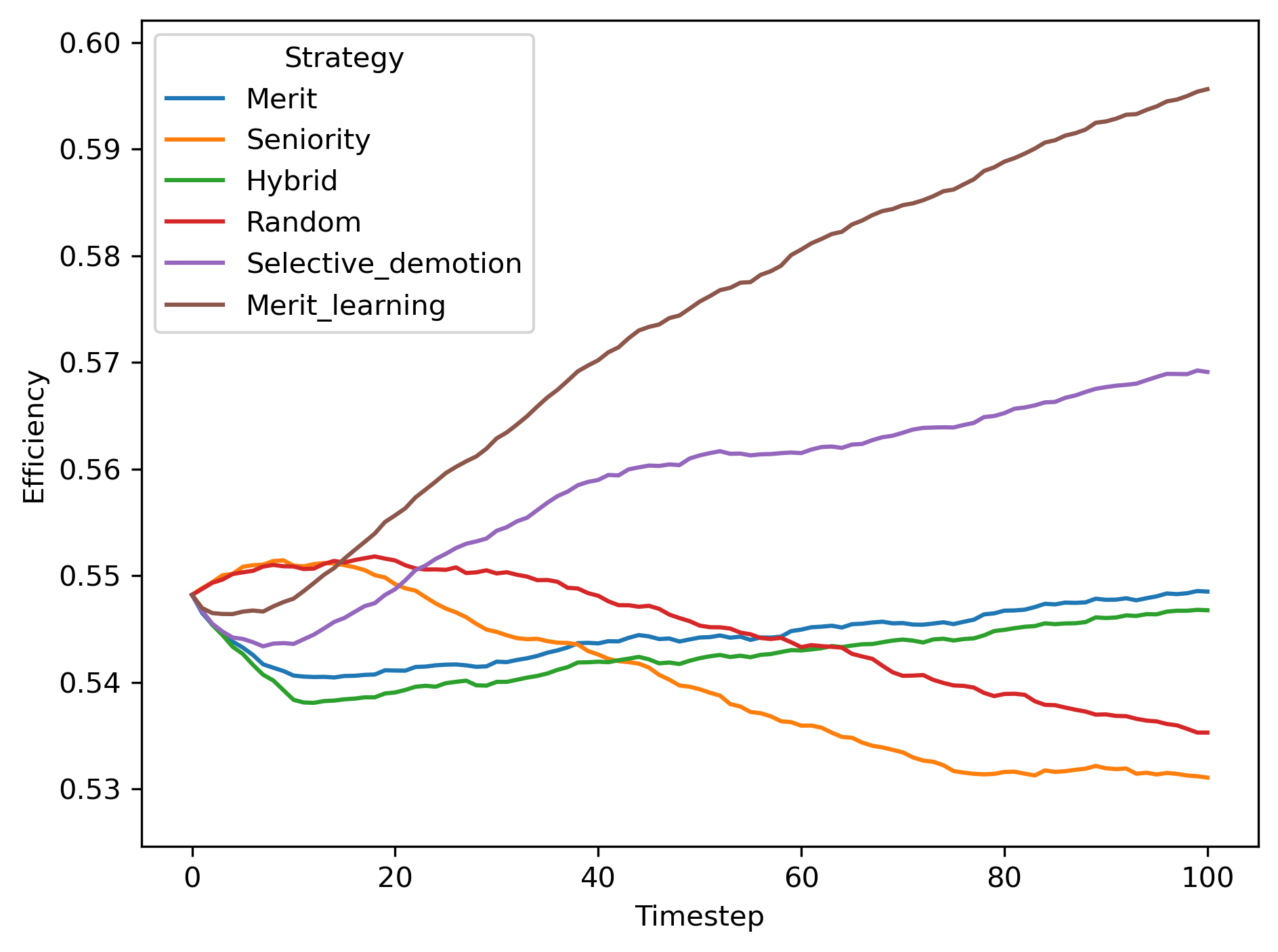}
    \caption{Organizational efficiency trajectories for high-mismatch regime under six promotion rules.}
    \label{fig:efficiency_high_mismatch}
\end{figure}

All strategies begin at $E_0 \approx 0.54820$. 

At $T=100$:
\begin{itemize}
  \item Merit: 0.54851 (+0.06\%),
  \item Seniority: 0.53106 (-3.13\%).
  \item Hybrid: 0.54676 (-0.26\%),
  \item Random: 0.53528 (-2.36\%),
  \item Selective Demotion: 0.56910 (+3.81\%),
  \item Merit+training: 0.59563 (+8.65\%),
\end{itemize}

\noindent\textbf{Merit.} The merit rule promotes the best \emph{in the current job}. In the high-mismatch setting this systematically elevates the ``best at Level~1'' - tech-heavy performers into L2/L3 roles which weighs management/compliance more. This results in many negative $\Delta P$ shocks. So we systematically lift the wrong skill mix and induce many negative $\Delta P$ shocks. That is why we see a drop in the first ~10–20 steps. Over time, the occasional naturally 'portable' agent (balanced skills) helps efficiency to recover, but there is no mechanism to prevent more misfits. This explains the slow crawl back to near-flat and plateauing.

We observed 96,059 promotions across 100 time-steps resulting in $\Delta P < 0$ (which is 88.1\% of all promotions) compared to only 12,941 promotions with $\Delta P > 0$.

\begin{figure}[H]
    \centering
    \begin{subfigure}[t]{0.48\linewidth}
        \centering
        \includegraphics[width=\linewidth]{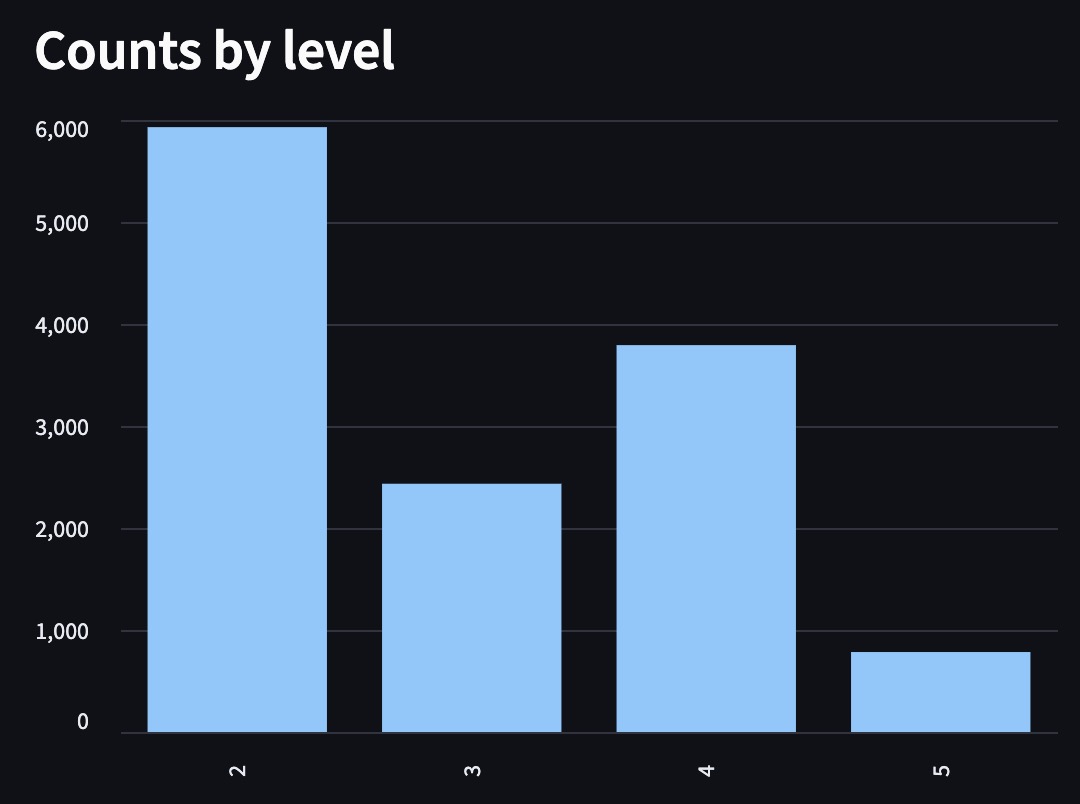}
        \caption{number of Promotions with $\Delta P > 0$ by level.}
        \label{fig:merit_pos}
    \end{subfigure}
    \hfill
    \begin{subfigure}[t]{0.48\linewidth}
        \centering
        \includegraphics[width=\linewidth]{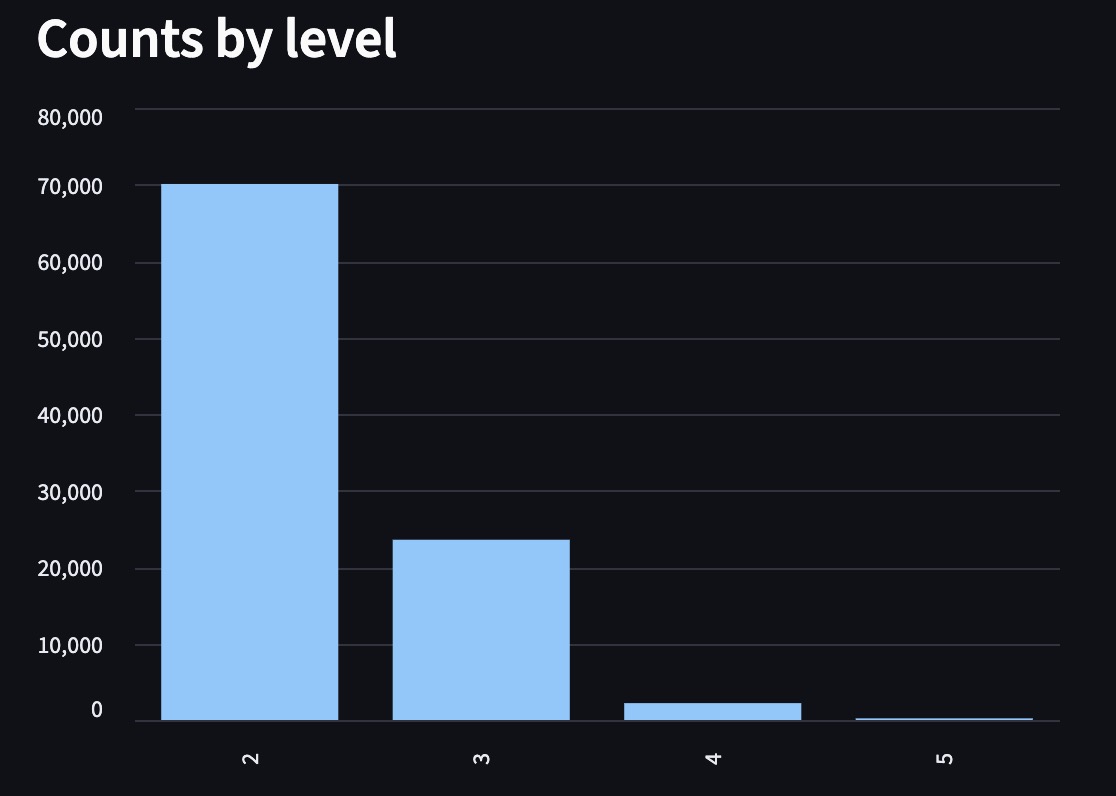}
        \caption{ number of Promotions with $\Delta P < 0$ by level.}
        \label{fig:merit_neg}
    \end{subfigure}
    \caption{Bar chart of promotion outcomes under merit-based promotion in the high-mismatch regime. 
The counts of agents with positive ($\Delta P>0$) and negative ($\Delta P<0$) performance changes are shown by level.}
    \label{fig:merit_deltas}
\end{figure}

\noindent When broken down by levels (Fig.~\ref{fig:merit_deltas}) these are the observed numbers for agents whose performance improves and for agents whose performance plummets:
\[
\begin{array}{lrr}
\toprule
\text{Level} & \Delta P > 0 & \Delta P < 0 \\
\midrule
\text{L2} & 5{,}930 & 70{,}070 \\
\text{L3} & 2{,}434 & 23{,}566 \\
\text{L4} & 3{,}793 & 2{,}207 \\
\text{L5} &   784   &   216 \\
\bottomrule
\end{array}
\]

\noindent\emph{Explanation:}
\begin{enumerate}[label=\arabic*.]
\item \textbf{L1\(\to\)L2:} weights change by \(\Delta w=(-0.4,+0.3,+0.0,+0.1)\) for (tech, mgmt, comp, soft). Under Merit at L1, promoted agents are typically tech-strong with comparatively lower management, so
\[
\Delta P \approx -0.4\,C_{\text{tech}} + 0.3\,C_{\text{mgmt}} + 0.1\,C_{\text{soft}}
\]
is usually negative. This matches the large imbalance (70{,}070 negative vs 5{,}930 positive).

\item \textbf{L2\(\to\)L3:} \(\Delta w=(-0.5,+0.2,+0.3,+0.0)\). Merit at L2 favors candidates with substantial tech and some management but does not screen on compliance (which carries zero weight at L2). When tech weight drops to zero at L3 and compliance becomes salient,
\[
\Delta P \approx -0.5\,C_{\text{tech}} + 0.2\,C_{\text{mgmt}} + 0.3\,C_{\text{comp}},
\]
the loss from tech commonly dominates, yielding many more negatives (23{,}566) than positives (2{,}434).

\item \textbf{L3\(\to\)L4:} \(\Delta w=(0.0,+0.2,-0.2,0.0)\). At L3, Merit ranks candidates primarily on management (tech has zero weight). The move to L4 increases the management weight and reduces the compliance weight, so for top-ranked L3 candidates who tend to have \(C_{\text{mgmt}}\) higher than \(C_{\text{comp}}\) then we have
\[
\Delta P \approx 0.2\,C_{\text{mgmt}} - 0.2\,C_{\text{comp}} > 0
\]
which is consistent with 3{,}793 positive vs 2{,}207 negative outcomes.

\item \textbf{L4\(\to\)L5:} \(\Delta w=(0.0,+0.1,0.0,-0.1)\). Merit at L4 emphasizes management; moving to L5 boosts management and trims soft skills. For many promoted L4 agents with \(C_{\text{mgmt}}>C_{\text{soft}}\),
\[
\Delta P \approx 0.1\,(C_{\text{mgmt}}-C_{\text{soft}}) > 0,
\]
aligning with 784 positives vs 216 negatives.

\end{enumerate}

\begin{figure}[H]
    \centering
    \includegraphics[width=\linewidth]{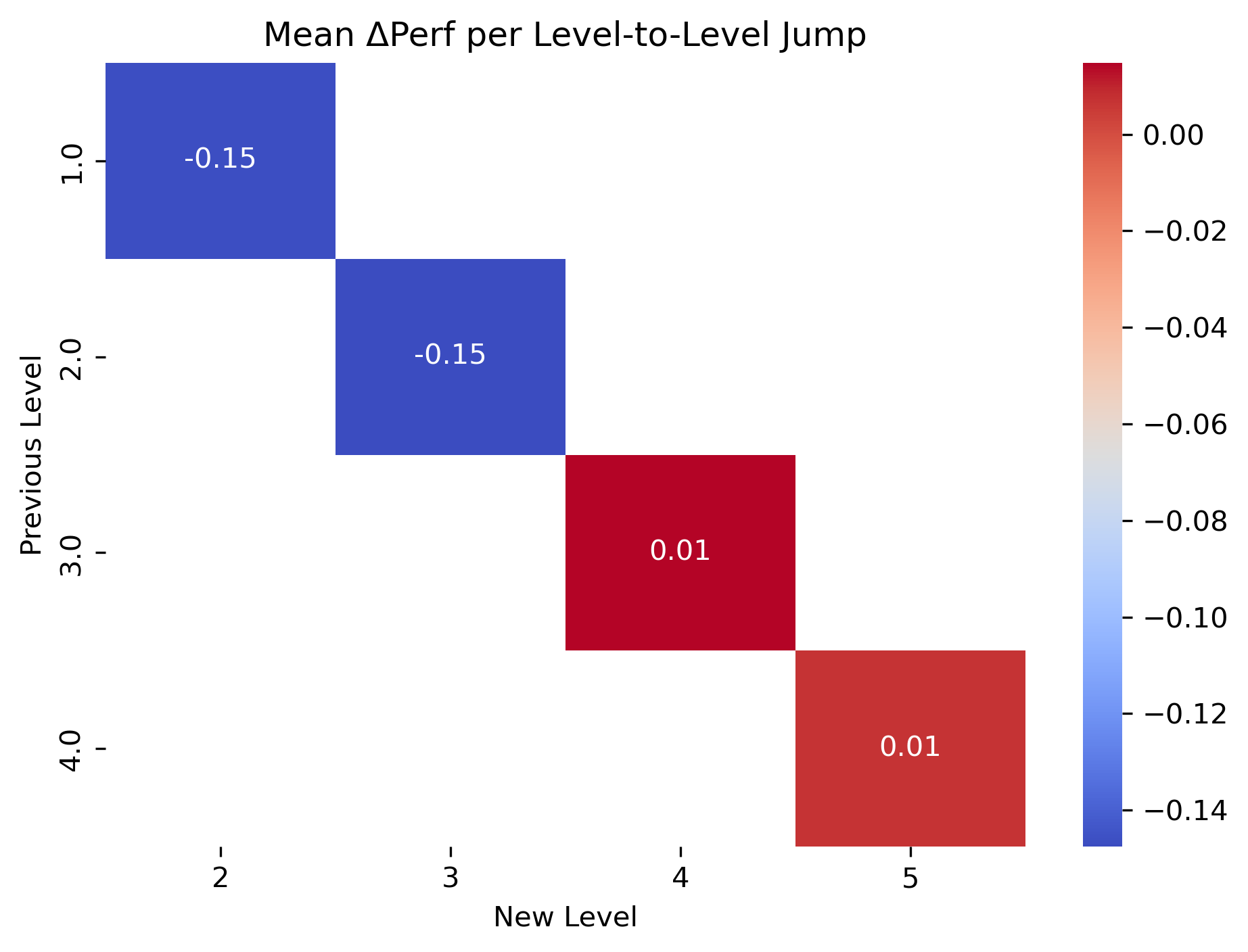}
    \caption{Mean $\Delta P$ per Level-to-Level jump for merit based promotion strategy.}
    \label{fig:merit_heatmap}
\end{figure}

\noindent The heatmap (Fig.~\ref{fig:merit_heatmap}) above also shows a similar trend. Specifically, the mean $\Delta P$ from L1$\to$L2: $-0.15$, L2$\to$L3: $-0.15$ are slightly negative due to the rapid change in role requirements and from L3$\to$L4: $+0.01$, L4$\to$L5: $+0.01$ is slightly positive as the roles requirements are very similar towards the top (high management).

\vspace{\baselineskip}
\noindent The patterns above provide \textbf{strong evidence} of the \textbf{Peter Principle} under \textbf{Merit}-based promotions: many agents' performance drop after promotion due to difference in role requirements.

\begin{figure}[H]
    \centering
    \includegraphics[width=\linewidth]{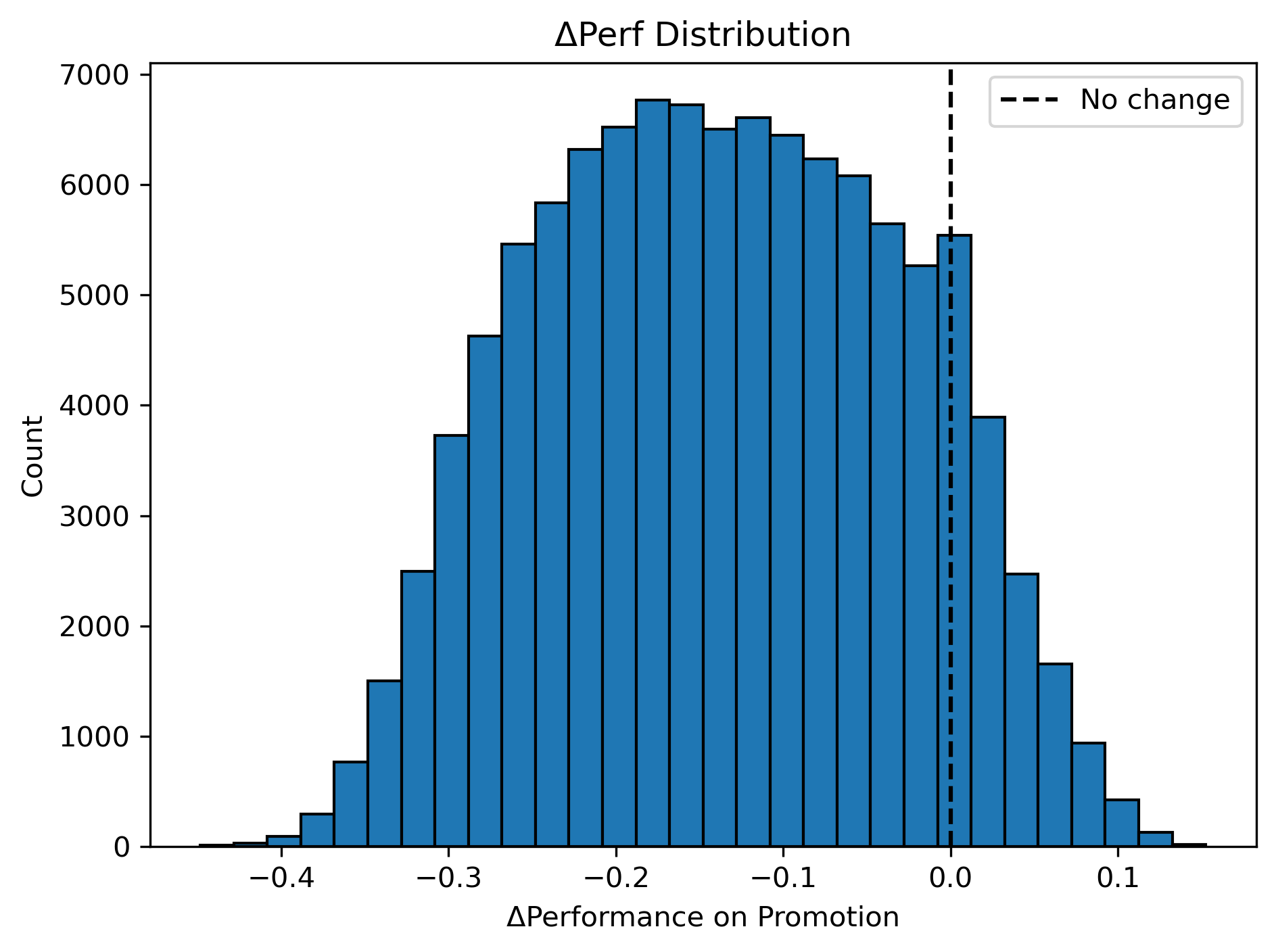}
    \caption{$\Delta P$ frequency distribution across all time-steps for merit based promotion.}
    \label{fig:merit_promotion_delta}
\end{figure}

\noindent The \emph{frequency distribution} of $\Delta P$ across all promotions (Fig.~\ref{fig:merit_promotion_delta}) is \textbf{heavily skewed to the left} around the \textbf{baseline $0$}, indicating \textbf{more promotions with $\Delta P<0$ than with $\Delta P>0$}. The average drop in performance is -0.136 and median drop is -0.138. If promotions were neutral on average, the mass would concentrate near $0$; the observed skew is \textbf{direct evidence} of the Peter Principle in this setting. Moreover, the distribution is \emph{asymmetric in magnitude}: the \textbf{negative tail} extends to about \(\mathbf{-0.448}\), whereas \textbf{positive tail} reaches \(\mathbf{+0.153}\) (1st/99th percentiles \(-0.350\) / \(+0.080\)), implying that losses are not only more frequent but also larger than gains.

\begin{figure}[H]
    \centering
    \includegraphics[width=\linewidth]{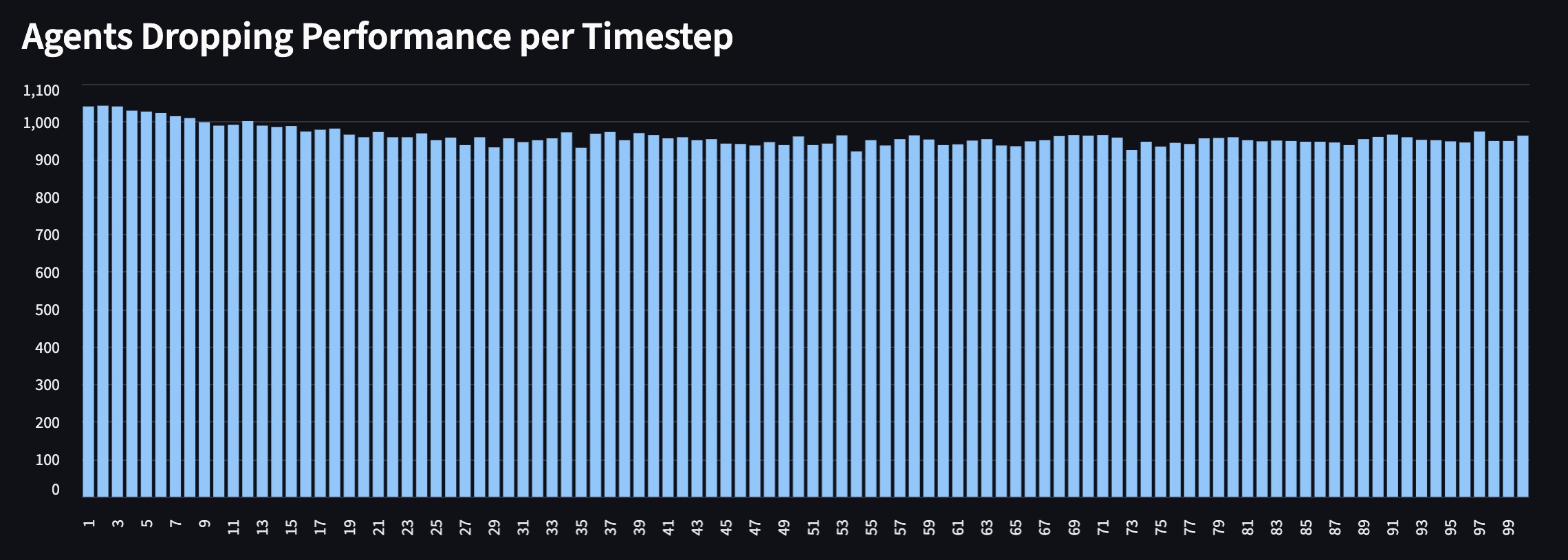}
    \caption{number of promotions with $\Delta P\!<\!0$ at each timestep for merit based promotion.}
    \label{fig:merit_neg_tseries}
\end{figure}

\noindent To complement the distributional view, Fig.~\ref{fig:merit_neg_tseries} plots the \emph{count} of promotions with $\Delta P<0$ at each timestep. The series is remarkably \textbf{flat}—roughly \textbf{$900$–$1{,}100$} negative shocks \emph{every} step, accumulating to \textbf{$96{,}059$} across $T{=}100$. This pattern shows that harmful moves are \textbf{persistent rather than episodic}: fresh L(i)$\!\to\!$L(i+1) misfits are produced each period under \textbf{Merit}. The steady inflow of losses helps explain the \emph{early dip} and \emph{prolonged plateau} in efficiency, gains from the minority of portable promotions are continuously offset by new negative shocks.


\vspace{\baselineskip}
\noindent Let us now take an example by going through journeys of two specific individuals. And see how they perform under the merit based promotion:

\paragraph{Case A — Portable excellence (Agent\#~138950).} Agent\# 138950 has the following \emph{skills:}
\begin{verbatim}
{
  "tech": 0.8621461961850508,
  "management": 0.9757467544033296,
  "compliance": 0.9716362302335507,
  "soft_skills": 0.7250920869880361
}
\end{verbatim}

\noindent The Agent joins the Organization at time-step 15 being hired at L1. The agent performs strongly at L1 ($P\!=\!0.8484$) and is promoted to L2 in the next timestep (t=16). At L2, performance \emph{improves} ($0.8688$): due to his high management skills and slightly strong soft-skills (as this level demands 30\% more management skills and 10\% more soft-skills) and is promoted to L3 in the following timestep (t=17). A further promotion to L3 raises performance again ($0.9244$): L3 emphasizes management and compliance which are both high for this agent. This leads to him getting promoted to L4.
The move to L4 is essentially neutral/positive ($0.9252$) as management weight rises further. The agent remains stagnant for long interval even due to his high performance due to scarcer vacancies and higher competition at the top, the agent reaches L5 at t=51 with the best score ($0.9503$).

\begin{figure}[H]
    \centering
    \includegraphics[width=\linewidth]{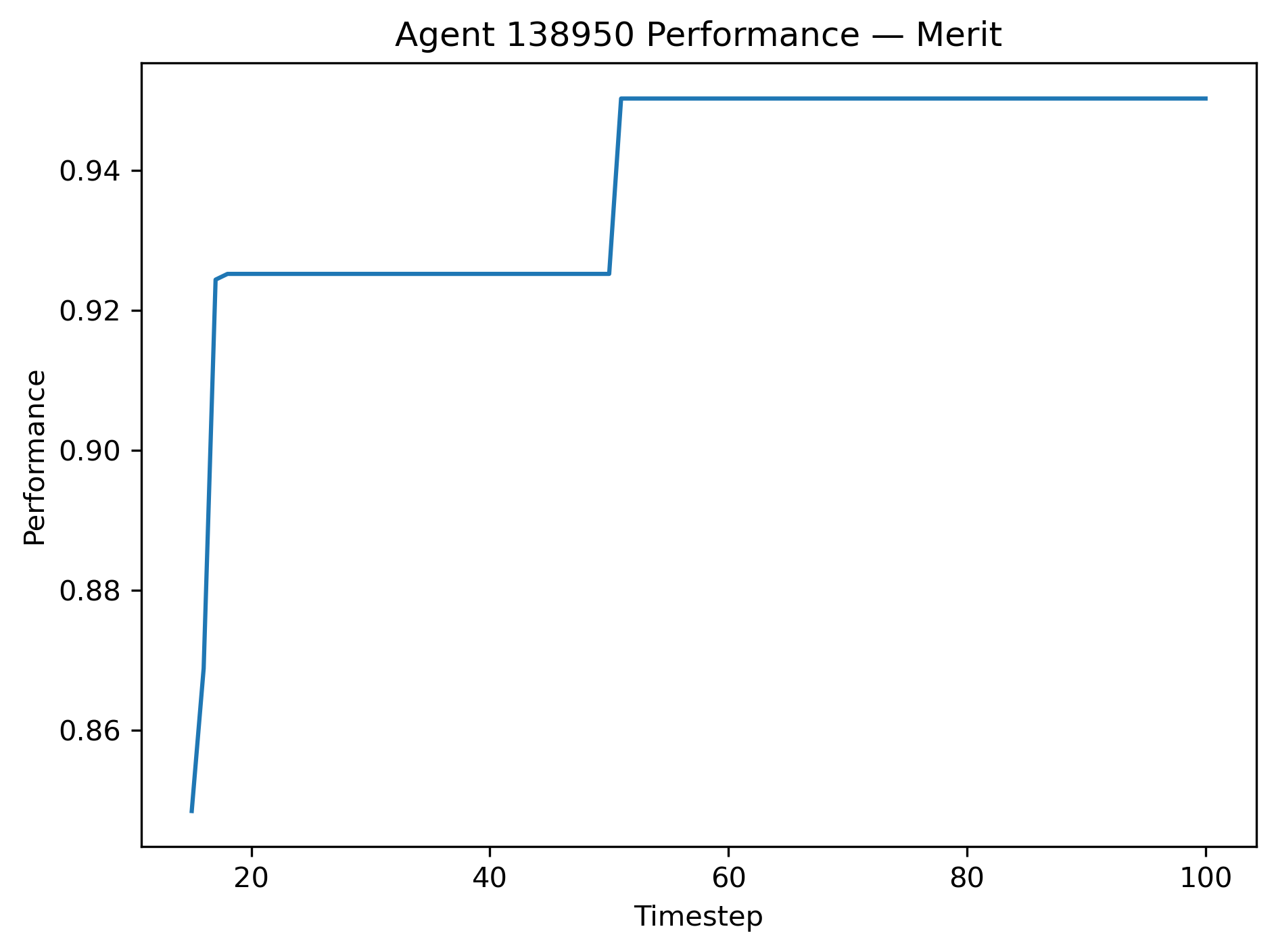}
    \caption{Agent 138950s performance over time for merit based promotion.}
    \label{fig:merit_agent_13950}
\end{figure}

\noindent \textbf{Takeaway:} When an agent’s competence vector already matches the next job’s weights, merit yields \emph{consistently positive} $\Delta P$ and rising performance. \textbf{But there is a bottleneck:} because early promotions are judged by the \emph{current} role’s objective (L1 is tech-heavy), an agent who would be excellent at L5 yet lacks standout \emph{tech} may \emph{never} clear the L1 screen—getting filtered out long before reaching the role where they would add the most value.

\paragraph{Case B — Classic misfit escalation (Agent\#~315680).} Agent\# 315680 has the following \emph{skills:} \begin{verbatim}
{
  "tech": 0.9810806939771801,
  "management": 0.6931592421945483,
  "compliance": 0.9239861161586994,
  "soft_skills": 0.922179441944758
}
\end{verbatim}
\noindent The Agent joins the Organization at time-step 79 being hired at L1. In the next timestep (t=80) due to excellent performance ($0.9752$) because of high tech skills and L1’s heavy tech weight, the agent is promoted to L2
After being promoted to L2, performance \emph{falls} to $0.8829$: because although L2 values management and soft, it still leans on tech ($50\%$), so the drop is modest. The performance even though is lower is still very high compared to other agents and that is why he is promoted to L3 in the next time-step (t=81). The next step to L3 produces another \emph{drop} to $0.8082$: this is because tech weight collapses to $0$, and while compliance becomes important, the loss of $0.5\times C_{\text{tech}}$ is not fully offset by the added management/compliance weight. The performance is still decently high at $0.8082$ which results in a further promotion to L4 \emph{reduces} performance again ($0.7620$): as L4 increases management and cuts compliance; because this agent’s $C_{\text{comp}} >> C_{\text{mgmt}}$, the compliance down-weighting hurts more than the management up-weighting helps. Therefore, no further promotion happens till the end of simulation

\begin{figure}[H]
    \centering
    \includegraphics[width=\linewidth]{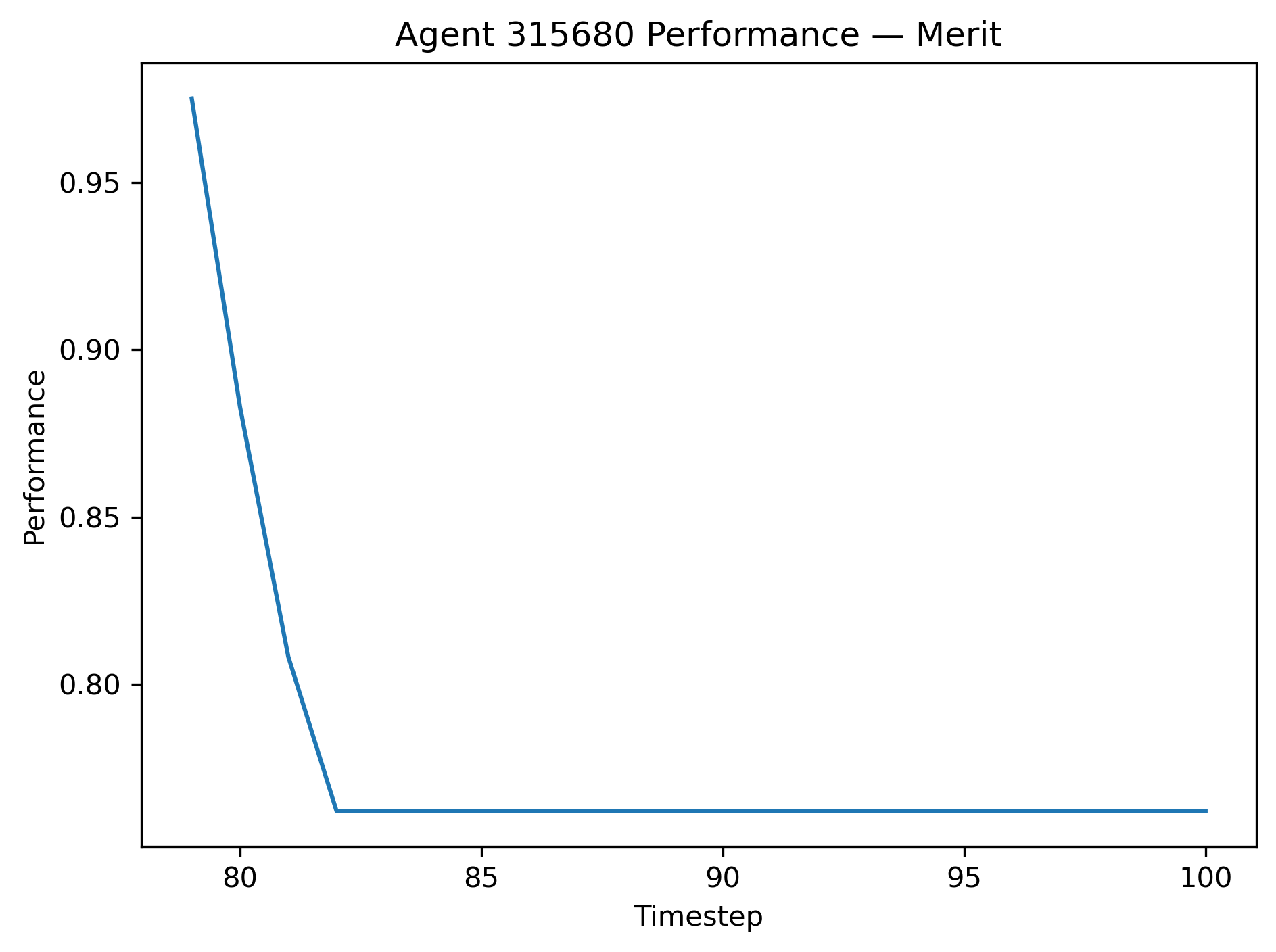}
    \caption{Agent 315680s performance over time for merit based promotion.}
    \label{fig:merit_agent_315680}
\end{figure}

\noindent \textbf{Takeaway:} repeated promotions selected on \emph{current-role} success shift the agent into profiles that down-weight their strongest contributor (tech), yielding \textbf{negative} $\Delta P$ at multiple steps which shows existence of the Peter Principle. He was promoted to his “level of incompetence” at L4 where there was no way for him to be promoted due to his poor performance. In this case, keeping the agent in a tech-weighted role (like L1) would likely have yielded more organizational value than pushing upward into increasingly management-weighted jobs.

\paragraph{Interpretation.}
Under \textbf{Merit}, the early transitions (L1$\!\to$L2, L2$\!\to$L3) reliably generate \textbf{Peter shocks}: destination roles reweight sharply \emph{away} from tech and \emph{toward} management/\allowbreak compliance, so success in the source job is \emph{not portable}. Near the top, the pattern softens: L4 and L5 are both \textbf{management-heavy}, so high performance at L4 tends to \emph{carry over} to L5, yielding small positive $\Delta P$ on average. Taken together with the left-skewed $\Delta P$ histogram and the steady stream of negative shocks each timestep, this level-specific profile is \textbf{consistent with the Peter Principle}: merit-based promotion systematically manufactures mismatches at the lower rungs even if the upper rungs look safer because the roles are similar. But we still see effects of Peter Principle in some agents transitioning form L4$\!\to$L5 although not very significant (e.g Agent\#~158384 dropped performance from $0.9226\!\to0.9154$). 

\vspace{\baselineskip}

\noindent\textbf{Seniority.} The seniority rule purely on \emph{organizational tenure}. In this model, that makes Seniority effectively \emph{skill-blind} and therefore qualitatively similar to \textbf{Random} - who goes up is largely independent of whether their competence vector matches the destination role. In the high-mismatch regime this produces a smooth, modest 
\textbf{drift downward} in efficiency (to $E_{100} \approx 0.531$), 
without the early crash seen under \textbf{Merit}. 
The reason is structural: because selection is \emph{uncorrelated} with the source job, 
we do not systematically lift tech-heavy Level~1 stars into 
management-weighted Level~2/3 roles (the mechanism that generates large negative $\Delta P$ under Merit). 
Instead, promotions are a \emph{mix} of good and bad matches. 
Because seniority promotions are unrelated to role fit, each step mixes helpful and harmful moves that mostly cancel out; the average promotion effect stays near zero, and efficiency drifts down slowly as small mismatches accumulate.

Across $T{=}100$ we observe $\mathbf{52{,}084}$ promotions with $\Delta P>0$ and $\mathbf{56{,}916}$ with $\Delta P<0$—a near balance, consistent with an essentially random mix of good and bad matches.

\begin{figure}[H]
    \centering
    \begin{subfigure}[t]{0.48\linewidth}
        \centering
        \includegraphics[width=\linewidth]{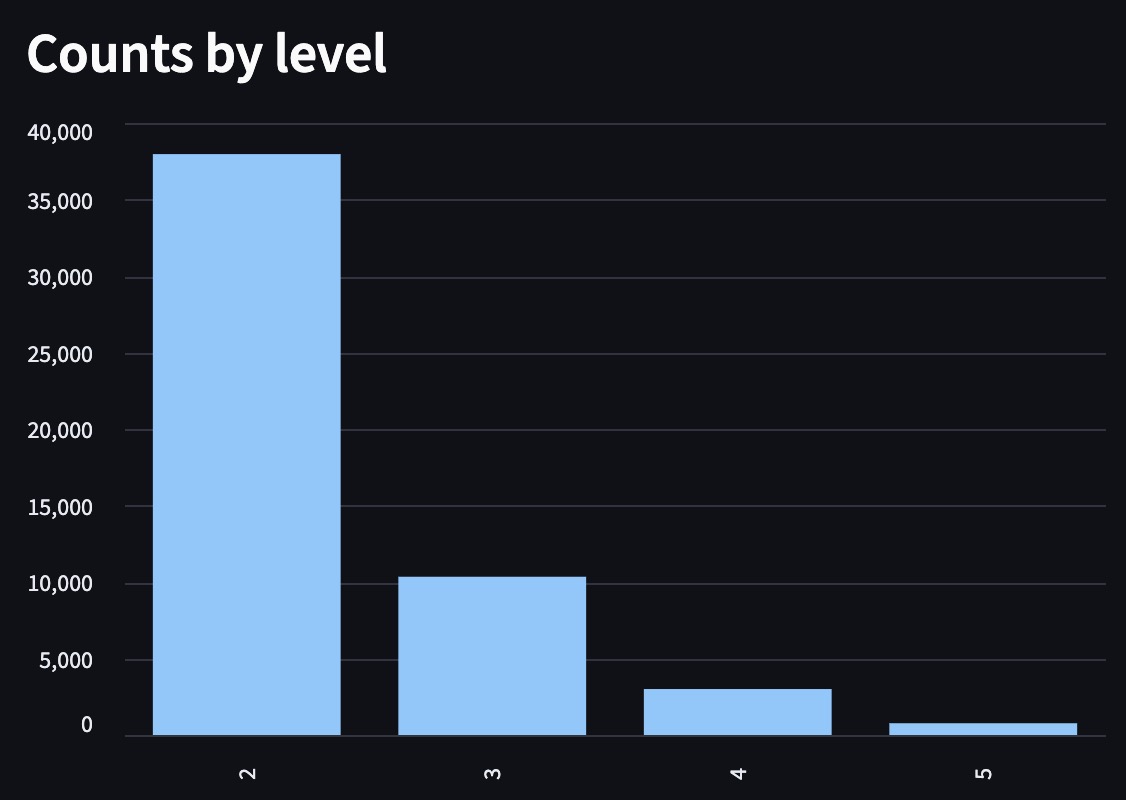}
        \caption{number of Promotions with $\Delta P > 0$ by level.}
        \label{fig:seniority_pos}
    \end{subfigure}
    \hfill
    \begin{subfigure}[t]{0.48\linewidth}
        \centering
        \includegraphics[width=\linewidth]{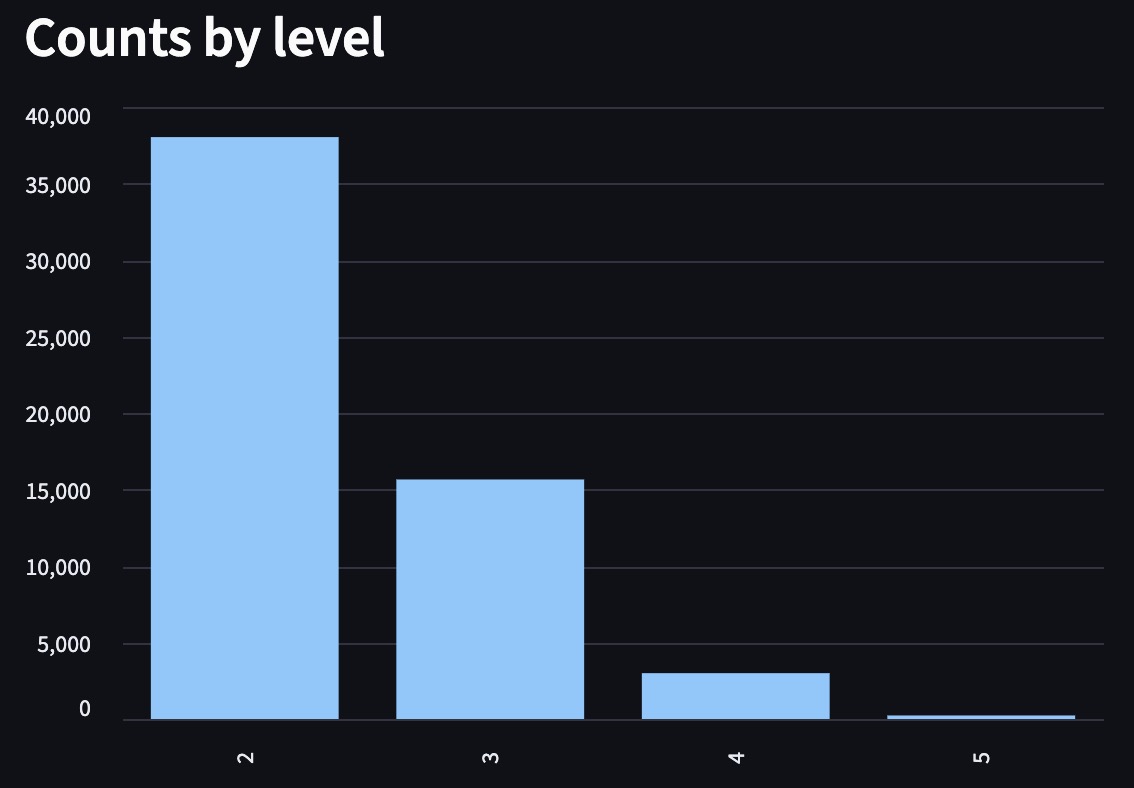}
        \caption{ number of Promotions with $\Delta P < 0$ by level.}
        \label{fig:seniority_neg}
    \end{subfigure}
    \caption{Bar chart of promotion outcomes under seniority-based promotion in the high-mismatch regime. 
The counts of agents with positive ($\Delta P>0$) and negative ($\Delta P<0$) performance changes are shown by level.}
    \label{fig:seniority_deltas}
\end{figure}

\noindent When broken down by levels (Fig.~\ref{fig:seniority_deltas}) these are the observed numbers for agents whose performance improves and for agents whose performance plummets:
\[
\begin{array}{lrr}
\toprule
\text{Level} & \Delta P > 0 & \Delta P < 0 \\
\midrule
\text{L2} & 37{,}967 & 38{,}033 \\
\text{L3} & 10{,}344 & 15{,}656 \\
\text{L4} & 3{,}004 & 2{,}996 \\
\text{L5} &   769   &   231 \\
\bottomrule
\end{array}
\]

\noindent\emph{Explanation:} Because \textbf{Seniority} promotes purely by \emph{time in the organization}, selection is essentially \emph{uncorrelated} with the destination role’s skill weights. This yields a near-random mix of helpful and harmful moves. Seniority does not ``favor'' any skill, but the \emph{level pools} it draws from are not neutral: 
the L2 pool is relatively tech-tilted (from our initialization), and the $L2 \to L3$ transition sharply de-weights technical skills and introduces compliance, creating a mild, systematic headwind there. 
By contrast, the top two levels are both management-heavy, so success at L4 carries over to L5. 
In short, helpful and harmful moves mostly cancel, with a small negative bias concentrated at the $L2 \to L3$ and a small positive bias at the $L4 \to L5$ path which consistent with a random-like policy interacting with non-neutral role reweightings, rather than with a strong Peter-Principle effect.

\begin{figure}[H]
    \centering
    \includegraphics[width=\linewidth]{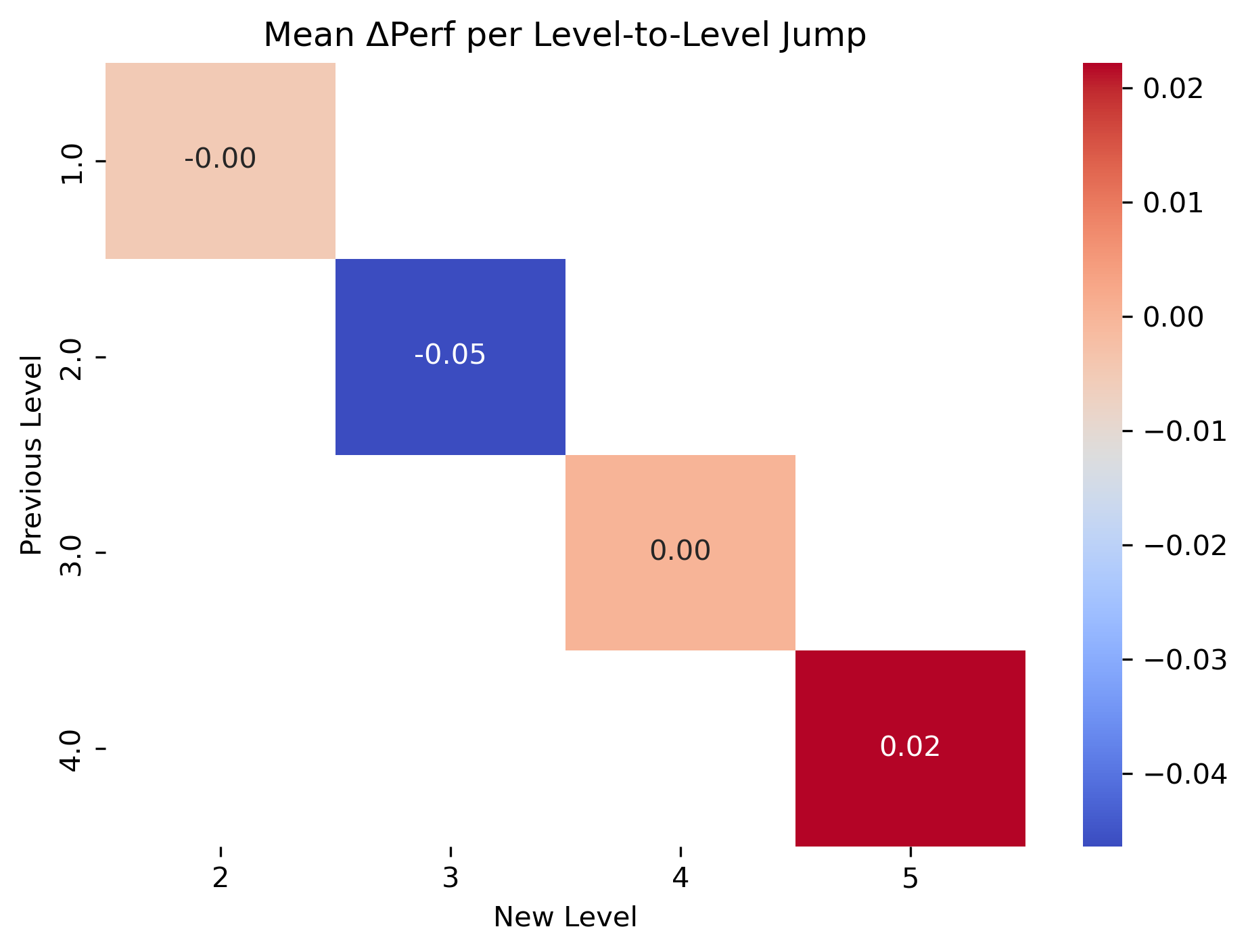}
    \caption{Mean $\Delta P$ per Level-to-Level jump for seniority based promotion strategy.}
    \label{fig:seniority_heatmap}
\end{figure}

\noindent The mean effects from the heatmap confirm this story (Fig.~\ref{fig:seniority_heatmap}): L1$\to$L2 $\approx \mathbf{0.00}$, L2$\to$L3 $\approx \mathbf{-0.05}$, L3$\to$L4 $\approx \mathbf{0.00}$, L4$\to$L5 $\approx \mathbf{+0.02}$. Early transitions include a mild negative due to the tech$\downarrow$ / compliance$\uparrow$ shift; near the top, roles are similar and the average impact is near zero or slightly positive.

\noindent The patterns above do not provide \textbf{enough evidence} of the \textbf{Peter Principle} under \textbf{Seniority}-based promotions: about equal number of people drop performance and improve performance upon promotion.

\begin{figure}[t!]
    \centering
    \includegraphics[width=\linewidth]{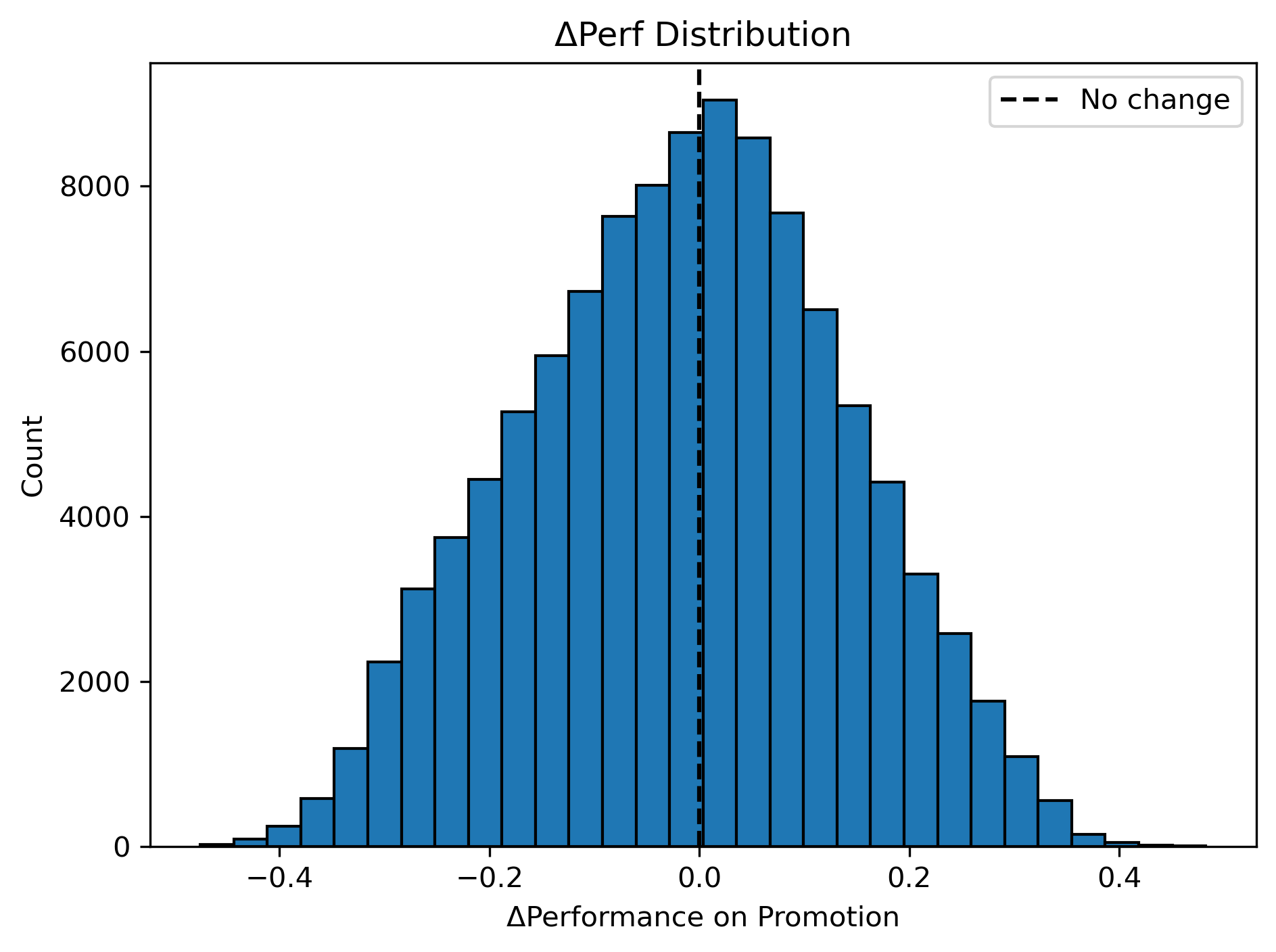}
    \caption{$\Delta P$ frequency distribution across all time-steps for seniority based promotion.}
    \label{fig:seniority_promotion_delta}
\end{figure}

\noindent The \emph{frequency distribution} of $\Delta P$ across all promotions under \textbf{Seniority} (Fig.~\ref{fig:seniority_promotion_delta}) shows a \textbf{marginally slight leftward skew} around the \textbf{baseline $0$}, indicating only a marginal excess of promotions with $\Delta P<0$ over $\Delta P>0$. Across $109{,}000$ promotions, the mean change is $-0.014$ (median $-0.009$), with $52.2\%$ of promotions showing a decrease. The \textbf{negative tail} extends to about \(\mathbf{-0.475}\) and \textbf{positive tail} to \(\mathbf{+0.482}\) (1st/99th percentiles \(-0.342\) / \(+0.311\)). A bootstrap 95\% confidence interval for the mean [$-0.0153,\,-0.0134$] and a one-sided Wilcoxon signed-rank test ($p<10^{-6}$) confirm detectability of a negative shift. However, the magnitude is negligible: relative to a random baseline, the standardized difference is trivial (Cohen’s $d\approx-0.02$), and the share of ``meaningful’’ drops ($\Delta P\leq-0.05$) is virtually unchanged from Random (difference $<\!1$ percentage point). Collectively, these results indicate that the observed dip is not practically significant. In other words, no practical Peter Principle is evident for \textbf{Seniority}; the pattern reflects a baseline step-change associated with role transition rather than a substantive degradation in person–role fit.

\begin{figure}[H]
    \centering
    \includegraphics[width=\linewidth]{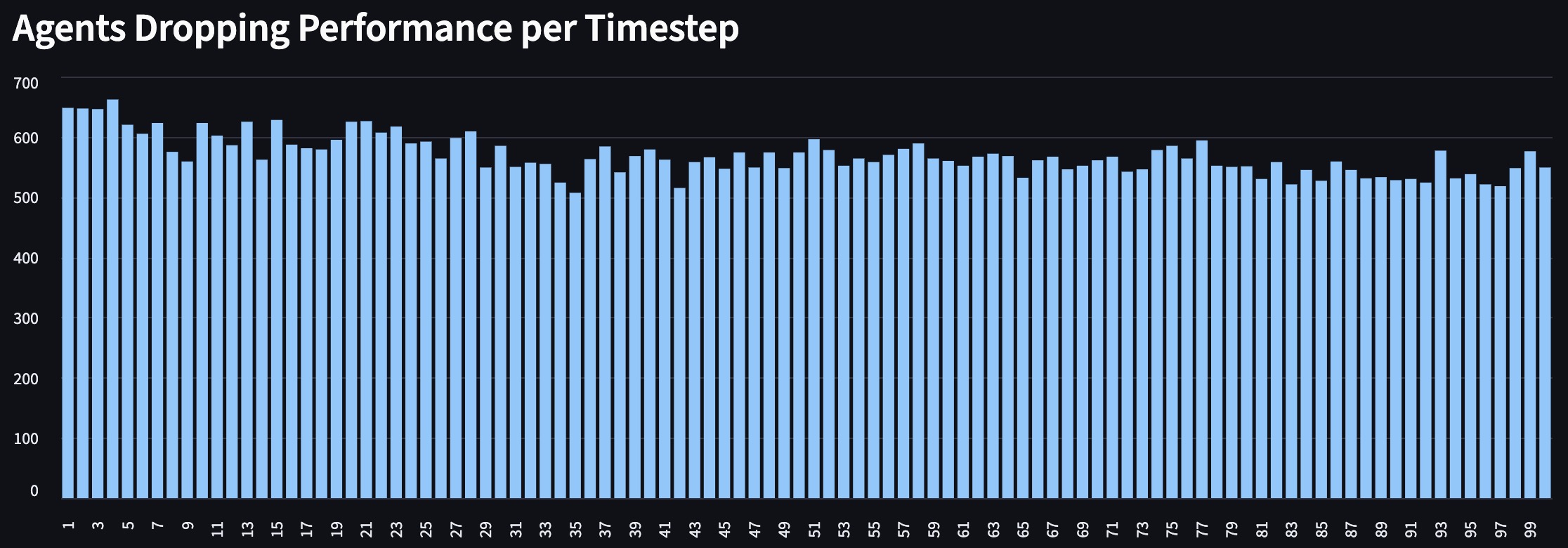}
    \caption{number of promotions with $\Delta P\!<\!0$ at each timestep for seniority based promotion.}
    \label{fig:seniority_neg_tseries}
\end{figure}

\noindent To complement the distributional view, Fig.~\ref{fig:seniority_neg_tseries} plots the \emph{count} of promotions with $\Delta P<0$ at each timestep. The series is remarkably \textbf{flat}—roughly \textbf{$500$–$700$} negative shocks \emph{every} step, accumulating to \textbf{$56{,}916$} across $T{=}100$. Because selection is unrelated to role fit, each period also generates a comparable number of gains (about \textbf{52{,}084} promotions with $\Delta P>0$), so the \emph{net} effect is small and efficiency declines only slightly. For context, the per-step volume of negative shocks under Seniority is well below Merit’s \textbf{900–1{,}100}. 


\paragraph{Interpretation.}
Under \textbf{Seniority}, selection is \emph{skill-blind}: tenure rather than fit to the destination role determines who moves. That makes promotions a near-random mix of matches and mismatches. In aggregate this yields (i) an almost even split of outcomes overall ($\Delta P > 0 \approx 52{,}084$ vs.\ $\Delta P < 0 \approx 56{,}916$), (ii) small mean step effects clustered near zero (heatmap: $L2 \to L3 \approx -0.05$, $L1 \to L2 \approx 0.00$, 
$L3 \to L4 \approx 0.00$, $L4 \to L5 \approx +0.02$), and (iii) a smooth, modest drift in efficiency to $E_{100} \approx 0.531$ without the early crash seen under \textbf{Merit}. The only consistent headwind appears at $L2\!\to\!L3$, and it follows directly from how agents are seeded: levels are initialized by thresholding on their role profiles (with relaxation), so $L2$ begins \emph{tech-tilted} (it screens on tech and not on compliance), whereas $L3$ swaps in compliance and downweights tech. When tenure alone drives movement, many over-tech/under-compliance incumbents at $L2$ rise to $L3$, producing slightly more losses than gains there. By contrast, $L4$ and $L5$ are both management-heavy, 
so performance tends to carry over, yielding mild positives. Together with the near-symmetric $\Delta P$ histogram, these patterns indicate \textbf{no practical Peter-Principle effect} for \textbf{Seniority} in this regime; the small dip reflects routine reweighting across roles rather than systematic misplacement.


\noindent\textbf{Hybrid ($\alpha{=}0.70$ on performance).}
Hybrid scores candidates with a 70–30 blend of \emph{current performance (Merit)} and \emph{tenure (Seniority)} (tenure normalized with a 12-year cap: $\,\mathrm{Tenure}_{\text{norm}}=\min(\tfrac{\text{years}}{12},\,1)\,$ so years beyond 12 do not further raise the seniority term). Because performance dominates, Hybrid behaves like “\emph{slightly-noisy Merit}.” it still pulls top performers in the source role upward. In the high-mismatch regime that means many L1 tech-strong agents are lifted into L2/L3 roles that emphasize management/compliance, so negative $\Delta P$ events persist. The tenure component, however, slightly \emph{dampens} the extremes (especially at lower levels where $T<12$ and the tenure term still varies; it occasionally advances long-tenured, mid-performing candidates and modestly slows the rapid ascent of very new top performers). This helps certain people who maybe have high tenure but poor technical skills at lower level to shine and reclaim higher positions where managerial skills matter the most. (e.g Agent\#~33705). The net effect is the same mechanism as in \emph{merit} but \emph{diluted}: slightly fewer large losses (Atleast for L1$\to$L2 transitions [details in the Heatmap]) with a trajectory that ends just below the start ($E_{100}\approx0.54676$ vs.\ $E_0\approx0.54820$). In short, Hybrid behaves pretty much like Merit-based in our simulation ($\alpha{=}0.7$). Theoretically, the outcome would behave more like seniority for smaller $\alpha$ and merit for larger $\alpha$. The condition that has to change is not to normalize tenure with a cap or increase the cap if seniority is prioritized, or cap it with the highest-tenure each time-step for more robust results and giving seniority some more weightage.

\emph{How the diagnostics line up with Merit.} 

Here are the numbers for the same:

\[
\begin{array}{lrr}
\toprule
\text{Level} & \Delta P > 0 & \Delta P < 0 \\
\midrule
\text{L2} & 8{,}250 & 67{,}480 \\
\text{L3} & 2{,}862 & 23{,}138 \\
\text{L4} & 3{,}808 & 2{,}192 \\
\text{L5} &   795   &   205 \\
\bottomrule
\end{array}
\]

\begin{figure}[H]
    \centering
    \begin{subfigure}[t]{0.48\linewidth}
        \centering
        \includegraphics[width=\linewidth]{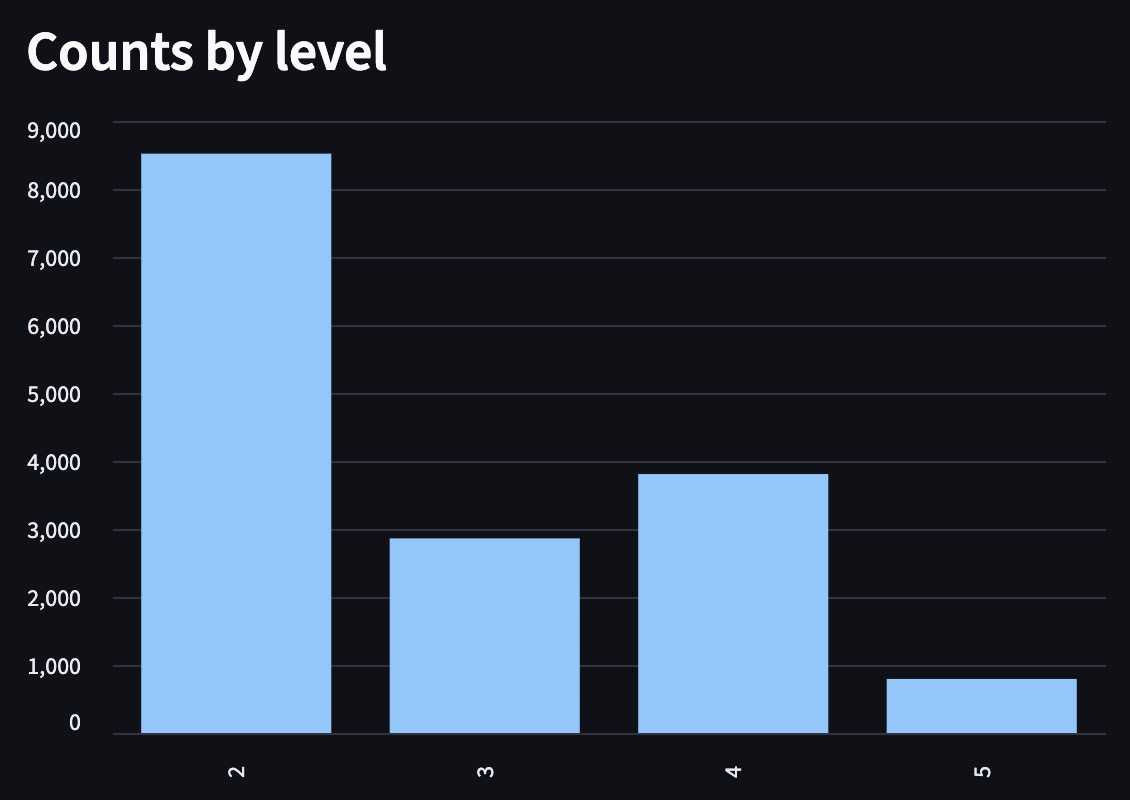}
        \caption{number of Promotions with $\Delta P > 0$ by level.}
        \label{fig:hybrid_pos}
    \end{subfigure}
    \hfill
    \begin{subfigure}[t]{0.48\linewidth}
        \centering
        \includegraphics[width=\linewidth]{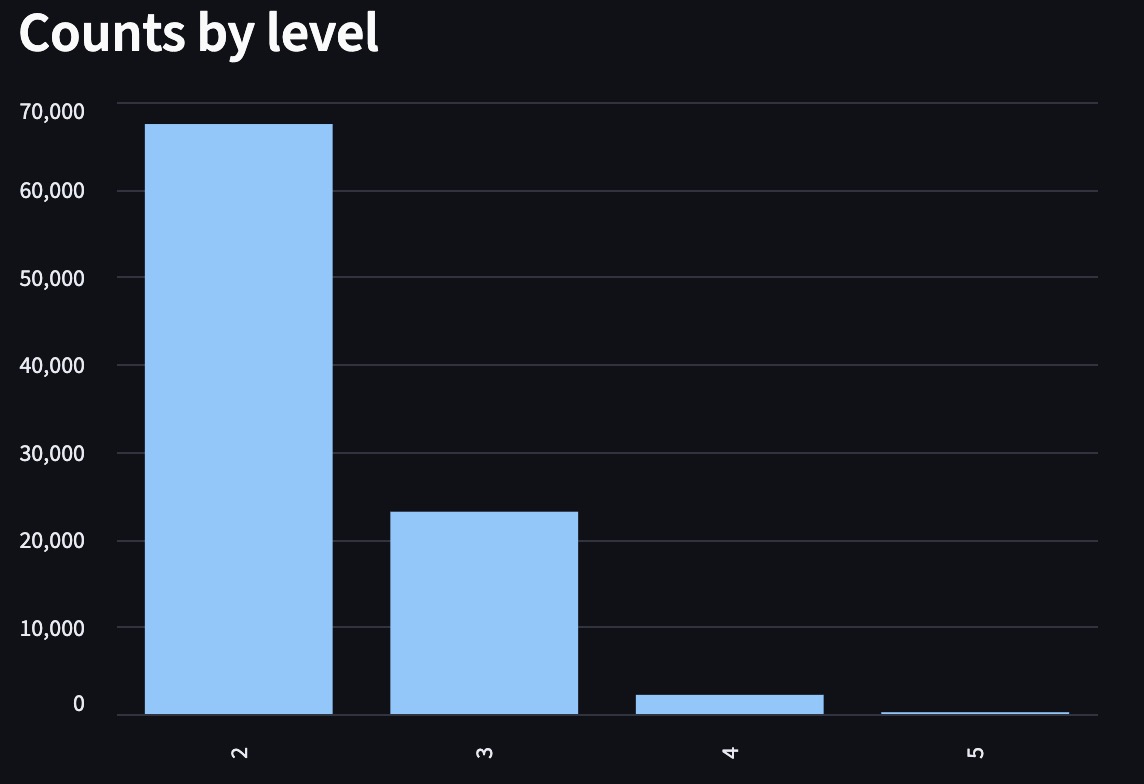}
        \caption{ number of Promotions with $\Delta P < 0$ by level.}
        \label{fig:hybrid_neg}
    \end{subfigure}
    \caption{Bar chart of promotion outcomes under hybrid promotion in the high-mismatch regime. 
The counts of agents with positive ($\Delta P>0$) and negative ($\Delta P<0$) performance changes are shown by level.}
    \label{fig:hybrid_deltas}
\end{figure}

\noindent \textbf{Hybrid} largely mirrors \textbf{Merit}: strong negatives at the bottom and modest positives at the top. The one consistent deviation is \textbf{L1$\to$L2}, where Hybrid softens early mismatches (Hybrid: $8{,}250$ $\Delta P{>}0$ vs.\ $67{,}480$ $\Delta P{<}0$; Merit: $5{,}930$ vs.\ $70{,}070$). This attenuation is exactly what the tenure term is expected to do at the bottom levels. Hybrid occasionally advances longer-tenured, mid-performing L1 candidates whose competence vector might match what is required at the level being promoted to. Beyond L2, the seniority term \emph{diminishes}: tenure is capped at 12 years and, given entry bands (L2: 2–5, L3: 4–7), agents quickly hit the cap, especially after $\gtrsim$10 steps. Once most candidates are saturated on tenure, the Hybrid score is effectively performance-driven and behaves like Merit.

\begin{figure}[H]
    \centering
    \includegraphics[width=\linewidth]{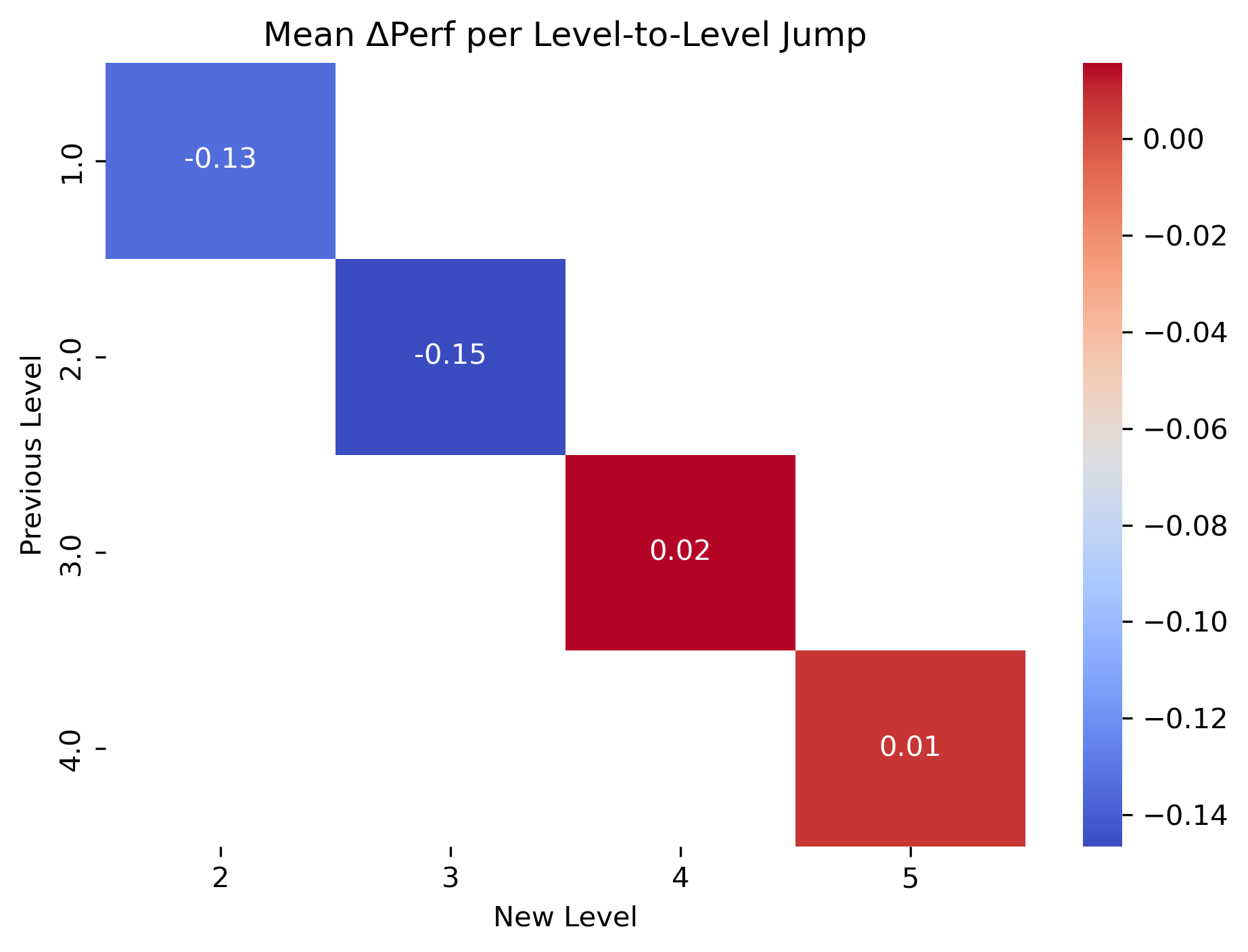}
    \caption{Mean $\Delta P$ per Level-to-Level jump for hybrid promotion strategy.}
    \label{fig:hybrid_heatmap}
\end{figure}

\noindent Similar data is shown in the heatmap, where L1$\to$L2 transition has lower $\Delta P$ value ($-0.13$ compared to $-0.15$ in merit).

\begin{figure}[H]
    \centering
    \includegraphics[width=\linewidth]{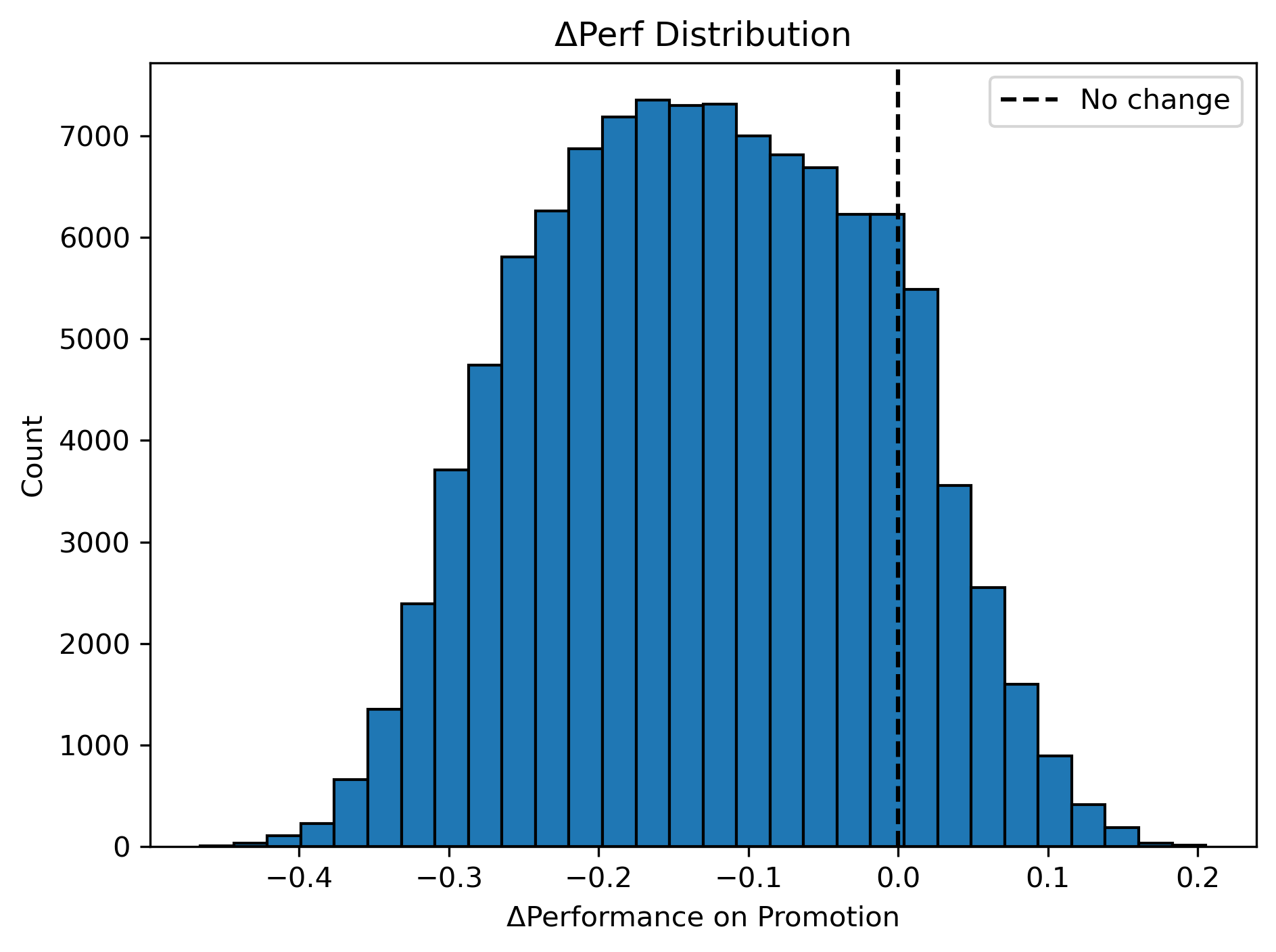}
    \caption{$\Delta P$ frequency distribution across all time-steps for hybrid promotion.}
    \label{fig:hybrid_promotion_delta}
\end{figure}

 \noindent The \textbf{$\Delta P$ histogram} is still \emph{left-skewed}, yet noticeably shallower than merit (mean = -0.128 (-0.136 for merit) and median=-0.129 (-0.138)). The \textbf{negative tail} extends to about \(\mathbf{-0.466}\) (-0.448 for merit) and \textbf{positive tail} to \(\mathbf{+0.205}\) (+0.153) (1st/99th percentiles \(-0.353\) / \(+0.103\)). Overall, the data show the same qualitative pattern as Merit.

\paragraph{Interpretation.} Hybrid \textbf{dampens} (but does not remove) the effects of Peter-Principle present under Merit especially for lower levels. Because performance still carries 70\% of the score and destination roles reweight away from tech, the early transitions remain the main source of losses; the tenure term mostly \emph{reduces} their frequency and severity rather than reversing them. At the same time, Hybrid sometimes lifts long-tenured agents who underperform in tech-weighted L1 work but have stronger management profiles, allowing them to claim higher positions where
managerial skills matter the most. One such example is of Agent\#~333705.

\begin{figure}[H]
    \centering
    \includegraphics[width=\linewidth]{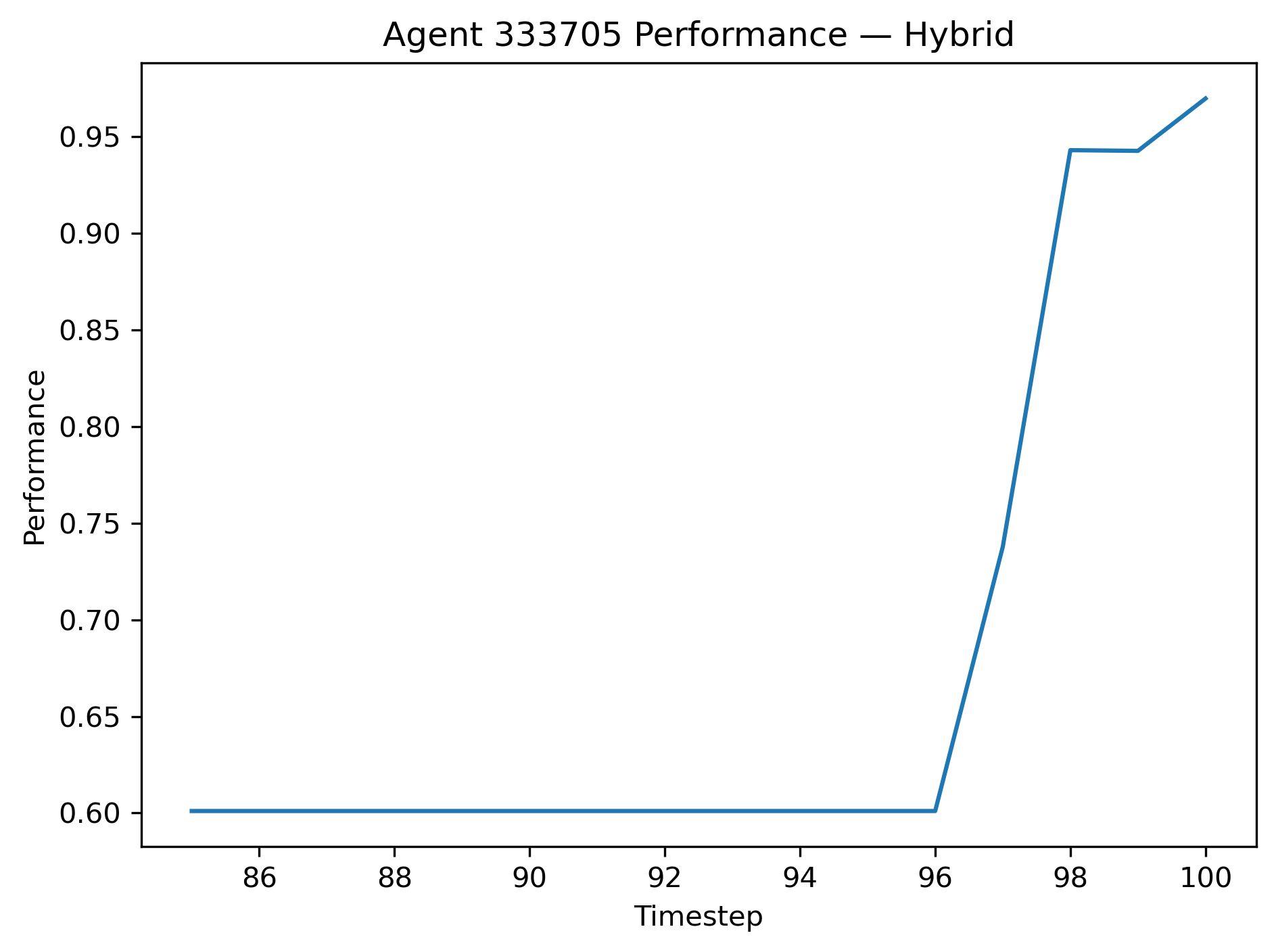}
    \caption{Agent 333705s performance over time for hybrid promotion.}
    \label{fig:hybrid_agent_13950}
\end{figure}

\noindent Agent\# 333705 has the following \emph{skills:}
\begin{verbatim}
{
  "tech": 0.5871914335043364,
  "management": 0.9967401028519282,
  "compliance": 0.9985460600423186,
  "soft_skills": 0.7257766012186696
}
\end{verbatim}

\noindent Agent\#~333705 joins the organization at time step $t{=}85$ with \textbf{low L1 performance} ($0.601$) due to weak technical ability. As a result, he spends a long time at \textbf{L1} and is finally promoted at \emph{$t{=}97$}, when his tenure reaches 12 years. The promotion is mainly the result of high tenure and relatively average L1 performance. At \textbf{L2}, where technical skills are less central and management matters more, his performance increases to $0.7378$. Due to this higher performance and his relatively high L2 tenure (13 years), he is promoted to \textbf{L3} at the next timestep (\emph{$t{=}98$}). \textbf{L3} emphasizes management and compliance which are both strong for this agent which raises his performance further ($0.9431$). From \textbf{L3} onward, the merit component dominates relative to tenure, and with this high performance he is promoted to \textbf{L4} at \emph{$t{=}99$}, where he again performs well ($0.9427$). Finally, at \emph{$t{=}100$} he is promoted to \textbf{L5}, where he records his best performance ($0.9698$). This type of transitions is what hybrid promotions unlock: under pure Merit, this agent would likely have stalled at L1 given his relatively low performance there, but tenure allowed him to surface and excel as the role mix shifted toward management.


\vspace{\baselineskip}
\vspace{\baselineskip}

\noindent\textbf{Random.}
Random draws promotees uniformly from each level’s candidate pool, fully \emph{skill-blind} and therefore, tracks Seniority closely. In the high-mismatch regime, efficiency follows the same \textbf{smooth, modest drift} as Seniority, ending at $E_{100}!\approx!0.5353$ (no early crash).
\par\medskip

\emph{How the diagnostics line up with Seniority.} 

\begin{figure}[H]
    \centering
    \begin{subfigure}[!t]{0.48\linewidth}
        \centering
        \includegraphics[width=\linewidth]{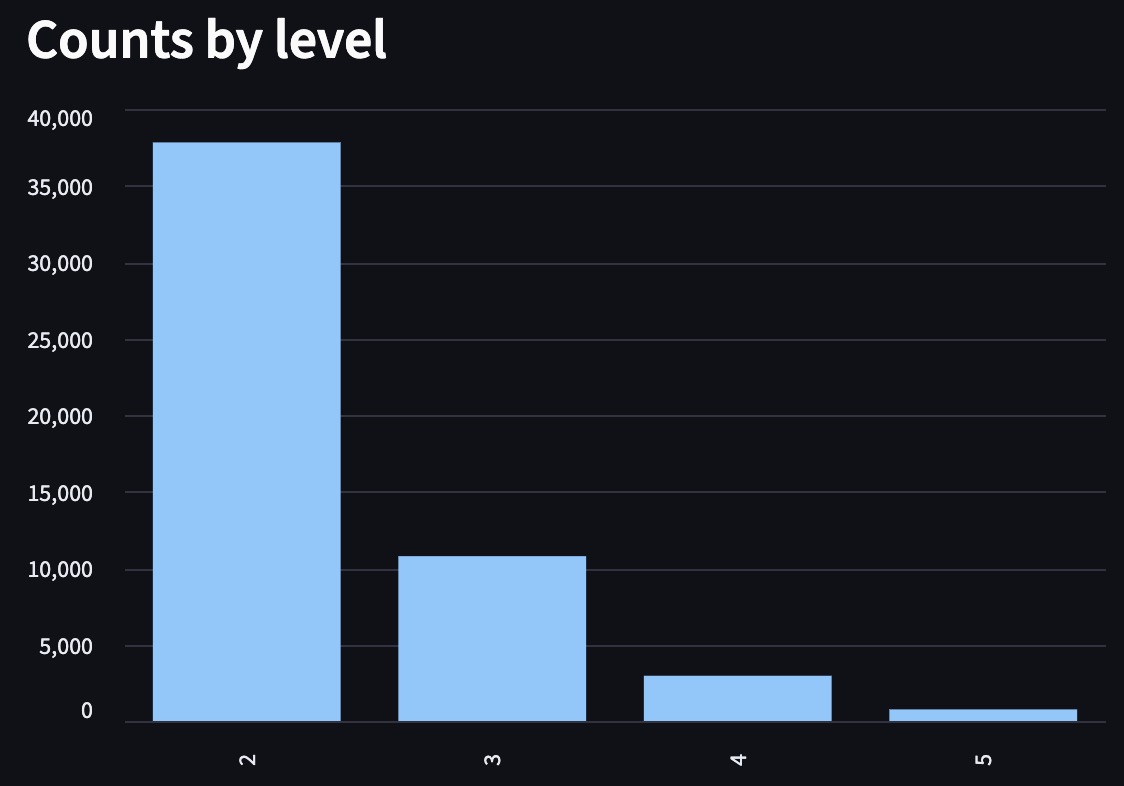}
        \caption{number of Promotions with $\Delta P > 0$ by level.}
        \label{fig:random_pos}
    \end{subfigure}
    \hfill
    \begin{subfigure}[!t]{0.48\linewidth}
        \centering
        \includegraphics[width=\linewidth]{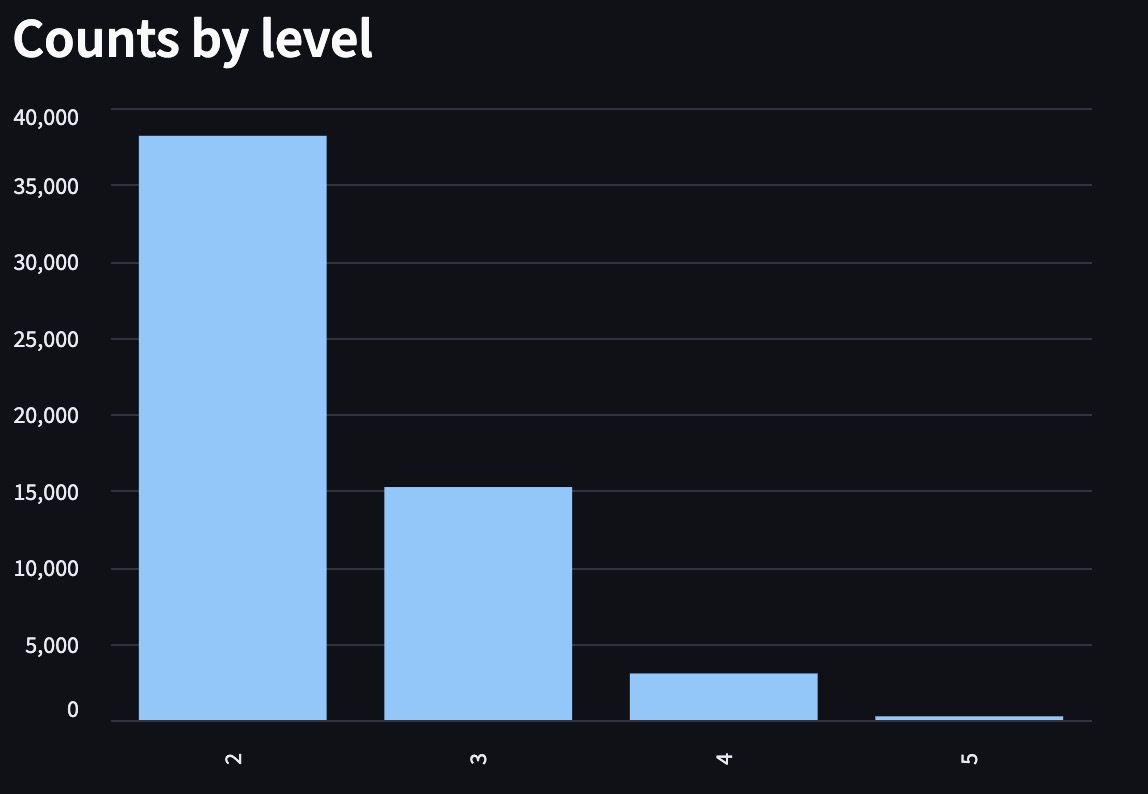}
        \caption{ number of Promotions with $\Delta P < 0$ by level.}
        \label{fig:random_neg}
    \end{subfigure}
    \caption{Bar chart of promotion outcomes under random promotion in the high-mismatch regime. 
The counts of agents with positive ($\Delta P>0$) and negative ($\Delta P<0$) performance changes are shown by level.}
    \label{fig:random_deltas}
\end{figure}

\par\medskip

Here are the numbers for the same:

\[
\begin{array}{lrr}
\toprule
\text{Level} & \Delta P > 0 & \Delta P < 0 \\
\midrule
\text{L2} & 37{,}817 & 38{,}183 \\
\text{L3} & 10{,}774 & 15{,}226 \\
\text{L4} & 2{,}958 & 3{,}042 \\
\text{L5} &   759   &   241 \\
\bottomrule
\end{array}
\]

\begin{figure}[H]
    \centering
    \includegraphics[width=\linewidth]{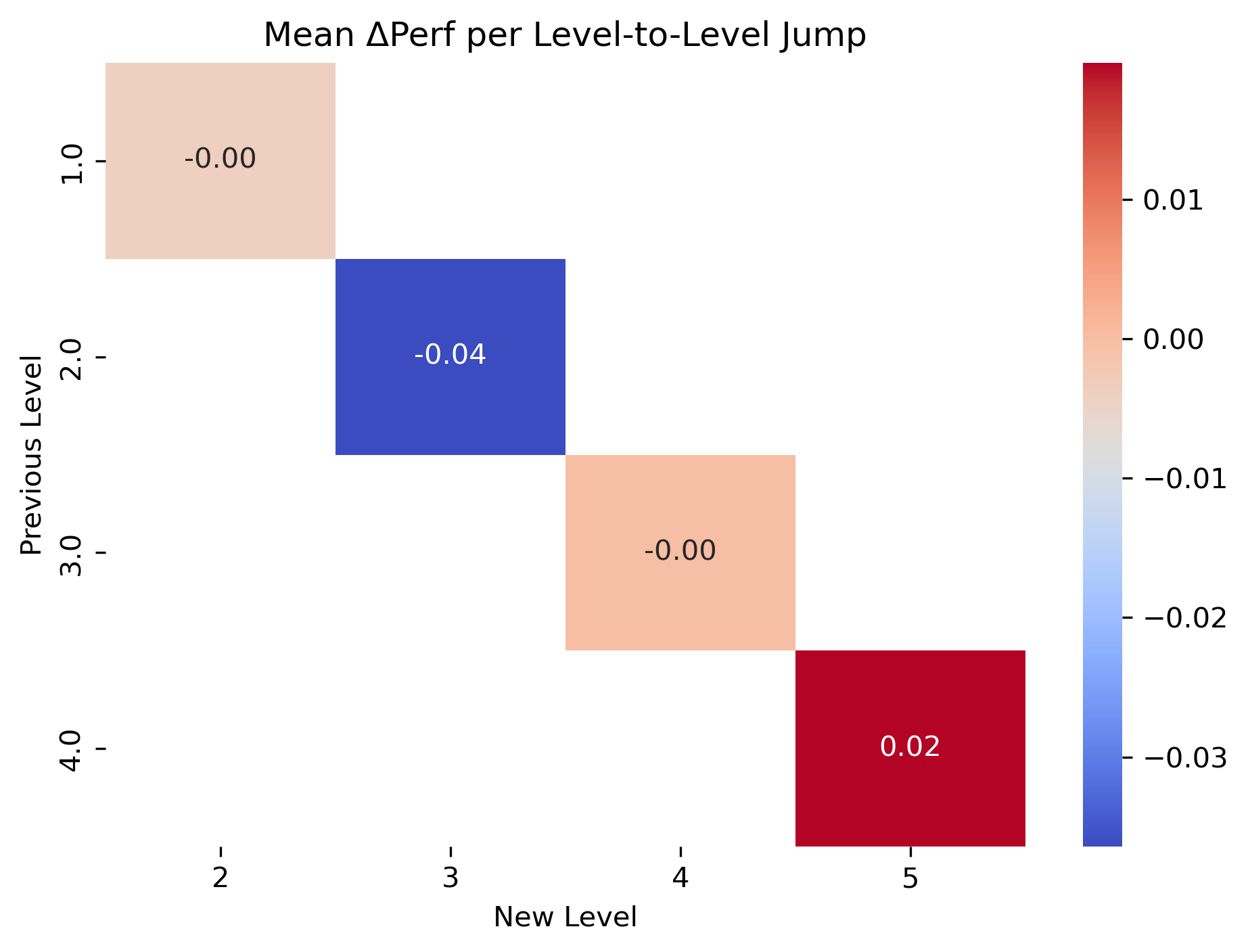}
    \caption{Mean $\Delta P$ per Level-to-Level jump for random promotion strategy.}
    \label{fig:random_heatmap}
\end{figure}
\par\medskip

\noindent The \textbf{bar charts} and the \textbf{mean $\Delta P$ heatmap} show an almost balanced mix on L1$\to$L2 and L3$\to$L4, a \emph{mild} tilt to negatives on L2$\to$L3 (tech weight $\downarrow$, compliance weight $\uparrow$), and slightly positive average shocks on L4$\to$L5 where roles are similar and management-heavy. Which is qualitatively same as seniority.
\par\medskip

\begin{figure}[H]
    \centering
    \includegraphics[width=\linewidth]{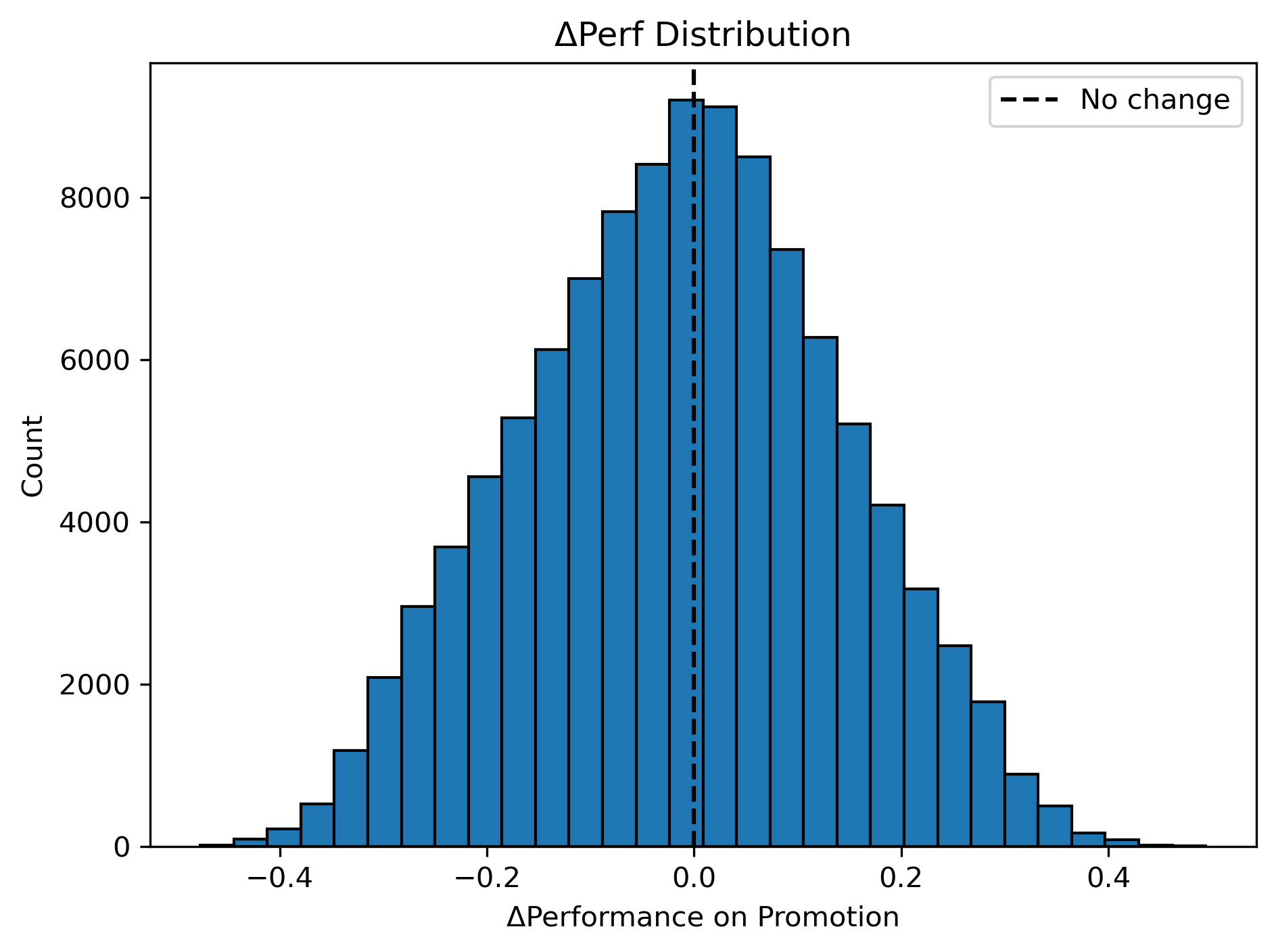}
    \caption{$\Delta P$ frequency distribution across all time-steps for random promotion.}
    \label{fig:random_promotion_delta}
\end{figure}
\par\medskip

\noindent The \textbf{$\Delta P$ histogram} is near-symmetric with only a negligibly small left tail (mean=-0.011, median=-0.008, \% of promotions with drop: 52.0\%). The \textbf{negative tail} extends to about \(\mathbf{-0.477}\) and \textbf{positive tail} to \(\mathbf{+0.494}\) (1st/99th percentiles \(-0.339\) / \(+0.319\)). 
\par\medskip

\paragraph{Interpretation.} With selection \emph{uncorrelated} to the destination profile, Random yields a near-50/50 mix of matches and mismatches. The slight negative bias arises from the role reweighting itself and initialization (especially L2$\to$L3), not from systematic misplacement. Consistent with Seniority, there is \textbf{no practical Peter-Principle signal} under Random; the observed dip reflects routine step changes in weights rather than a directional selection error.
\par\medskip

\vspace{\baselineskip}

\noindent \textbf{Selective Demotion.} (Mitigation; see Sec.~\ref{sec:mitigations}.) This rule promotes the best performer from the source level but \emph{demotes} anyone at the destination level whose immediate post-move performance falls below a fixed threshold (here, ($\Delta P \le -\tau$, $\tau=0.05$); the vacated seat is refilled from the level below by merit, while the demoted agent is blacklisted from re-promotion to avoid making the same move again. In the \emph{high-mismatch} regime, this filter converts the large L1$\to$L2 and L2$\to$L3 Peter shocks into short-lived blips instead of persistent mismatches, so the organization avoids the prolonged drag seen under pure Merit. The efficiency path shows a brief early dip while misfits are identified and rolled back (minimum $E_t \approx 0.543$ at $t=7$ from $E_0 \approx 0.548$), followed by a steady rise to a higher plateau by $T=100$ ($E_{100} \approx 0.569$, $+3.8\%$ vs.\ baseline). Mechanistically, demotion curbs the accumulation of low-fit promotions that drag efficiency under pure Merit; the blacklist-and-refill step shifts subsequent selections toward more “portable” competence profiles (higher management/compliance relative to tech), reducing both the \emph{frequency} and \emph{magnitude} of negative $\Delta P$ events and allowing aggregate performance to recover and surpass baseline rather than plateau near it. 
\par\medskip

\noindent
We observed \(97{,}648\) promotions over \(T=100\), with \(60{,}333\) having \(\Delta P < 0\) (\(61.8\%\)) and \(37{,}315\) having \(\Delta P > 0\) (\(38.2\%\)). Relative to \emph{Merit} in the same regime (\(88.1\%\) negatives), selective demotion reduces the negative share by \(\approx 26.3\) percentage points \(\bigl(\tfrac{26.3}{88.1} \approx 30\%\ \text{relative drop}\bigr)\).
\par\medskip

\begin{figure}[H]
    \centering
    \begin{subfigure}[t]{0.48\linewidth}
        \centering
        \includegraphics[width=\linewidth]{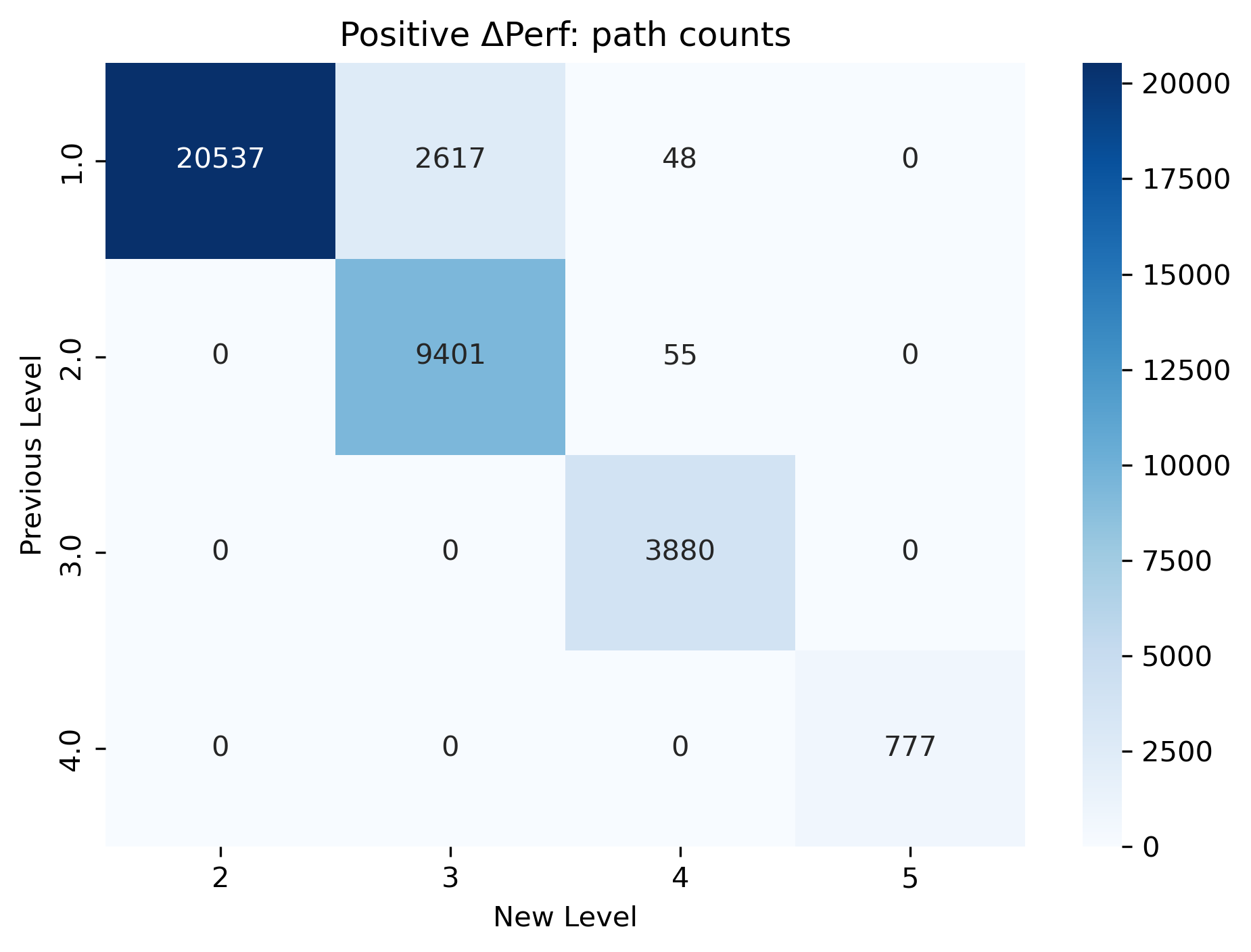}
        \caption{number of Promotions with $\Delta P > 0$ by level.}
        \label{fig:selective_demotion_pos}
    \end{subfigure}
    \hfill
    \begin{subfigure}[t]{0.48\linewidth}
        \centering
        \includegraphics[width=\linewidth]{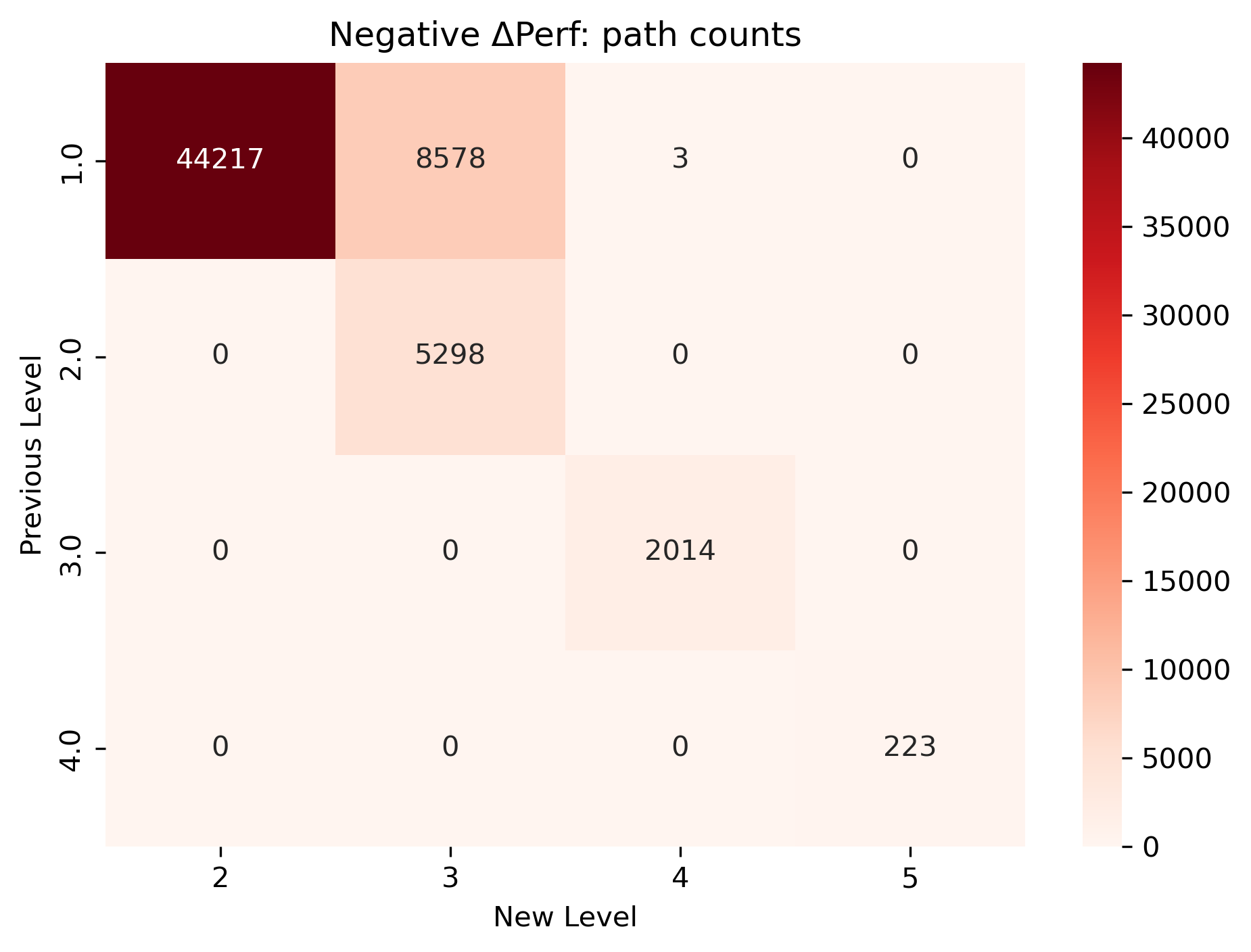}
        \caption{ number of Promotions with $\Delta P < 0$ by level.}
        \label{fig:selective_demotion_neg}
    \end{subfigure}
    \caption{Heatmap of promotion outcomes under selective demotion in the high-mismatch regime.}
    \label{fig:selective_demotion_deltas}
\end{figure}

\noindent\emph{Heatmaps (Fig.~\ref{fig:selective_demotion_deltas}).} Under \textbf{selective demotion} in the \emph{high-mismatch} regime. At lower levels, we have more promotions with $\Delta P < 0$ than $\Delta P > 0$. This is due to sharp reweighting away from tech at lower levels (L1$\to$L2) and our threshold ($\tau = 0.05$) reversing big drops but keeping small losses. As we move higher, L2$\to$L3 where roles are more similar, the counts of positive and negative shocks are more balanced. This is because the \emph{demotion+blacklist} removes poorly matched promotees early, so by the time agents reach L3/L4 the pool is enriched for management/compliance-heavy profiles which is rewarded at higher levels. At higher levels (L3$\to$L4 and L4$\to$L5), roles are even more similar and management-heavy, and due to early filtering the counts of positive are more compared to negatives.

\begin{figure}[H]
    \centering
    \includegraphics[width=\linewidth]{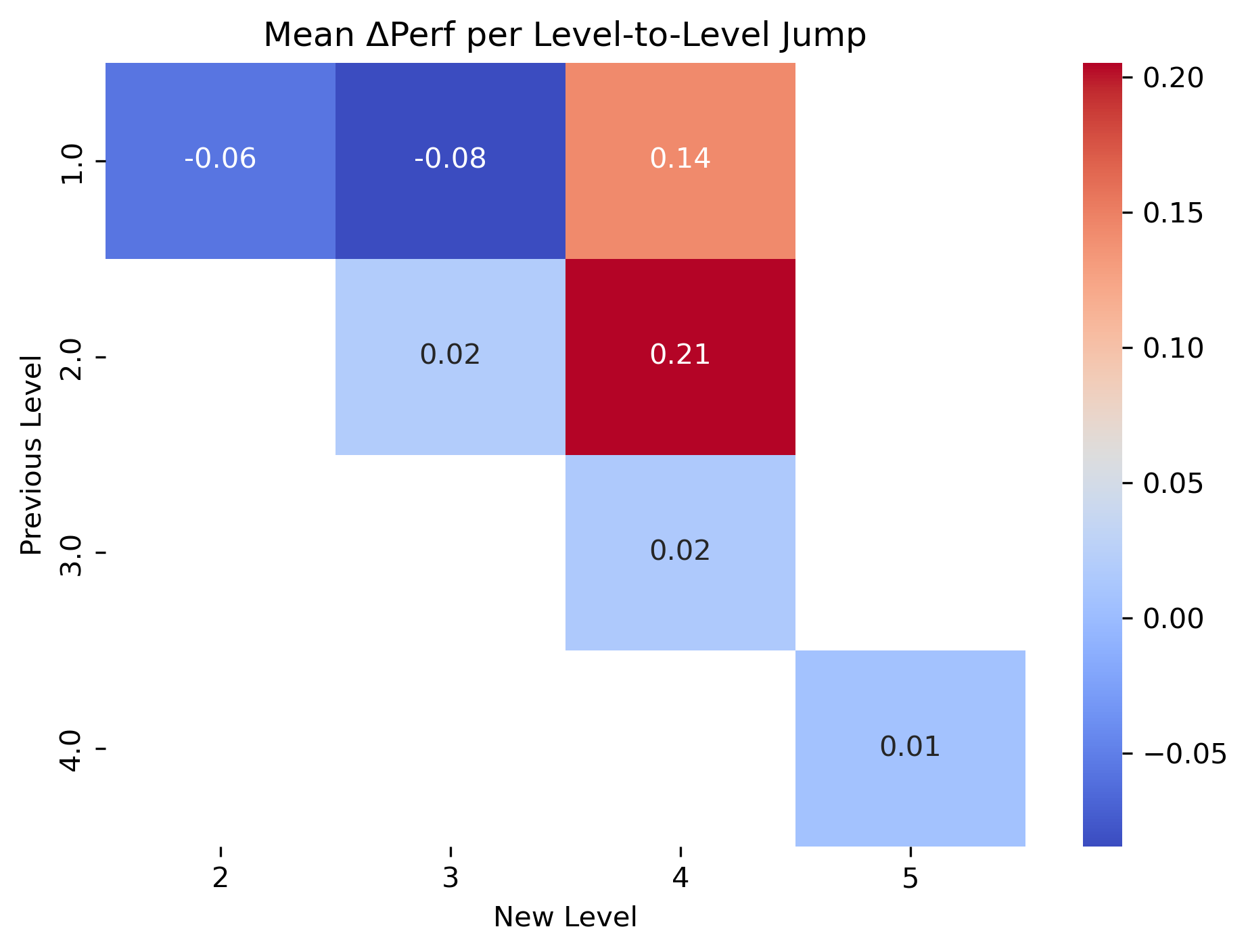}
    \caption{Mean $\Delta P$ per Level-to-Level jump for selective demotion promotion strategy.}
    \label{fig:selective_demotion_heatmap}
\end{figure}

\noindent The \textbf{mean $\Delta P$ heatmap} (Fig.~\ref{fig:selective_demotion_heatmap}) shows a similar pattern: large negative shocks at L1$\to$L2 ($-0.06$) due to tech$\downarrow$/management$\uparrow$ reweighting and threshold keeping small drops, but the magnitude is smaller than pure Merit ($-0.15$). At L2$\to$L3 we see a small positive ($0.02$). This is because the selective demotion already filtered for high compliance and management compared to high tech in the L1$\to$L2 promotion, so this promotion is slightly positive because the role profiles for L2 and L3 are very similar. Through similar logic, L3$\to$L4 and L4$\to$L5 are also slightly positive ($0.02$ and $0.01$ respectively).

\noindent We also see L1$\to$L3, L1$\to$L4 and L2$\to$L4 transitions because of the demotion and refill mechanism. These ``skip-level'' transitions are a direct consequence of the selective-demotion cascade. In one timestep, we (i) fill attrition vacancies top--down, (ii) demote any just-promoted misfits \((\Delta P \leq -\tau)\), which creates new vacancies, and then (iii) immediately refill those new vacancies from the current lower level. Because the refill ranks candidates using their post-promotion performance at the new level, someone who just moved \(L1 \rightarrow L2\) can be chosen again for \(L2 \rightarrow L3\) in the same tick. And our logs only record only \emph{prev-level $\to$ end-of-step level}, so it appears as two level jumps. Rare three-step jumps (e.g., L1$\to$L4) happen when multiple upper tiers open simultaneously and the candidate is already well aligned with top-heavy weights. That is the reason why L2$\to$L4 and L1$\to$L4 has more positive shocks. Due to selection of management and compliance heavy profiles at lower levels, these agents are well suited for higher levels which are also management and compliance heavy. But L1$\to$L3 transitions still have a negative mean $\Delta P$ because of the tech$\downarrow$/compliance$\uparrow$ reweighting at lower levels.

\begin{figure}[H]
    \centering
    \includegraphics[width=\linewidth]{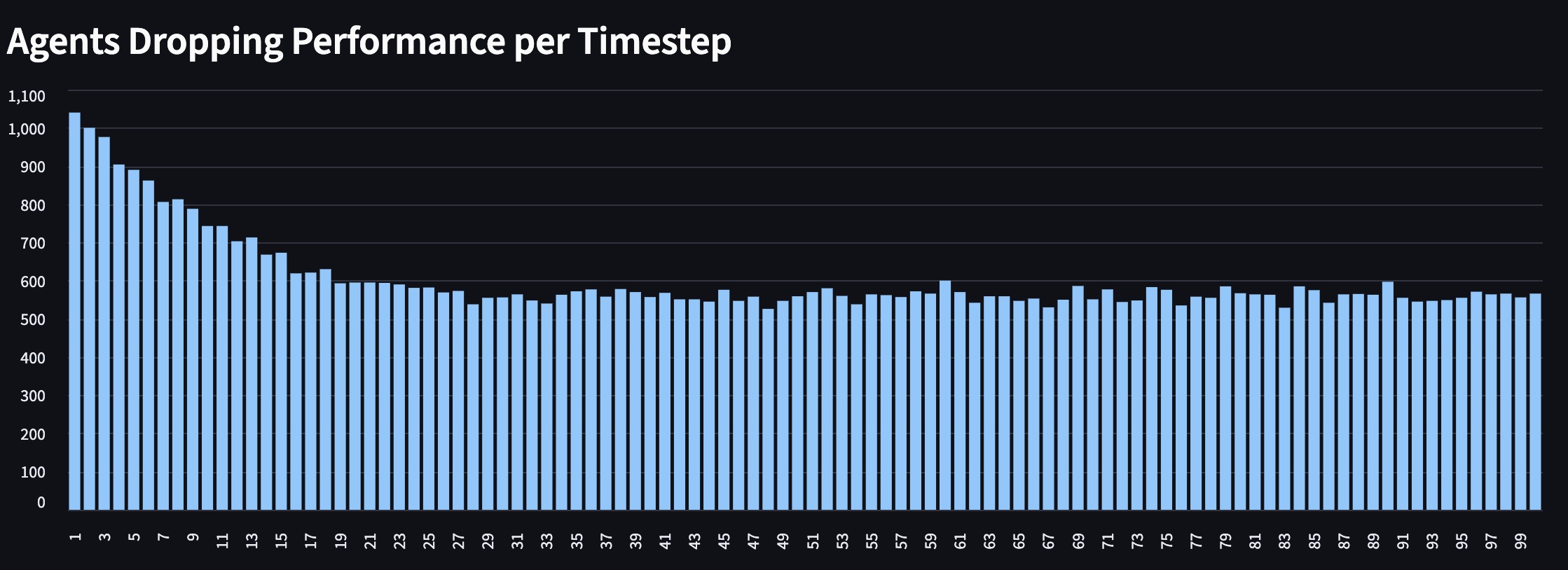}
    \caption{$\Delta P$ Number of promotions with $\Delta P<0$ at each timestep.}
    \label{fig:selective_demotion_neg_tseries}
\end{figure}

\noindent The \textbf{negative shocks over time histogram} shows the bigger picture. We see during earlier timsteps that roughly $\sim\!1000$ promotions per step result in drop in performance which reduces drastically every timestep and after about timestep 20 it stabilizes to $\sim\!500-600$ agents and remains constrant throughout. This reflects the immediate testing of many tech-heavy L1 performers against more managerial roles (the classic Peter shock) during early timesteps. As the \emph{selective demotion} filter reverses the largest drops and blacklists those agents from re-promotion into the same vacancy, the candidate pool becomes progressively better aligned with destination weights. Consequently, resulting in the per-step count of negatives declining rapidly and stabilizing at $\sim\!500-600$ per step. This floor is \emph{expected}. First, fresh L1 hires enter every timestep and some will experience small mismatches on their first move. Second, the tolerance $\tau=0.05$ only reverses \emph{large} drops; modest declines (e.g., $-0.01$ to $-0.04$) persist by design. Together, these forces explain the stable tail of $\sim\!500-600$ negatives per step after the initial adjustment period.
\par\medskip

\begin{figure}[H]
    \centering
    \includegraphics[width=\linewidth]{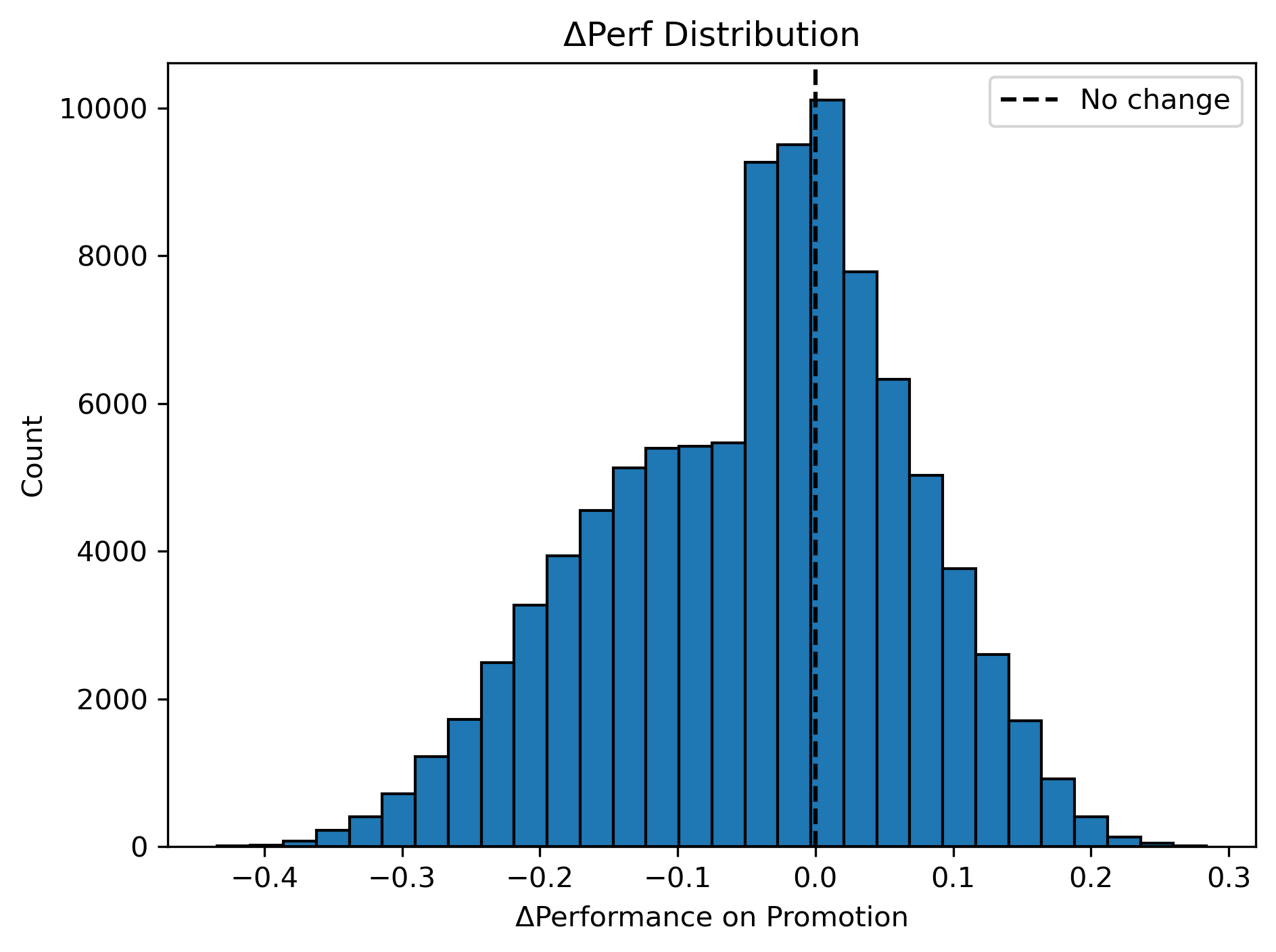}
    \caption{$\Delta P$ frequency distribution across all time-steps for selective demotion strategy.}
    \label{fig:selective_demotion_promotion_delta}
\end{figure}
\par\medskip

\noindent The \textbf{$\Delta P$ histogram} under \emph{selective demotion} is \emph{left-skewed} with a small negative central tendency, but noticeably shallower than pure Merit (mean \(-0.043\)  vs.\(-0.136\) for Merit; median = \(-0.028\) vs.\(-0.138\); \% of promotions with drop: 61.8\% vs.\ 88.1\%). The \textbf{negative tail} extends to about \(\mathbf{-0.434}\) while the \textbf{positive tail} tops out near \(\mathbf{+0.284}\) (1st/99th percentiles \(-0.305\) / \(+0.176\)).Compared with pure Merit, the selective-demotion threshold trims the \emph{magnitude} of large losses and pulls mass toward mild negatives (e.g., \(-0.01\) to \(-0.04\)) that survive the \(\tau=0.05\) filter. The asymmetry arises from the sharp tech\(\downarrow \to\)management/compliance\(\uparrow\) reweighting at lower levels (creating a longer left tail).

Let's take example of 2 agents to understand the impact of selective demotion:
\par\medskip

\paragraph{Case A — Skip-level gain (Agent\#~100462).} Agent\# 100462 has the following \emph{skills:}
\begin{verbatim}
{
  "tech": 0.7604235466122206,
  "management": 0.9511536650823397,
  "compliance": 0.6392829972727467,
  "soft_skills": 0.9765699963299141
}
\end{verbatim}
\par\medskip

\noindent The Agent begins in the Organization at time-step 0 at L1. The agent's performance is mildly strong at L1 ($P\!=\!0.782$). Due to this the agent is not promoted to L2 early. But due to selective demotions and blacklisting of other agents, at timestep 4 the agent is promoted to L2. But immidiately due to skills aligning well with L2 (management$\uparrow$, softskills$\uparrow$) and vacancies at L3 the agent is directly promoted to L3 in the \emph{apply\_dra\_demotion\_and\_refill} step which in our code happens after \emph{promote\_agents}. At L3, performance \emph{improves} ($0.8627$): due to his high management skills, strong soft-skills and slightly strong compliance(as this level demands 50\% more management skills, 20\% more soft-skills and 30\% more compliance). He stays in this level for a few more timesteps as the performance is not enough to get promoted to L4. At timestep 11, he gets promoted to L4 due to more selective demotions at L3$\to$L4. The move to L4 is positive ($0.925$) as management weight rises further and compliance drops. The agent remains at this level till the end of the simulation as his performance is not enough to get promoted to L5.
\par\medskip

\begin{figure}[H]
    \centering
    \includegraphics[width=\linewidth]{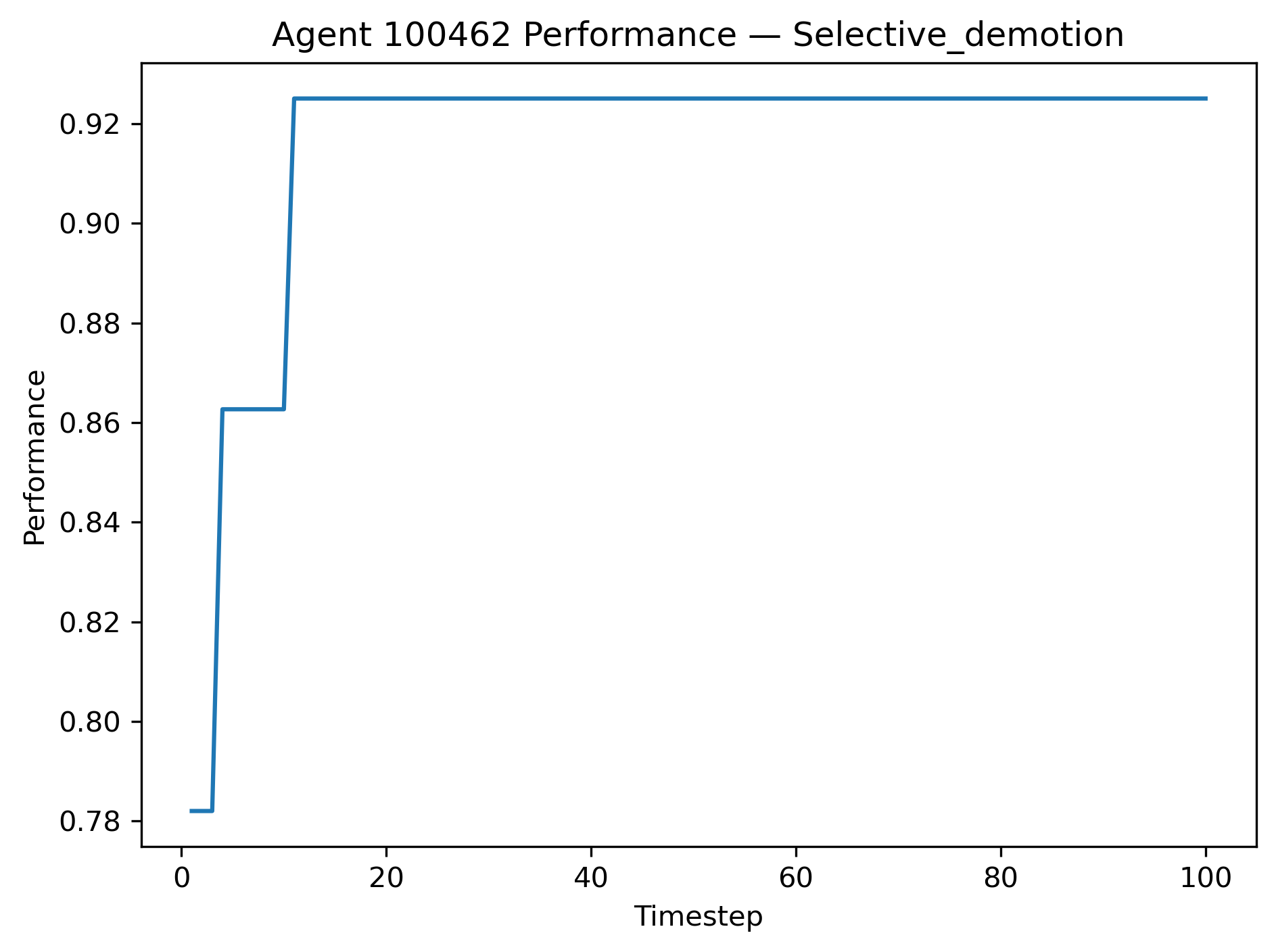}
    \caption{Agent 100462's performance over time for selective demotion strategy .}
    \label{fig:selective_demotion_agent_100462}
\end{figure}
\par\medskip

\noindent \textbf{Takeaway:} Selective demotion’s ``trial--then--confirm'' logic lets \emph{portable} profiles surface quickly even if they  weren’t standout fits for L1’s tech-heavy screen. Once this agent cleared L1$\to$L2, their  \emph{post-promotion L2 performance} was immediately strong, so the cascade refill selected them again  (L2$\to$L3) in the same tick. The later L3$\to$L4 move is again positive because upper levels are management-heavy, matching the agent’s strengths; the eventual stall at \(L4\) reflects scarcity/competition rather than mismatch. In short, the mitigation simultaneously prunes big misfits and \emph{accelerates} well-aligned candidates to levels where they create more value.
\par\medskip

\paragraph{Case B — Skip-level loss (Agent\#~100114).} \label{para:case100114_merit}Agent\# 100114 has the following \emph{skills:}

\begin{verbatim}
{
  "tech": 0.8808329324958385,
  "management": 0.7831324047407058,
  "compliance": 0.036530444057425115,
  "soft_skills": 0.6639328247375101
}
\end{verbatim}
\par\medskip

\noindent The Agent begins in the Organization at time-step 0 at L1. The agent's performance is strong at L1 ($P\!=\!0.8591$). Due to this the agent is promoted to L2. But immidiately due to skills aligning well with L2 (management$\uparrow$, softskills$\uparrow$) and vacancies at L3 the agent is directly promoted to L3 in the \emph{apply\_dra\_demotion\_and\_refill} step which in our code happens after \emph{promote\_agents}. At L3, performance \emph{decreases drastically} ($0.5353$): due to his very low compliance(as this level demands 30\% more compliance). Due to this big drop the agent is not promoted to L4 and remains at L3 for rest of the simulation till timestep 99 where he is attrited. 
\par\medskip

\begin{figure}[H]
    \centering
    \includegraphics[width=\linewidth]{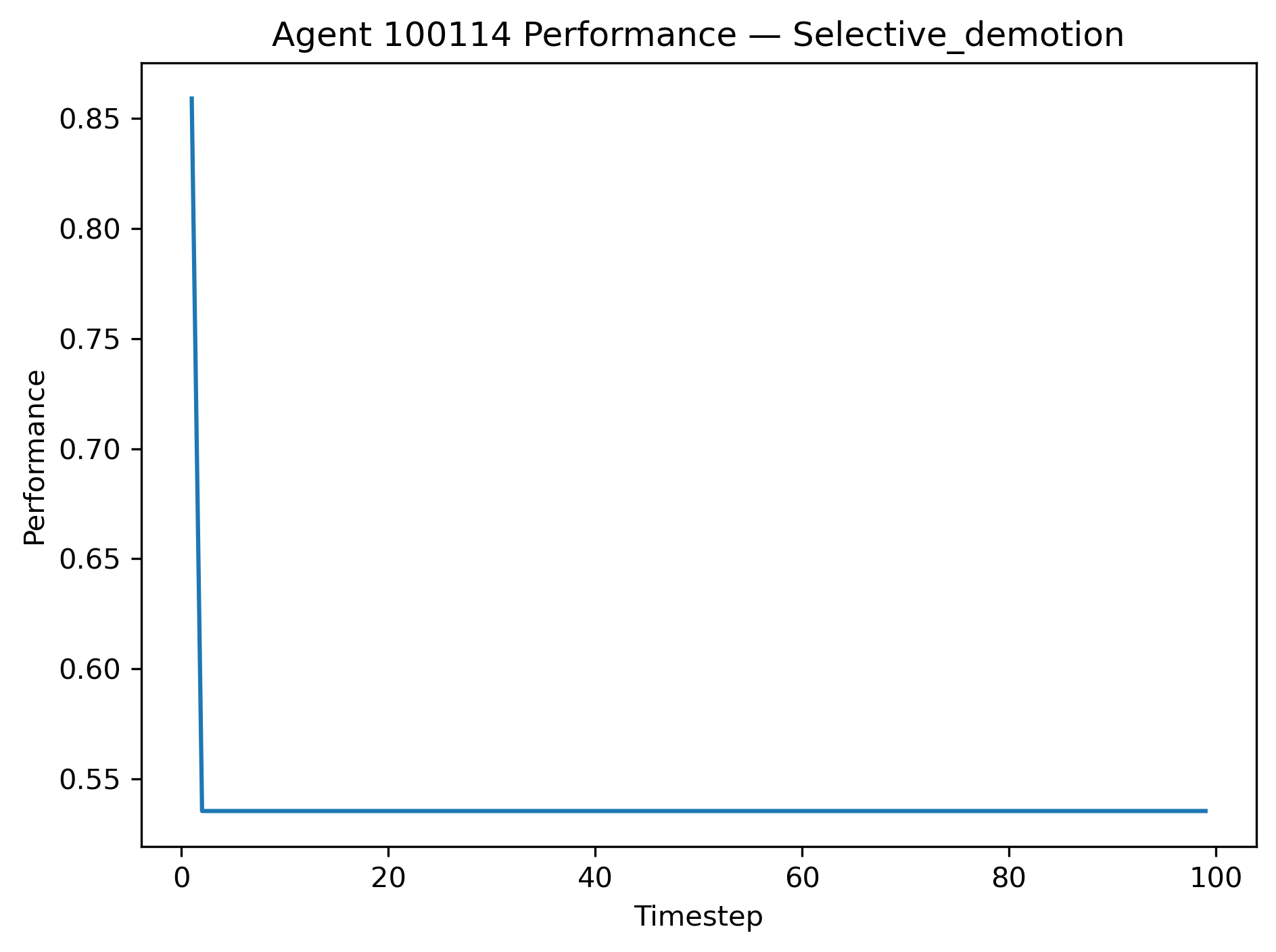}
    \caption{Agent 100114's performance over time for selective demotion strategy .}
    \label{fig:selective_demotion_agent_100114}
\end{figure}
\par\medskip

\noindent The agent isn't demoted back to L2 even with such a huge drop in performance because of our implementation in which demotion checks are executed once per timestep—prior to the backfill stage—so promotions arising during the backfill are not subjected to the same-tick demotion test. This design choice preserves simplicity and speed but allows a small number of large, skip-level losses to persist for one tick. But these cases are rare and have minimal impact on overall efficiency and efficiently explains the outcomes of this strategy. 
\par\medskip

\noindent \textbf{Takeaway:} This is the mitigation’s ``edge case'': a tech/\hspace{0pt} management strong agent with near-zero compliance is fast-tracked to \(L3\) by the cascade and then suffers a large negative shock (\(\Delta P \approx -0.324\)) because \(L3\) weights compliance heavily (30\%). Given our one-sweep implementation, the same-tick demotion test occurs before the refill, so this skip-level loss can persist for a single timestep; on the next step the agent simply remains at L3 due to low performance. Empirically, such cases are rare and account for part of the residual left tail and the \(\sim 500\)--\(600\) negatives/step floor.
\par\medskip

\paragraph{Interpretation.} Selective demotion \textbf{mitigates} the Peter-\hspace{0pt} Principle drag seen under pure Merit by identifying and rolling back large mismatches early. The threshold-and-blacklist mechanism curbs the accumulation of low-fit promotions that depress efficiency, allowing the organization to recover and surpass baseline rather than plateau near it. The diagnostics show that selective demotion reduces both the \emph{frequency} and \emph{magnitude} of negative $\Delta P$ events, but some mismatches persist at lower levels (due to fresh hires and the tolerance $\tau$), their impact is substantially softened. At higher levels, due to filtering and blacklisting, the candidate pool is enriched for management/compliance-heavy profiles which results in more positive shocks. The number of negative drops also reduces drastically over time as the demotion black lists agents who drop in performance and the candidate pool becomes more aligned with the destination roles. Overall, this mitigation yields a smoother efficiency trajectory with a higher terminal value. \emph{Compared with Merit—where early L1$\to$L2/L2$\to$L3 moves manufacture persistent misfits, Selective Demotion prunes large losses early and lets portable profiles advance, raising upper-path means and the terminal efficiency.}
\par\medskip


\vspace{\baselineskip}

\noindent \textbf{Merit-with-training.} (Mitigation; see Sec.~\ref{sec:mitigations}.) Promotions are by \emph{merit} (current-role performance), and each 
just-promoted agent receives a \emph{one-shot training update} on tech and management:
\[
C_{\text{new}} = \min\bigl(1,\; C_{\text{old}} + \text{learning\_rate}\bigr),
\quad 
\text{with} \quad \text{learning\_rate} = C \,(1-C).
\]
In the \emph{high-mismatch} ladder, this post-move training partially repairs misalignment created by reweighting tech$\downarrow$/management$\uparrow$ at L1$\to$L2 but it cannot fix the L2$\to$L3 jump where tech drops to zero and compliance (untrained) becomes salient. Accordingly, the efficiency path shows only a \emph{shallow} early dip from \(E_{0} \approx 0.548\) to a minimum \(E_{t=4} \approx 0.546\) (\(\sim 0.33\%\) down), then a \emph{steady rise} as trained L2 agents carry higher management into upper levels and management-heavy moves (L3$\to$L4, L4$\to$L5) become mildly positive with training, reaching \(E_{100} \approx 0.596\) (\(+8.7\%\) over baseline). Because training is one-shot and targets only tech/management, gains taper (diminishing \(C(1-C)\)) and the curve settles into a higher plateau 
rather than compounding indefinitely.

\noindent We observe 109{,}000 promotions over \(T=100\): 69{,}080 with \(\Delta P < 0\) (63.38\%) and 39{,}920 with \(\Delta P > 0\) (36.62\%). Compared to pure Merit in the same regime (88.1\% negatives), training cuts the negative share by \(\approx 24.7\) percentage points.

\begin{figure}[H]
    \centering
    \begin{subfigure}[t]{0.48\linewidth}
        \centering
        \includegraphics[width=\linewidth]{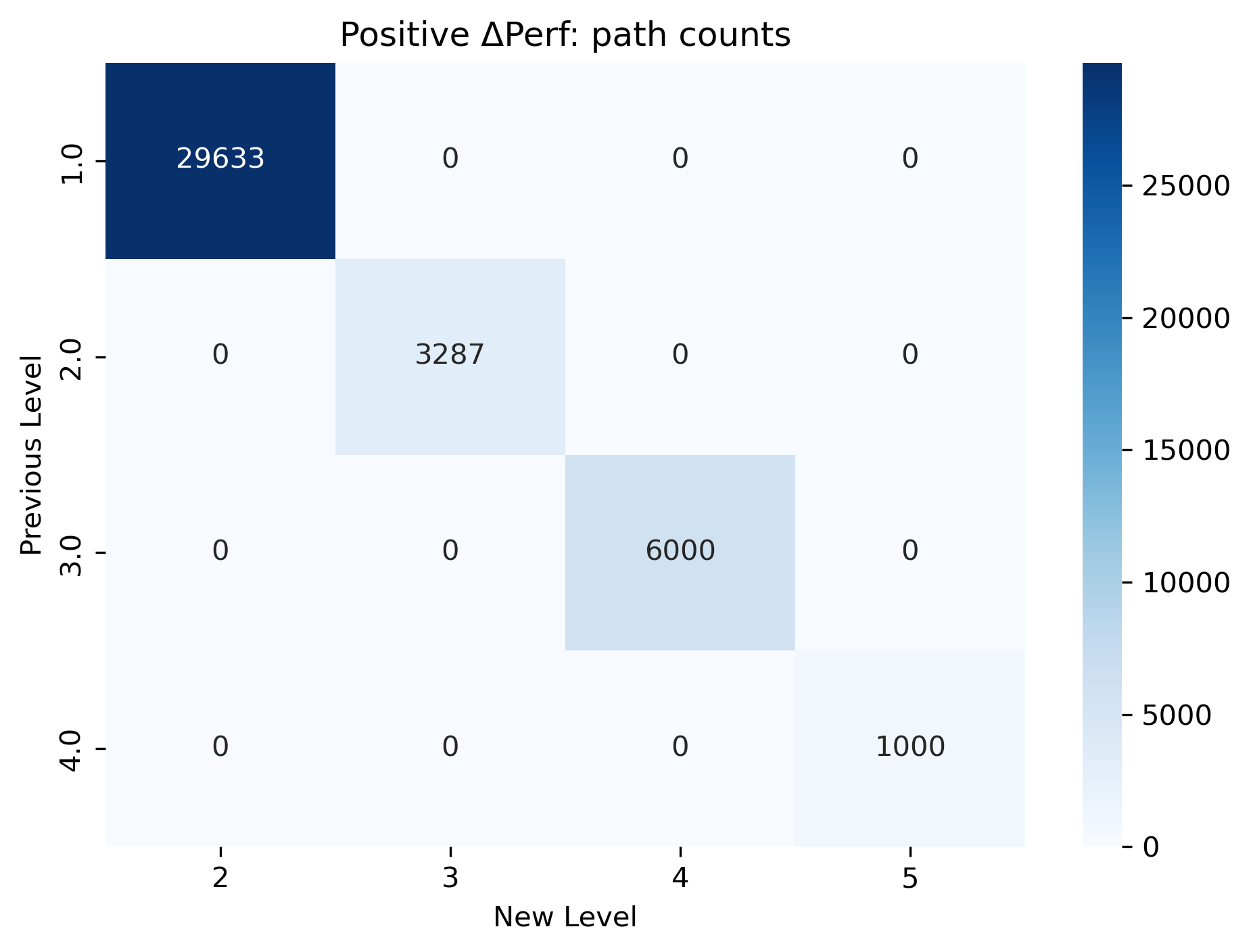}
        \caption{number of Promotions with $\Delta P > 0$ by level.}
        \label{fig:merit_training_pos}
    \end{subfigure}
    \hfill
    \begin{subfigure}[t]{0.48\linewidth}
        \centering
        \includegraphics[width=\linewidth]{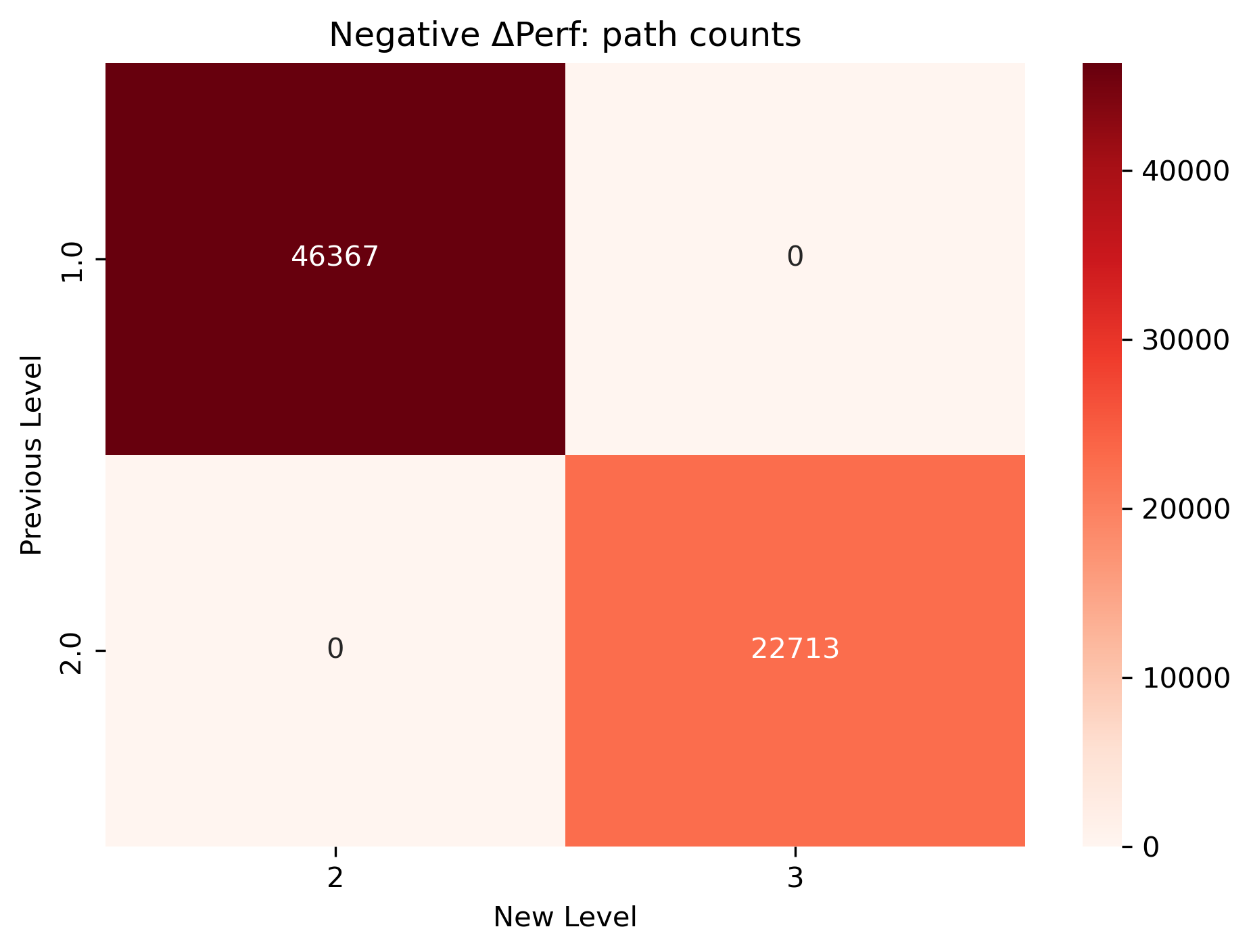}
        \caption{ number of Promotions with $\Delta P < 0$ by level.}
        \label{fig:merit_training_neg}
    \end{subfigure}
    \caption{Heatmap of promotion outcomes under merit-with-training in the high-mismatch regime.}
    \label{fig:merit_training_deltas}
\end{figure}

\noindent\emph{Heatmaps (Fig.~\ref{fig:merit_training_deltas}).} Under \textbf{merit-with-training} in the \emph{high-mismatch} regime. At lower levels, we have more promotions with $\Delta P < 0$ than $\Delta P > 0$. This is due to sharp reweighting away from tech at L1$\to$L2 (\(\Delta P>0=29{,}633\) vs.\ \(\Delta P<0=46{,}367\); 39.0\% positive) and our training only partially repairing the mismatch which trims losses relative to pure Merit, but the tech\(\downarrow\downarrow\)/management\(\uparrow\uparrow\) reweighting still leaves many small dips because we weigh in on \emph{tech} when we select agents for promotion. As we move higher, L2$\to$L3 we see the largest imbalance (\(\Delta P>0=3{,}287\) vs.\ \(\Delta P<0=22{,}713\); 12.6\% positive) because the \(L2 \rightarrow L3\) jump zeros out tech and \emph{introduces compliance} (0.3) that training does not affect; extra management from training only partially offsets this shift. By contrast, the upper levels are management-heavy and more similar in profile: moves into \textbf{L4} (6{,}000 vs.\ 0) and \textbf{L5} (1{,}000 vs.\ 0) are uniformly positive in this run. Two forces drive this: (i) training delivers a small, well-timed bump to management exactly when its weight increases (\(L3 \rightarrow L4\), \(L4 \rightarrow L5\)), and (ii) survivorship---agents reaching L3/L4 are already enriched for management/compliance, so their performance is more portable at the top.

\begin{figure}[H]
    \centering
    \includegraphics[width=\linewidth]{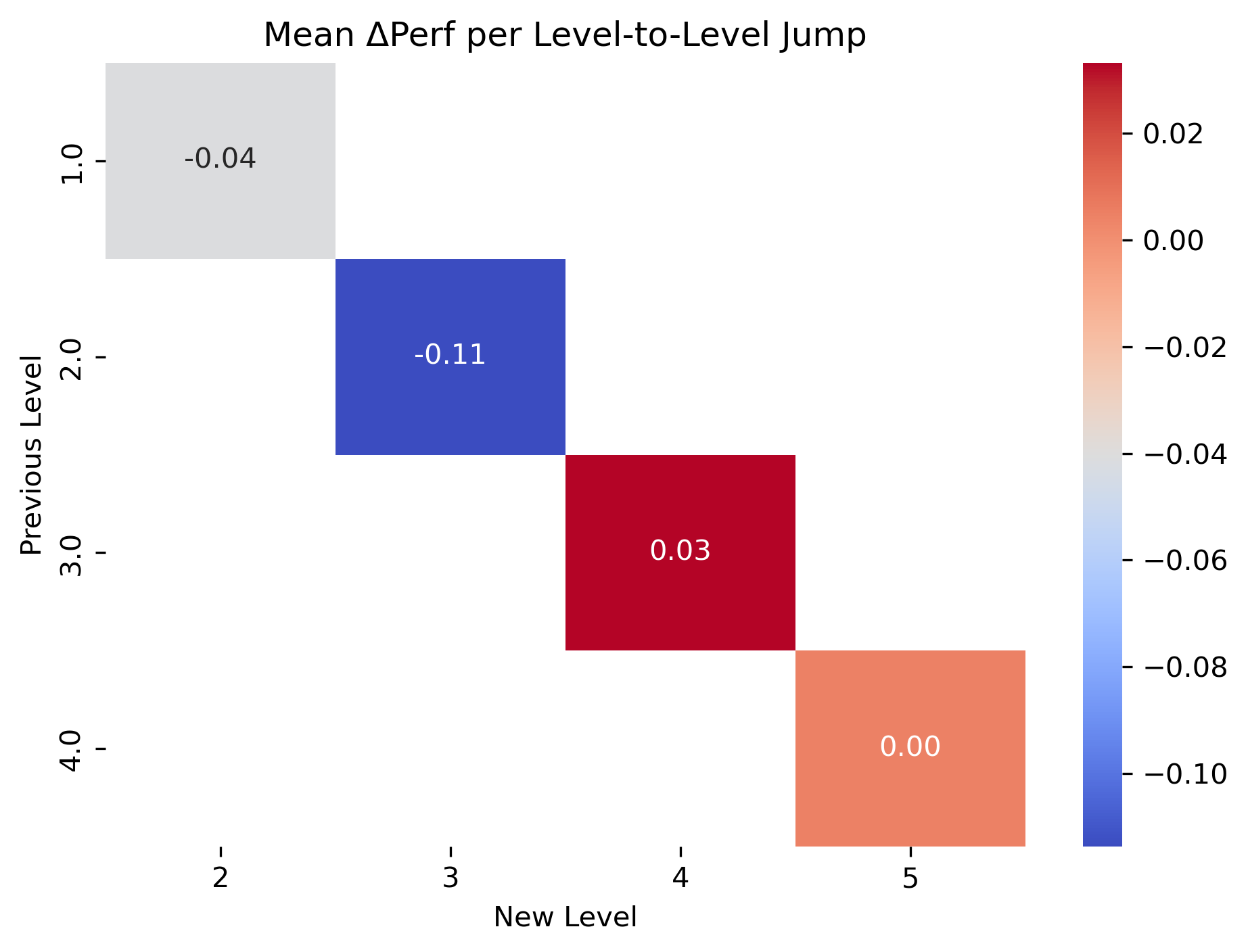}
    \caption{Mean $\Delta P$ per Level-to-Level jump for merit-with-training promotion strategy.}
    \label{fig:merit_training_heatmap}
\end{figure}

\noindent The \textbf{mean $\Delta P$ heatmap} (Fig.~\ref{fig:merit_training_heatmap}) shows a similar pattern that \emph{training softens but does not remove} early Peter shocks. We observe a modest loss in \(L1 \to L2\) jump (\(\bar{\Delta P} \approx -0.04\)), much smaller than under pure Merit (\(\sim -0.15\)), because the one-shot update boosts \emph{management} precisely when its weight rises. The main bottleneck remains \(L2 \to L3\) (\(\bar{\Delta P} \approx -0.11\)): tech drops to zero and \emph{compliance} (0.3) becomes salient, yet training does not touch compliance, so added management only partially offsets the shift. Upper-level moves turn mildly positive---\(L3 \to L4 \approx +0.03\) and \(L4 \to L5 \approx +0.005\)---as roles are management-heavy and the training bump aligns with their weights; with diminishing \(C(1-C)\) increments, these gains taper, yielding a higher but stable plateau.

\begin{figure}[H]
    \centering
    \includegraphics[width=\linewidth]{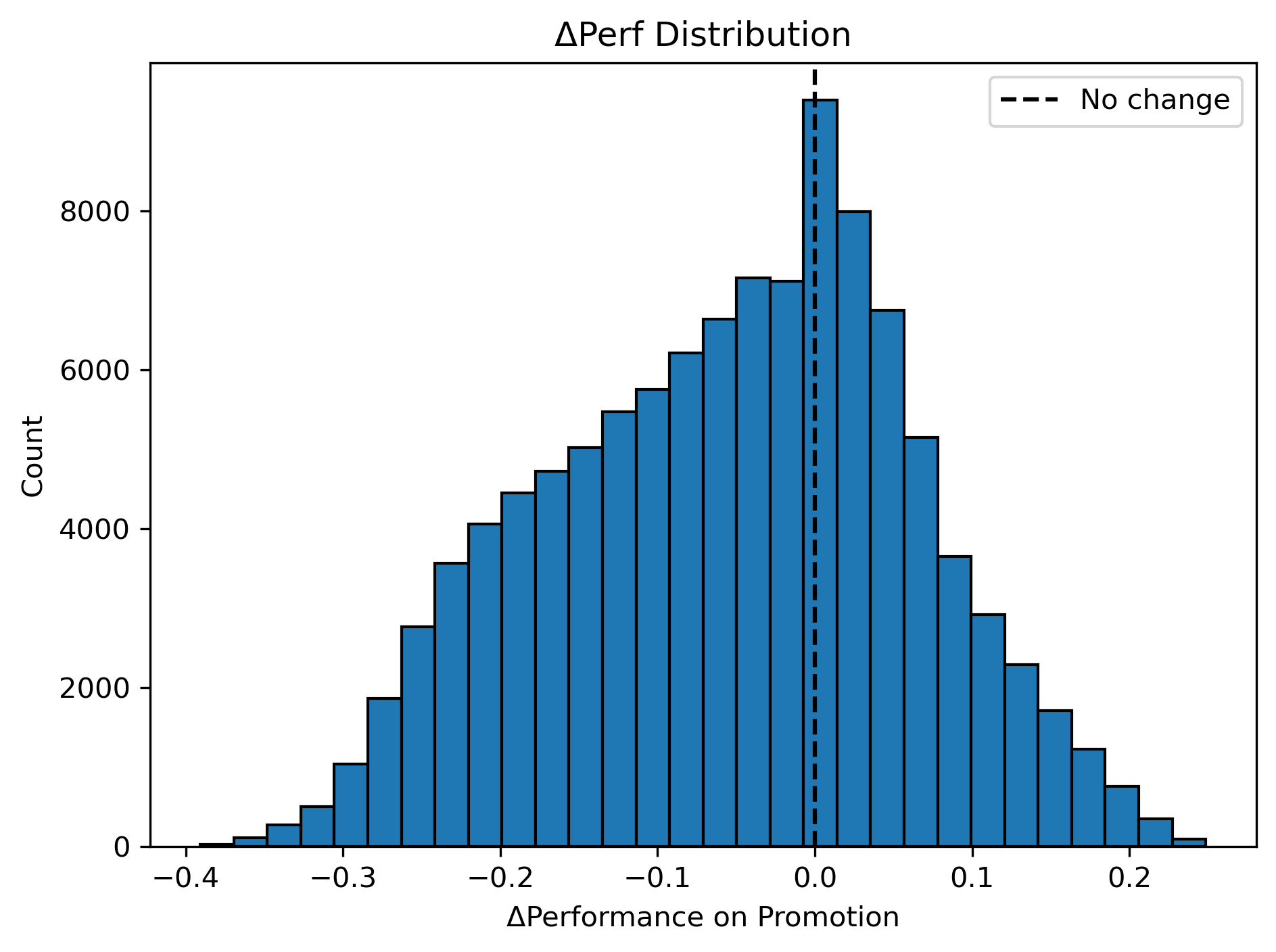}
    \caption{$\Delta P$ frequency distribution across all time-steps for merit-with-training strategy.}
    \label{fig:merit_training_promotion_delta}
\end{figure}

\noindent\textbf{$\Delta P$ histogram} under \emph{Merit-with-training} is \emph{left-skewed} with a modest negative center: mean \(-0.054\) vs. \(-0.136\) for merit, median \(-0.044\) vs. \(-0.138\) for merit; \(N=109{,}000\)). Losses remain more frequent than gains (63.38\% vs.\ 36.62\%), but the mass has shifted toward \emph{smaller} drops relative to pure Merit (mean \(-0.1365\); 88.13\% negatives). The \textbf{negative tail} now reaches \(\mathbf{\approx -0.391}\) (vs.\ \(-0.448\) under Merit), while the \textbf{positive tail} extends to \(\mathbf{\approx +0.249}\) (vs.\ \(+0.153\) under Merit); the 1st/99th percentiles are \(-0.300 / +0.187\). This pattern reflects the mechanism: the one-shot post-promotion update boosts management (and a bit of tech), trimming \(L1 \rightarrow L2\) losses and enhancing upper-level gains, but it does not touch \emph{compliance}, so the \(L2 \rightarrow L3\) shift continues to generate many negatives---keeping the overall histogram left-heavy even as its left tail becomes noticeably shallower. The positive tail thickens because training is applied \emph{at} promotion and targets exactly the skills that upper rungs reward. First, the one-shot bump to \emph{management} (and a smaller boost to \emph{tech}) makes \(L3 \rightarrow L4\) and \(L4 \rightarrow L5\) moves more positive---\(L4\) increases the weight on management and \emph{decreases} the weight on compliance, so trained, management-heavy profiles can register unusually large \(\Delta P > 0\). Second, \(C(1-C)\) is biggest around mid-skill values, so agents with moderate management receive relatively large training increments right before moving into management-heavy roles, pushing some outcomes far into the right tail. Third, survivorship concentrates candidates with adequate compliance and rising management at the upper levels; repeated promotions mean repeated one-shot updates, compounding portability and yielding rarer but larger positive shocks.

\begin{figure}[H]
    \centering
    \includegraphics[width=\linewidth]{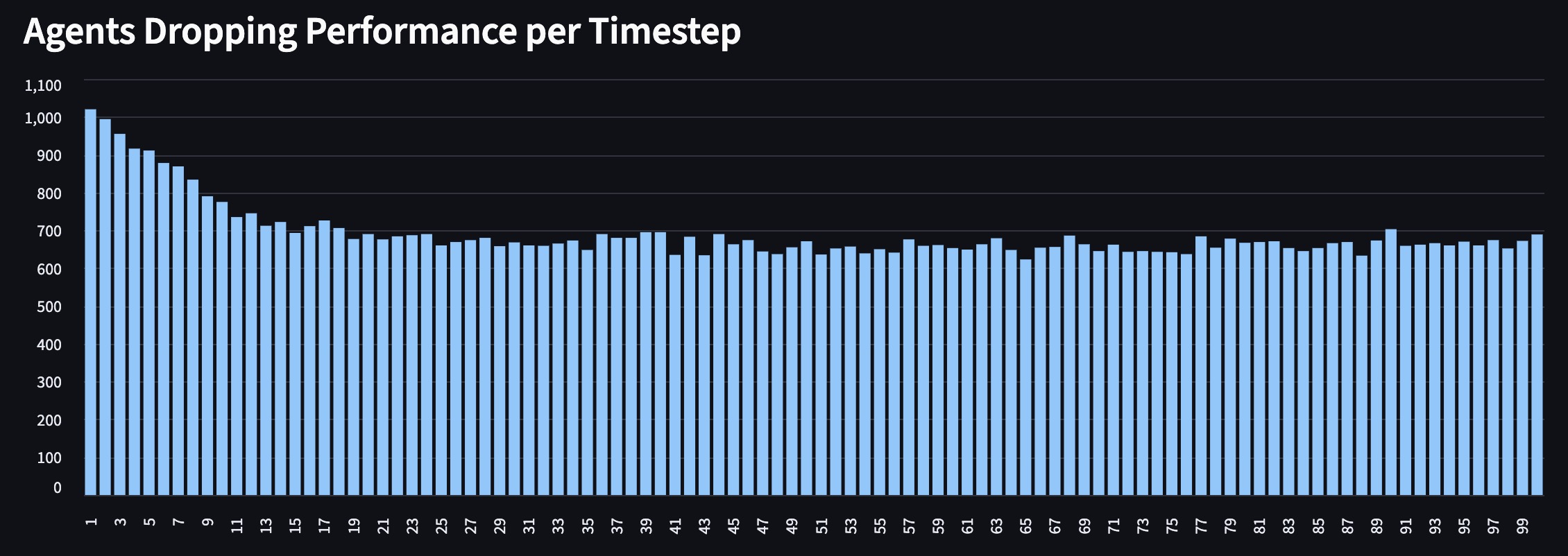}
    \caption{$\Delta P$ Number of promotions with $\Delta P<0$ at each timestep.}
    \label{fig:merit_training_neg_tseries}
\end{figure}

\noindent The \textbf{negative shocks over time histogram} shows the bigger picture. We see during earlier timsteps that roughly $\sim\!1000$ promotions per step result in drop in performance which reduces drastically every timestep and after about timestep 20 it stabilizes to $\sim\!600-700$ agents and remains constrant throughout. This reflects the immediate testing of many tech-heavy L1 performers against more managerial roles (the classic Peter shock) during early timesteps. During early timesteps all flows are active (per step: 760 \(L1 \rightarrow L2\), 260 \(L2 \rightarrow L3\), 60 \(L3 \rightarrow L4\), 10 \(L4 \rightarrow L5\)). The first two paths are strongly Peter-prone in the high-mismatch ladder (tech\(\downarrow\downarrow\) at \(L1 \rightarrow L2\); tech\(\rightarrow 0\) and compliance\(\uparrow\uparrow\) at \(L2 \rightarrow L3\)), so the initial waves are heavily negative. Because training is applied \emph{post-move} promotion and boosts only tech/management, it trims those losses but cannot fix the L2$\to$L3 compliance requirement, so the initial counts remain high ($\approx1000\to1100$). As the pipeline warms, the feeder pools change: the very tech-skewed L1 backlog is depleted; newly promoted L2 agents carry a management bump from training; and candidates without enough latent compliance fail to advance to L3. Consequently, the per-step negatives fall and settle by about \(t \approx 20\) into a stable band of \(\sim 600{-}700\) negatives per step.

\noindent\emph{Why it never goes to zero.} There is a structural floor: (a) continuous L1 hiring ensures some first-move mismatches every step; (b) training is \emph{one-shot} and targets only tech/management, so the \(L2 \rightarrow L3\) compliance bottleneck persists; and (c) the update \(C(1-C)\) has diminishing returns, so later cohorts receive smaller gains. Together these forces keep a stable tail of negatives even as the early spike subsides and the efficiency curve transitions from a shallow dip to a steady rise.

Let's take example of 2 agents to understand the impact of merit-with-training:

\paragraph{Case A — Residual Peter shock despite training (Agent\#~100114).} Agent\# 100114 has the following \emph{skills} at t = 0:

\begin{verbatim}
{
  "tech": 0.8808329324958385,
  "management": 0.7831324047407058,
  "compliance": 0.036530444057425115,
  "soft_skills": 0.6639328247375101
}
\end{verbatim}

\noindent The Agent begins in the Organization at time-step 0 at L1. The agent's performance is strong at L1 ($P\!=\!0.8591$). Due to this the agent is promoted to L2 at timestep 4. At L2, performance \emph{increases} further due to added training step ($0.9116$) his skills become:

\begin{verbatim}
{
  "tech": 0.9857992100224586,
  "management": 0.9529684461264509,
  "compliance": 0.036530444057425115,
  "soft_skills": 0.6639328247375101
}
\end{verbatim}

The agent is promoted to L3 immidiately due to his excellent performance at L2, but at L3, performance \emph{decreases drastically} ($0.6437$): due to his very low compliance(as this level demands 30\% more compliance). Due to this big drop the agent is not promoted to L4 and remains at L3 for rest of the simulation till the rest of the simulation.

His skills at L3 become:

\begin{verbatim}
{
  "tech": 1,
  "management": 1,
  "compliance": 0.036530444057425115,
  "soft_skills": 0.6639328247375101
}
\end{verbatim}

\begin{figure}[H]
    \centering
    \includegraphics[width=\linewidth]{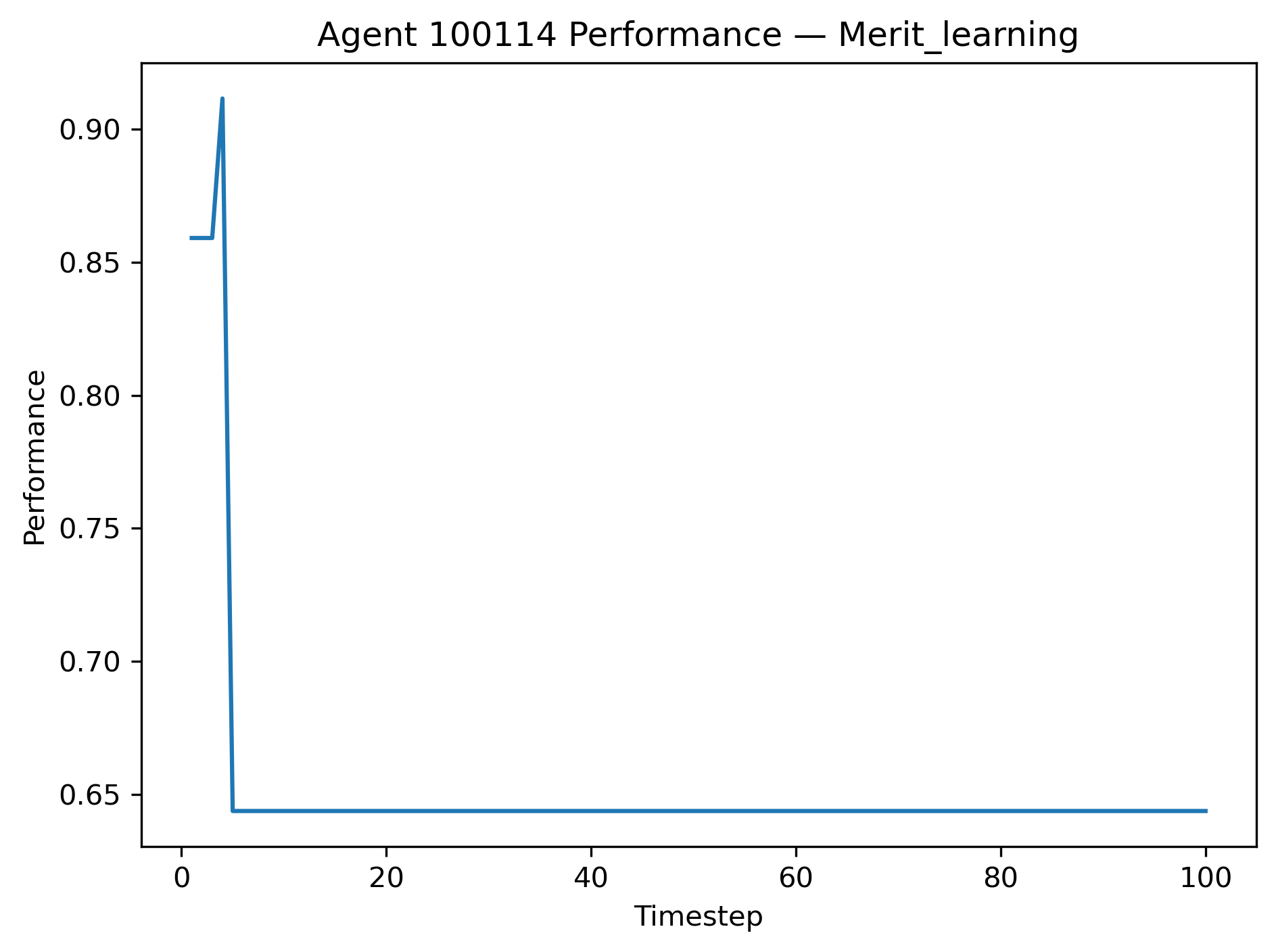}
    \caption{Agent 100114's performance over time for merit-with-training strategy .}
    \label{fig:merit_training_agent_100114}
\end{figure}

\noindent \textbf{Takeaway:} This case illustrates the mitigation’s limitation: the one-shot training update boosts \emph{tech} and \emph{management}, but does not affect \emph{compliance}.\footnote{We model training as a short, on-promotion intervention aimed at skills that are plausibly upskilled via workshops/bootcamps (tech, management). Compliance and soft skills, while trainable, typically accrue over longer horizons (standards mastery, audits, documentation culture), which our one-step update does not simulate. Extending the model to multi-period training for compliance/soft skills is straightforward and left for future work.} As a result, the agent still suffers a large negative shock (\(\Delta P \approx -0.2679\)) at the \(L2 \rightarrow L3\) jump because \(L3\) demands significant compliance (0.3) that the agent lacks. The training helps somewhat (the drop is smaller than under pure Merit (Performance at L3 0.5353)), but it cannot fully offset the mismatch created by the sharp reweighting and new compliance requirement. This example highlights that while training can mitigate some Peter shocks, it cannot eliminate all mismatches, especially when unaddressed skills become critical at higher levels. his mirrors real organizations: e.g., a high-performing \emph{engineering manager} at a \emph{medical-device} firm (excellent architecture and people leadership) promoted into a \emph{Regulatory/Quality lead} role may struggle without deep familiarity with FDA 21 CFR Part 820, ISO 13485, design-control documentation, and audit cadence; short onboarding improves managerial portability but cannot substitute for accumulated regulatory practice. Hence, even strong technical/managing profiles can plateau when a compliance bottleneck becomes binding.

\paragraph{Case B — Portable excellence with training (Agent\#~101433).} Agent\# 101433 has the following \emph{skills} at t = 0:

\begin{verbatim}
{
  "tech": 0.6931240193169654,
  "management": 0.6155462990198542,
  "compliance": 0.8866214862660718,
  "soft_skills": 0.9326679807251644
}
\end{verbatim}

\noindent The Agent begins in the Organization at time-step 0 at L1. The agent's performance is above average at L1 ($P\!=\!0.7171$). This is due to about average technical skills. Due to this the agent remains in L1 for a long time. At timestep 29 he is promoted to L2 and recieves his first training. At L2, performance \emph{increases} further due to added training step ($0.8951$) his skills become:

\begin{verbatim}
{
  "tech": 0.9058271324798258,
  "management": 0.8521953518026687,
  "compliance": 0.8866214862660718,
  "soft_skills": 0.9326679807251644
}
\end{verbatim}

This massive increase is due to our training function which gives higher boost to mid-level skills. The agent is promoted to L3 immidiately due to his excellent performance at L2, and at L3, performance \emph{increases} further ($0.9525$): due to his high compliance(as this level demands 30\% more compliance). And his skills at L3 become:

\begin{verbatim}
{
  "tech": 1,
  "management": 1,
  "compliance": 0.8866214862660718,
  "soft_skills": 0.9326679807251644
}
\end{verbatim}

The agent is promoted to L4 immidiately due to his excellent performance at L3, and at L4, performance \emph{increases} further ($0.9752$): due to his high compliance(as this level demands 20\% compliance). And his skills at L4 remain the same as the limit for tech and management hits. The agent is promoted to L5 immidiately due to his excellent performance at L4, and at L5, performance again \emph{increases} slightly ($0.9819$). The agent remains at L5 for rest of the simulation till timestep 85 when he is attrited.

\begin{figure}[H]
    \centering
    \includegraphics[width=\linewidth]{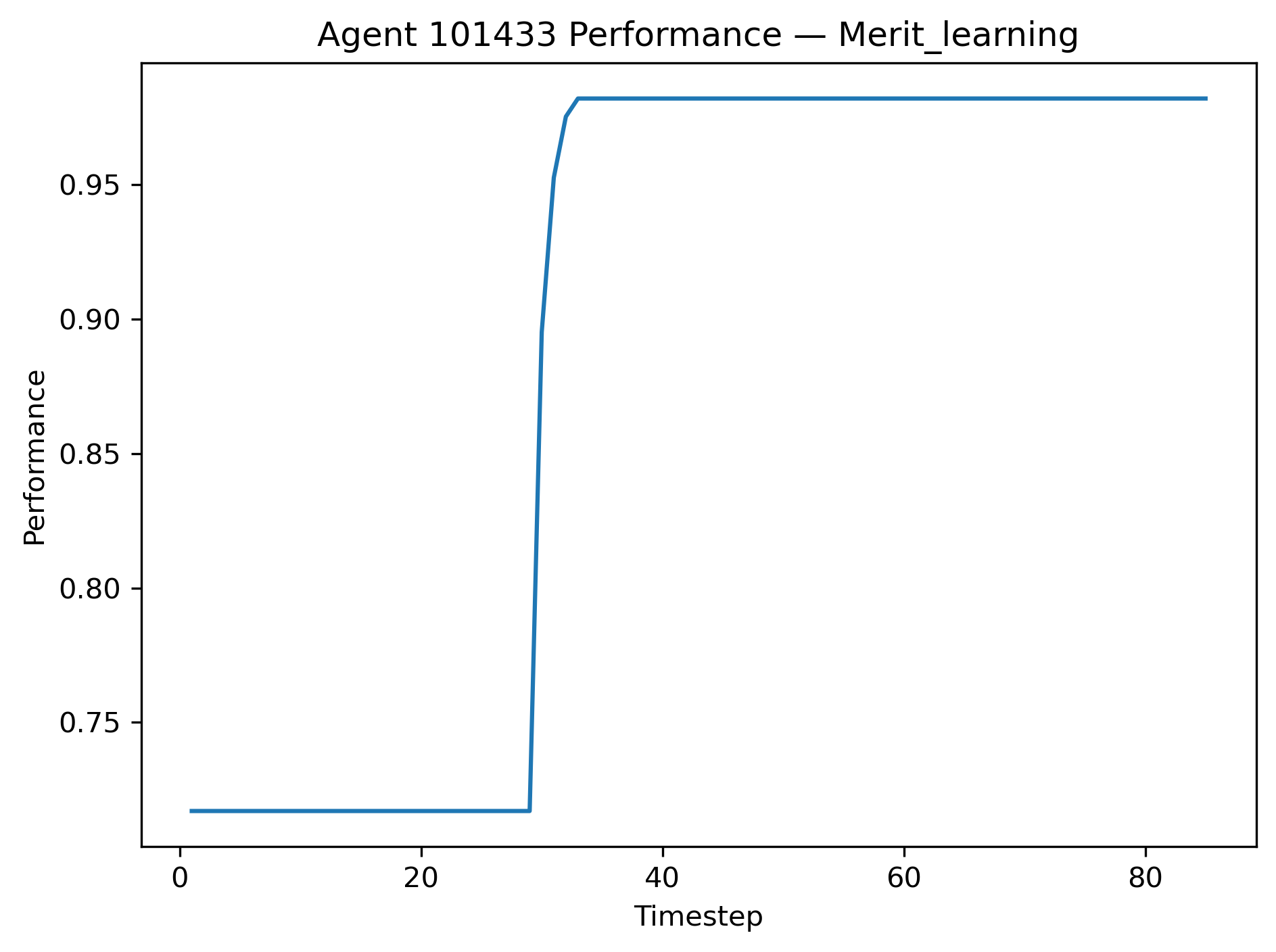}
    \caption{Agent 101433's performance over time for merit-with-training strategy .}
    \label{fig:merit_training_agent_101433}
\end{figure}

\noindent\textbf{Takeaway.} This is the upside of \emph{Merit-with-training}: a mid-skill profile can be transformed into ``portable excellence.'' Starting with only moderate \emph{tech} (0.69) and \emph{management} (0.62), the agent receives a large one-shot boost at L2 because the update \(C(1-C)\) is largest for mid-range \(C\). That jump (0.7171\(\rightarrow\)0.8951) immediately aligns the agent with L2’s management-heavier weights, and---crucially---his already-strong \emph{compliance} (0.89) makes the \(L2 \rightarrow L3\) shift \emph{positive} rather than punitive. From there, promotions stack small, well-timed gains in management into increasingly management-heavy roles (\(L3 \rightarrow L4 \rightarrow L5\)), producing a clean ascent with \(\Delta P > 0\) at every step (0.8951\(\rightarrow\)0.9525\(\rightarrow\)0.9752\(\rightarrow\)0.9819). In organizational terms, targeted post-promotion training can unlock latent capacity in ``solid, not standout'' performers, converting them into high performers once the role mix pivots toward management \footnote{or any other trainable skill}---provided the hard-to-train bottlenecks (here, compliance) are already in place. This mirrors real organizations: e.g., a competent \emph{software engineer} with solid but not exceptional coding and management skills, when promoted into a \emph{team lead} role and given targeted management training, can leverage their growing leadership abilities and existing compliance/soft skills to excel in increasingly managerial roles, ultimately reaching senior leadership positions.

\subsection{Transferable-skills Regime}

\begin{figure}[H]
    \centering
    \includegraphics[width=\linewidth]{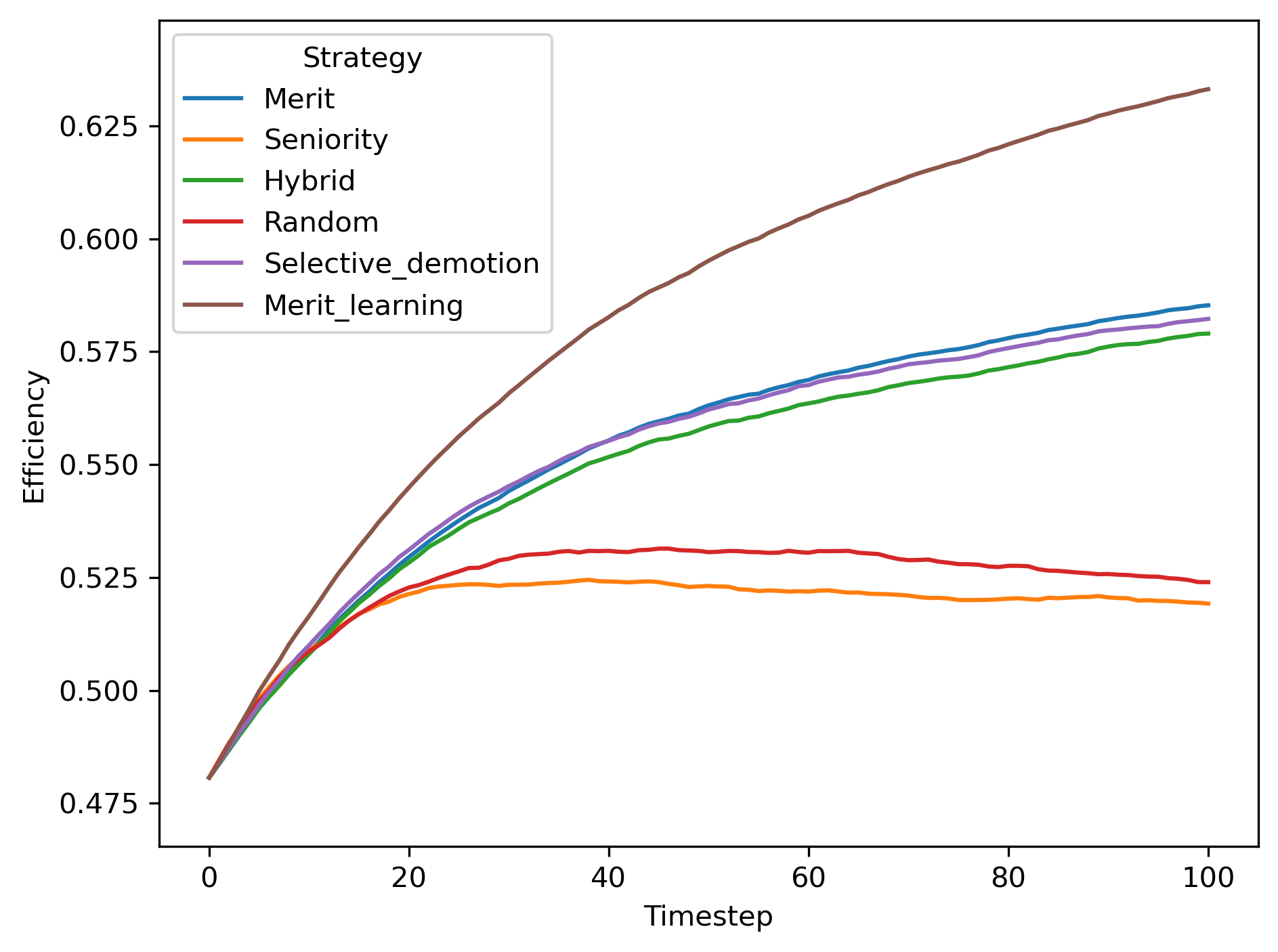}
    \caption{Organizational efficiency trajectories for transferable-skills regime under six promotion rules.}
    \label{fig:efficiency_skills-transfer}
\end{figure}

All strategies begin at $E_0 \approx 0.48069$. 

At $T=100$:
\begin{itemize}
  \item Merit: 0.58525 (+21.75\%),
  \item Seniority: 0.51923 (+8.02\%),
  \item Hybrid: 0.57900 (+20.45\%),
  \item Random: 0.52397 (+9.00\%),
  \item Selective Demotion: 0.58225 (+21.13\%),
  \item Merit+training: 0.63310 (+31.71\%),
\end{itemize}

\noindent\textbf{Merit.} As before, the rule promotes the best \emph{in the current job}. The difference here is the job ladder itself: L1 and L2 remain tech-dominant (0.9 and 0.8 on \textsf{tech}, with only a small bump in \textsf{management} at L2); L3 introduces modest \textsf{management} and \textsf{compliance} (0.65/0.15/0.10/0.10); and only at L4--L5 do weights pivot meaningfully toward \textsf{management}/\textsf{compliance} (L4: 0.4/0.2/0.2/0.2; L5: 0.2/0.4/0.3/0.1). Because adjacent levels ask for \emph{nearby} skills, the top L1 performers that merit elevates are not flung into a qualitatively different role. The immediate mismatch that drove negative $\Delta P$ shocks in the high-mismatch regime largely disappears; promotions tend to preserve fit, so early efficiency losses are rare.

\noindent Empirically (this run), efficiency rises \emph{monotonically} from $E_0 \approx 0.4807$ with no initial dip, reaching $E_{100}=0.5853$ ($+21.8\%$ vs.\ $E_0$). Gains are front-loaded and then taper: the series crosses $+5\%$ by $t\approx 9$, $+10\%$ by $t\approx 20$, $+15\%$ by $t\approx 38$, and $+20\%$ by $t\approx 78$.

\noindent\emph{Why the curve is concave (and plateaus).} We have wo bottlenecks:
\begin{enumerate}
    \item \textbf{Compliance} is negligible at L1--L2 but matters at L4--L5. Since merit at lower levels mostly selects on \textsf{tech} (the dominant weight there), the pipeline under promotes strong \textsf{compliance} profiles. By the time upper roles demand it, many promoted agents are relatively weaker on \textsf{compliance}, so marginal gains shrink and the series bends over (gradual plateau) rather than continuing linear growth.

    \item With the level mix fixed at $(0.40, 0.25, 0.20, 0.10, 0.05)$, even saturating L2--L5 with highly capable individuals leaves $40\%$ of the workforce at L1. L1 has no promotion-gated skill filtering at hire, so its competence distribution is effectively uniform. Under the L1 profile (0.9 \textsf{tech} $+$ 0.1 \textsf{soft}), the expected L1 performance hovers near $\sim 0.5$ and is constantly refreshed by higher attrition and backfills (exit $=5\%$ at L1 vs.\ $0.2\%\text{--}2\%$ above). As L2--L5 approach their stationary, high-fit composition, the organization’s efficiency becomes a weighted average dominated by this persistent L1 ``floor.'' Intuitively, even if upper tiers stabilize near strong performance, the $0.40\times$ L1 block fed by uniform entrants and higher turnover prevents the aggregate efficiency from climbing much beyond the observed $\sim 0.58\text{--}0.59$ plateau.
\end{enumerate}

\noindent\emph{Net effect.} Relative to the high-mismatch ladder, merit improves because promotions are \emph{adjacent} (fewer $\Delta P<0$ shocks), yielding a smooth rise. But two ceilings kick in: (i) late-stage \textsf{compliance} requirements the pipeline did not prioritize earlier, and (ii) the structural L1 share with uniform inflow and the highest exit rate. Together they explain the concave trajectory and the steady, modest plateau without any explicit training.

\noindent We observed 96,807 promotions across 100 time-steps resulting in $\Delta P < 0$ (which is 88.8\% of all promotions) compared to only 12,193 promotions with $\Delta P > 0$.

\begin{figure}[H]
    \centering
    \begin{subfigure}[t]{0.48\linewidth}
        \centering
        \includegraphics[width=\linewidth]{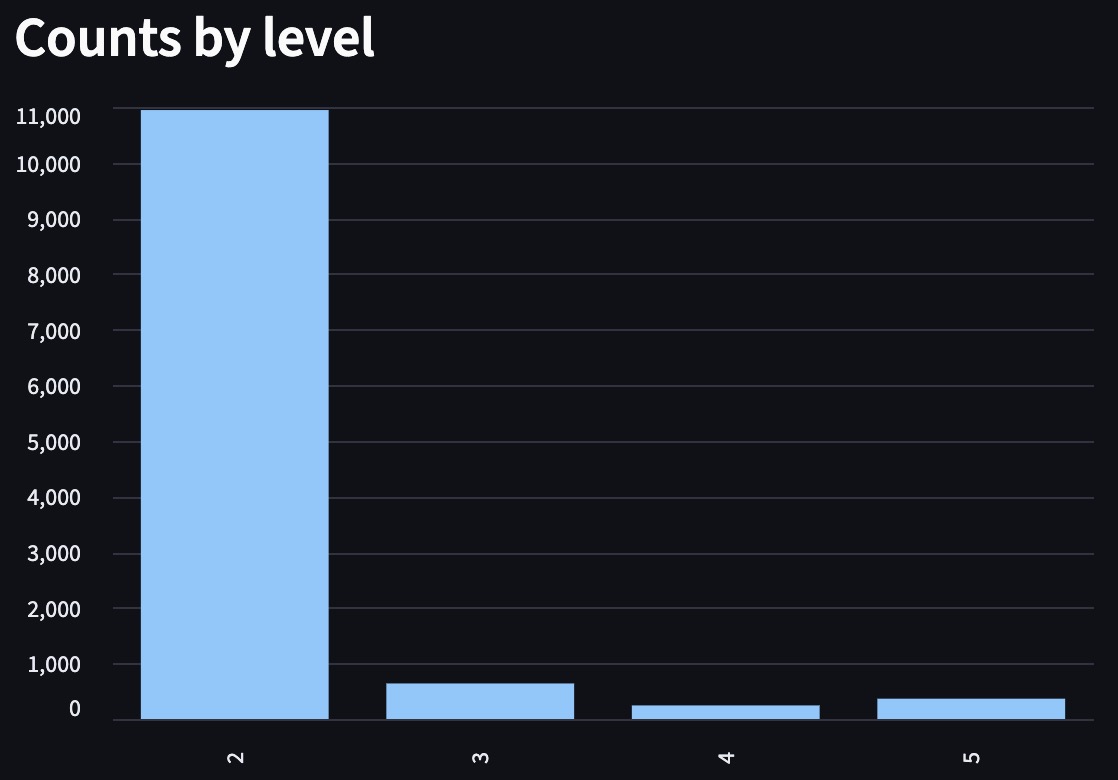}
        \caption{number of Promotions with $\Delta P > 0$ by level.}
        \label{fig:merit_pos_transfer}
    \end{subfigure}
    \hfill
    \begin{subfigure}[t]{0.48\linewidth}
        \centering
        \includegraphics[width=\linewidth]{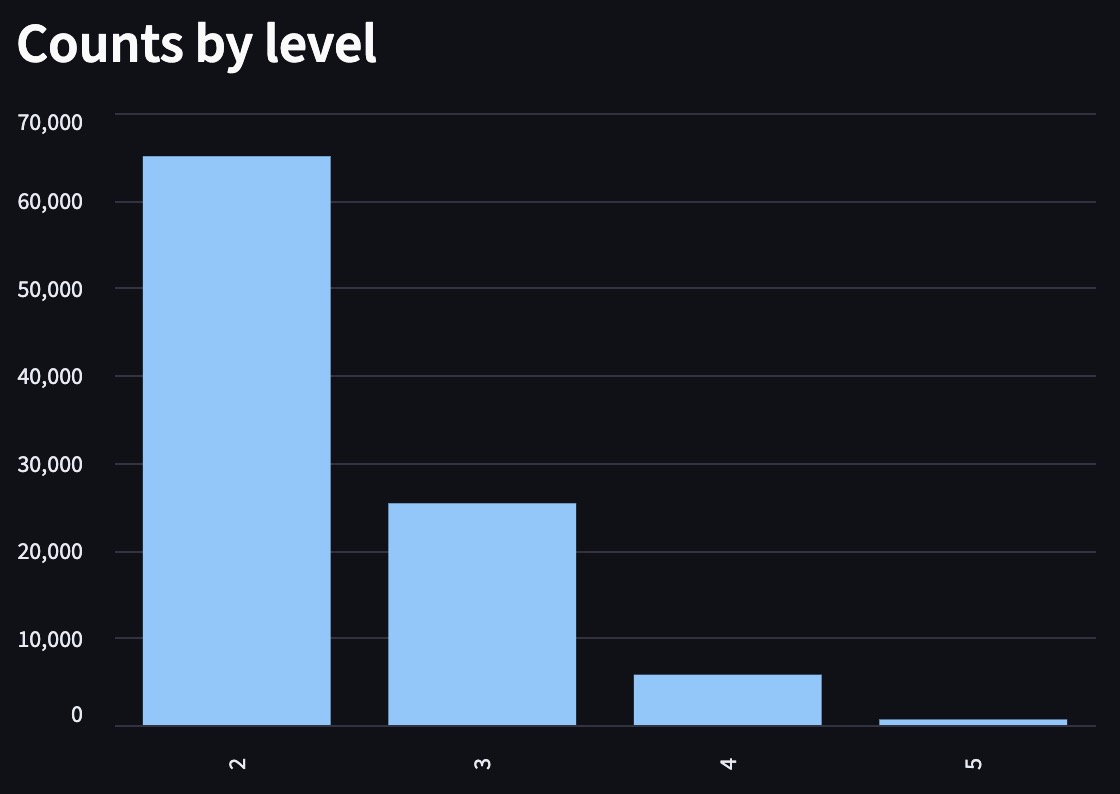}
        \caption{ number of Promotions with $\Delta P < 0$ by level.}
        \label{fig:merit_neg_transfer}
    \end{subfigure}
    \caption{Bar chart of promotion outcomes under merit-based promotion in the transferable-skills regime. 
The counts of agents with positive ($\Delta P>0$) and negative ($\Delta P<0$) performance changes are shown by level.}
    \label{fig:merit_deltas_transfer}
\end{figure}

\noindent When broken down by levels (Fig.~\ref{fig:merit_deltas_transfer}) these are the observed numbers for agents whose performance improves and for agents whose performance plummets:
\[
\begin{array}{lrr}
\toprule
\text{Level} & \Delta P > 0 & \Delta P < 0 \\
\midrule
\text{L2} & 10{,}945 & 65{,}055 \\
\text{L3} & 639 & 25{,}361 \\
\text{L4} & 244 & 5{,}756 \\
\text{L5} &   365   &   635 \\
\bottomrule
\end{array}
\]

\noindent\emph{Explanation:} Our first guess was that the \emph{average gain} per positive promotion would be larger than the \emph{average loss} per negative promotion, explaining the net efficiency rise despite many more losses. However, the data show the opposite: The \emph{average gain} per positive promotion is $+0.0099$, while the \emph{average loss} per negative promotion is $-0.0498$. Thus a typical negative event is about $5\times$ larger in magnitude than a typical positive one.\footnote{Overall counts: $109{,}000$ promotions, with $12{,}193$ ($11.2\%$) $\Delta P>0$ and $96{,}807$ ($88.8\%$) $\Delta P<0$. Overall mean $\Delta P=-0.043$; median $-0.043$;}

\begin{table}[h]
\centering
\caption{Promotion outcomes by transition (transferable-skills regime). ``Pos.\ Share'' is the fraction with $\Delta P>0$. The fourth column is the mean gain conditional on $\Delta P>0$, the fifth is the mean absolute loss conditional on $\Delta P<0$, and the last column is their ratio.}
\label{tab:transitions-adjacent}
\begin{threeparttable}
\begin{tabular}{lrrrrr}
\toprule
\textbf{Transition} & \textbf{Count} & \textbf{Pos.\ Share} & $\mathbf{\Delta P\mid(\Delta P>0)}$ & $\mathbf{\lvert\Delta P\rvert\mid(\Delta P<0)}$ & \textbf{Ratio} \\
\midrule
L1$\to$L2 & 76{,}000 & 0.144 & +0.00982 & 0.04374 & 0.225 \\
L2$\to$L3 & 26{,}000 & 0.025 & +0.00676 & 0.06366 & 0.106 \\
L3$\to$L4 & 6{,}000  & 0.041 & +0.01080 & 0.05911 & 0.183 \\
L4$\to$L5 & 1{,}000  & 0.365 & +0.01768 & 0.03061 & 0.578 \\
\midrule
\textbf{Overall} & 109{,}000 & 0.112 & +0.00992 & 0.04979 & 0.199 \\
\bottomrule
\end{tabular}
\end{threeparttable}
\end{table}

\noindent Therefore the hypothesis “per–positive promotion gain $>$ per–negative promotion loss” is \textbf{not supported}. Across all transitions—and especially at the high-volume steps (L1$\to$L2, L2$\to$L3)—negative events are significantly larger in magnitude. The only relatively “closer” case is L4$\to$L5 (ratio $=0.58$), but promotions at that step are \emph{few} (1{,}000 total), so they contribute little to the aggregate.

\medskip
\noindent\textbf{Why efficiency still rises more than in the high-mismatch ladder.} The key difference is not the \emph{share} of negatives (similar across the two ladders), but the \emph{size and placement} of shocks:

\medskip
\noindent\textbf{Shocks are shallower on the busy steps.} 
The lower transitions dominate volume (L1$\to$L2 $= 760$ and L2$\to$L3 $= 260$ per step; $\sim\!94\%$ of all promotions). 
On these steps the transferable-skills ladder has much \emph{smaller} average losses than high-mismatch:
L1$\to$L2: $-0.036$ vs.\ $-0.146$; 
L2$\to$L3: $-0.062$ vs.\ $-0.148$. 
This “softening” where promotions are most frequent is the main reason the transferable-skills trajectory climbs higher.

\medskip
\noindent\textbf{Top-end moves are mild and low-volume.} 
At the top the per-move effects are small and the counts are tiny (L3$\to$L4 $= 60$; L4$\to$L5 $= 10$ per step), so they contribute little to the aggregate. 
In fact, the signs differ across ladders:
L3$\to$L4 mean: $-0.056$ (transferable-skills) vs.\ $+0.015$ (high-mismatch);
L4$\to$L5 mean: $-0.013$ (transferable-skills) vs.\ $+0.007$ (high-mismatch).
Even though positives at L4$\to$L5 are larger when they occur in transferable-skills ($+0.0177$ for $\Delta P>0$) than in high-mismatch ($+0.0100$), they are \emph{less frequent} there (pos.\ share $36.5\%$ vs.\ $78.4\%$), yielding a tiny negative mean overall. Either way, because these steps are few, they do not drive the outcome.

\medskip
\noindent Therefore after weighting each transition’s mean by its per-step volume, the\textbf{net effect per period is far less harmful.} That results in a steady rise in efficiency, with no initial dip and a higher plateau than in the high-mismatch regime.

\begin{figure}[H]
    \centering
    \includegraphics[width=\linewidth]{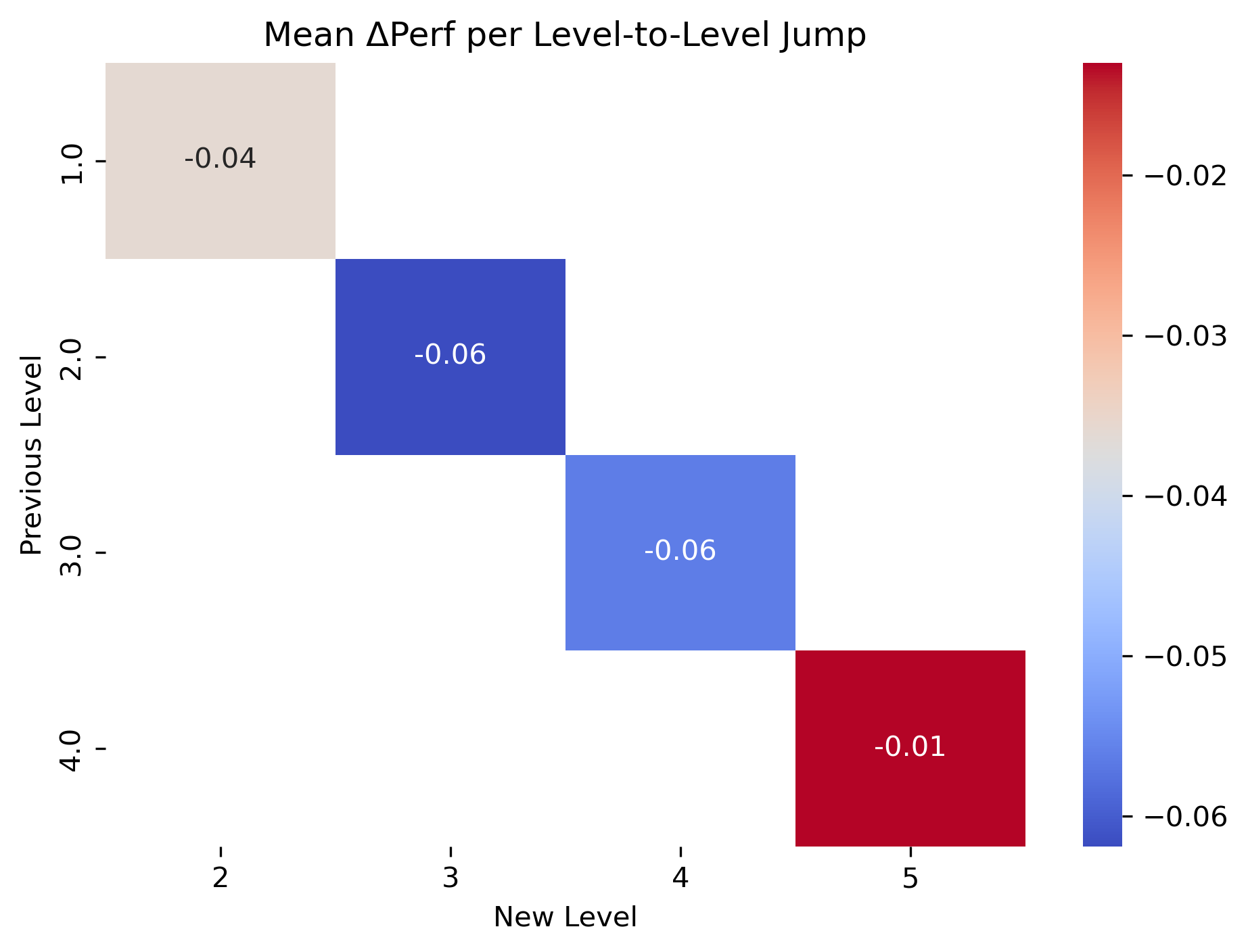}
    \caption{Mean $\Delta P$ per Level-to-Level jump for merit based promotion strategy.}
    \label{fig:merit_heatmap_transfer}
\end{figure}

\noindent The heatmap (Fig.~\ref{fig:merit_heatmap_transfer}) above also shows a similar trend. Specifically, the mean $\Delta P$ from L1$\to$L2: $-0.04$, L2$\to$L3: $-0.06$, L3$\to$L4: $-0.06$ and L4$\to$L5: $-0.01$. The drop is highest in the L2$\to$L3 transition which is expected as the role requirements change the most here (tech drops by 0.15 and compliance is introduced). Similarly, the drop from L3$\to$L4 can be explained. The drop from L1$\to$L2 is also expected as the role requirements change slightly (tech drops by 0.1 and management is introduced). And the drop is least in the L4$\to$L5 transition as the role requirements change the least (no new skills are introduced and there is modest increase in other skills).

\vspace{\baselineskip}
\noindent The patterns above provide \textbf{strong evidence} of the \textbf{Peter Principle} under \textbf{Merit}-based promotions: most promotions reduce the promoted agent’s performance because the destination role reweights skills relative to the source (even if by a little bit).

\begin{figure}[H]
    \centering
    \includegraphics[width=\linewidth]{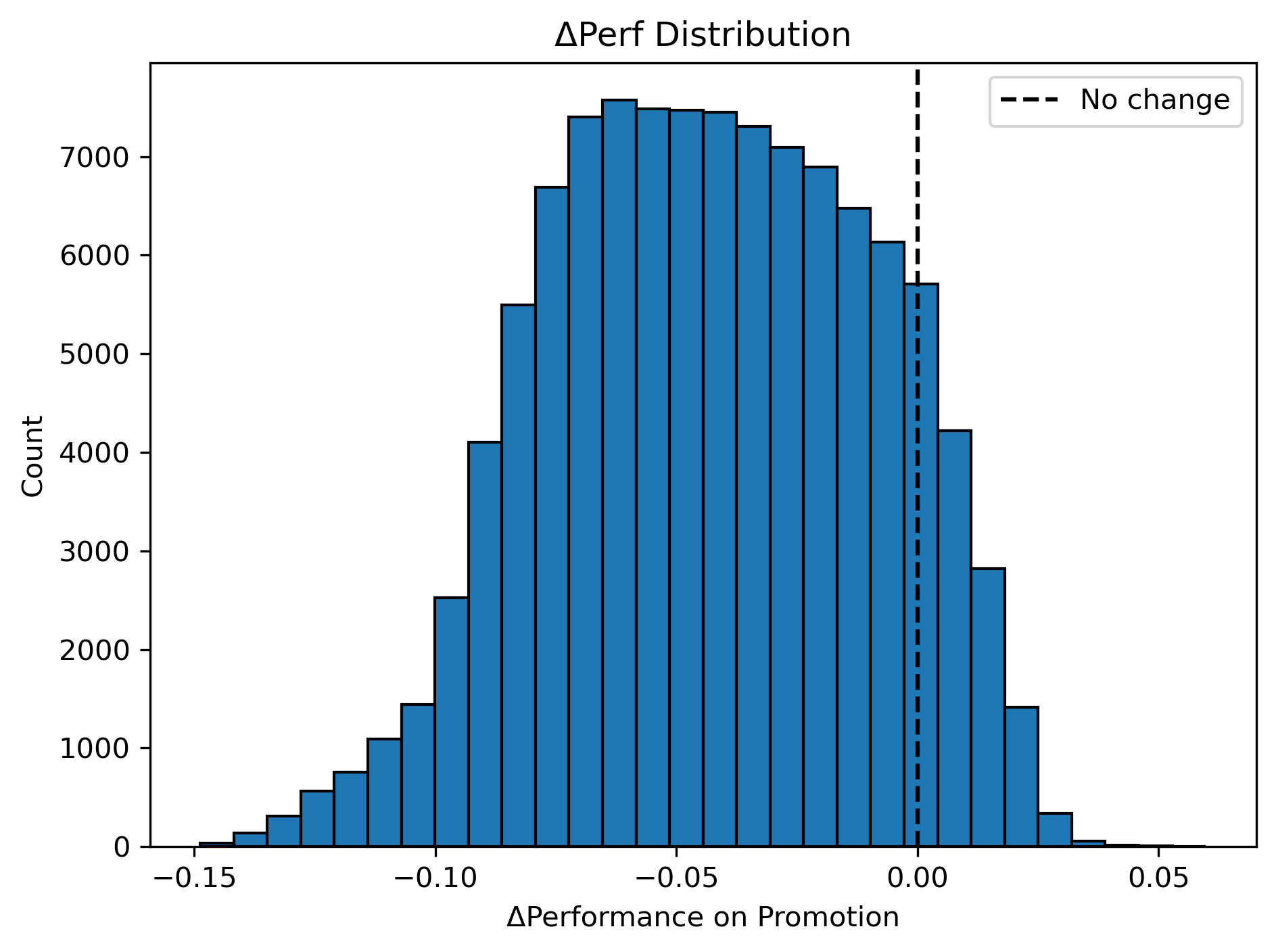}
    \caption{$\Delta P$ frequency distribution across all time-steps for merit based promotion.}
    \label{fig:merit_promotion_delta_transfer}
\end{figure}

\noindent The \emph{frequency distribution} of $\Delta P$ (Fig.~\ref{fig:merit_promotion_delta_transfer}) is \textbf{left-skewed} around $0$, again indicating \textbf{more promotions with $\Delta P<0$ than $\Delta P>0$}. 
The overall average shock is $-0.043$ and median is $-0.043$ (compared to $-0.136$ and $-0.138$ for high-mismatch regime), with a \textbf{shallow negative tail} extending to $-0.149$ (about $-0.4$ for high-mismatch) and a \textbf{short positive tail} extending to $+0.06$ ($+0.153$ for high-mismatch) (1st/99th percentiles \(-0.120\) / \(+0.021\)). 
Compared to the high-mismatch ladder (mean $\Delta P \approx -0.136$), the losses here are \emph{much smaller} in magnitude. That is due to the \emph{adjacent role weights}\footnote{By “adjacent role weights,” we mean that the change in required skill weights between consecutive levels is small} of the transferable-skills regime, which softens the shocks even though the Peter-Principle mechanism remains active.

\begin{figure}[H]
  \centering
  \includegraphics[width=\linewidth]{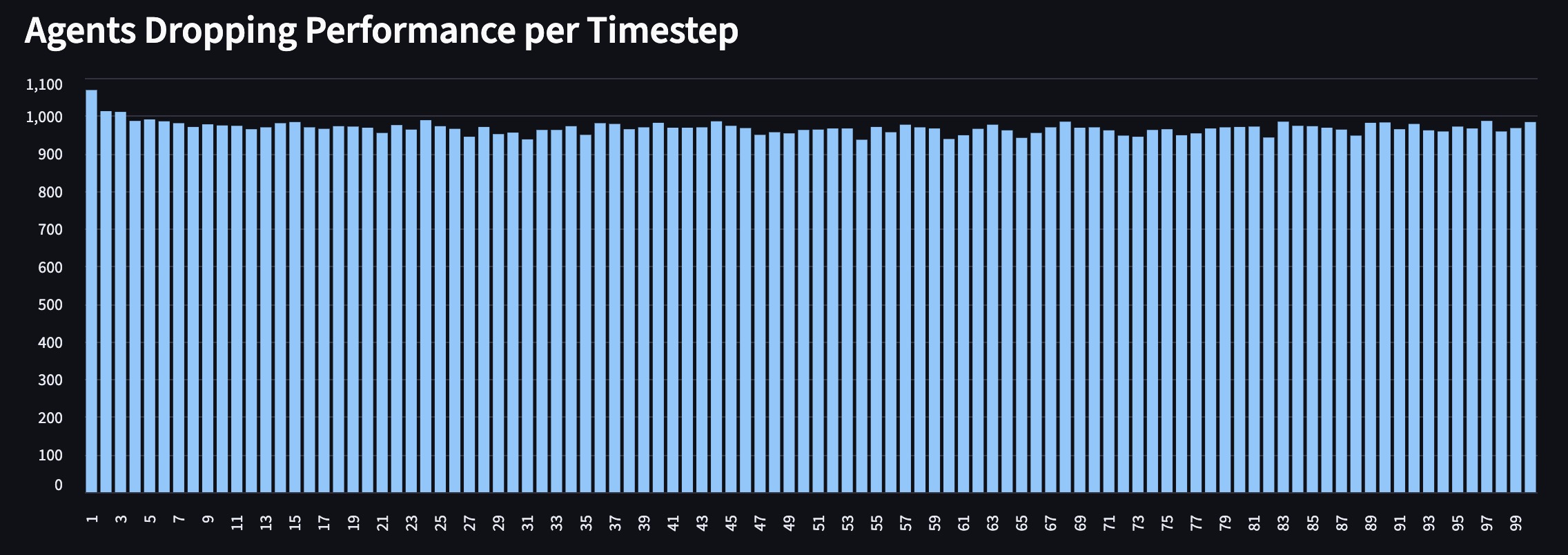}
  \caption{Number of promotions with $\Delta P<0$ at each timestep (transferable-skills regime).}
  \label{fig:merit_neg_tseries_transfer}
\end{figure}

\noindent Complementing the distributional view, Fig.~\ref{fig:merit_neg_tseries_transfer} shows the \emph{count} of $\Delta P<0$ promotions each step. 
The series is \textbf{flat}—about \textbf{900--1{,}100} negative shocks \emph{every} timestep—accumulating to \textbf{96{,}807} across $T{=}100$. 
This again indicates that misaligned moves are \textbf{persistent rather than episodic} under \textbf{Merit}.

\vspace{\baselineskip}
\noindent\emph{Why efficiency still rises here.} 
Despite the Peter-Principle signal (most promotions are negative), the \textbf{losses are substantially smaller} in this ladder—especially at the high-volume steps L1$\to$L2 and L2$\to$L3—so the promotion-induced “drag” per period is far lower than in high-mismatch. 
With adjacent roles requiring \emph{nearby} skills, success is more portable upward; the organization therefore climbs smoothly to a higher terminal efficiency ($0.5852$ vs.\ $0.5485$ in high-mismatch) before leveling off due to the fixed level mix and the L1 baseline.

\paragraph{Interpretation.}
Under \textbf{Merit}, selection rewards success in the \emph{current} role. In the transferable-skills ladder, where adjacent levels reweight skills in small steps, this produces a \emph{weaker} Peter-Principle pattern: most promotions still reduce performance ($96{,}807$ of $109{,}000$; $88.8\%$), but the average loss per move is much smaller than in the high-mismatch ladder (mean $\Delta P \approx -0.043$ here vs.\ $-0.136$ there). Transition means and volumes explain the shape: the two busy steps (L1$\!\to$L2 and L2$\!\to$L3; $\sim\!94\%$ of promotions) are only mildly harmful on average ($-0.04$, $-0.06$); the upper moves are rare and near-neutral on net (L3$\!\to$L4 $\approx -0.06$, L4$\!\to$L5 $\approx -0.01$); and the typical positive shock is smaller than the typical negative one, especially at the lower levels. If you multiply each step’s average impact by how many of those promotions happen, the total “drag” from promotions each period is much smaller here than in high-mismatch. Because promotions hurt less overall,so efficiency rises smoothly without an early dip and ends higher ($E_{100}\approx 0.5852$ vs.\ $0.5485$). The concave plateau follows naturally once L2–L5 settle into high-fit mixes and the large L1 block (40\% of seats, near-uniform entrants) anchors the organizational average.

\vspace{\baselineskip}

\noindent\textbf{Seniority.} As before, the rule promotes purely on \emph{organizational tenure}. Due to that the model makes makes Seniority effectively \emph{skill-blind} and therefore qualitatively similar to \textbf{Random} - who goes up is largely independent of whether their competence vector matches the destination role. In the transferable-skills regime graph climbs quickly starting from $E_{0} \approx 0.4807$ in the first 10 steps then it plateaus out and then decreases slightly to end at $E_{100} \approx 0.5192$. 
The reason is structural: because selection is \emph{uncorrelated} with the source job, promotions are a \emph{mix} of good and bad matches. 
Because seniority promotions are unrelated to role fit, each step mixes helpful and harmful moves that mostly cancel out; the average promotion effect stays near zero, but at lower level due to our initialization\footnote{We draw agents with random skills and fill fixed level caps by assigning each agent to the highest role they reasonably fit, loosening the bar only if a level can’t be filled. This makes upper pools roughly aligned to their role profiles from day one, while L1 remains a broad, near-uniform mix. That asymmetry helps explain the small early lift under Seniority when adjacent role weights make success more portable.} and \emph{adjacent role weights} transitions are mildly helpful and as they account for majority of promotions the efficiency increases at first and then falls when L2–L5 approach their steady mix and randomness takes over, pushing the efficiency to $E_{100} \approx 0.5192$.

Across $T{=}100$ we observe $\mathbf{58{,}753}$ promotions with $\Delta P>0$ and $\mathbf{50{,}247}$ with $\Delta P<0$ which is a near balance, consistent with an essentially random mix of good and bad matches with slightly more positive promotions (due to close similarity in skills across levels and our initialization).

\begin{figure}[H]
    \centering
    \begin{subfigure}[t]{0.48\linewidth}
        \centering
        \includegraphics[width=\linewidth]{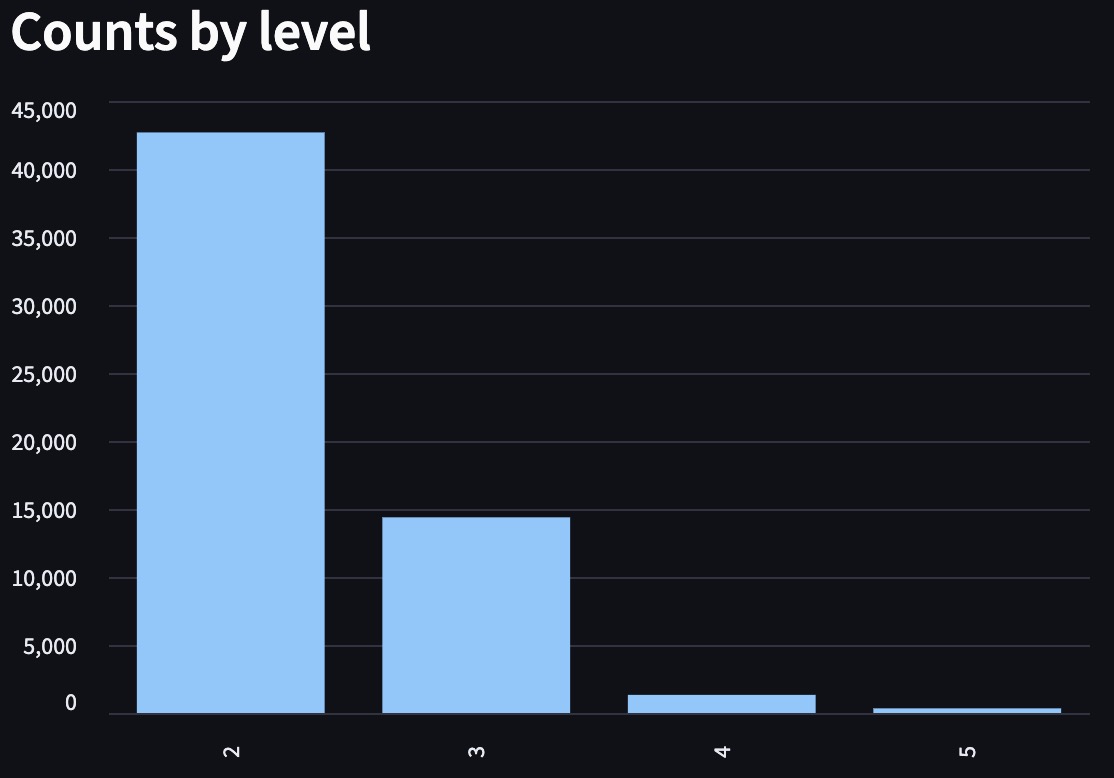}
        \caption{number of Promotions with $\Delta P > 0$ by level.}
        \label{fig:seniority_pos_transfer}
    \end{subfigure}
    \hfill
    \begin{subfigure}[t]{0.48\linewidth}
        \centering
        \includegraphics[width=\linewidth]{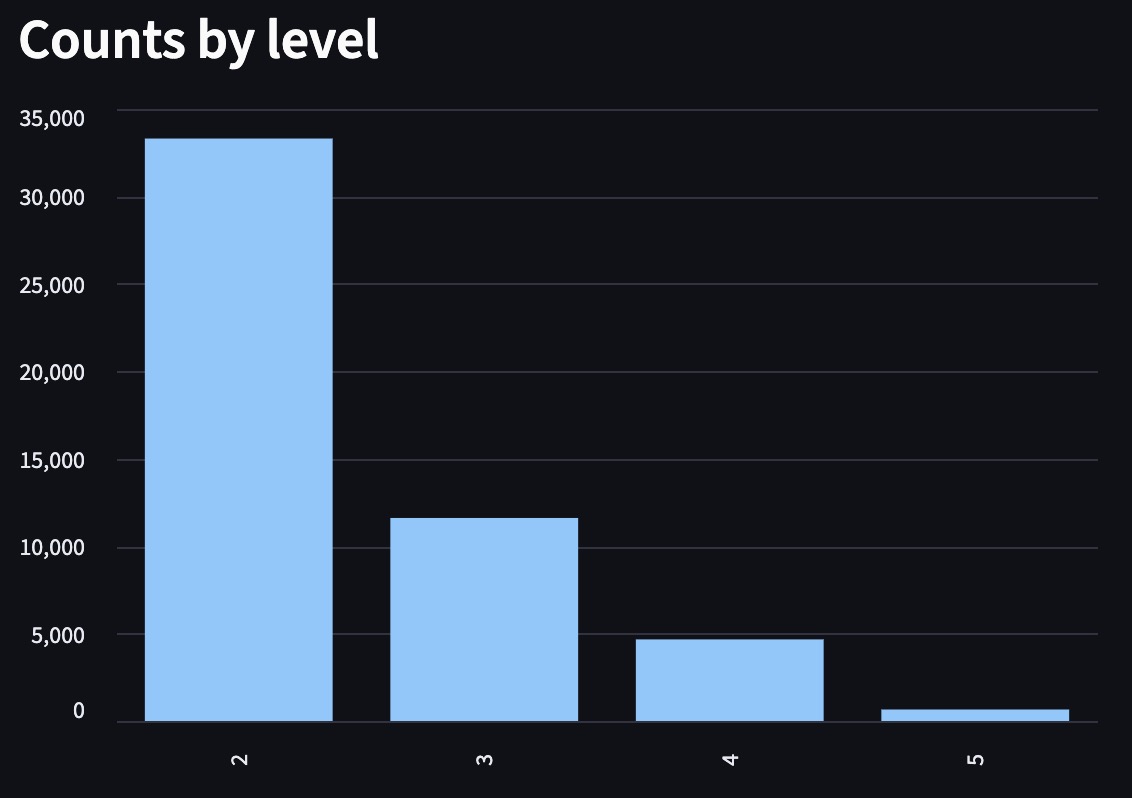}
        \caption{ number of Promotions with $\Delta P < 0$ by level.}
        \label{fig:seniority_neg_transfer}
    \end{subfigure}
    \caption{Bar chart of promotion outcomes under seniority-based promotion in the transferable-skills regime. 
The counts of agents with positive ($\Delta P>0$) and negative ($\Delta P<0$) performance changes are shown by level.}
    \label{fig:seniority_deltas_transfer}
\end{figure}

\noindent When broken down by levels (Fig.~\ref{fig:seniority_deltas}) these are the observed numbers for agents whose performance improves and for agents whose performance plummets:
\[
\begin{array}{lrr}
\toprule
\text{Level} & \Delta P > 0 & \Delta P < 0 \\
\midrule
\text{L2} & 42{,}689 & 33{,}311 \\
\text{L3} & 14{,}388 & 11{,}612 \\
\text{L4} & 1{,}334 & 4{,}666 \\
\text{L5} &   342   &   658 \\
\bottomrule
\end{array}
\]

\noindent\emph{Explanation:} Because \textbf{Seniority} promotes purely by \emph{time in the organization}, selection is essentially \emph{uncorrelated} with the destination role’s skill weights. This yields a near-random mix of helpful and harmful moves. Seniority does not ``favor'' any skill, but the \emph{level pools} it draws from are not neutral: L2 is tech-tilted (from our initialization), and the $L2 \to L3$ transition is \emph{mildly} tech-to-management, so it yields a small positive mean shock. And due to the same reasoning $L3 \to L4$ is more management heavy and hence the mean shock is slightly negative. Finally, $L4 \to L5$ is again more management heavy but the change is small and hence the mean shock is slightly negative. When we said that L1 is near-uniform it is not entirely true - it is slightly tech heavy due to the initialization as from a uniform distribution the agents who are eligible for $L2 \to L5$ are already selected out and the remaining agents are slightly similar to L1 profile. This explains the small positive mean shock from L1 to L2. But these shocks are only present in early timesteps as the levels approach their steady mix and then randomness takes over and therefore mean shock is very \emph{small} and close to zero.

\begin{figure}[H]
    \centering
    \includegraphics[width=\linewidth]{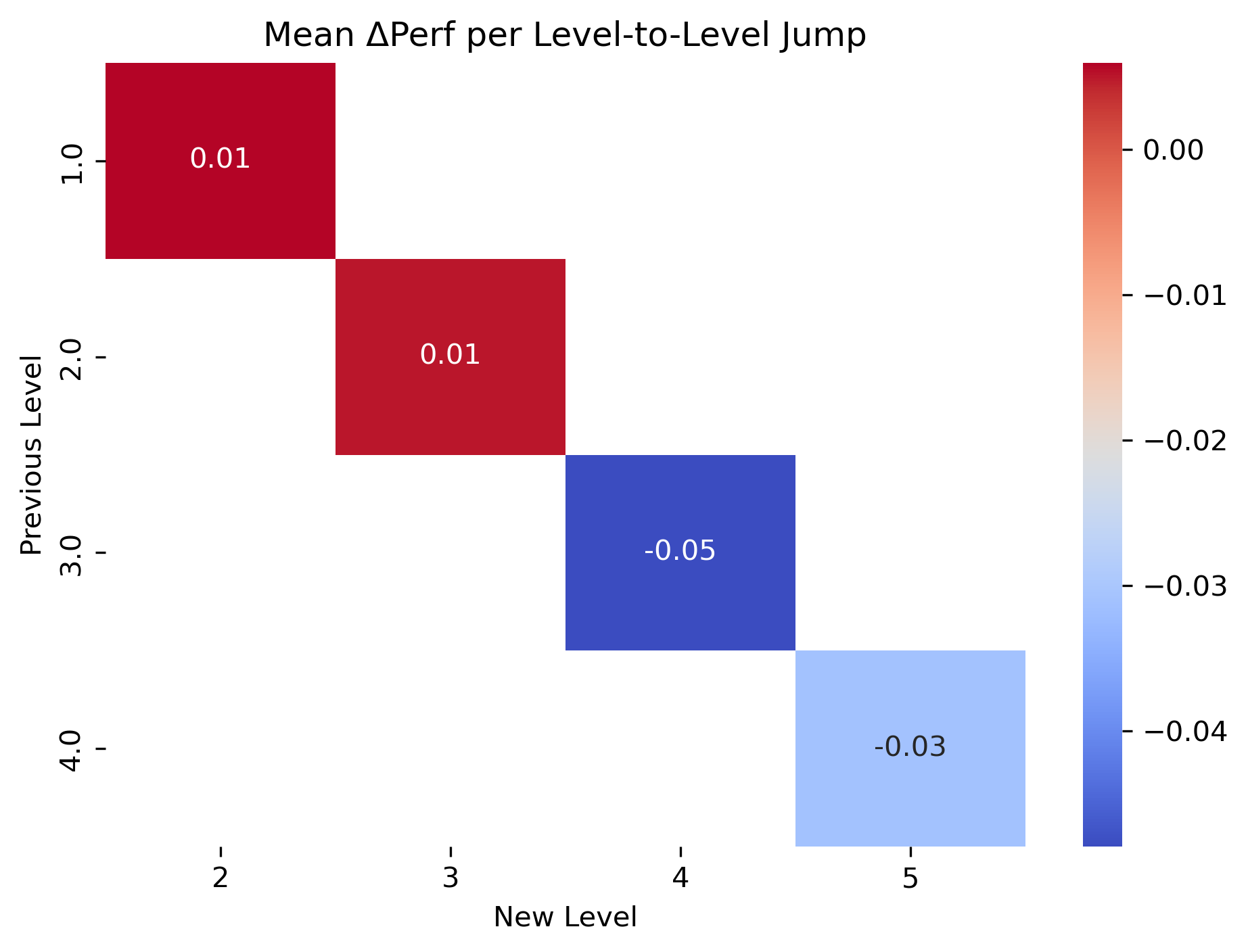}
    \caption{Mean $\Delta P$ per Level-to-Level jump for seniority based promotion strategy.}
    \label{fig:seniority_heatmap_transfer}
\end{figure}

\noindent The mean effects from the heatmap confirm this story (Fig.~\ref{fig:seniority_heatmap}): L1$\to$L2 $\approx \mathbf{0.01}$, L2$\to$L3 $\approx \mathbf{0.01}$, L3$\to$L4 $\approx \mathbf{-0.05}$, L4$\to$L5 $\approx \mathbf{-0.03}$. Early transitions include a mild positive due to high tech and similar skills requirements; near the top, roles become more different and the average impact is slightly negative.

\begin{figure}[H]
    \centering
    \includegraphics[width=\linewidth]{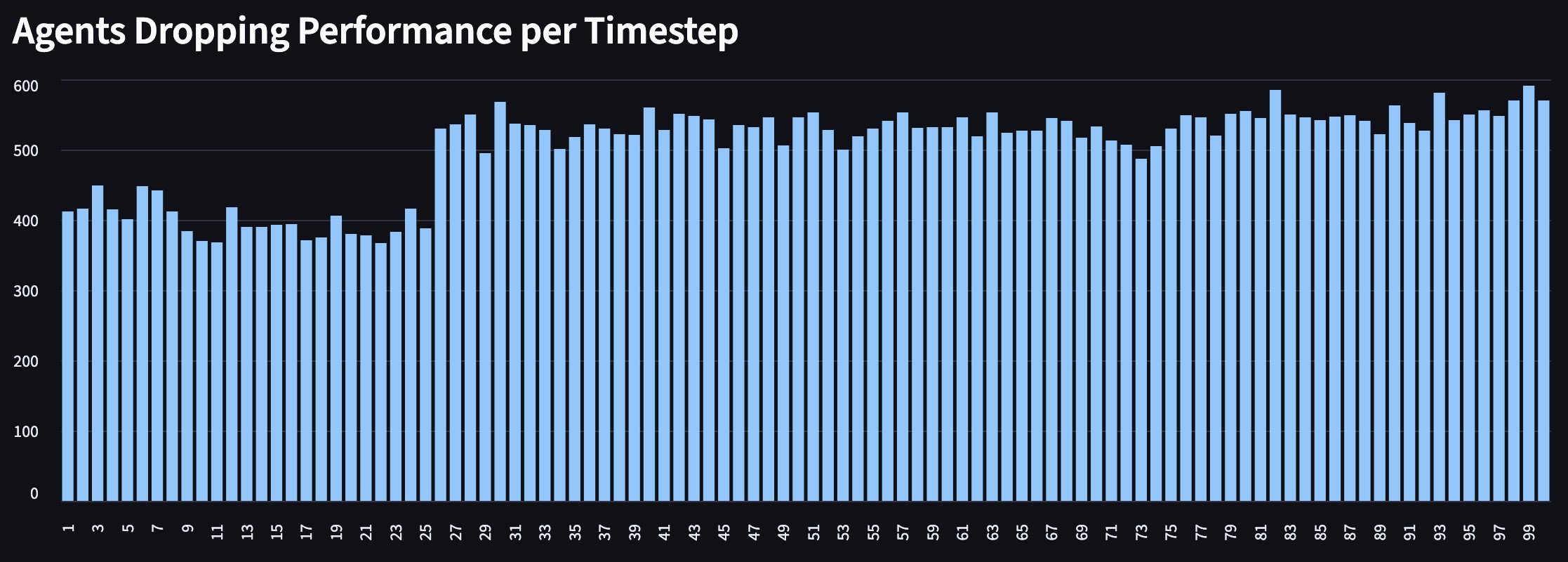}
    \caption{number of promotions with $\Delta P\!<\!0$ at each timestep for seniority based promotion.}
    \label{fig:seniority_neg_tseries_transfer}
\end{figure}

This is evident from the time series of number of promotions with $\Delta P < 0$ (Fig.~\ref{fig:seniority_neg_tseries_transfer}) which is \textbf{not flat} like previously. In the early timesteps (1-25) we have about \textbf{$350$-$450$} negative shocks. And after that (i.e $t > 25$) we have about \textbf{$500$-$600$} negative shocks. This accumulates to \textbf{50{,}247} across $T{=}100$. This indicates that during earlier timesteps the levels are filled with agents who are slightly similar to the role requirements and therefore the promotions which dispropotionately come $L1\to L2$ and $L2\to L3$ tend to be mildly positive and offset the small negatives from $L3\to L4$ and $L4\to L5$. Therefore we see less negative shocks in the early timesteps. But as the levels approach their steady mix the randomness takes over and therefore the number of negative shocks increases.


\begin{figure}[H]
    \centering
    \includegraphics[width=\linewidth]{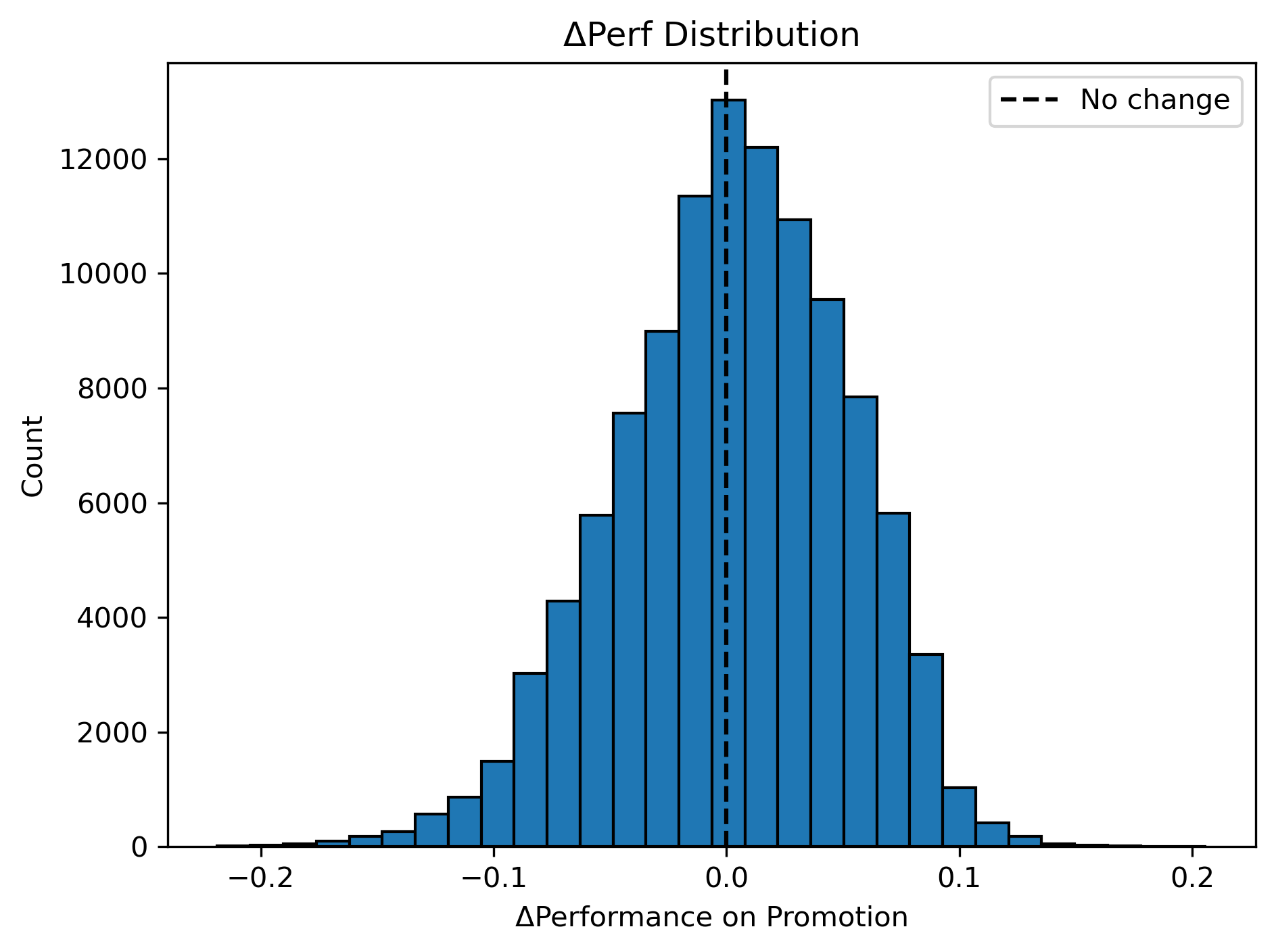}
    \caption{$\Delta P$ frequency distribution across all time-steps for seniority based promotion.}
    \label{fig:seniority_promotion_delta_transfer}
\end{figure}

\noindent The \emph{frequency distribution} of $\Delta P$ across all promotions under \textbf{Seniority} (Fig.~\ref{fig:seniority_promotion_delta_transfer}) is \textbf{right-shifted} but \textbf{left-skewed}: more promotions help than hurt (58{,}753 vs 50{,}247; 53.9\% positive), yet the negative tail is longer. Across $109{,}000$ promotions, the mean change is $0.002$ (median $0.005$), with $46.1\%$ of promotions showing a decrease. The \textbf{negative tail} extends to about \(\mathbf{-0.219}\) and \textbf{positive tail} to \(\mathbf{+0.206}\) (1st/99th percentiles \(-0.122\) / \(+0.099\)). The average change is $+0.0024$ with a bootstrap 95\% CI $[+0.0021,\,+0.0027]$; a one-sided Wilcoxon test confirms a detectable positive shift ($p \ll 10^{-6}$).
However, the magnitude is small in practice: relative to a random baseline the effect size is trivial 
(Cohen’s $d \approx 0.07$), and the share of ``meaningful'' declines ($\Delta P \le -0.05$) 
is slightly \emph{lower} than Random (14.8\% vs.\ 16.0\%). Together, these results explain the shape of the efficiency curve: a modest early lift (small positive bias in the high-volume lower steps) followed by a plateau and slight decline (as levels approach their steady mix and randomness takes over).

\paragraph{Interpretation.}
Under \textbf{Seniority}, selection is \emph{skill-blind}: tenure, not destination fit, determines who moves. In the transferable-skills ladder, where adjacent levels reweight skills only in small steps, that policy yields an almost random mix of matches and mismatches with a slight early positive bias. In aggregate we see (i) more improvements than declines overall ($\Delta P>0 \approx 58{,}753$ vs.\ $\Delta P<0 \approx 50{,}247$), (ii) small mean step effects consistent with locally aligned roles (heatmap, Fig.~\ref{fig:seniority_heatmap_transfer}: L1$!\to$L2 $\approx {+}0.01$, L2$!\to$L3 $\approx {+}0.01$, L3$!\to$L4 $\approx {-}0.05$, L4$!\to$L5 $\approx {-}0.03$), and (iii) a shape that rises quickly and then flattens, ending at $E_{100}\!\approx\!0.5192$. The time-series of negative promotions (Fig.~\ref{fig:seniority_neg_tseries_transfer}) explains the dynamics: early steps ($t{=}1$–25) feature fewer declines (about $350$–$450$ per step) because initialization leaves level pools roughly aligned with their roles and most moves come from the near-neutral lower transitions; as pools converge to their steady mix, selection becomes effectively random with respect to fit and the count of declines rises to $500$–$600$ per step. The histogram (Fig.~\ref{fig:seniority_promotion_delta_transfer}) is accordingly \emph{right-shifted but left-skewed}: many small gains, fewer but larger losses; the shift is statistically detectable (mean $\approx {+}0.0024$) but small in practice, so the early lift dissipates and the series plateaus once L2-L5 stabilize and the large L1 block anchors the average.

\vspace{\baselineskip}

\noindent\textbf{Hybrid ($\alpha{=}0.70$ on performance).}
Hybrid scores candidates with a 70–30 blend of \emph{current performance (Merit)} and \emph{tenure (Seniority)} (tenure normalized with a 12-year cap). Because performance dominates, Hybrid behaves like “\emph{slightly-noisy Merit}.” it still lifts strong current performers, but the tenure term occasionally advances longer-tenured, mid-performing agents and modestly slows the fastest ascents. In the transferable-skills ladder, where adjacent levels only \emph{lightly} reweight skills—this produces a Peter-Principle signal that is present but milder than in high-mismatch. The tenure component, however, slightly \emph{dampens} the effect (especially at lower levels where $T<12$, similar to the high-mismatch regime). We have similar graph as Merit which climbs quickly from $E_{0} \approx 0.4807$ in the first 10 steps then it plateaus out at $E_{100} \approx 0.5790$ ($+20.45\%$ vs.\ $E_0$). The concave plateau follows for the same reasons as Merit: (i) late-stage \textsf{compliance} requirements the pipeline did not prioritize earlier, and (ii) the structural L1 share with uniform inflow and the highest exit rate.

\emph{How the diagnostics line up with Merit.} 

\begin{figure}[H]
    \centering
    \begin{subfigure}[t]{0.48\linewidth}
        \centering
        \includegraphics[width=\linewidth]{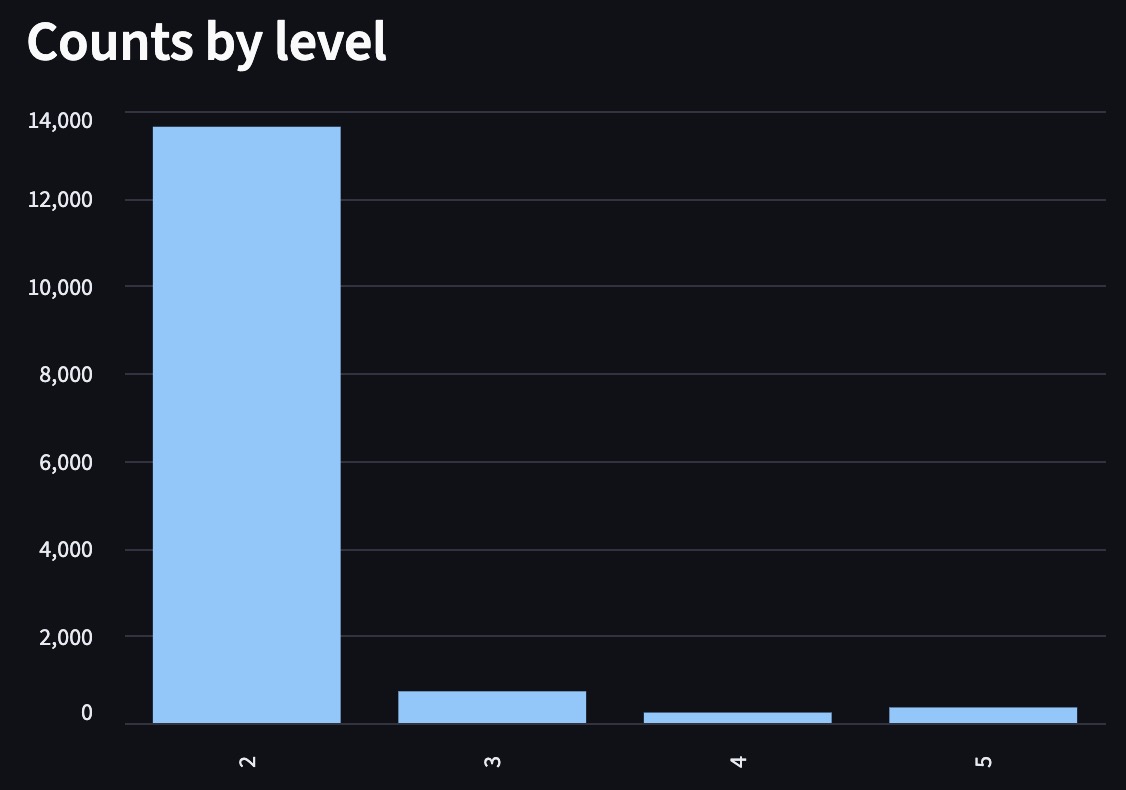}
        \caption{number of Promotions with $\Delta P > 0$ by level.}
        \label{fig:hybrid_pos_transfer}
    \end{subfigure}
    \hfill
    \begin{subfigure}[t]{0.48\linewidth}
        \centering
        \includegraphics[width=\linewidth]{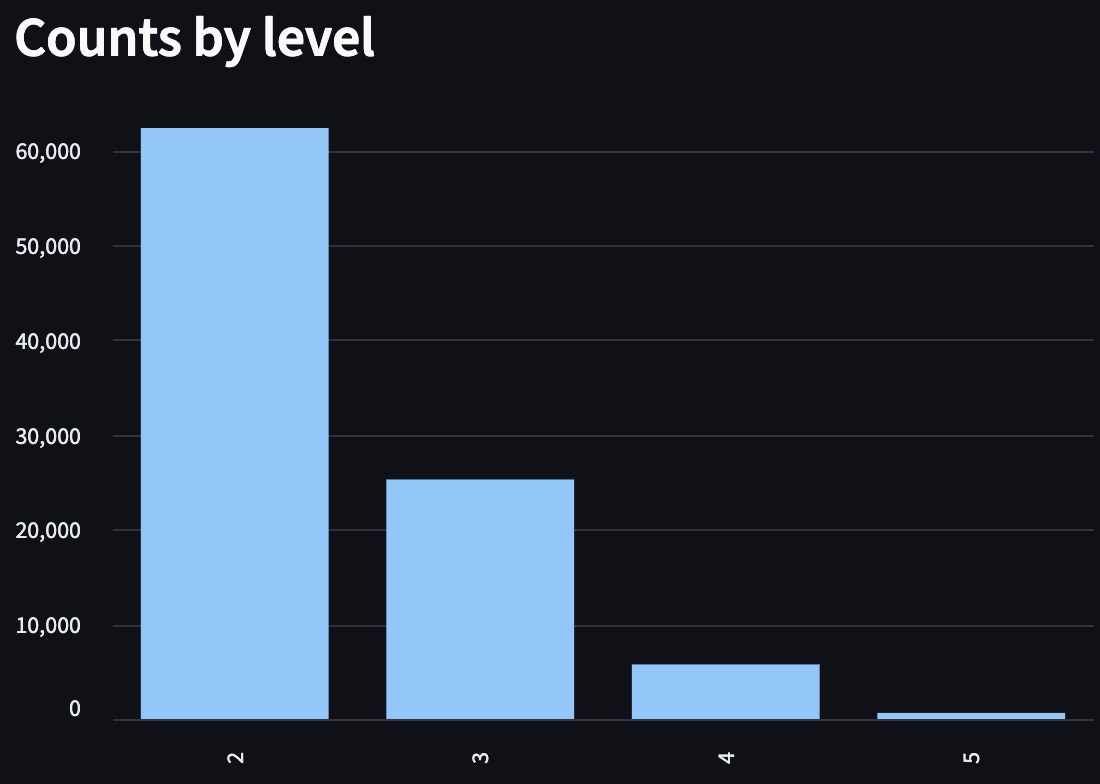}
        \caption{ number of Promotions with $\Delta P < 0$ by level.}
        \label{fig:hybrid_neg_transfer}
    \end{subfigure}
    \caption{Bar chart of promotion outcomes under hybrid promotion in the transferable-skills regime. 
The counts of agents with positive ($\Delta P>0$) and negative ($\Delta P<0$) performance changes are shown by level.}
    \label{fig:hybrid_deltas_transfer}
\end{figure}

Here are the numbers for the same:

\[
\begin{array}{lrr}
\toprule
\text{Level} & \Delta P > 0 & \Delta P < 0 \\
\midrule
\text{L2} & 13{,}640 & 62{,}360 \\
\text{L3} & 725 & 25{,}275 \\
\text{L4} & 239 & 5{,}761 \\
\text{L5} &   356   &   644 \\
\bottomrule
\end{array}
\]

\noindent \textbf{Hybrid} largely mirrors \textbf{Merit} in the transferable-skills ladder: performance still drives most moves, and the tenure term acts as gentle noise. The one consistent deviation is at \textbf{L1$\!\to$L2}, where Hybrid softens early mismatches compared with Merit (Hybrid: $13{,}640$ with $\Delta P>0$ vs.\ $62{,}360$ with $\Delta P<0$; Merit: $10{,}945$ vs.\ $65{,}055$). A smaller, similar softening appears at \textbf{L2$\!\to$L3} (Hybrid: $725$ vs.\ $25{,}275$; Merit: $639$ vs.\ $25{,}361$), while the top transitions are nearly identical (L3$\!\to$L4: $239$ vs.\ $5{,}761$; L4$\!\to$L5: $356$ vs.\ $644$). This is exactly what the tenure term is expected do at the bottom: occasionally advance longer-tenured, mid-performing L1/L2 candidates whose profiles are “close enough,” and slightly delay very new stars. Once most candidates accumulate tenure (the 12-year cap saturates quickly mid-ladder), the Hybrid score is effectively performance-driven and behaves like Merit. 

\noindent \textbf{Net effect:} the diagnostics line up with Merit but with a modestly higher share of positives overall (Hybrid: $14{,}960$ positives, $94{,}040$ negatives; $13.7\%$ positive vs.\ Merit’s $11.2\%$), yielding the same smooth, concave trajectory that rises quickly and plateaus slightly \emph{below} Merit ($E_{100}\approx 0.5790$ vs.\ $0.5853$), for the same reasons—late \textsf{compliance} becoming binding and the large L1 block anchoring the aggregate.

\begin{figure}[H]
    \centering
    \includegraphics[width=\linewidth]{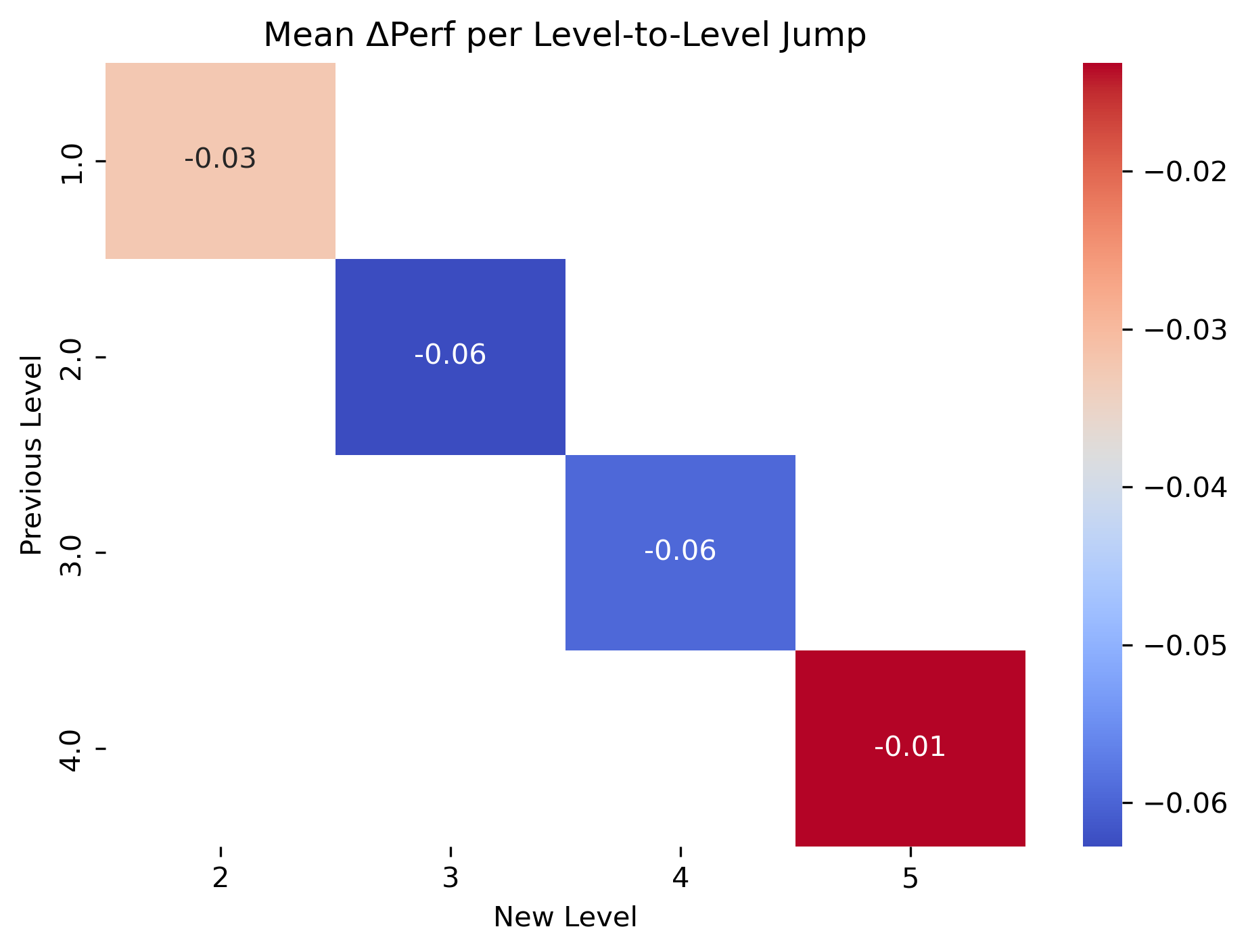}
    \caption{Mean $\Delta P$ per Level-to-Level jump for hybrid promotion strategy.}
    \label{fig:hybrid_heatmap_transfer}
\end{figure}

\noindent Similar data is shown in the heatmap, where L1$\to$L2 transition has lower $\Delta P$ value ($-0.03$ compared to $-0.04$ in merit).

\begin{figure}[H]
    \centering
    \includegraphics[width=\linewidth]{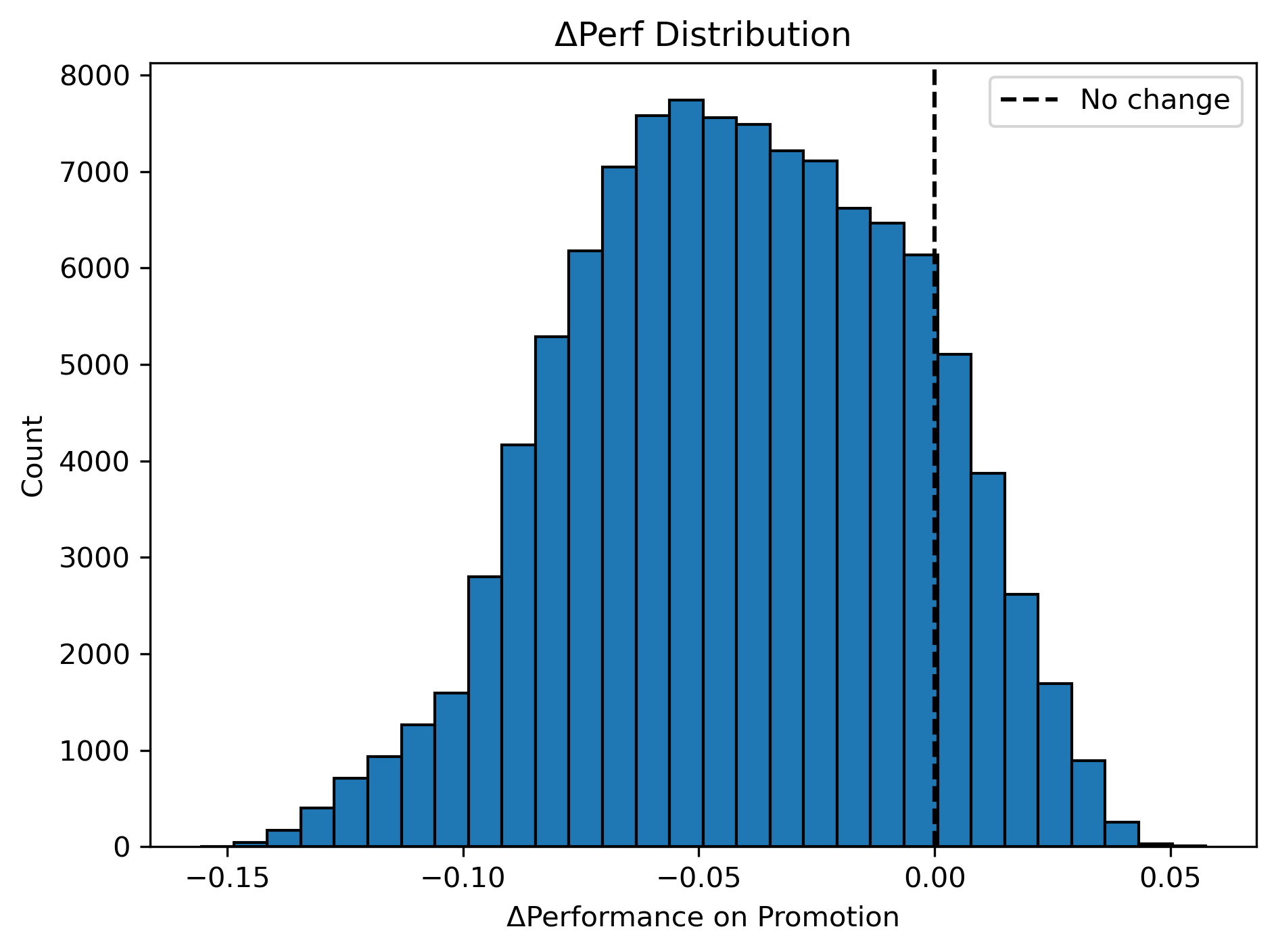}
    \caption{$\Delta P$ frequency distribution across all time-steps for hybrid promotion.}
    \label{fig:hybrid_promotion_delta_transfer}
\end{figure}

\noindent The \textbf{$\Delta P$ histogram} is still \emph{left-skewed}, with a mean of $-0.041$ and median $-0.041$, similar to Merit ($-0.043$ and $-0.043$). The \textbf{negative tail} extends to about \(\mathbf{-0.156}\) and \textbf{positive tail} to \(\mathbf{+0.057}\) (1st/99th percentiles \(-0.122\) / \(+0.030\)). Overall, the data show the same qualitative pattern as Merit. 

\paragraph{Interpretation.} Similar to high-mismatch regime, Hybrid \textbf{dampens} (but does not remove) the effects of Peter-Principle present under Merit especially for lower levels. {Hybrid ($\alpha{=}0.7$).}
Selection is ``\emph{slightly noisy Merit}'': current performance still drives most moves, while tenure occasionally nudges up longer-tenured mid-performers and slows very new stars. In the transferable-skills ladder---where adjacent levels reweight skills only a little---this yields the same basic mechanism as Merit but \emph{milder}: most promotions still reduce performance (left-skewed $\Delta P$ with mean $-0.041$ and median $-0.041$), yet the losses at the busy lower steps are a touch softer (heatmap: $L_{1}\to L_{2} \approx -0.03$ vs.\ $-0.04$ in Merit; $L_{2}\!\to\! L_{3}$ remains $\approx -0.06$). Diagnostics line up with this story: Hybrid has more positives than Merit at the bottom (e.g., $L_{1}\!\to\! L_{2}$: $13{,}640$ positives vs.\ $62{,}360$ negatives, vs.\ Merit’s $10{,}945/65{,}055$), with nearly identical outcomes higher up ($L_{3}\!\to\! L_{4}$: $239/5{,}761$; $L_{4}\!\to\! L_{5}$: $356/644$). Weighting each step’s mean by its promotion volume yields a slightly \emph{smaller} per-period drag than Merit (Hybrid $\approx -44.6$ units vs.\ Merit $\approx -47.0$), so the efficiency path looks the same shape but sits a little lower and climbs a bit more slowly: a smooth rise from $E_{0}\approx 0.4807$ to a concave plateau at $E_{100}\approx 0.5790$ (vs.\ $0.5853$ for Merit). As with Merit, the plateau is explained by two ceilings: compliance becomes more salient near the top (not prioritized early in the pipeline), and the large $L_{1}$ block ($40\%$) with near-uniform inflow anchors the organizational average.

\vspace{\baselineskip}

\noindent\textbf{Random.}
Random draws promotees uniformly from each level’s candidate pool, fully \emph{skill-blind} and therefore, tracks Seniority closely. In the transferable-skills regime, the graph follows a similar pattern as Seniority which climbs quickly from $E_{0} \approx 0.4807$ in the first 10 steps then it plateaus out and then decreases slightly to end at $E_{100} \approx 0.5240$. The reason is structural: because selection is \emph{uncorrelated} with the source job, promotions are a \emph{mix} of good and bad matches. And most of the helpful and harmful promotions cancel each other out; the average promotion effect stays near zero, but at lower level due to our initialization and \emph{adjacent role weights} transitions are mildly helpful and as they account for majority of promotions the efficiency increases at first and then falls when L2–L5 approach their steady mix and randomness takes over, pushing the efficiency to $E_{100} \approx 0.5240$.

\emph{How the diagnostics line up with Seniority.} 

\[
\begin{array}{lrr}
\toprule
\text{Level} & \Delta P > 0 & \Delta P < 0 \\
\midrule
\text{L2} & 40{,}563 & 35{,}437 \\
\text{L3} & 12{,}860 & 13{,}140 \\
\text{L4} & 1{,}568 & 4{,}432 \\
\text{L5} &   337   &   663 \\
\bottomrule
\end{array}
\]

\begin{figure}[H]
    \centering
    \begin{subfigure}[t]{0.48\linewidth}
        \centering
        \includegraphics[width=\linewidth]{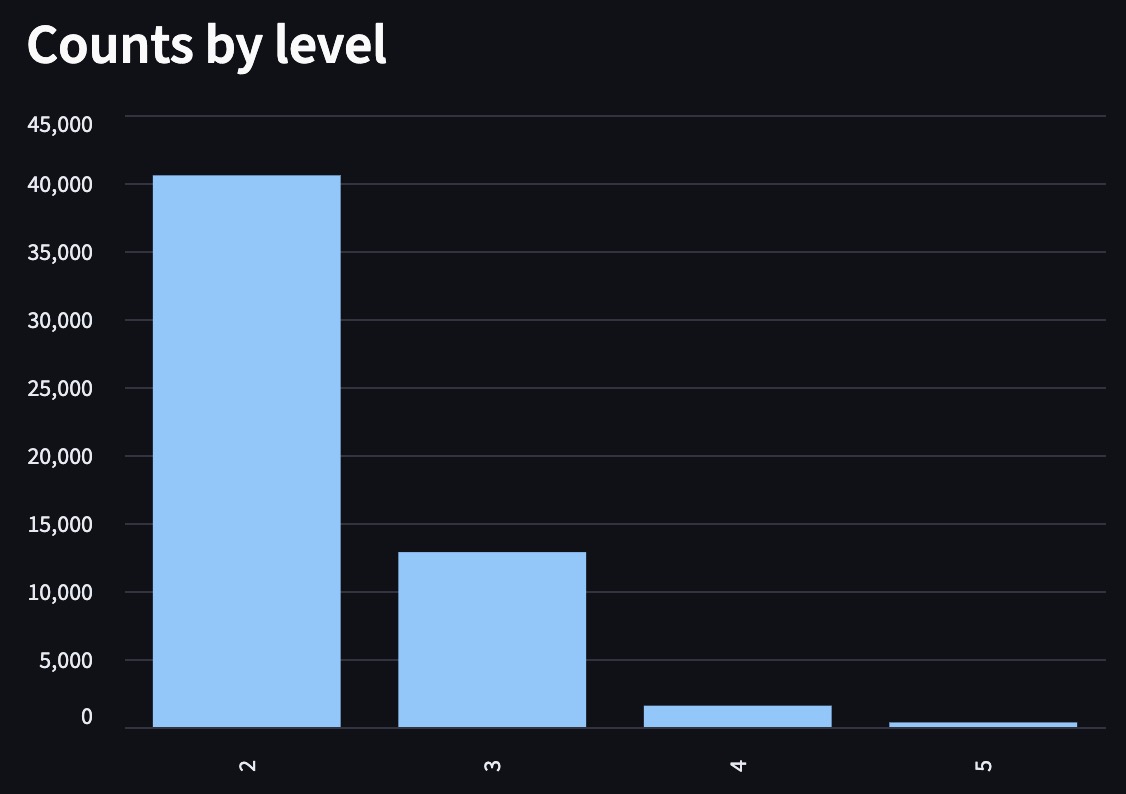}
        \caption{number of Promotions with $\Delta P > 0$ by level.}
        \label{fig:random_pos_transfer}
    \end{subfigure}
    \hfill
    \begin{subfigure}[t]{0.48\linewidth}
        \centering
        \includegraphics[width=\linewidth]{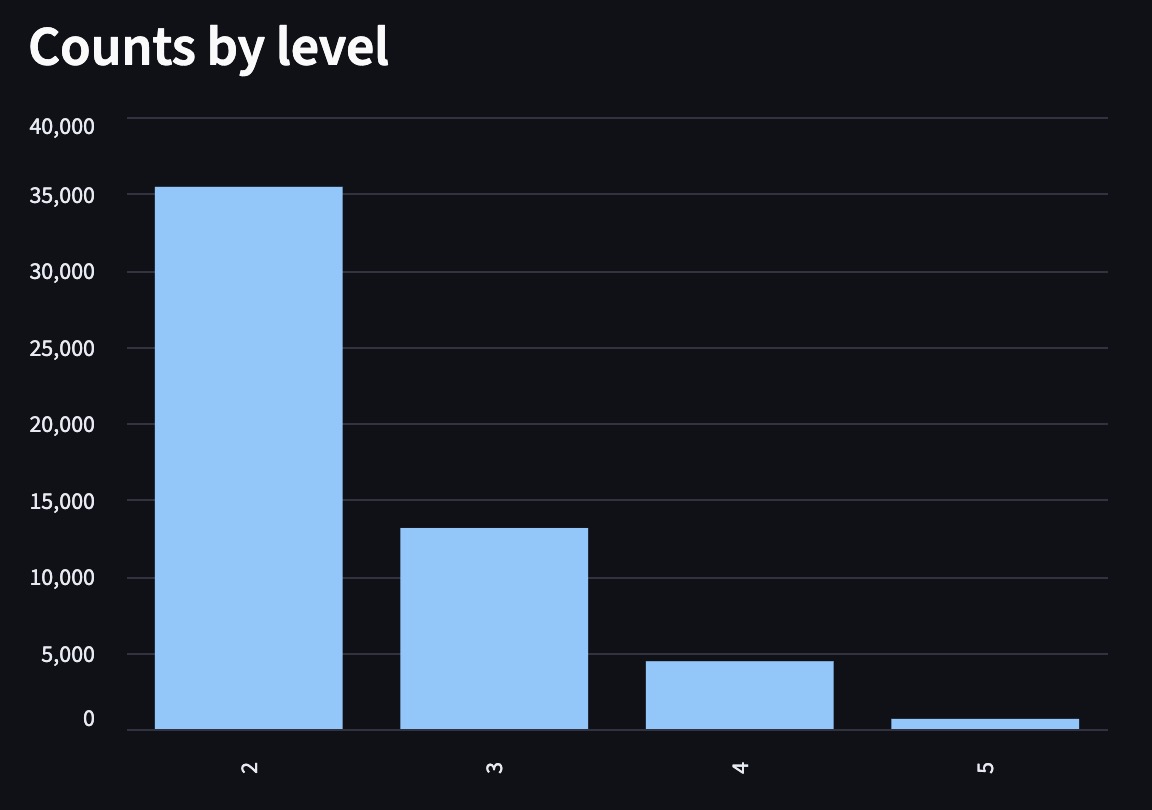}
        \caption{ number of Promotions with $\Delta P < 0$ by level.}
        \label{fig:random_neg_transfer}
    \end{subfigure}
    \caption{Bar chart of promotion outcomes under random promotion in the transferable-skills regime. 
The counts of agents with positive ($\Delta P>0$) and negative ($\Delta P<0$) performance changes are shown by level.}
    \label{fig:random_deltas_transfer}
\end{figure}

\begin{figure}[H]
    \centering
    \includegraphics[width=\linewidth]{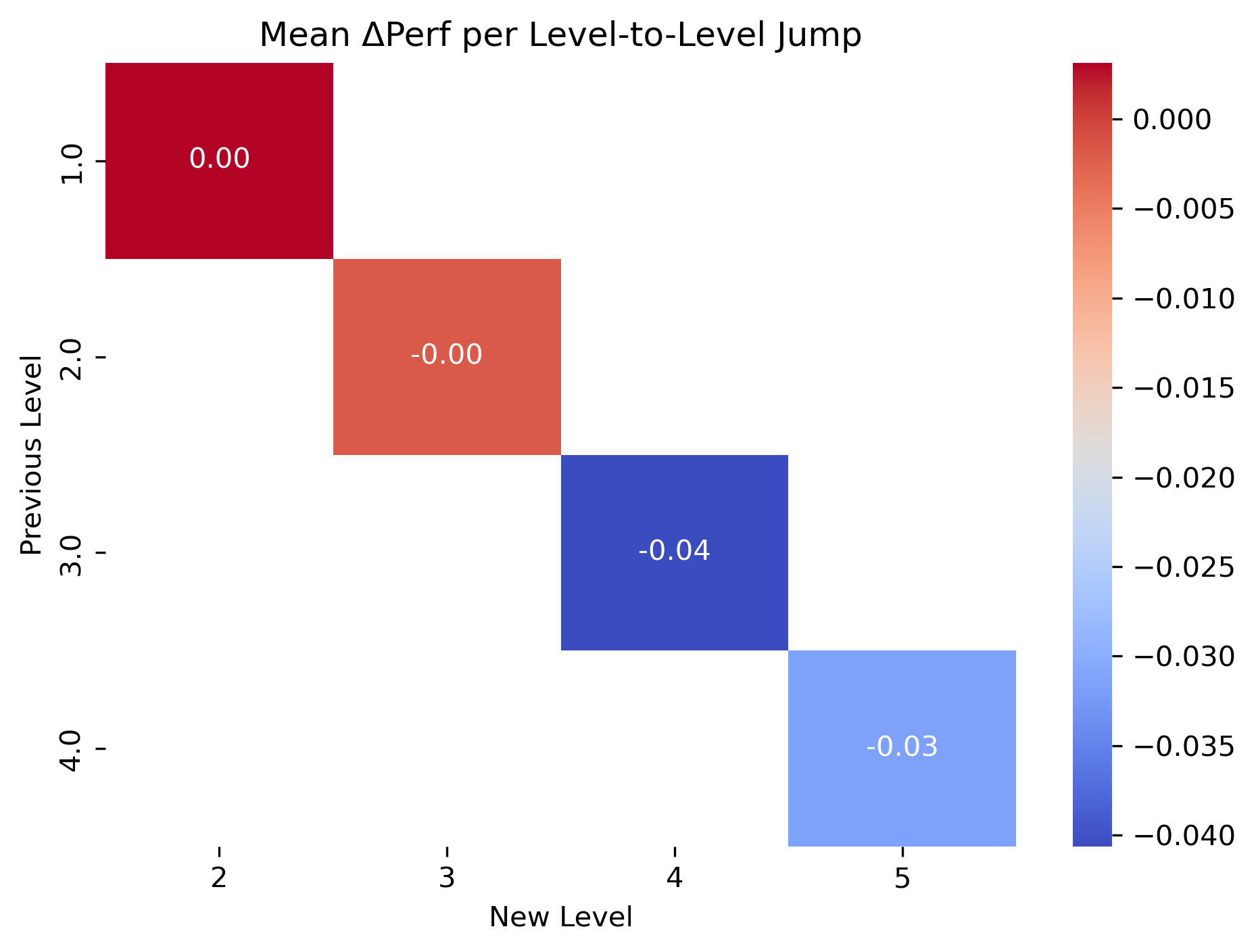}
    \caption{Mean $\Delta P$ per Level-to-Level jump for random promotion strategy.}
    \label{fig:random_heatmap_transfer}
\end{figure}

\noindent The \textbf{bar charts} and the \textbf{mean $\Delta P$ heatmap} show an almost balanced mix on L1$\to$L2 and L2$\to$L3 (compared to slight positive bias in Seniority), with a mild negative tilt at the top (L3$\to$L4: $-0.04$; L4$\to$L5: $-0.03$), again consistent with Seniority.

\begin{figure}[H]
    \centering
    \includegraphics[width=\linewidth]{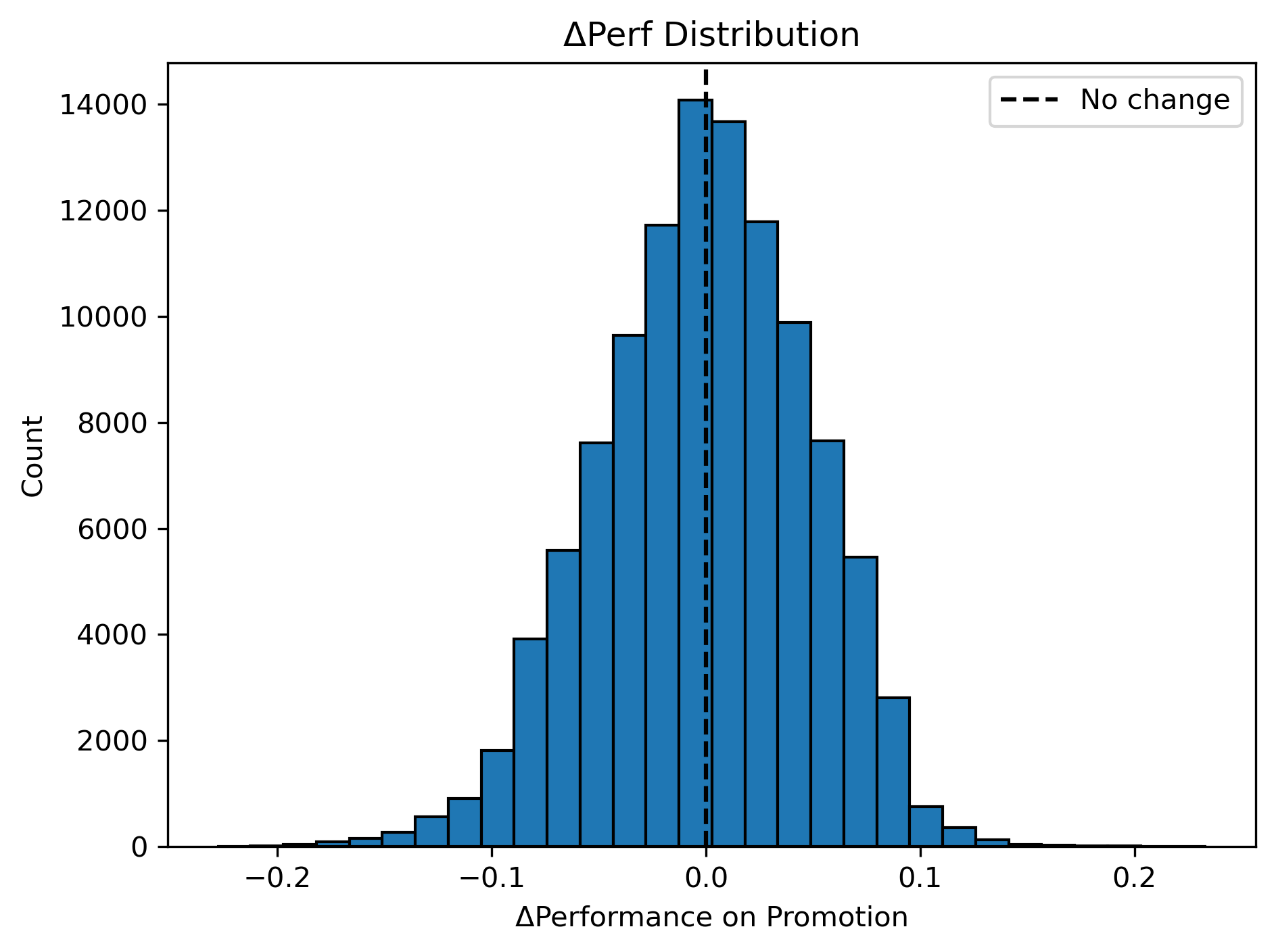}
    \caption{$\Delta P$ frequency distribution across all time-steps for random promotion.}
    \label{fig:random_promotion_delta_transfer}
\end{figure}

\noindent The \textbf{$\Delta P$ histogram} is near-symmetric with only a negligibly small left tail (mean=-0.001, median=0.001, \% of promotions with drop: 49.2\%). The \textbf{negative tail} extends to about \(\mathbf{-0.228}\) and \textbf{positive tail} to \(\mathbf{+0.234}\) (1st/99th percentiles \(-0.121\) / \(+0.098\)).

\paragraph{Interpretation.}
Under \textbf{Random}, selection is fully \emph{skill-blind}, so promotions are an almost even mix of matches and mismatches. In the transferable-skills ladder---where adjacent roles reweight skills only slightly---this yields near-neutral moves at the busy lower steps: $L_{1}\!\to\!L_{2}$ is modestly favorable ($40{,}563$ with $\Delta P>0$ vs.\ $35{,}437$ with $\Delta P<0$), $L_{2}\!\to\!L_{3}$ is essentially balanced ($12{,}860$ vs.\ $13{,}140$), while the top steps carry a mild negative tilt ($L_{3}\!\to\!L_{4}$: $1{,}568$ vs.\ $4{,}432$; $L_{4}\!\to\!L_{5}$: $337$ vs.\ $663$), consistent with the heatmap's small negative means at the top (Fig.~\ref{fig:random_heatmap_transfer}). The histogram (Fig.~\ref{fig:random_promotion_delta_transfer}) is correspondingly near-symmetric (mean $\approx -0.001$, median $\approx 0.001$, $49.2\%$ with $\Delta P<0$), indicating no practical Peter-Principle effect. Put together, these near-zero per-promotion effects explain the shape of the efficiency path: a quick early rise (driven by the high-volume, near-neutral lower transitions under our initialization and adjacent role weights), followed by a long plateau and slight late decline, settling at $E_{100}\approx 0.5240$. Random thus tracks Seniority closely but ends a touch higher, in line with its slightly more balanced lower-pipeline outcomes.

\vspace{\baselineskip}

\noindent \textbf{Selective Demotion.} (Mitigation; see Sec.~\ref{sec:mitigations}.) This rule promotes the best performer from the source level but \emph{demotes} anyone at the destination level whose immediate post-move performance falls below a fixed threshold (here, ($\Delta P \le -\tau$, $\tau=0.05$); the vacated seat is refilled from the level below by merit, while the demoted agent is blacklisted from re-promotion to avoid making the same move again. In the \emph{transferable-skills} regime, where adjacent roles are more compatible and competence carries upward, this filter rarely needs to fire ($\sim$one in six promotions), serving mainly to trim a small set of outlier misfits rather than to undo large cohorts. As a result, the efficiency path is \emph{monotone and concave} as seen under Merit: starting at \(E_{0} \approx 0.4807\), \(E_{t}\) rises every step with no early dip, capturing roughly half of the total gain by \(t \approx 21\) and \(\sim 80\%\) by \(t \approx 50\), then flattening toward a higher plateau of \(E_{100} \approx 0.5823\) (\(+21.1\%\)). Mechanistically, gentle inter-level reweighting keeps most \(\Delta P\) near zero or slightly positive; selective demotion removes the few large negatives and, via blacklist+refill, incrementally enriches upper tiers with ``portable'' profiles---yielding steady aggregate improvement that slows only as the pool saturates. Selective Demotion performs slightly worse than Merit (0.5822 vs.\ 0.5853). The small gap is expected: the demotion+refill cascade reduces large losses (share of \(\Delta P \leq -0.05\) falls from \(\sim 43\%\) to \(\sim 17\%\)) but also (i) yields \emph{fewer total promotions} over \(T\) (94.3k vs.\ 109k), slowing the upward diffusion of portable talent, and (ii) introduces some \emph{skip-level} moves (e.g., \(L1 \rightarrow L3\), \(L1 \rightarrow L4\)) whose average shocks are more negative than adjacent steps (mean \(\Delta P \approx -0.079\) and \(-0.098\)), marginally offsetting the variance reduction. Net: the mechanism mirrors Merit’s shape but lands a bit lower because it trades a bit of throughput for robustness.

\noindent We observed 94{,}302 promotions over \(T=100\), with 72{,}801 having \(\Delta P < 0\) (77.2\%) and 21{,}501 having \(\Delta P > 0\) (22.8\%). Relative to \emph{Merit} in the same regime (109{,}000 promotions with 96{,}807 negatives; 88.8\% negative), selective demotion lowers the negative share by \(\approx 11.6\) percentage points (\(\sim 13.1\%\) relative drop) and dramatically trims \emph{large} losses (\(\Delta P \leq -0.05\)) from 43.3\% to 16.6\% (\(-26.7\) pp; \(\sim 61.7\%\) relative cut).

\begin{figure}[H]
    \centering
    \begin{subfigure}[t]{0.48\linewidth}
        \centering
        \includegraphics[width=\linewidth]{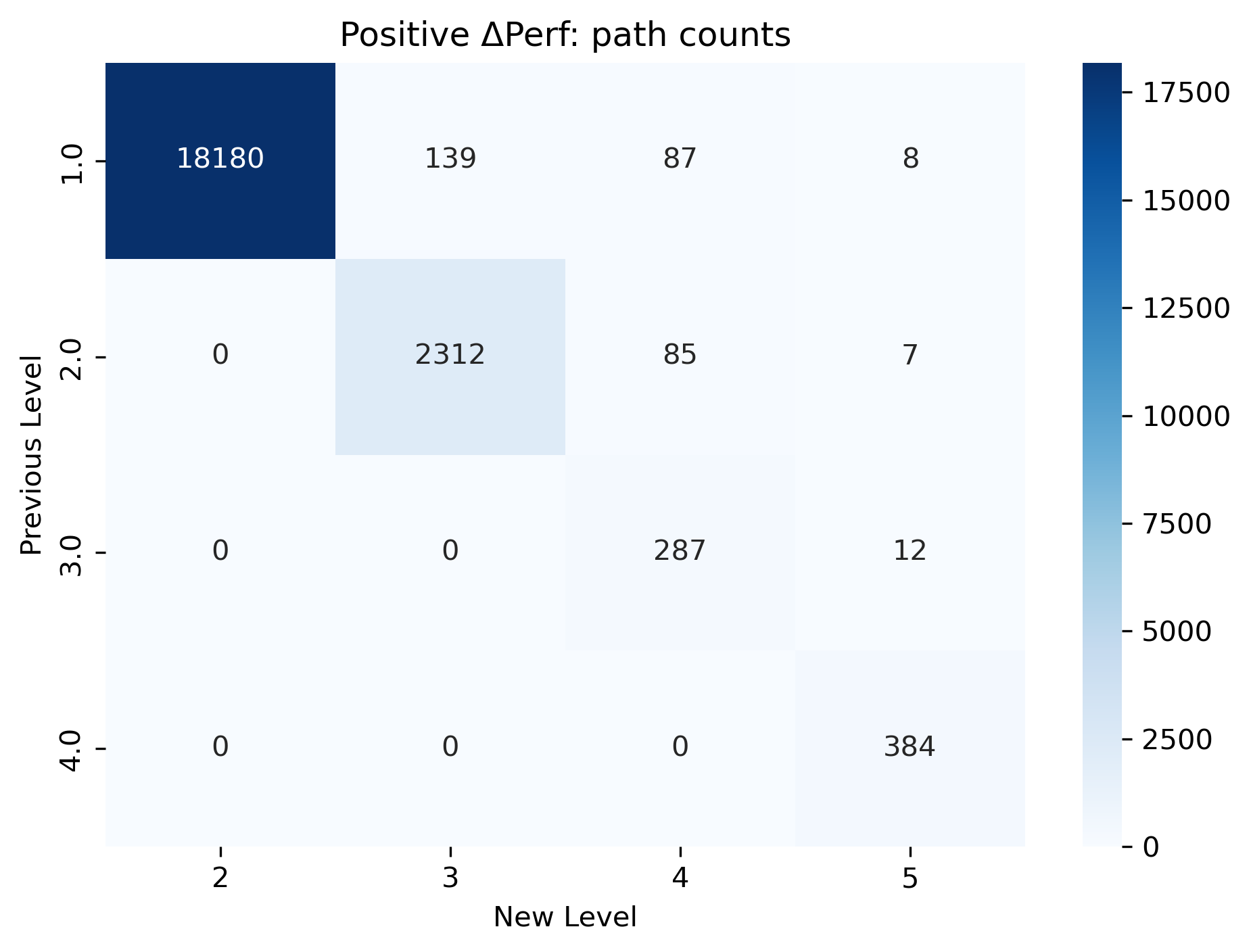}
        \caption{number of Promotions with $\Delta P > 0$ by level.}
        \label{fig:selective_demotion_pos_transfer}
    \end{subfigure}
    \hfill
    \begin{subfigure}[t]{0.48\linewidth}
        \centering
        \includegraphics[width=\linewidth]{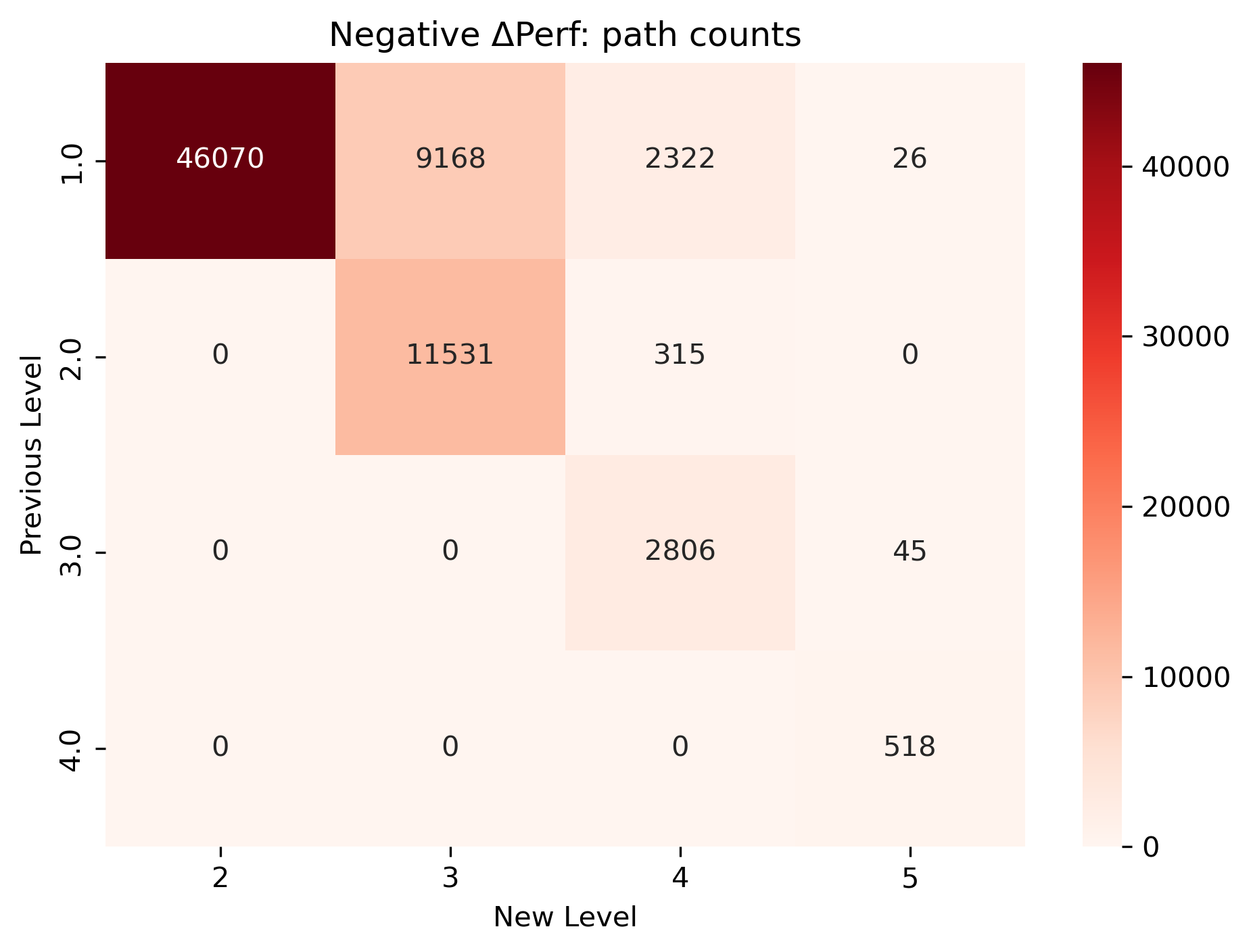}
        \caption{ number of Promotions with $\Delta P < 0$ by level.}
        \label{fig:selective_demotion_neg_transfer}
    \end{subfigure}
    \caption{Heatmap of promotion outcomes under selective demotion in the transferable-skills regime.}
    \label{fig:selective_demotion_deltas_transfer}
\end{figure}

\noindent\emph{Heatmaps (Fig.~\ref{fig:selective_demotion_deltas_transfer}).} Under \textbf{selective demotion} in the \emph{transferable-skills} regime. At lower levels, we still have more promotions with \(\Delta P < 0\) than \(\Delta P > 0\). This pattern persists up the ladder except the \(L2 \to L5\) path. This pattern arises because (i) adjacent roles reweight gently, so a large fraction of first moves incur \emph{small} declines that the \(\tau = 0.05\) screen is designed to tolerate, leaving a persistent excess of mild negatives; (ii) the demotion+blacklist fires less often than in high-mismatch, so it trims the tail without \emph{strongly} reshaping the pool toward management/compliance-heavy ``portable'' profiles, and the negative majority never flips at the top; and (iii) the refill cascade generates \emph{more skip-level jumps} (up to \(L1 \rightarrow L5\)), because adjacent moves tend to be ``good enough'' to merit immediate re-selection the same tick, yet those multi-step arrivals land in more management-weighted roles without another same-tick demotion test. In high-mismatch, heavy pruning at lower tiers flipped upper-tier counts toward positives; here, with softer reweighting and lighter pruning, the mitigation mainly reduces variance, so negatives remain dominant even at the higher tiers of the organizational ladder.

\begin{figure}[H]
    \centering
    \includegraphics[width=\linewidth]{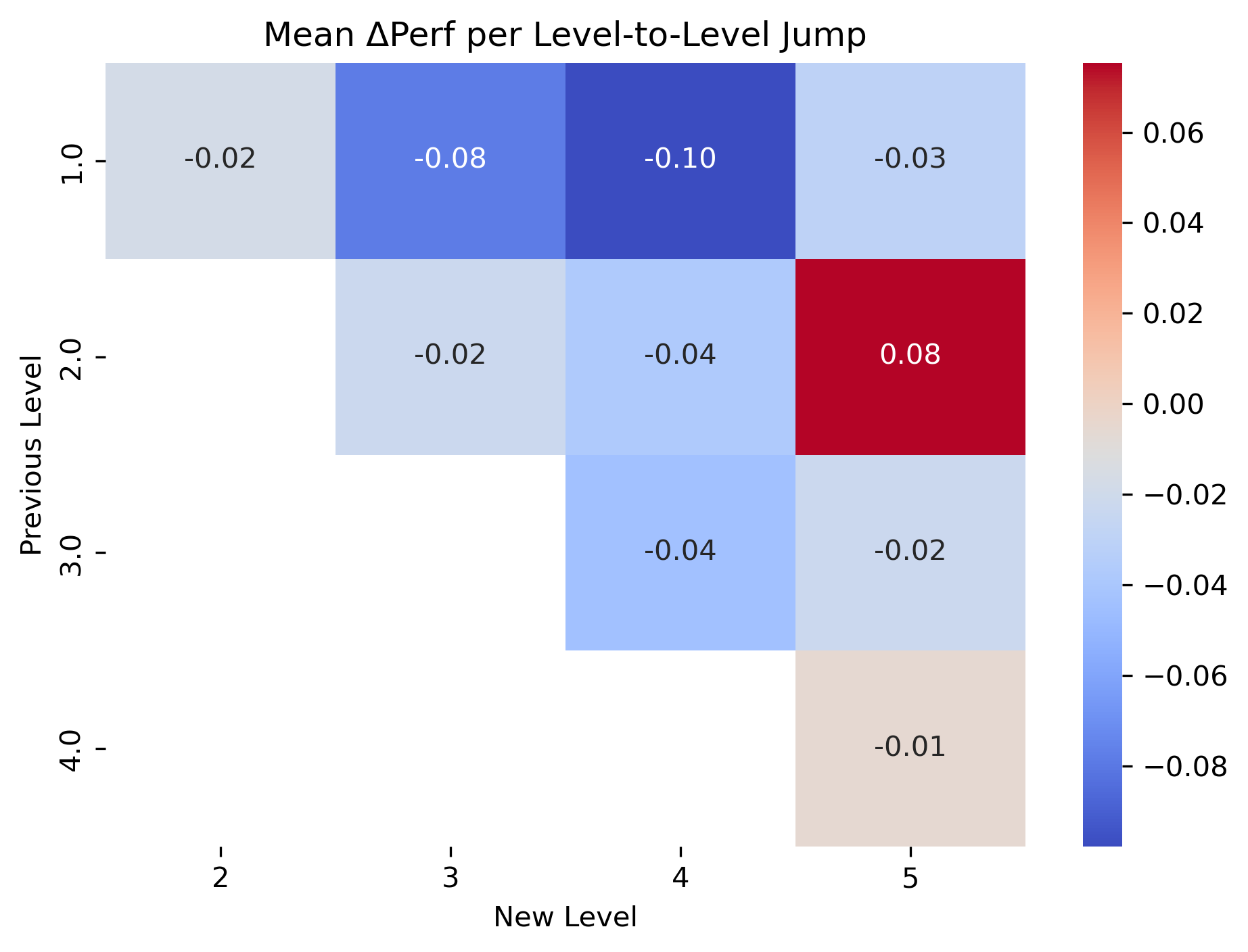}
    \caption{Mean $\Delta P$ per Level-to-Level jump for selective demotion promotion strategy.}
    \label{fig:selective_demotion_heatmap_transfer}
\end{figure}

\noindent\emph{Mean $\Delta P$ heatmap (Fig.~\ref{fig:selective_demotion_heatmap_transfer}).} The pattern under \textbf{selective demotion} in the \emph{transferable-skills} regime the outcomes tend to be negative (left of zero), but only slightly so with no extreme collapses, just mild declines persisting throughout the levels. The adjacent moves post a string of small negatives---\(L1 \rightarrow L2 \approx -0.02\), \(L2 \rightarrow L3 \approx -0.02\), \(L3 \rightarrow L4 \approx -0.04\), and a near-neutral \(L4 \rightarrow L5 \approx -0.01\) this is because inter-level reweighting is gentle and the \(\tau = 0.05\) screen is designed to reverse only \emph{large} drops, allowing the many mild declines to persist. Unlike high-mismatch (where heavy pruning flipped some upper paths positive), the filter here fires less often, so the pool is only modestly reweighted toward management/compliance and adjacent path means stay slightly negative through \(L4\). Skip-level means are more negative overall---\(L1 \rightarrow L3 \approx -0.08\), \(L1 \rightarrow L4 \approx -0.10\), \(L3 \rightarrow L5 \approx -0.02\)---because cascade promotions can push candidates into more managerial roles without another same-tick demotion test, compounding small misalignments in a single step. The one clear exception is the rare \(L2 \rightarrow L5\) jump (\(\approx +0.08\)), which is positive precisely because only highly ``portable'' profiles are selected for such a leap in the refill cascade. Net: selective demotion trims the tail but, in transferable-skills, leaves most path means slightly negative (with \(L4 \rightarrow L5\) nearly neutral), explaining why efficiency tracks Merit’s shape closely while landing just below it.

\begin{figure}[t!]
    \centering
    \includegraphics[width=\linewidth]{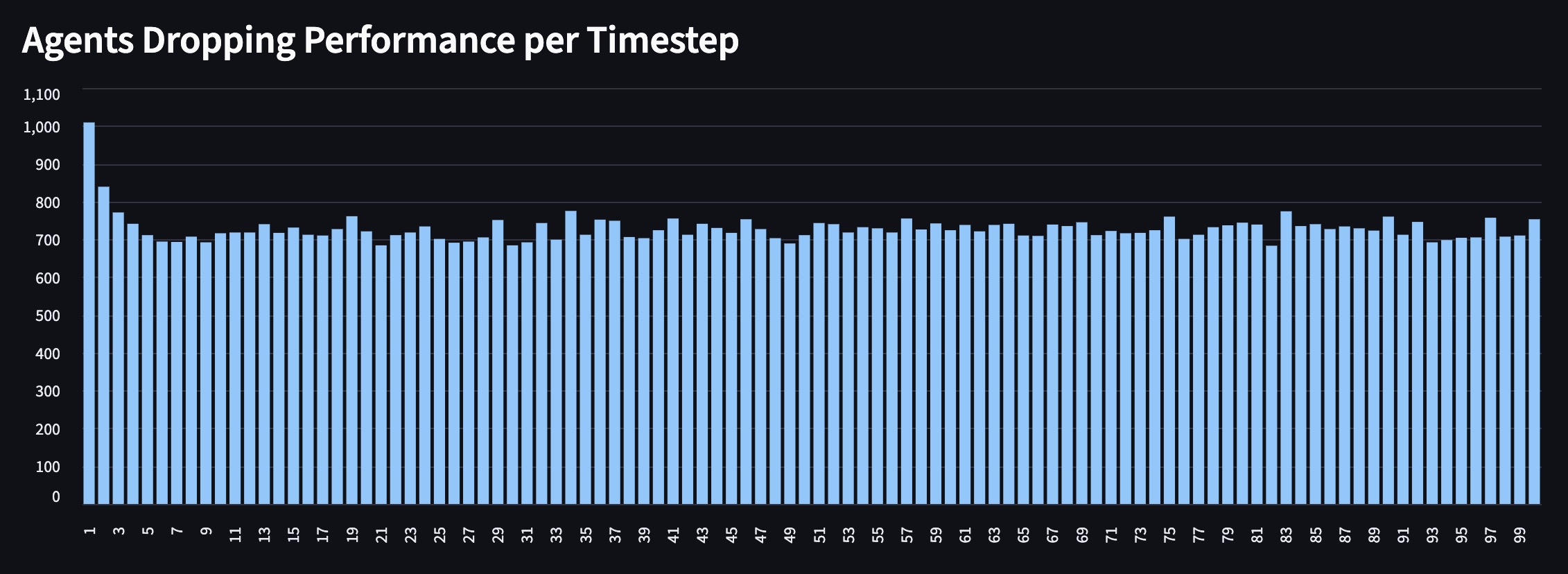}
    \caption{$\Delta P$ frequency distribution across all time-steps for selective demotion promotion.}
    \label{fig:selective_demotion_neg_tseries_transfer}
\end{figure}

\noindent\textbf{Negative shocks over time histogram} shows the bigger picture. At \(t=1\) there's an early spike with 1{,}009 promotions with \(\Delta P < 0\), but by \(t \approx 3\) the count has already fallen into a stable \(\sim 700{-}800\) band that persists through \(T=100\). This pattern reflects the \emph{transferable-skills} mechanics: adjacent roles are compatible, so most first moves produce \emph{small} declines that the \(\tau = 0.05\) screen intentionally lets pass, while the filter quickly removes the relatively few large misses in the first couple of ticks. After that ``quick cleanup'', three forces maintain the flat ``floor'': (i) a continual inflow of new L1 hires, a fraction of whom take small first-move losses; (ii) the tolerance only reverses large drops, so modest declines (\(-0.01\) to \(-0.04\)) persist by design; and (iii) cascade refill creates additional skip-level arrivals (e.g., \(L1 \rightarrow L4/L5\)) that aren’t re-tested for demotion in the same tick, adding steady trickle negatives. In contrast, under \emph{high-mismatch} the per-step negatives decline more gradually and settle lower (\(\sim 500{-}600\) after \(\sim 20\) steps) because there are many more large mismatches to prune, and sustained demotions progressively reshape the pool; in transferable-skills, fewer large errors and more mild ones mean the series stabilizes earlier but at a higher plateau.

\begin{figure}[H]
    \centering
    \includegraphics[width=\linewidth]{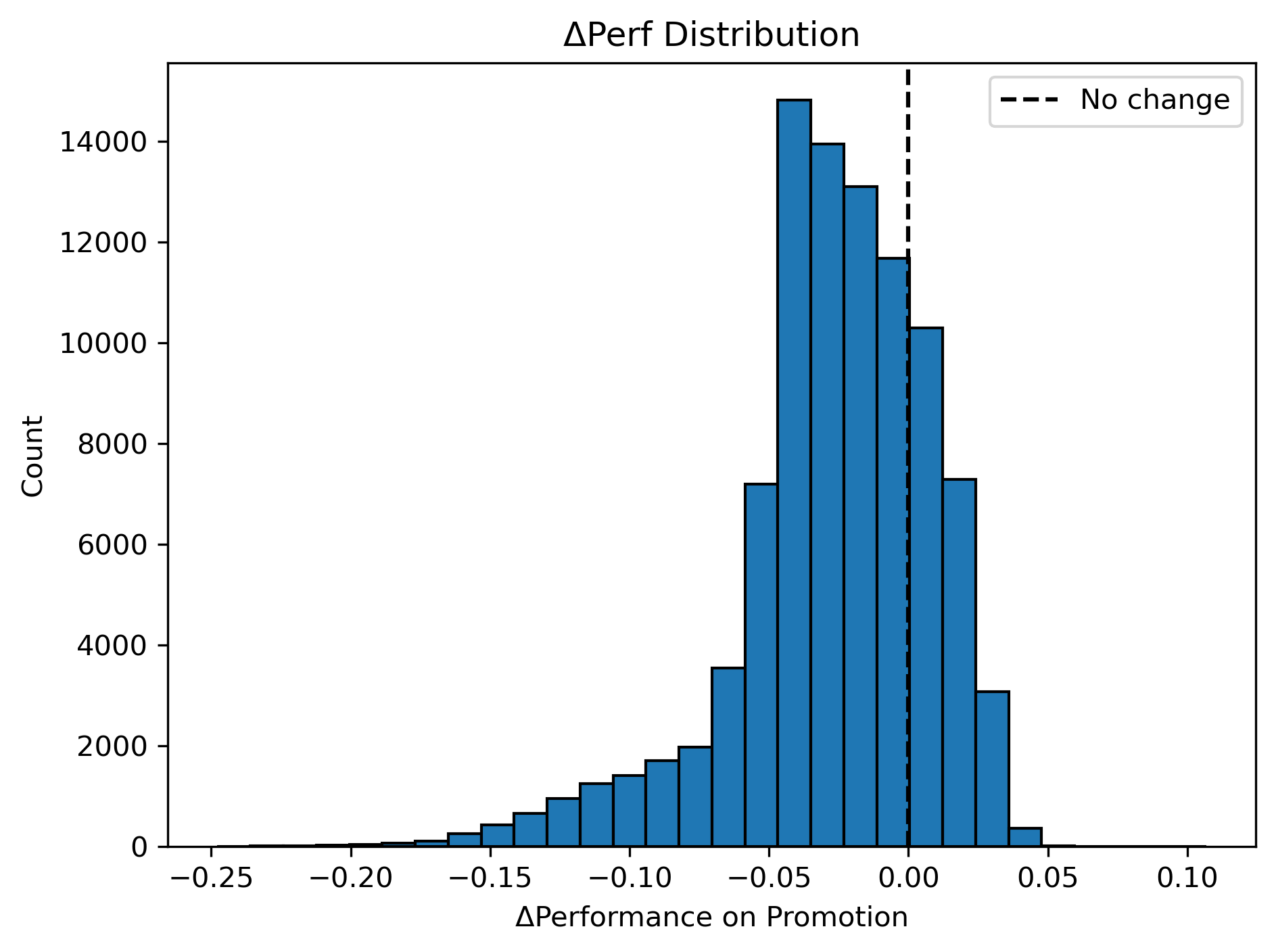}
    \caption{$\Delta P$ frequency distribution across all time-steps for selective demotion strategy.}
    \label{fig:selective_demotion_promotion_delta_transfer}
\end{figure}

\noindent The \textbf{$\Delta P$ histogram} under \emph{selective demotion} in \emph{transferable-skills} is \emph{left-skewed} but shallow, with a small negative center (mean \(-0.027\), median \(-0.024\)) and a high share of mild declines: 77.2\% of promotions are negative, yet only 16.6\% are ``large'' (\(\Delta P \leq -0.05\)), so most mass sits in \(-0.01\) to \(-0.04\). The tails are tight (1st/99th percentiles \(-0.142 / +0.032\); min/max \(-0.248 / +0.107\)), reflecting gentle inter-level reweighting and the \(\tau = 0.05\) screen trimming outliers. Compared with \emph{Merit} in the same regime (mean \(-0.043\), median \(-0.043\), 88.8\% negatives, with 43.3\% large losses), selective demotion shifts probability mass from severe losses into mild ones, producing a shallower left tail and a modest right tail. Relative to \emph{high-mismatch} selective demotion (which showed a more pronounced left tail and rarer but larger extremes), the transferable-skills histogram is tighter and closer to zero throughout which is consistent with smaller weight shocks and less frequent demotion.

Let's take an example of an agent to to understand the impact of selective demotion:

\paragraph{Case — Gentle cascade with small losses (Agent\#~272388).} Agent\#~272388 has the following skills:

\begin{verbatim}
{
  "tech": 0.9809747076184643,
  "management": 0.8367465850953305,
  "compliance": 0.8045381174430996,
  "soft_skills": 0.8610815684776533
}
\end{verbatim}

\noindent The agent joins the organization at t=63 at L1 and due to his high tech and soft-skills he performs very well (0.969) and is promoted to L2 at next timestep (t=64). But immidiately due to skills aligning well with L2 due to transferable-skills regime (tech 0.8, management 0.1, soft-skills 0.1) and vacancies at L3 the agent is directly promoted to L3. At L3 the agent's performance drops to 0.9297 due to increased management and compliance requirements. Which is lower but it is not filtered by our tollerence $\tau=0.05$ . Due to this impressive performance at L3, he is promoted to L4 at the next timestep (t=65) where again performance drops to 0.8929. Again this is not filtered by our tollerence $\tau=0.05$. Finally at t=67 the agent is promoted to L5 where his performance drops even more to 0.8584. Again the drop is less than our tollerence $\tau=0.05$ and therefore the agent is remains at L5 till the end of simulation (t=100).

\begin{figure}[H]
    \centering
    \includegraphics[width=\linewidth]{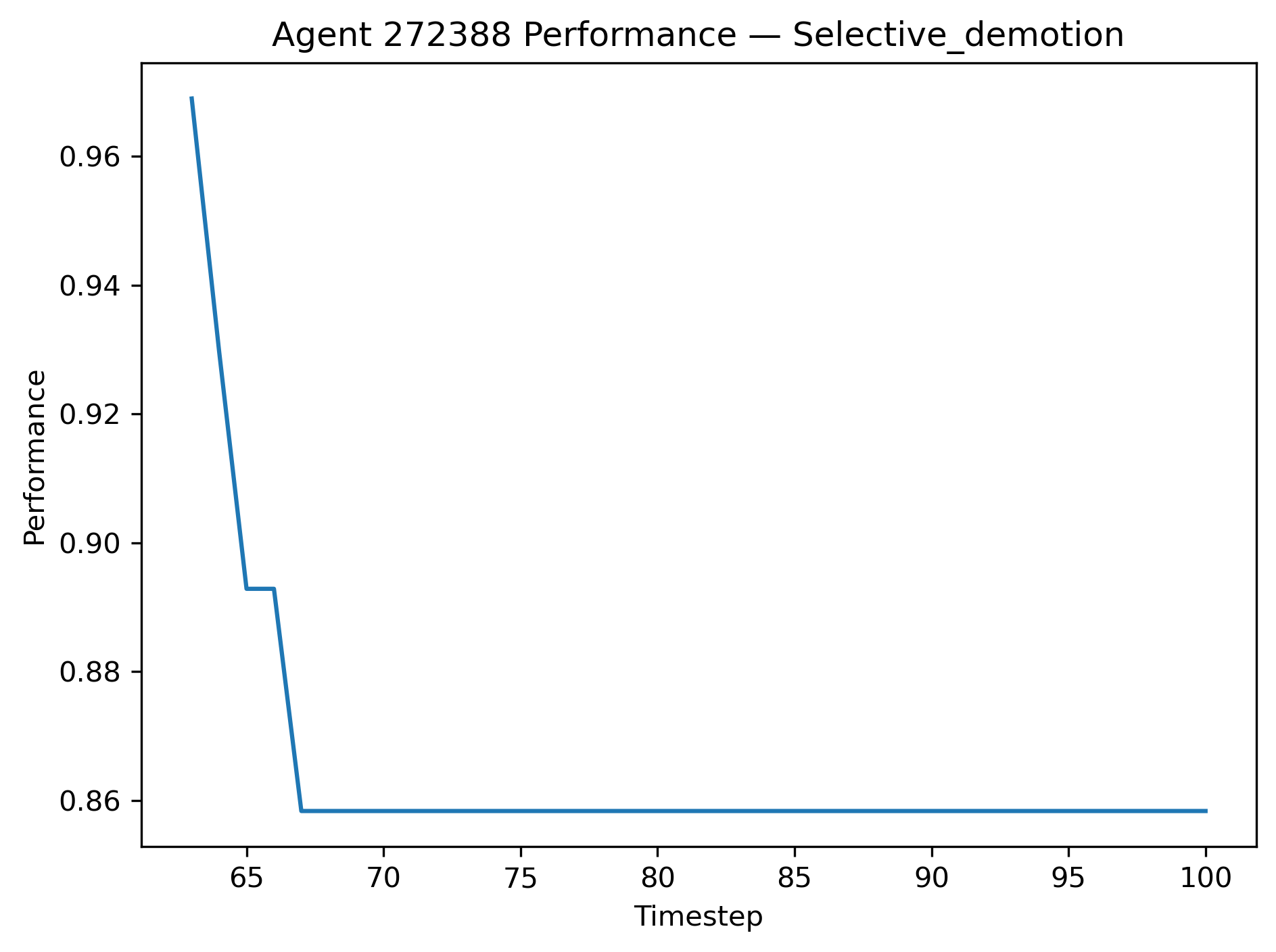}
    \caption{Agent 272388's performance over time for selective demotion strategy .}
    \label{fig:selective_demotion_agent_272388}
\end{figure}

\noindent \textbf{Takeaway: } Agent~272388 is a compact example of the \emph{transferable-skills + selective-demotion} diagnostics taken together. His path (\(L1 \rightarrow L3 \rightarrow L4 \rightarrow L5\)) includes a same-tick skip and a sequence of \emph{small} declines (\(\Delta P \approx -0.039, -0.037, -0.035\)), which matches the \emph{by-level counts heatmap} where negatives dominate even at the higher tiers of the organizational ladder, driven largely by \emph{mild} drops rather than catastrophic ones. It also mirrors the \emph{mean $\Delta P$ heatmap}, where adjacent steps are slightly negative and skip-levels are more fragile, yet here the losses stay sub-threshold. It aligns well with the \emph{histogram’s} shallow left mass (\(-0.01\) to \(-0.04\)) created by the \(\tau = 0.05\) rule pruning only the tail of large misses. His strong management/compliance (0.84/0.80) makes him ``portable'' enough to be re-selected in the cascade (explaining the logged \(L1 \rightarrow L3\) skip), but the tech\(\downarrow \rightarrow\)management/compliance\(\uparrow\) reweighting ensures each jump is a small net loss. In short, this case ties the figures together: selective demotion curbs extremes but tolerates small declines in transferable-skills regime explaining \emph{left-skewed} but shallow, with a small negative center $\Delta P$ histogram and only slightly negative heatmap of Mean $\Delta P$.

\paragraph{Interpretation.} In \emph{transferable-skills} regime, selective demotion behaves as a \emph{variance trimmer} rather than a mean shifter: it removes severe misfits but tolerates the many \emph{small} mismatches induced by gentle reweighting due to this regime. This explains why negatives outnumber positives at nearly every destination (heatmap counts), adjacent-path means remain slightly below zero. The per-step negatives rapidly stabilize at a higher plateau (\(\sim 700{-}800\)) after an early cleanup compared to \emph{high-mismatch} regime (\(\sim 500{-}600\)). The histogram’s shallow left skew (\(|\Delta P| < 0.05\) mass) is exactly what the \(\tau\)-filter is designed to produce. Mechanistically, demotion is executed \emph{before} refill, so cascade promotions that follow are not re-tested the same tick; this yields more skip-level arrivals (including \(L1 \rightarrow L5\)), preserves mild losses, and leaves a pocket of negatives. The lone systematic exception, \(L2 \rightarrow L5\) (positive on average), reflects strong selection on \emph{post-move} performance at \(L2\), which enriches for portability prior to the leap due to already selecting for high-management/compliance. Net effect: a controlled, shallow Peter signal---efficiency follows Merit’s smooth, concave rise but lands slightly lower because the rule trades throughput (fewer promotions and some riskier skips) for robustness (far fewer large drops). This mitigation is less critical in transferable-skills, where the Peter Principle is milder, and yields a slightly poorer efficiency than Merit due to lower number of promotions.

\noindent \textbf{Merit-with-training.} (Mitigation; see Sec.~\ref{sec:mitigations}.) Promotions are by \emph{merit} (current-role performance), and each 
just-promoted agent receives a \emph{one-shot training update} on tech and management: 
\nolinebreak
\[
C_{\text{new}} = \min\{1, \; C_{\text{old}} + \ell(C_{\text{old}})\}.
\]
where $\ell(C)$ is learning rate given by:
\nolinebreak
\[
\ell(C) = k \, C \, (1 - C),
\]
\nolinebreak

\noindent In the \emph{transferable-skills} regime, this post-move training amplifies upward portability: skills required are similar across levels, so boosting tech/management immediately after a merit-based promotion helps the agent adapt to the new role’s slightly reweighted demands. The efficiency path is a smooth, concave rise with \emph{no} early dip, increasing from $E_0 = 0.4807$ to $E_{100} =0.6331$. Half of the total gain is reached by $t \approx 26$ and $\sim 80\%$ by $t \approx 58$, then the curve flattens as $C(1-C)$ naturally tapers at higher skill levels and the pool saturates. Mechanistically, training boosts management (and tech) just as its weight increases at higher levels, enhancing portability and reducing mismatch. Compared to \emph{Merit} in the same regime ($E_{100}=0.5853$), merit-with-training delivers $+46\%$ \emph{more} total gain (0.1524 vs.\ 0.1046), reflecting that small, well-timed boosts to management (and some tech) are rewarded repeatedly as roles become more managerial with altitude. The plateau occurs due to three ceilings: compliance (untrained) becomes more salient near the top, the learning rate $C(1-C)$ naturally tapers at higher skill levels, and the large $L_{1}$ block ($40\%$) with near-uniform inflow anchors the organizational average.

\noindent We observe 109{,}000 promotions over ($T=100$): 36{,}972 with \(\Delta P < 0\) (33.9\%) and 72{,}028 with \(\Delta P > 0\) (66.1\%). Compared to pure Merit in the same regime (88.8\% negatives), training reduces the negative share by \(\approx 54.9\) percentage points (\(\sim 61.8\%\) relative drop).

\begin{figure}[H]
    \centering
    \begin{subfigure}[t]{0.48\linewidth}
        \centering
        \includegraphics[width=\linewidth]{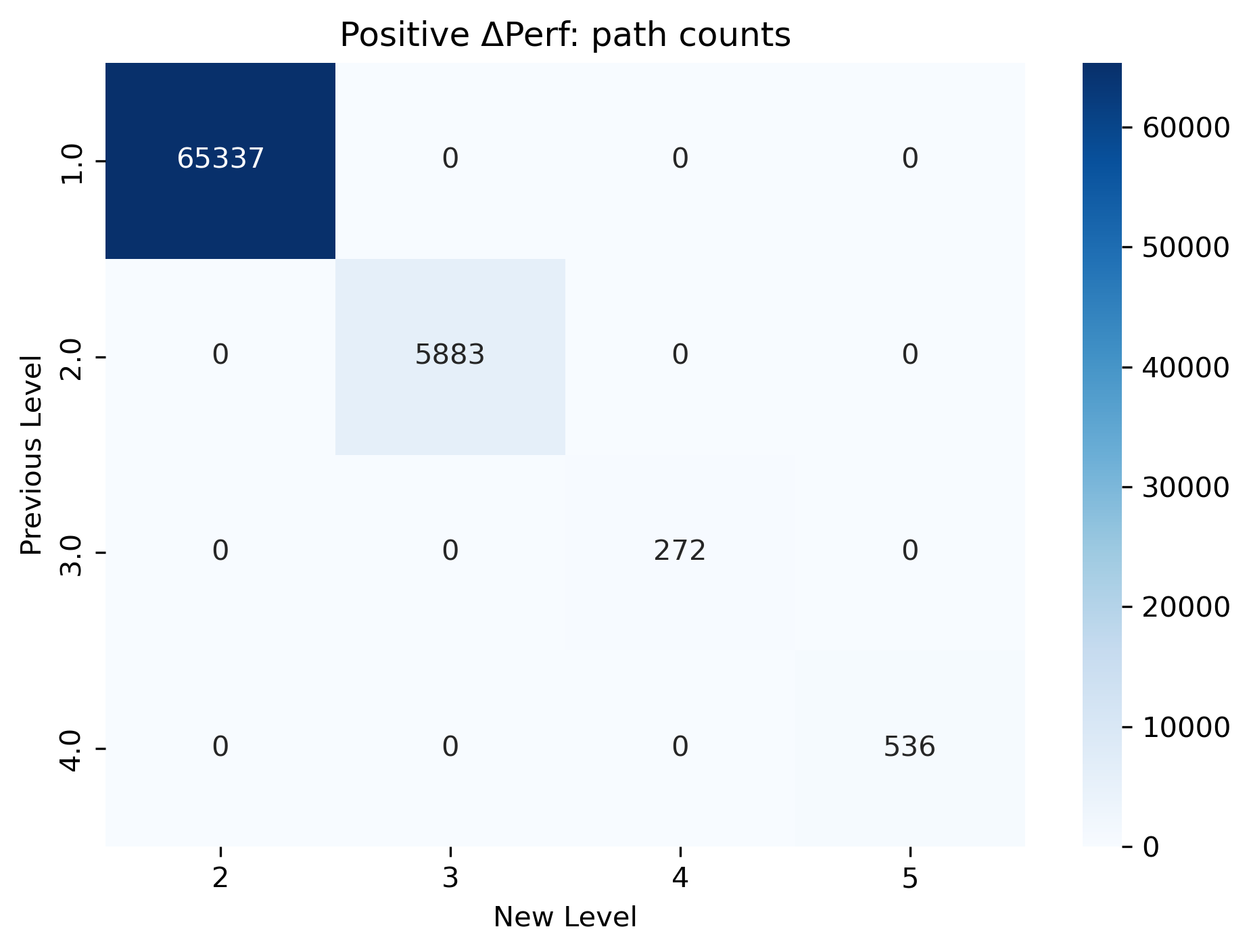}
        \caption{number of Promotions with $\Delta P > 0$ by level.}
        \label{fig:merit_training_pos_transfer}
    \end{subfigure}
    \hfill
    \begin{subfigure}[t]{0.48\linewidth}
        \centering
        \includegraphics[width=\linewidth]{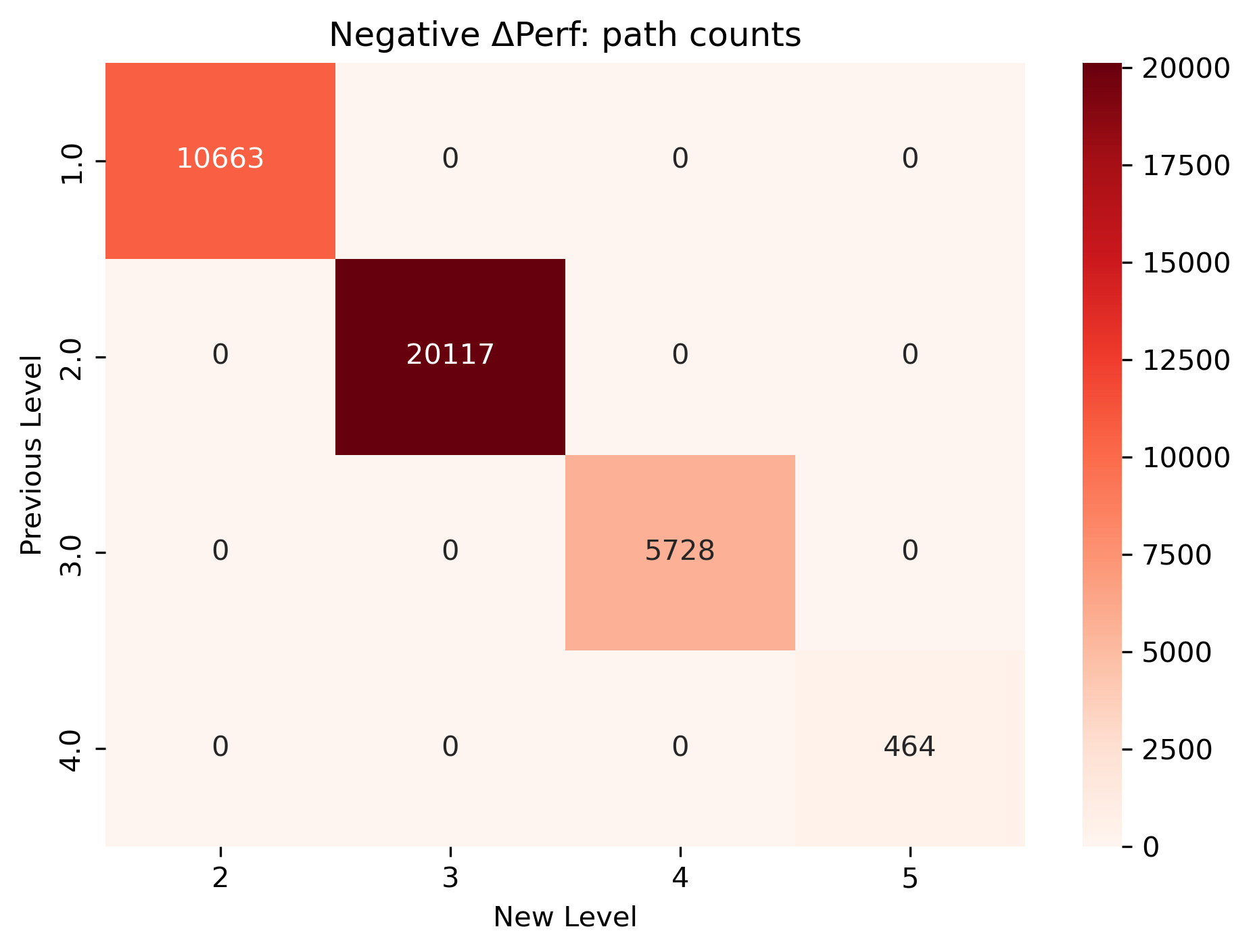}
        \caption{ number of Promotions with $\Delta P < 0$ by level.}
        \label{fig:merit_training_neg_transfer}
    \end{subfigure}
    \caption{Heatmap of promotion outcomes under merit-with-training in the transferable-skills regime.}
    \label{fig:merit_training_deltas_transfer}
\end{figure}

\noindent\emph{Heatmaps (Fig.~\ref{fig:merit_training_deltas_transfer}).} Under \textbf{merit-with-training} in the \emph{transferable-skills} regime. At lower levels (L1$\to$L2), we have more promotions with $\Delta P > 0$ than $\Delta P < 0$. This is due to the transferable-skills: tech remaining important (0.8 instead of 0.9) and training improves the tech skills of promoted agents which helps them perform well in the new role. Along with that training also boosts management which is slightly more important at L2 (0.1) but this early management boost will help them more at higher levels when management becomes crucial. As we move higher, L2$\to$L3 we now have more promotions with $\Delta P < 0$ than $\Delta P > 0$. This is due to added compliance (0.1) which training does not affect. And even with a small extra management weight (0.15) many profiles take mild losses. The imbalance is strongest at \(L3 \rightarrow L4\) (272 vs.\ 5{,}728), where tech drops further (0.65\(\rightarrow\)0.40) and compliance doubles (0.10\(\rightarrow\)0.20), overwhelming the modest management gain (0.15\(\rightarrow\)0.20) and the training boost. At the top step \(L4 \rightarrow L5\), counts approach balance with a slight positive edge (536 vs.\ 464): management weight jumps (0.20\(\rightarrow\)0.40) and the on-promotion training directly augments it, while survivorship leaves a pool with better compliance, offsetting the additional compliance demand (0.20\(\rightarrow\)0.30). Net: training flips the first move positive, but mid-ladder compliance remains the bottleneck; by \(L5\), the large management increase and selection effects restore near parity.

\begin{figure}[H]
    \centering
    \includegraphics[width=\linewidth]{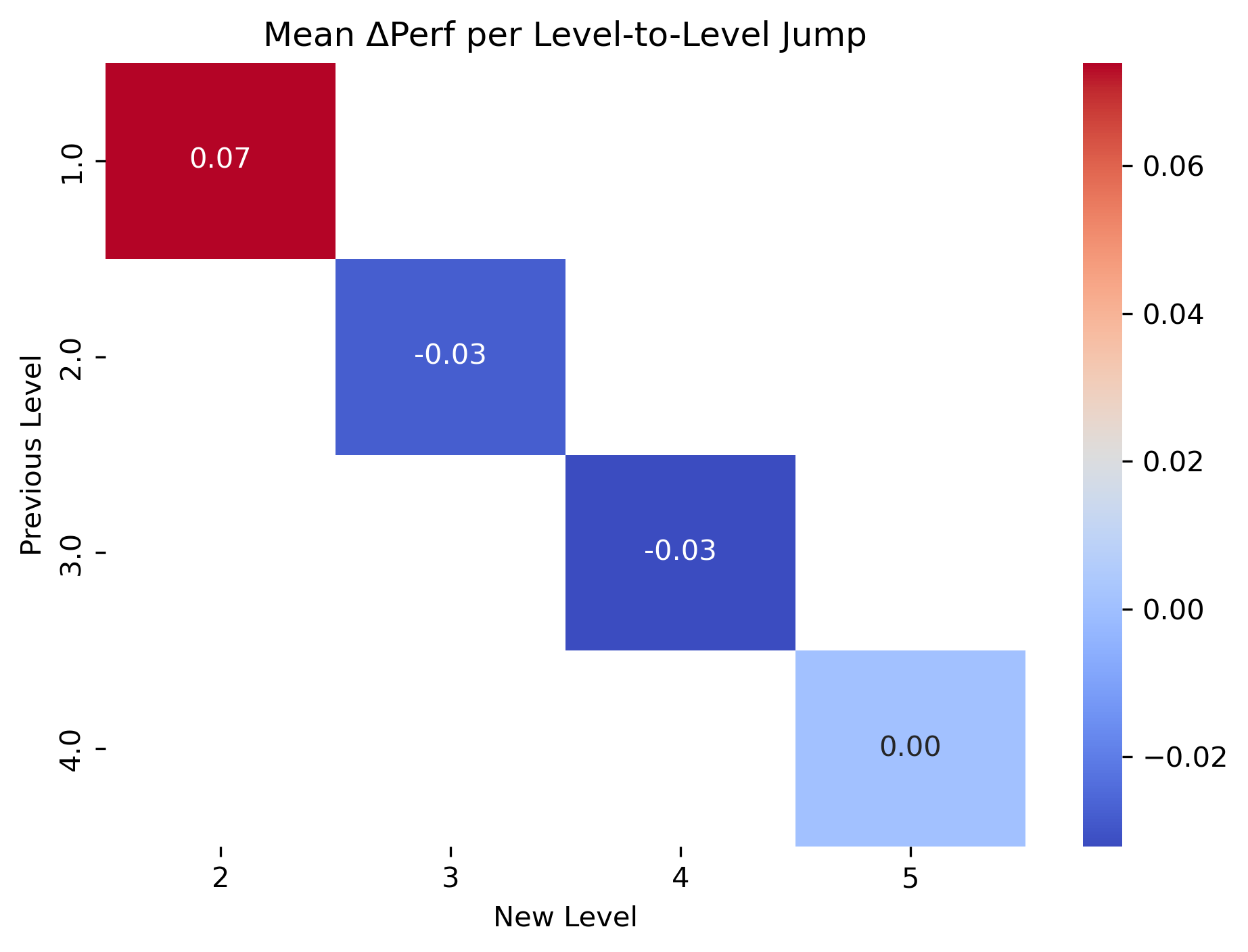}
    \caption{Mean $\Delta P$ per Level-to-Level jump for merit-with-training promotion strategy.}
    \label{fig:merit_training_heatmap_transfer}
\end{figure}


\noindent The \textbf{mean $\Delta P$ heatmap} (Fig.~\ref{fig:merit_training_heatmap_transfer}) shows a similar pattern that \emph{training softens and even reverses} early Peter shocks in the \emph{transferable-skills} regime. The entry step flips \emph{positive} from \(L_1 \to L_2\) (\(\bar{\Delta P} \approx 0.07\) vs.\ -0.04 in Merit), because the one-shot update boosts \emph{management} precisely when its weight rises (\(0.0 \to 0.1\)) and keeps \emph{tech} high (\(0.90 \to 0.80\)). In the mid-levels, residual negatives persist where \emph{compliance} tightens and training does not act on it—\(L_2 \to L_3\) (\(\bar{\Delta P} \approx -0.03\) vs.\ -0.06 for Merit) as compliance appears (\(0.10\)) and tech falls (\(0.80 \to 0.65\)). The average loss is smaller than under the high-mismatch regime as well (\(-0.11\)), as the tech and management weights are closer to the previous levels. \(L_3 \to L_4\) (\(\bar{\Delta P} \approx -0.03\) vs.\ -0.06 for Merit) as tech drops again (\(0.65 \to 0.40\)) and compliance doubles (\(0.10 \to 0.20\)). At the top, the big management jump (\(0.20 \to 0.40\)), along with selection effects and the training boost, nearly cancels the tech downshift and higher compliance burden: \(L_4 \to L_5\) (\(\bar{\Delta P} \approx +0.001\), \(+0.00\) in the figure, vs.\ -0.01 for Merit). Net: training turns the first move into a gain and halves the size of mid-ladder losses; with diminishing \(C(1-C)\) increments, the advantages taper, yielding a smooth concave rise that outperforms Merit overall.

\begin{figure}[H]
    \centering
    \includegraphics[width=\linewidth]{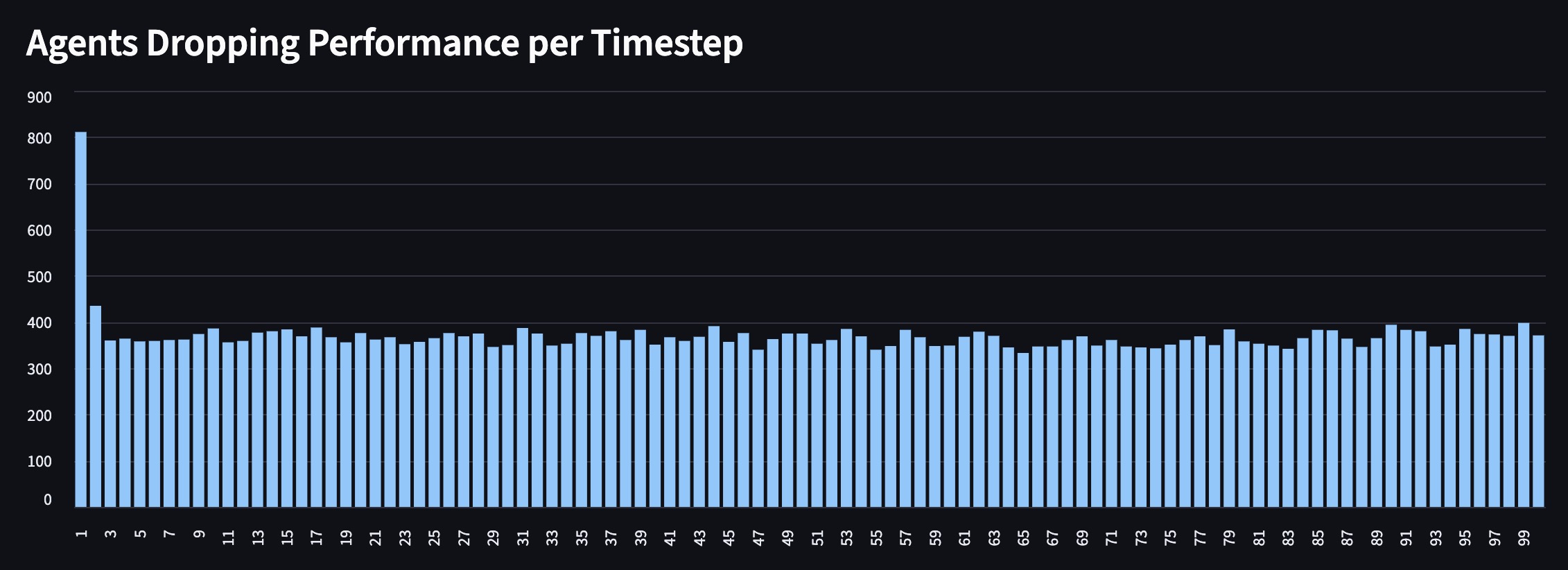}
    \caption{$\Delta P$ Number of promotions with $\Delta P<0$ at each timestep.}
    \label{fig:merit_training_neg_tseries_transfer}
\end{figure}

\noindent The \textbf{negative shocks over time histogram} shows the bigger picture. We see brief spike followed by very fast stabilization: at \(t=1\) there are 811 promotions with \(\Delta P < 0\), at \(t=2\) this already falls to 435, and from \(t \approx 3\) onward it settles into a flat \(\sim 300{-}400\) band through \(T=100\). Mechanistically, training flips a large share of \(L1 \rightarrow L2\) moves \emph{positive} on impact (management boost where its weight first appears, tech still high), so the initial wave of negatives is smaller and burns off quickly. What remains are mostly \emph{mild} losses from mid-ladder steps where compliance rises and training does not act (\(L2 \rightarrow L3\), \(L3 \rightarrow L4\)). The one-shot update also has diminishing returns (\(C(1-C)\) shrinks as skills approach 1), which prevents the floor from collapsing to zero even as large drops become rare. 

\begin{figure}[H]
    \centering
    \includegraphics[width=\linewidth]{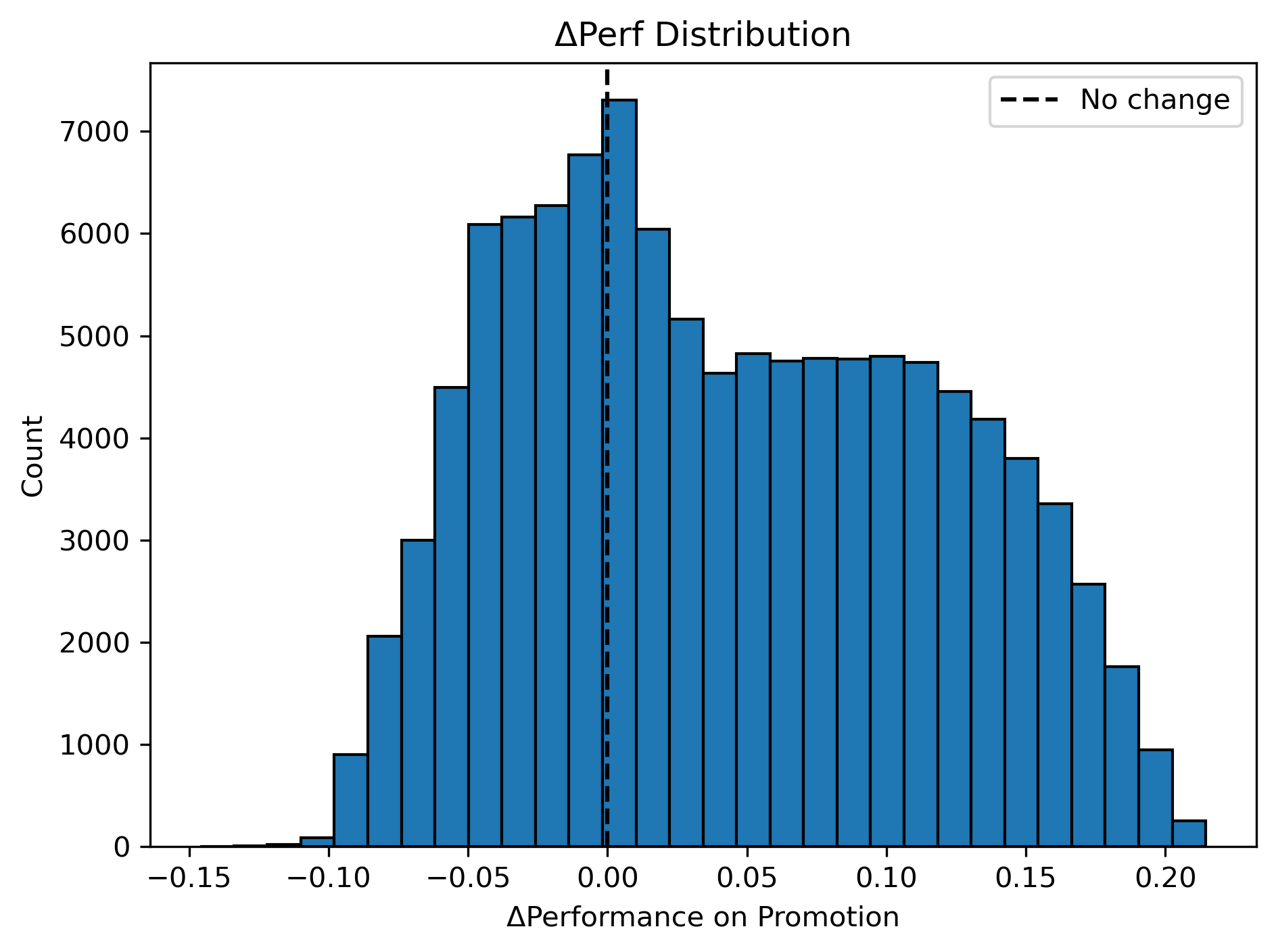}
    \caption{$\Delta P$ frequency distribution across all time-steps for merit-with-training strategy.}
    \label{fig:merit_training_promotion_delta_transfer}
\end{figure}

\noindent\textbf{$\Delta P$ histogram} under \emph{Merit-with-training} in \emph{transferable-skills} regime is \emph{right-skewed} with a positive center: Mean \(\approx +0.043\), median \(\approx +0.035\), mildly right-skewed, and shows a truncated left tail (only 9.6\% with $\Delta P \leq -0.05$); by contrast, \emph{Merit} shows mean \(-0.043\), median \(-0.043\), left-skewed, with 88.8\% negatives, and 43.3\% with \(\Delta P \leq -0.05\). This histogram is also \emph{thick on the right} (large gains $\geq$ 0.05 = 44.6\% vs.\ 0.008\% under Merit), while tails overall stay tight (1st/99th percentiles \(=-0.085/+0.192\); min/max \(=-0.146/+0.215\)). Mechanistically, the one-shot update---applied \emph{at} promotion---boosts tech and especially management right where those weights remain high or rise with levels, flipping \(L1 \rightarrow L2\) to positive and cushioning many subsequent mid-level moves; residual left mass comes from mid-ladder steps where \emph{compliance} grows and training does not act. Repeated promotions mean repeated updates resulting in higher tech and management skills, and survivorship concentrates portability, pushing more mass into the moderate right tail while truncating severe losses.

\paragraph{Interpretation.} In \emph{transferable-skills} regime, merit-with-training behaves as a \emph{mean and variance improver}: the one-shot update boosts tech and management just as their weights rise, enhancing upward portability and flipping the first move positive. Due to this training, the efficiency curve climbs smoothly with no early dip ending up higher than Merit by \(\sim 46\%\) in total gain. The by-level counts and mean-heatmap confirm the mechanism: mid-ladder paths (\(L2 \rightarrow L3\), \(L3 \rightarrow L4\)) remain negative-heavy because \emph{compliance} grows in salience and training does not touch it, but the average losses are small---consistent with the histogram’s positively centered, mildly right-skewed shape and truncated left tail (few large drops). The time-series shows the same story dynamically: an early, modest spike in negatives turns to a low, flat floor (\(\sim 300{-}400\) per step) by \(t \approx 3\) which is sustained by continuous L1 inflow, compliance bottlenecks, and diminishing \(C(1-C)\) increments---not by accumulating misfits. Selection further amplifies portability at the top (\(L4 \rightarrow L5\)): survivors arrive with higher management and adequate compliance, so the large management weight jump nearly neutralizes the tech downshift and compliance burden. Overall, training shifts the whole distribution of promotion shocks to the right and compresses the left tail; it does not eliminate the mid-ladder compliance drag, but it prevents persistent Peter accumulation and delivers a faster, smoother, higher efficiency path than Merit in this regime. In transferable-skills, where role demands shift gently, the merit-with-training mechanism is especially effective in mitigating the Peter Principle. The first step turns positive and severe misfits are rare, so there’s no accumulating drag. A small, residual Peter effect remains mid-ladder (\(L2 \rightarrow L3\), \(L3 \rightarrow L4\)) because compliance rises and training doesn’t touch it, but it shows up as brief, mild dips rather than a persistent stall.

\section{Mitigations}\label{sec:mitigations}

The Peter Principle posits that in hierarchical organizations, employees tend to be promoted based on success in their current role until they reach a level at which they are no longer competent. In other words, a star performer in one job may become an ineffective manager when promoted, because the skills that made them excel in the old role are different from those needed in the new role. The Peter Principle presents a fundamental challenge to organizational design: how can firms recognize and reward high performers without systematically placing them in roles where they underperform? This section first reviews existing literature on strategies to mitigate Peter Principle effects, then presents empirical results from our two proposed interventions—selective demotion and merit-with-training—across both regime types.

\subsection{Literature Review: Strategies for Mitigating the Peter Principle}

The economics, management, and organizational psychology literature has proposed
several approaches to address promotion-induced inefficiencies, spanning
organizational design, selection mechanisms, and post-promotional support. The literature contains a mix of empirical papers, theoretical treatments, meta-analyses of human resource strategies, and computational studies that address ways to lessen promotion incongruence. In what follows, we review these strategies with an eye toward developing our own mitigation strategies.

The economics literature offers several strategies for mitigating meritocratic promotion inefficiencies. Lazear \cite{lazear2004peter} suggests that firms set a higher bar for promotion than just meeting the minimum threshold. That is, companies should inflate their promotion criteria to compensate for the expected drop due to the regression-to-the-mean effect. Fairburn and Malcomson \cite{fairburn2001performance} go a step further by arguing that firms should rely on subjective promotion strategies. These strategies foster the promotion of people who are well suited for the job by focusing on managerial skills, communication, decision-making capacity, and leadership qualities. They also propose creating institutions to limit influence-seeking and politics in the promotion process. These institutions prevent poorly suited candidates from lobbying for promotions for which they lack competency. Finally, both Lazear \cite{lazear2004peter} and Fairburn and Malcomson \cite{fairburn2001performance} propose separating compensation and promotion decisions.

Grabner and Moers \cite{grabner2013managers} also emphasize the importance of selection and promotion decision design in mitigating inefficiency stemming from meritocratic promotion. They find that having firms shift promotion weights from current-job performance to subjective assessments of potential mitigates inefficiencies when tasks differ. Similarly, Benson et al. \cite{benson2019promotions} suggest decoupling rewards and promotions as a mitigation strategy. In addition, they suggest using trial promotions and dual career ladders for more deliberate talent planning.

Meta-analysis studies provide two distinct mitigation strategies for inefficiencies resulting from promotional design. Hermelin et al. \cite{hermelin2007validity} show that assessment centers have moderate predictive validity for supervisory or managerial performance, and their predictive power is stronger for measures of potential than for simple promotions-to-role outcomes. Lacerenza et al. \cite{lacerenza2017leadership} use meta-analyses of leadership training to find substantial positive effects on learning, transfer, and results when programs use best practices. Similarly, coaching has positive effects on performance and skill development across organizational settings.

Computational studies such as those conducted by Pluchino et al. \cite{pluchino2010peter} and Farias et al. \cite{farias2021peter} model promotion systems and find unexpected results: in some stylized settings, randomized promotion or promoting not strictly the top performers can outperform meritocratic promotion if the skill sets required at the next level are uncorrelated with current performance.

\subsection{Selective demotion}
\paragraph{Introduction.} \textbf{Selective demotion} implements a ``trial--then--confirm'' approach to promotions by monitoring immediate post-promotion performance and reversing moves that result in substantial performance degradation. The strategy operates through a four-step mechanism:

\begin{enumerate}
    \item \textbf{Promotion:} Candidates are initially promoted using standard merit-based selection.
    \item \textbf{Performance Assessment:} Immediate change in post-promotion performance is evaluated using the following metric
    \[
        \Delta P = P_{\text{post}} - P_{\text{pre}},
    \]
    holding the agent's competence vector fixed while applying the new role's demand weights.
    \item \textbf{Demotion Trigger:} If
    \(\Delta P \leq -\tau\) (with \(\tau = 0.05\) in our implementation), the agent is immediately demoted back to their previous level.
    \item \textbf{Blacklisting and Refill:} Demoted agents are permanently blacklisted from re-promotion to prevent repeated failed attempts, and vacated positions are refilled using merit-based selection from the level below.
\end{enumerate}

\noindent This mechanism operationalizes the Peter Principle's core insight that competence at one level does not guarantee competence at the next, providing a structural ``safety valve'' against persistent mismatches.

\paragraph{Motivation.} Selective demotion addresses the fundamental limitation of traditional promotion systems: they lack mechanisms to reverse decisions when role-fit proves poor. As established in the introduction, the Peter Principle warns that promoting top performers can systematically place employees in roles where their competence no longer translates, thereby degrading overall efficiency. Traditional merit-based promotion, while effective at identifying current-role excellence, cannot predict performance when skill requirements shift substantially between levels.

\noindent The strategy emerged from our empirical observation that under merit-based promotion in high-mismatch regimes, 88.1\% of promotions result in immediate performance decreases (\(\Delta P < 0\)), with mean performance drops of \(-0.136\). Rather than accepting these mismatches as inevitable, selective demotion provides a corrective mechanism that preserves merit-based selection while protecting against its systematic failures.

\noindent The approach mirrors real-world practices such as probationary periods, trial appointments, and acting roles, where organizations test fit before confirming permanent placements. Our blacklisting mechanism prevents the ``thrashing'' that could occur from repeated failed promotion attempts, ensuring that agents settle at their level of competence.

\paragraph{Why it works.} Selective demotion mitigates the Peter Principle by introducing a feedback loop that corrects misaligned promotions before they can degrade organizational performance. It works through three complementary mechanisms that address different aspects of the Peter Principle:

\begin{enumerate}
    \item \textbf{Immediate Mismatch Correction.} By monitoring performance immediately after promotion, the strategy identifies and reverses the most severe cases of role-misfit before they can contribute to sustained organizational inefficiency. In our high-mismatch regime, this mechanism reduces the share of promotions with $\Delta P \leq -0.05$ from $43.3\%$ under pure merit to just $16.6\%$ under selective demotion.

    \item \textbf{Peter Mismatch Elimination.} The strategy directly addresses Peter Principle dynamics by removing agents who cannot perform adequately in higher roles. For example, when a role demands more management skills but the agent lacks sufficient management competence, they are demoted back to their previous level where their skills are better aligned with role requirements.

    \item \textbf{Blacklist Prevention of Repeated Failures.} The blacklisting mechanism prevents the same agent from being re-promoted to the same vacancy, ensuring that once an agent demonstrates poor fit (e.g., low management skills for a management-heavy role), they cannot repeatedly fail in the same transition. This eliminates cycling and ensures agents remain at levels where their competence vectors align with role demands.
\end{enumerate}

\noindent The strategy's effectiveness stems from its ability to distinguish agents who are at their ``level of incompetence'' and ensuring they return to their ``level of competence''. It simply removes Peter mismatches by identifying cases where the same skills that drove success at level~$\ell$ actively impede performance at level~$\ell{+}1$.

\paragraph{Results.} Our simulation results demonstrate that selective demotion significantly attenuates Peter Principle effects across both regime types, with particularly strong performance in high-mismatch environments:

\paragraph*{High-Mismatch Regime (Regime B)}
\begin{itemize}
    \item \textbf{Efficiency Improvement:} Achieves $E_{100} = 0.569$ ($+3.8\%$ vs.\ baseline), compared to Merit’s $E_{100} = 0.549$ ($+0.06\%$).
    \item \textbf{Promotion Quality:} Reduces negative promotions from $88.1\%$ to $61.8\%$, representing a $26.3$ percentage point improvement.
    \item \textbf{Severe Mismatch Reduction:} Cuts large performance drops ($\Delta P \leq -0.05$) from $43.3\%$ to $16.6\%$, a $61.7\%$ relative reduction.
    \item \textbf{Trajectory Pattern:} Shows brief early dip (minimum $E_{7} = 0.543$) followed by steady recovery, avoiding the prolonged efficiency plateau seen under pure Merit.
    \item \textbf{Temporal Dynamics:} Negative promotions decrease from $\sim 1000$ per timestep initially to a stable floor of $500$--$600$ per timestep after period 20, reflecting the combined effects of pool enrichment through blacklisting and continuous new hiring.
\end{itemize}

\paragraph*{Transferable-Skills Regime (Regime A)}
\begin{itemize}
    \item \textbf{Efficiency Performance:} Reaches $E_{100} = 0.582$ ($+21.1\%$ vs.\ baseline), slightly below Merit’s $E_{100} = 0.585$ ($+21.8\%$).
    \item \textbf{Promotion Quality:} Reduces negative promotions from $88.8\%$ to $77.2\%$, representing an $11.6$ percentage point improvement.
    \item \textbf{Severe Mismatch Reduction:} Dramatically trims large performance drops ($\Delta P \leq -0.05$) from $43.3\%$ to $16.6\%$, a $26.7$ percentage point reduction.
    \item \textbf{Path-Specific Effects:} Adjacent-level moves (L1$\to$L2, L2$\to$L3) show mean $\Delta P$ values of $-0.02$, compared to more severe drops under pure Merit (mean $-0.043$ overall).
    \item \textbf{Skip-Level Dynamics:} Cascade refill mechanism generates beneficial skip-level promotions (e.g., L2$\to$L5 with mean $\Delta P = +0.08$) for highly portable profiles, though most skip-level moves remain slightly negative due to gentle reweighting.
    \item \textbf{Temporal Dynamics:} Negative promotions stabilize more quickly at $700$--$800$ per timestep after early cleanup, higher than the high-mismatch regime due to gentler filtering requirements.
\end{itemize}

\paragraph*{Mechanism Effects Across Both Regimes}
The strategy operates as a \emph{variance trimmer} in transferable-skills environments and a \emph{mean shifter} in high-mismatch environments. In high-mismatch settings, heavy pruning at lower tiers progressively enriches upper organizational levels with management-and-compliance-heavy profiles, eventually flipping upper-tier promotion outcomes toward positive. In transferable-skills settings, the mechanism removes severe outliers while tolerating mild performance decreases, maintaining Merit’s efficiency trajectory shape while providing robustness against catastrophic mismatches.

\paragraph{Real-Life Implications.} Selective demotion offers several practical advantages for organizational implementation by mimicking \emph{probationary periods} which are already common in many firms. We now outline how a firm would implement this in practice:

\begin{enumerate}
    \item \textbf{Probationary implementation.} Selective demotion directly parallels common organizational practices such as probationary periods, trial appointments, and acting roles. In our model, the $\tau=0.05$ threshold functions as an objective performance benchmark that determines whether a "trial promotion" is confirmed or reversed. This mirrors real-world 90-day reviews, interim appointments, and acting manager positions where organizations test fit before permanent placement.

    \item \textbf{Technology firms and sales organizations.}  In environments with sharp skill transitions (IC-\allowbreak to-\allowbreak manager, sales-\allowbreak to-\allowbreak management), selective demotion provides critical protection against the well-documented pattern where top individual contributors become ineffective managers. Our results suggest that implementing trial periods with clear performance thresholds could prevent approximately 27\% of promotion-induced performance declines.

    \item \textbf{Stigma-mitigating design choices.} Rather than formal ``demotion,'' organizations can implement functionally equivalent approaches that maintain employee dignity while optimizing role fit:
    \begin{itemize}
        \item Acting appointments with confirmation contingent on performance metrics.
        \item Dual-track career progression allowing high performers to advance without management responsibilities.
        \item Pay-protected reassignments that maintain compensation while optimizing role fit---recognizing that extraordinary performance benefits the company economically even if the employee is not fit for the next hierarchical level.
        \item Enhanced compensation packages (monetary or non-monetary benefits such as flexible work arrangements, professional development opportunities, or specialized project assignments) for exceptional performers who excel in their current role but are not suited for promotion.
    \end{itemize}

    \item \textbf{Threshold calibration.} The tolerance parameter $\tau$ offers organizations a tunable lever for managing promotion risk:
    \begin{itemize}
        \item \textbf{Smaller $\tau$:} Increases demotion rate and early organizational churn, but more aggressively eliminates persistent mismatches. Higher risk of reacting to performance noise, potentially demoting agents experiencing temporary adjustment challenges.
        \item \textbf{Larger $\tau$:} Reduces demotion rate and organizational disruption, but allows more embedded mismatches to persist. Efficiency gains shrink, particularly in transferable\allowbreak -skills environments where modest performance drops might indicate normal role transition.
        \item \textbf{Optimal Calibration:} The trade-off represents classic Type-I versus Type-II error management in misfit detection. Risk-averse organizations might prefer higher $\tau$ to minimize false positives (demoting good fits), while performance-driven organizations might choose lower $\tau$ to aggressively eliminate misfits despite occasional false positives.
        \item \textbf{Level-Specific Thresholds:} Organizations can implement differentiated tolerance parameters across organizational levels to account for varying transition difficulties. For example:
        \begin{itemize}
            \item \textbf{Lower thresholds for sharp transitions:} $\tau = 0.03$ for L1→L2 moves where technical skills must translate to management responsibilities, requiring more aggressive mismatch detection.
            \item \textbf{Higher thresholds for similar roles:} $\tau = 0.07$ for L3→L4 transitions where role requirements are more aligned, allowing for normal adjustment variation.
            \item \textbf{Rationale:} This approach recognizes that some organizational transitions inherently involve larger skill reweighting (high\allowbreak-mismatch characteristics) while others represent incremental progression (transferable\allowbreak -skills characteristics), even within the same organizational hierarchy.
        \end{itemize}

    \end{itemize}

    \item \textbf{Implementation timing and scale.} Our results indicate that selective demotion is most valuable during organizational growth phases when promotion volume is high. The mechanism's effectiveness increases with scale, as larger promotion cohorts provide more opportunities for improved selection through the blacklisting mechanism.
\end{enumerate}

\noindent Overall, selective demotion is practically deployable because it formalizes what many organizations already do informally: confirm fit with transparent criteria, reverse when necessary, and recognize high performance even when optimal contribution occurs at the current level.

\subsection{Merit-with-training}
\paragraph{Introduction.}
Merit-with-training combines standard merit-based promotion selection with immediate post-promotion skill development on trainable dimensions. The strategy operates through a two-phase process:

\begin{enumerate}
    \item \textbf{Selection:} Candidates are ranked and promoted using traditional merit-based criteria, selecting those with highest current-role performance.
    \item \textbf{Training Intervention:} Immediately upon promotion, agents receive a targeted, one-shot competency boost on technical and management skills using a logistic-derivative learning function:
    \[
    \ell(C) = k \cdot C (1-C), \quad \text{where } k=1,
    \]
    \[
    C_{\text{new}} = \min\{1,\, C_{\text{old}} + \ell(C_{\text{old}})\}.
    \]
\end{enumerate}

\noindent This learning rate function derives from the derivative of a sigmoid (logistic) learning curve, as established in Section~\ref{interventions}. The underlying learning dynamics follow a bounded S-shaped competence trajectory:
\[
    C(t) = \frac{C_{max}}{1 + e^{-k(t - t_{0})}},
\]
where $C_{max} = 1$ represents the performance ceiling and $k = 1$ normalizes the growth rate. Differentiating this sigmoid function yields the instantaneous learning rate
\[
    \frac{dC}{dt} = kC(1-C),
\]
which captures the empirically observed pattern that learning progress is driven by successful task executions rather than raw trial counts~\cite{leibowitz2010sigmoid}.

\noindent The logistic-derivative form $\ell(C) = C(1-C)$ ensures that agents with mid-level competence ($C \approx 0.5$) benefit most from training (receiving the maximum increment of $0.25$), while novices ($C \approx 0$) and near-experts ($C \approx 1$) show minimal improvement due to diminishing returns near the performance boundaries. This mirrors real-world training effectiveness where foundational scaffolding is required for beginners, peak performers approach natural ceilings, and intermediate-level agents experience the steepest learning gains from structured interventions.

\noindent The training scope is deliberately restricted to technical and management skills only. We hold compliance and soft skills fixed because, in our organizational setting, they reflect ingrained regulatory knowledge and interpersonal dispositions that change slowly relative to the immediate post-promotion onboarding window. By contrast, technical and managerial competencies are directly teachable through targeted workshops, coaching sessions, and structured practice exercises\allowbreak — precisely the interventions that organizations can implement immediately after promotion to accelerate role transitions and close emerging skill gaps.

\paragraph{Motivation.}
Merit-with-training addresses the core tension in organizational promotion systems: the need to maintain performance-based advancement while mitigating the skill misalignments that create Peter Principle effects. Our model analysis revealed that merit-based promotion generates substantial post-promotion performance drops (mean $\Delta P = -0.136$ in high-mismatch regimes) precisely because the skills that drive excellence at level $\ell$ may not align with requirements at level $\ell+1$.

\noindent Rather than abandoning merit-based selection, this strategy augments it with targeted intervention at the moment of greatest vulnerability—immediately after promotion when agents face new role demands. The approach was motivated by three key insights from our analysis:

\begin{enumerate}
    \item \textbf{Timing Criticality:} Performance drops occur immediately upon promotion as the same competence vector encounters different role weights, suggesting that early intervention could be highly effective.
    \item \textbf{Trainable Skill Selection:} We restrict training to technical and management skills based on practical implementation constraints rather than empirical transferability. Compliance and soft skills are held fixed because they reflect ingrained regulatory knowledge and interpersonal dispositions that develop slowly over longer horizons, whereas technical and managerial competencies can be directly enhanced through targeted workshops, coaching, and structured practice—precisely the interventions organizations can feasibly implement in immediate post-promotion onboarding windows.
    \item \textbf{Learning Dynamics:} The logistic-derivative form captures the realistic constraint that training effectiveness depends on baseline competence, preventing unrealistic skill gains while maximizing impact for agents at optimal learning points.
\end{enumerate}

\noindent The strategy reflects common organizational practices such as management boot camps, technical certifications, leadership development programs, and structured onboarding, providing a formal framework for optimizing these investments.

\paragraph{Why it works.}
Merit-with-training operates through three synergistic mechanisms that directly counter Peter Principle dynamics:

\begin{enumerate}
    \item \textbf{Immediate Gap Closure:} The post-promotion training boost targets the specific skills that become more heavily weighted at higher levels. In our high-mismatch regime, management weight increases from $0.0$ at L1 to $0.3$ at L2, and the training update provides precisely the management skill enhancement needed to improve fit with the new role profile.
    \item \textbf{Competence-Adaptive Learning:} The $\ell(C) = C(1-C)$ function reflects the natural learning dynamics where training effectiveness varies with baseline competence. This isn't a deliberate choice to favor intermediate performers, but rather a consequence of how skill acquisition fundamentally operates: agents with management competence $C = 0.5$ naturally receive the maximum boost of 0.25 because they are at the steepest part of the learning curve, while agents with very low ($C \approx 0$) or very high ($C \approx 1$) competence experience diminishing returns due to the inherent constraints of the learning process. Novices lack the foundational scaffolding to absorb advanced training effectively, while near-experts approach natural performance ceilings where marginal improvements become increasingly difficult to achieve. This pattern emerges organically from the logistic learning model and irrors real-world training effectiveness where mid-level performers show the greatest improvement.


    \item \textbf{Multiplicative Benefit Accumulation:} In the transferable-skills regime, where technical and management competencies remain valuable across multiple levels, the training boost compounds across subsequent promotions. An agent receiving management training at L2 benefits not only in their current role but also in future moves to L3, L4, and L5 where management weight continues to increase.
\end{enumerate}

\noindent The strategy's effectiveness stems from its ability to provide immediate, targeted skill enhancement precisely when agents face new role demands, while preserving the merit-based selection that rewards current performance. The choice to focus training on technical and management skills reflects practical implementation constraints—these dimensions can be meaningfully improved through short-term intensive interventions, whereas compliance and soft skills require longer development horizons. The core mechanism that addresses the Peter Principle problem is the timing and magnitude of the skill boost, not the specific dimensions chosen: by enhancing competence vectors immediately after promotion, the approach reduces the misalignment between agent capabilities and new role demands. Unlike corrective approaches, it invests in capability building before problems manifest rather than addressing failures after they occur.

\paragraph{Results.}
Merit-with-training demonstrates consistently strong performance across both regime types, with particularly impressive results in environments where technical and management skills maintain relevance across organizational levels:

\paragraph*{High-Mismatch Regime (Regime B)}
\begin{itemize}
    \item \textbf{Efficiency Achievement:} Reaches $E_{100} = 0.596$ ($+8.7\%$ vs.\ baseline), substantially outperforming pure Merit ($E_{100} = 0.549$).
    \item \textbf{Promotion Quality Transformation:} Reduces negative promotions from $88.1\%$ to $63.4\%$, representing a $24.7$ percentage point improvement.
    \item \textbf{Path-Specific Improvements:} L1$\to$L2 transitions show dramatically reduced mean performance drops ($\Delta P = -0.04$ vs.\ $-0.15$ under pure Merit).
    \item \textbf{Upper-Level Performance:} L3$\to$L4 and L4$\to$L5 moves become consistently positive ($+0.03$ and $+0.005$ respectively) due to training effects and role similarity.
    \item \textbf{Temporal Dynamics:} Shows smooth, monotonic efficiency improvement with no early dip phase. Brief spike in negative promotions at $t=1$ (811 agents) rapidly stabilizes to $300$--$400$ per timestep by $t=3$.
\end{itemize}

\paragraph*{Transferable-Skills Regime (Regime A)}
\begin{itemize}
    \item \textbf{Efficiency Excellence:} Achieves $E_{100} = 0.633$ ($+31.7\%$ vs.\ baseline), delivering $46\%$ more total gain than pure Merit.
    \item \textbf{Promotion Outcome Reversal:} Flips promotion outcomes to $66.1\%$ positive vs.\ $33.9\%$ negative, a complete reversal from Merit’s $88.8\%$ negative rate.
    \item \textbf{First-Move Success:} L1$\to$L2 transitions become net positive (mean $\Delta P = +0.07$) compared to Merit’s $-0.04$.
    \item \textbf{Sustained Benefits:} Maintains positive trajectory through mid-career with L4$\to$L5 moves nearly neutral ($\Delta P \approx 0.001$).
    \item \textbf{Learning Distribution Effects:} Agents with mid-range technical and management skills ($C \approx 0.4$--$0.6$) receive the largest training boosts and subsequently become the most successful at higher levels.
\end{itemize}

\paragraph*{Edge Cases and Limitations.}
In very high-mismatch environments or for agents with low baseline competence in trainable dimensions, training yields modest gains and some performance shocks persist. Most critically, the strategy cannot address compliance gaps that become binding at higher levels, creating structural bottlenecks where L2$\to$L3 transitions in high-mismatch regimes continue to generate negative outcomes despite management training. This limitation is exemplified by Agent~100114 (see \nameref{para:case100114_merit} for merit-with-training), who, despite receiving substantial technical and management training (skills reaching $C=1.0$ for both dimensions), suffers a large negative performance shock ($\Delta P=-0.268$) at the L2$\to$L3 transition due to very low compliance competence ($C=0.037$) when L3 demands $30\%$ compliance weight. As noted in our case study analysis, ``training helps somewhat \dots{} but it cannot fully offset the mismatch created by the sharp reweighting and new compliance requirement,'' highlighting that while training can mitigate some Peter shocks, it cannot eliminate all mismatches when unaddressed skills become critical at higher levels.

\paragraph{Real-Life Implications.}
In real organizations, merit-with-training mirrors practices commonly found in leadership development pipelines across sectors. After a merit-based promotion, firms operationalize this strategy by immediately enrolling newly promoted managers or specialists into role-specific, tightly sequenced onboarding and training programs. Typically, this involves:

\begin{enumerate}
    \item \textbf{Implementation in Practice:} Within days of promotion, companies enroll new managers in intensive 2--4 week workshops covering budgeting, performance reviews, and project coordination. Simultaneously, they deploy targeted technical modules through learning management platforms and assign experienced mentors for 60--90 days of coached feedback. This front-loaded approach concentrates the highest-impact training in the first month---precisely when our model predicts the largest performance gaps.

    \item \textbf{Sector-Specific Variations:} Technology firms focus these bootcamps heavily on the IC-to-manager transition, emphasizing management skills over technical depth. Healthcare organizations blend administrative training with clinical currency maintenance for newly promoted physician-administrators. Professional services firms combine enhanced technical expertise with business development skills for partner-track promotions. Academia and R\&D environments emphasize strategic thinking and advanced technical competencies that build upon existing knowledge bases.

    \item \textbf{System Integration:} Companies can embed these programs into existing HR workflows by using promotions as automatic triggers for training enrollment, deploying targeted curricula based on destination roles, and incorporating training readiness into succession planning decisions. Performance management systems track completion, and learning platforms deliver role-specific content based on the promotee's new organizational level and function.

    \item \textbf{ROI Justification:} Organizations can justify substantial training investments using our efficiency improvements as benchmarks ($+8.7\%$ in high-mismatch environments, $+31.7\%$ in transferable-skills environments), which provide clear performance improvements and reduced promotion failure rates that translate directly to measurable business value.

    \item \textbf{Recognized Limitations:} Practitioners acknowledge that one-shot training cannot address all skill gaps. Compliance requirements and soft skill development are typically handled through separate, longer-term programs, as these domains require extended development horizons beyond the immediate post-promotion window. The strategy works best when foundational interpersonal and regulatory competencies are already established.
\end{enumerate}

\noindent The strategy's strength lies in its preventive rather than corrective approach---investing in capability building before problems manifest rather than addressing failures after they occur. This aligns with organizational preferences for positive development interventions while delivering measurable efficiency gains through better role-person fit optimization.

\section{Discussion \& Conclusion}\label{sec:discussion}





\noindent Our comprehensive agent-based modeling investigation provides the first systematic, quantitative exploration of \emph{when} and \emph{why} the Peter Principle manifests in organizational hierarchies, revealing that its emergence depends critically on the alignment between promotion criteria and inter-level skill requirements. The central finding of this study is that the Peter Principle is not universal—it emerges predictably under specific organizational conditions that we can now characterize precisely.

\paragraph{The Fundamental Mechanism.}
The Peter Principle occurs when the most common \emph{merit-based promotion} systematically elevates agents whose competence vectors perform well under current role weights but are poorly aligned with destination role demands. Fundamentally, the Peter Principle arises when promotion criteria (in merit-based selection) are misaligned with the skill requirements of higher-level roles. In our model, performance is the dot product of an agent’s static competence vector with the role’s weight profile. Promoting the top performer at level $\ell$ means selecting an individual whose competence is finely tuned to the weight profile of that level. Unless the next level demands a similar mix of skills, that same competence vector will be a poor fit for the new role. In other words, when the best-performing agents at one level are elevated, their strengths under the old role weights often become weaknesses or insufficiencies under the new weights, leading inevitably to a performance drop. Our simulations confirm this mechanistic misalignment effect: holding an agent’s abilities constant and only changing the role demands produces an immediate performance shock upon promotion. In practical terms, employees are ``promoted to their level of incompetence''—the very skills that drove success in the prior job can hinder performance in the next.

\paragraph{} 
\noindent Crucially, the severity of Peter Principle effects depends on the organization’s \emph{role-profile regime}, i.e., how dramatically skill priorities shift between levels. In a high-mismatch regime, where each promotion entails a sharp reweighting of skills (e.g., technical skills give way to management and compliance in moving from engineer to manager), the misfit is extreme. For example, in our high-mismatch scenario the first promotion (Level~1$\to$2) caused an average performance drop of $\sim -0.15$, with 43\% of promotions resulting in a large ($>$5\%) performance decline. By contrast, in a transferable-skills regime (gradual skill shifts across levels), the same merit-based promotions yielded a much smaller average drop ($\sim -0.04$) and far fewer severe declines. Skill portability thus acts as a buffer: when higher roles still value the lower-role skills, promoted individuals remain competent. Nevertheless, even in the transferable-skills regime we observed a majority of promotions ($\approx 88\%$) causing at least a slight performance decrease. This underscores a key insight: whenever there is any change in role skill weights, a top performer’s current competence profile is unlikely to perfectly match the new role, making some drop virtually mathematically inevitable. The larger the shift in required skills, the more pronounced the Peter Principle effect. Another structural factor is promotion volume: organizations that promote many people rapidly (e.g., large cohorts moving up from entry levels) will see more aggregate efficiency loss simply because they generate more opportunities for mismatches. 

\paragraph{} 
\noindent In sum, the Peter Principle emerges as a systemic consequence of multi-level skill misalignment—especially acute in environments with sharp skill transitions between roles and high rates of upward mobility.

\subsection{Why Avoiding the Peter Principle Doesn't Guarantee Superior Efficiency}
A counterintuitive finding is that strategies which successfully avoid Peter Principle mismatches do not necessarily achieve superior organizational efficiency. One might assume that if promotions did not cause performance drops, the organization would perform better. Counterintuitively, avoiding Peter Principle mismatches does not guarantee higher efficiency. We found that promotion strategies explicitly designed to sidestep the Peter Principle---notably seniority-based promotions and random promotions---indeed eliminate the systematic post-promotion drop\footnote{\textbf{What we mean by “systematic” drops.} A drop is \emph{systematic} when the promotion rule itself makes negative post-promotion changes more likely for the promoted cohort than for the eligible pool; formally, when $E[\Delta P \mid \text{promoted}] < E[\Delta P \mid \text{eligible}]$ because selection is based on performance under level $\ell$ weights while the destination level uses different weights. Under \emph{random} or \emph{seniority} promotion, selection is (approximately) independent of destination-role fit, so the distribution of $\Delta P$ among promotees mirrors that of the pool (typically roughly symmetric, i.e., $\Pr[\Delta P<0]\approx \Pr[\Delta P>0]$), with any residual mean driven by the regime’s structure rather than the rule. By contrast, \emph{merit} selects agents optimized for level $\ell$; when weights shift at $\ell{+}1$, this induces a selection bias toward $\Delta P<0$—hence “systematic” drops.} in performance, yet they plateau at a lower overall efficiency than merit-based promotion. Under seniority or random rules, the organization avoids the drastic mismatches seen under pure merit selection, but it also forgoes the benefits of meritocratic sorting. 

For example, in the high-mismatch regime, a pure merit system suffered frequent drops ($88.1\%$ of promotions reduced performance) yet still ended up about breaking even in efficiency ($+0.06\%$ over baseline), whereas seniority-only promotions led to a $-3.1\%$ efficiency decline and random promotions to $-2.4\%$ (even though neither produced an early performance crash). Similarly, in the transferable-skills regime, merit-based policy achieved about $21.8\%$ efficiency gain ($88.8\%$ of promotions reduced performance), far outpacing seniority ($+8.0\%$) or random ($+11.2\%$) promotions. This \emph{``Seniority and Random Paradox''}---that ostensibly safer promotion methods underperform---stems from trade-offs in allocation and motivation.

\paragraph{} 
The resolution to this paradox lies in understanding that merit-based selection, despite creating Peter mismatches and generating many misaligned promotions, systematically elevates individuals with higher overall competence. These high performers bring substantial strengths (e.g., exceptional technical or soft skills) that continue to add value at the higher position, partially offsetting their weaker areas.  While these agents may experience performance drops when role weights shift unfavorably, their superior baseline competencies in at least some dimensions (which resulted in higher performance in the first place) continue to contribute positively to organizational performance. In contrast, seniority or random promotions disregard competence, meaning they often advance mediocre or low-skill individuals who contribute little in any role. Thus, even misaligned high performers often outcontribute randomly selected low performers. The distribution of talent thus favors the meritocratic system: a few brilliantly successful promotions can contribute outsized gains that outweigh the numerous smaller losses from misfires. 

\paragraph{}
The mathematical insight is that organizational efficiency depends on the distribution of performance contributions, not just the frequency of promotion successes. Even if most promotions create individual performance drops, the positive contributions from successful promotions can dominate the aggregate calculation if they are sufficiently large. This results in merit-based systems achieving higher overall efficiency despite more frequent mismatches.

\paragraph{} 
Moreover, merit-based systems preserve incentives for employees to work hard and develop skills---knowing that good performance will be rewarded---which bolsters overall productivity (a motivation effect absent in random or tenure-based systems). These factors explain why avoiding Peter Principle promotions does not necessarily maximize organizational efficiency. A promotion policy that is too cautious (ignoring merit) sacrifices the dynamic benefits of recognizing and leveraging talent. The optimal approach for efficiency lies in harnessing meritocratic selection while managing its downsides rather than abandoning merit altogether.

\subsection{Efficiency Outcomes by Regime and Promotion Strategy}

Our findings paint a nuanced efficiency landscape across different organizational regimes and promotion strategies. High-mismatch versus transferable-skills regimes exhibit distinctly different performance trajectories. In high-mismatch environments, a merit-based promotion strategy triggers an early efficiency drop---a sharp dip as many newly promoted stars falter---followed by a slow recovery and plateau. Efficiency under pure merit in this regime reached a modest equilibrium around $E_{100} \approx 0.549$ (essentially no net gain over the start) after an initial trough (minimum efficiency $\sim 0.541$). By contrast, in the transferable-skills regime (where each promotion entails only a gentle change in required skills), merit promotions yield steady, monotonic improvements in performance with no significant initial dip, ultimately achieving a much higher efficiency ($E_{100} \approx 0.585$, about $+21.8\%$ from baseline). Gentle skill transitions make any Peter Principle effects milder and more transient.

\paragraph{} 
Importantly, promotion strategies that avoid merit showed lower peak performance in both regimes. Under both seniority-based and random promotion policies, the high-mismatch case saw no dramatic drop at the outset, but the efficiency leveled off at a lower value (around $0.531$--$0.535$) and never caught up to the meritocratic strategy. A hybrid strategy (e.g., $70\%$ merit $+$ $30\%$ tenure) landed in between---in high-mismatch conditions it nearly matched pure merit’s eventual efficiency ($0.547$ vs.\ $0.549$) while softening the initial performance shock. In the transferable-skills regime, seniority and random promotions achieved some growth (single-digit percentage efficiency gains), but still considerably lagged the merit-based approach in the long run. The trade-off is clear: purely merit-based promotion maximizes performance growth in the favorable (transferable-skills) scenario and remains competitive even in hostile (high-mismatch) scenarios, whereas seniority and random rules produce more stable but ultimately weaker performance outcomes. Hybrid promotions offer a middle ground, blunting the worst merit-based mismatches at the cost of slight talent misallocation. Overall, we observed that meritocratic promotion drives higher peak efficiencies in both regime types, confirming that the benefits of selecting the most competent candidates often outweigh the costs of the Peter Principle---especially when skill carryover between roles is high.

\paragraph{} 
To put the promotion strategies in perspective: Merit-based promotion yields the highest organizational performance in the long run, but at the risk of many individual post-promotion failures in a high-mismatch context. Seniority-based promotion avoids abrupt performance drops by design (since promotions are uncorrelated with skill), yet it consistently results in under-utilization of talent and lower overall efficiency. Random promotion similarly disperses opportunity without regard to competence; it produces a roughly even mix of successes and failures (about $50/50$), avoiding systematic mismatches but also forgoing any intentional skill-based allocation. Both seniority and random strategies plateau at mediocre efficiency levels (roughly $5$--$12\%$ below the merit strategy outcomes, depending on regime). The hybrid strategy (merit plus a tenure component) demonstrates intermediate characteristics---it still leverages performance to pick many promotions, but the tenure quota introduces some randomness that occasionally places a solid veteran in a role even if they were not the top performer. This tends to dampen the extreme Peter Principle cases (fewer dramatic crashes) at a small cost to peak efficiency. In high-mismatch settings our $70/30$ hybrid nearly matched the merit system’s efficiency by the final time-step, indicating that a bit of randomness/tenure can buffer the system without much sacrifice. In transferable-skill settings, hybrid promotions also perform well, though pure merit was still best. 

\paragraph{}
This paper set out to answer a fundamental question: \emph{Does the Peter Principle exist? If so does it render hierarchical organizations inherently inefficient?} Using an agent-based model that isolates role–skill misalignment from confounds (e.g., learning-by-doing, politics), we show that Peter effects emerge predictably when promotion criteria and destination-role weights diverge. Yet, across regimes, \emph{meritocratic promotion}—despite producing many incipient misfits—consistently delivers the strongest or competitive \emph{system-level} efficiency, whereas strategies that suppress Peter Principle \emph{incidence} (seniority or random) sacrifice allocation and incentive gains and plateau at lower performance. Thus, “removing” Peter signals is \emph{not} sufficient to maximize efficiency. The right design question becomes: \emph{How do we keep the benefits of merit while attenuating its misfit costs?} We therefore evaluate two complementary levers that target the \emph{severity} and \emph{persistence} of misalignment rather than its mere existence: a corrective \emph{undo} policy (Selective Demotion) and a preventive \emph{prepare} policy (Merit-with-Training). This reframes the Peter Principle as a systemic feature to be \emph{managed}, not eliminated.

\paragraph{} 
Finally, we examined two mitigation strategies: \emph{Selective Demotion} (a merit-based scheme augmented with a policy to undo failed promotions) and \emph{Merit-with-Training} (merit-based promotion plus immediate post-promotion training). Both of these outperformed the unaugmented merit system in the \textbf{high-mismatch} regime, demonstrating that targeted interventions can recover much of the efficiency lost to Peter Principle effects. In the \textbf{transferable-skills} regime, merit-with-training produced the strongest results overall. By contrast, selective demotion offered little headroom there and performed on par with or slightly \emph{below} pure merit. We next discuss these mitigation strategies in depth.

\subsection{Mitigation Strategies: Reducing Peter Principle Inefficiencies}

Organizational policies can mitigate the efficiency loss due to Peter Principle effects by either correcting misallocations after they occur or preemptively equipping promotees for their new roles. Our study evaluated two such mechanisms and found each to be effective in different ways:

\subsubsection{Selective Demotion (Corrective).}
This strategy monitors each promotion and reverses ``mistakes''---if an employee’s performance drops by more than a set threshold (e.g., $5\%$) after promotion, they are demoted back to their previous position at the next opportunity. The demoted individual is also blacklisted from filling that same higher-level vacancy immediately, ensuring the organization tries a different candidate. The effect is to create a ``trial promotion'' period that tests fitness for the new role. 

Selective demotion proved highly effective at pruning out the worst mismatches. In the \textbf{high-mismatch} regime, it cut the incidence of large performance drops ($\Delta P \leq -0.05$) by about $60\%$ (from $43.3\%$ of promotions under pure merit down to $16.6\%$). By quickly removing those who truly cannot perform in the higher role, the policy boosted overall efficiency from $E_{100} \approx 0.549$ under pure merit to $0.569$ under demotion. We observed only a brief initial dip in efficiency followed by a healthy recovery and growth, instead of the prolonged stagnation seen with unmitigated merit promotions. In the \textbf{transferable-skills} regime, selective demotion offers little additional headroom. Because role-weight shifts are mild, relatively few promotions breach the demotion tolerance (i.e., $\Delta P \leq -\tau$), so the policy rarely fires. As a result, terminal efficiency is \emph{on par with or slightly below} pure merit (e.g., $E_{100}\!\approx\!0.582$ vs.\ $0.585$), reflecting occasional rollback churn and the lost compounding benefits from otherwise portable promotions.

The mechanism works largely by cleansing the upper ranks of misfits: over time, the higher levels become populated by individuals whose skill profiles better match the managerial demands (since those who were promoted but performed poorly get filtered out). A trade-off is that demotion introduces some turbulence---repeated churn in roles until a good match is found---which could harm morale or stability if not managed well. In practice, organizations can reap the benefits of this strategy by implementing it as a probationary period or acting assignment for promotions, so that reversal carries less stigma. When carefully designed (e.g., using ``acting'' titles or lateral reassignment with pay protection), selective demotion can correct poor promotion decisions with minimal political fallout. The key advantage of this approach is its insurance against the worst cases: it does not prevent all performance declines, but it prevents those declines from persisting. An incompetent promotee is promptly returned to a role where they are competent, rather than remaining in over their head. This keeps the overall efficiency trajectory on track by removing outliers that would otherwise drag it down.

\subsubsection{Merit-with-Training (Preventive).} 
This strategy preserves merit-based promotion but immediately follows it with targeted training for the promoted employee. In our implementation, each promotee received a one-time skill boost (computed via a logistic learning function) on \emph{trainable} dimensions —\textbf{technical} and \textbf{managerial}— right after promotion. The intuition is to close the skill gap before it can manifest as failure---essentially updating the agent’s competence vector to better align with the new role’s weight profile. 

This preventive approach proved extremely powerful, especially in regimes with moderate skill continuity. In the transferable-skills regime, adding post-promotion training led to the highest efficiency levels observed in any scenario: final efficiency reached $E_{100} \approx 0.633$, a $31.7\%$ improvement from baseline (about $46\%$ more total gain than even the pure merit policy). Remarkably, training flipped the usual pattern of mostly-negative promotion outcomes to two-thirds positive---under merit-with-training, $66\%$ of promotions actually increased the agent’s performance, a complete reversal of the Peter Principle trend seen under merit alone. 

Even in the high-mismatch regime, training significantly blunted the Peter effect: the share of promotions causing performance drops fell from $88\%$ to about $63\%$, and the average performance loss per promotion shrank dramatically (mean $\Delta P \approx -0.05$ with training vs.\ $-0.14$ without). The mechanism here is straightforward: by investing in employees’ capabilities at the moment of transition, the organization makes them more adaptable and better suited for the demands of the higher role. Essentially, it treats the cause of the Peter Principle (skill mismatch) rather than the symptoms, by proactively realigning the competence of promoted staff to their new responsibilities. 

The trade-offs for merit-with-training involve resource cost and recognition of its limits. Training programs require time, money, and effort---organizations must dedicate onboarding curricula, mentoring, or formal courses for newly minted managers, which is a non-trivial investment. However, our results suggest the return on investment can be very high, with substantial efficiency gains to be had (e.g., $+8$--$32\%$ improvements) if training is well-targeted. Another limitation is that not all skill gaps can be closed instantly. In our model, we only trained technical and managerial skills; if a higher-level role suddenly demands something like compliance knowledge or soft skills that the individual lacks, a brief training burst may not fully compensate. Indeed, we noted cases of residual Peter Principle effects where an agent with a very low compliance ability still struggled at a level requiring compliance, despite receiving the maximum technical/management training. Real-world leadership development programs recognize this and often combine immediate training with longer-term development for complex competencies. Despite these caveats, merit-based promotion with training is a potent preventive remedy---it maintains the motivational and sorting benefits of meritocracy while alleviating misalignment by actively grooming employees for their new roles. In contrast to the corrective nature of demotion, training is a positive intervention that organizations generally find palatable and supportive of employees. Where feasible, it allows companies to have their cake and eat it too: reward high performers with advancement and set them up to succeed, thereby reducing the inefficiency traditionally associated with the Peter Principle.

\paragraph{} 
It is worth noting that these two strategies are not mutually exclusive. In fact, they can complement each other. An organization in a high-mismatch context (e.g., a tech firm where technical experts are often promoted to people managers) might implement both policies: provide training to all promotees to boost their managerial skills and use a safety net demotion policy to catch the ones who still cannot perform. Training will elevate the general success rate of promotions by addressing common skill gaps, while selective demotion will handle the outliers by undoing the occasional failures that slip through. In a transferable-skills context, the need for demotion is minimal---training alone may suffice to almost eliminate Peter Principle losses, since mismatches are rarer and smaller. 

In practice, choosing the right mitigation (or combination) will depend on the organizational culture and the nature of role transitions. Selective demotion provides a corrective backstop against bad promotions, whereas merit-with-training is a preventive boost to help people grow into new positions. Both approaches demonstrated that the Peter Principle, while pervasive, can be substantially managed through thoughtful policy design rather than simply being an unavoidable fate.




\section*{Future Research Directions}

Our agent-based model opens several avenues for future investigation to deepen and broaden the understanding of promotions and the Peter Principle. It serves as a starting outline and further modifications can be made to the model to explore a range of organizational, behavioral, and policy complexities. Many promising extensions have already been hinted at by our findings and could be pursued in subsequent research:

\begin{enumerate}
    \item \textbf{Expanded Skill and Organizational Scope:} One extension is to increase the dimensionality and realism of the competence and role profiles. We used four skill dimensions; future models could incorporate additional skills or traits (e.g., creativity, emotional intelligence, domain-specific knowledge) or sub-skills (e.g., technical $\to$ programming + system design + troubleshooting). This might help to calibrate the model to specific industries or job families. 
    \item \textbf{Dynamic Skill Evolution and Learning:} In our model, agents’ competences were mostly static aside from post-promotion training spurts. Future work should explore dynamic skill evolution---agents continuously learning or depreciating in skills over time, even without formal training interventions. This would allow modeling of experience accumulation, on-the-job learning, or skill atrophy, and how those influence promotion outcomes. This would also allow us to update Seniority based promotion where agents gain mild skills (e.g., compliance or soft-skills) upon reaching certain tenure to make sure it is not just mimicking the Random promotion strategy.
    \item \textbf{External and Lateral Hiring Dynamics:} Our current model focuses on internal promotions in an hierarchy; by allowing lateral role transitions (not just upward promotions) could capture scenarios where employees move to different departments or specialties at the same level to better match their skills. We could also consider external hiring by simulating external labour markets, where new employees enter at various levels with different competence profiles, to see how this impacts overall efficiency and promotion dynamics. Note: Cost considerations would be crucial here (external hires are often more expensive than internal promotions even if they promise better overall outcomes)
    \item \textbf{Continuous and Multi-Period Training Interventions:} We modeled training as a one-time boost immediately after promotion. Future work could explore more realistic training regimes, such as multi-stage or ongoing development programs that unfold over time, or providing training for all employees periodically (not just new promotees). This would generally result in better outcomes, but at a higher cost. Future work could explore the cost-benefit tradeoffs of different training cadences and intensities.
    \item \textbf{Organizational Politics and Social Networks:} Our current model is intentionally meritocratic and skill-focused, but real organizations have political behaviors, cultural norms, and coalition dynamics that affect promotions. Future research could integrate factors like office politics, favoritism/nepotism, influence networks, or cultural attitudes toward promotions and demotions. Agent-based models could simulate scenarios where promotions require not just performance but also political support, or where demotions carry stigma that impacts morale. Incorporating such social dynamics would provide a richer picture of promotion systems, albeit with added complexity. We could consider this as a \emph{Game-Theory} style simulation where we can figure out optimal strategies for agents to get promoted (e.g., \citet{lazear2004peter} demonstrates how the fundamental principle that different reward structures cause behavioral changes in agents in promotion contexts.) We could also model ``Who you know vs. what you know'' dynamics. It is evident from literature network position, sponsorship, and brokerage strongly shape promotion odds independent of human-capital controls(cite literature)
    \item \textbf{Threshold Sensitivity and Adaptive Parameters:} We could systematically vary key parameters such as the demotion threshold $\tau$ to identify the effects and figure out optimal tolerance levels. Further we could also change attrition rates, level distribution, role profiles, or hybrid weightings to see how sensitive outcomes are to these design choices. This would help identify robust policies that perform well given the conditions. We could also consider defining other promotion strategies and see their effects, comparing them to the ones we have already defined. These parameters could be tuned to a specific organization or industry to simulate realistic scenarios.
    \item \textbf{Resource Constraints and Optimization:} We could introduce budget constraints for training programs, external hiring or limits on how many demotions can occur in a period, to simulate real-world resource trade-offs. This would allow us to study how organizations can optimally allocate limited resources between promotion, training, and retention to maximize efficiency. We could implement prioritization heuristics for allocating capacity among training, demotions, and hires to optimize training ROI.
    \item \textbf{Empirical Validation and Cross-Industry Studies:} To connect these modeling insights with the real world, more empirical research is needed. One direction is to analyze longitudinal organizational data---tracking companies that use different promotion strategies (strict meritocracy vs.\ seniority, or those that have training programs, etc.) and measuring their performance over time to see if patterns align with our simulated results. Natural experiments could be used, for example if a firm changes its promotion policy, to observe before-and-after efficiency changes. Cross-industry comparisons would also be enlightening: our model suggests that industries like tech\allowbreak /sales (high mismatch) are prone to strong Peter effects, while academia\allowbreak /research (more gradual skill progression) fare better, but systematic data across many firms and sectors would validate these regime characterizations. Validating the model’s predictions with real promotion outcomes (such as the fraction of promoted managers who fail and either leave or get demoted, as studied in \citet{benson2019promotions} for sales organizations) would strengthen confidence in the theory. It could also reveal aspects the model misses---for instance, maybe some firms already effectively mitigate Peter effects through informal practices.
\end{enumerate}

\section*{Limitations}

While our agent-based model provides a powerful tool for understanding promotion dynamics, it is built on several simplifications that warrant caution. First, we assumed a fixed four-dimensional competence vector for each agent (apart from discrete post-promotion training updates). This means we did not model gradual development or decay of skills over time---in reality, employees learn from experience, and performance in a new role might improve after an initial adjustment period even without formal training. Our model’s agents either stayed static or received one-shot training, which likely overstates the permanence of initial promotion shocks. 

Second, the skill set and role profiles, though multi-\linebreak dimensional, are still limited. We focused on technical, managerial, compliance, and soft skills, but other qualities like creativity, leadership charisma, or network influence were not captured. We also treated the organization in a vacuum, ignoring interpersonal factors (mentoring, team synergy, morale effects) that could moderate promotion outcomes. These simplifications were deliberate to keep the model tractable and the results interpretable. Indeed, there is a complexity--transparency trade-off at play: a more intricate model with many variables might mimic reality more closely, but it would be harder to isolate cause and effect or to generalize findings. Our streamlined design sacrifices some realism (no emotional intelligence, no cultural nuances, etc.) in exchange for clarity in analyzing the core mechanism of the Peter Principle. 

Finally, as with any simulation, our quantitative results (exact percentages, etc.) are not meant to be precise forecasts for actual organizations. They depend on the chosen parameters (e.g., distribution of agent skills, magnitude of weight changes, promotion rates) and initial conditions. We aimed for realistic orders of magnitude, but real-world organizations vary widely. The value of the model is in the qualitative insights and comparative outcomes, not the specific numeric efficiency values.

\section*{Broader Implications and Conclusion}

Our study suggests that the Peter Principle is not a sign of dysfunctional management, but rather a nearly unavoidable byproduct of meritocratic promotion in multi-level organizations. In essence, whenever people are promoted based on performance in their current job, and the next job is materially different, some degree of performance decline is a systemic feature. Recognizing this reality is the first step toward managing it. 

Organizations should not overreact by rejecting merit-based advancement altogether; as we showed, meritocratic systems, for all their Peter Principle flaws, often still outperform the alternatives because they motivate employees and allocate talent better overall. Instead, the goal should be to manage the Peter Principle effect, not eliminate it. This means designing promotion systems and career structures that anticipate performance drops and cushion their impact, through measures like training, trial periods (demotions if needed), or even redesigning roles to reduce skill mismatches. 

Our findings have several concrete implications. For high-mismatch industries (think fast-growing tech firms or sales organizations), it is critical to provide new managers with substantial training and/or to use provisional promotions so that top technical specialists are not permanently trapped in roles where they underperform. For more skill-continuous fields (academia, research), a straightforward merit-based system with light-touch interventions may suffice. In any case, organizations that understand these dynamics can strategically harness promotions as a tool for human capital optimization rather than a gamble. By measuring post-promotion performance (e.g., tracking $\Delta P$ as a diagnostic) and responding with smart policies, firms can sustain high performance and employee engagement at the same time.

\paragraph{}
In conclusion, the Peter Principle emerges from a fundamental tension in organizational design---the very act of rewarding people for excellence can inadvertently move them into roles where that excellence disappears. Our agent-based model provided a theoretical lens to see when and why this happens, and it demonstrated that while you cannot completely eliminate the Peter Principle in a merit-driven hierarchy, you can certainly mitigate its impact. 

Far from rendering merit promotions ``wrong,'' the Peter Principle calls for nuanced management: it reminds us that every promotion is both an opportunity and a risk. Wise organizations will continue to promote talent, but will do so with eyes open---deploying training, probation, or alternative career paths to ensure that the rise of an employee is aligned with their ability to contribute. In this way, the Peter Principle can be tamed and turned from a purely negative phenomenon into a manageable aspect of workforce strategy. 

Ultimately, treating the Peter Principle as a systemic reality to be managed rather than a failure to be rid of will lead to more resilient and efficient organizations---ones that reward merit, support employee development, and optimize the fit between people and positions for the benefit of all.


\bibliographystyle{model1-num-names}
\bibliography{peter} 

@book{peter1969peter,
  author    = {Peter, Laurence J. and Hull, Raymond},
  title     = {The Peter Principle: Why Things Always Go Wrong},
  publisher = {William Morrow and Company},
  address   = {New York},
  year      = {1969}
}

@article{kane1970promotion,
  author    = {Kane, Edward J.},
  title     = {Dynamics of the Peter Principle},
  journal   = {Management Science},
  volume    = {16},
  number    = {12},
  pages     = {B-800--B-811},
  year      = {1970},
  publisher = {INFORMS},
  doi       = {10.1287/mnsc.16.12.B800}
}

@article{lazear2004peter,
  author    = {Lazear, Edward P.},
  title     = {The Peter Principle: A Theory of Decline},
  journal   = {Journal of Political Economy},
  volume    = {112},
  number    = {S1},
  pages     = {S141--S163},
  year      = {2004},
  publisher = {The University of Chicago Press},
  doi       = {10.1086/379943}
}

@article{pluchino2010peter,
  author    = {Pluchino, Alessandro and Rapisarda, Andrea and Garofalo, Cesare},
  title     = {The Peter Principle Revisited: A Computational Study},
  journal   = {Physica A: Statistical Mechanics and its Applications},
  volume    = {389},
  number    = {3},
  pages     = {467--472},
  year      = {2010},
  publisher = {Elsevier},
  doi       = {10.1016/j.physa.2009.09.045}
}

@article{benson2019promotions,
  author    = {Benson, Alan and Li, Danielle and Shue, Kelly},
  title     = {Promotions and the Peter Principle},
  journal   = {The Quarterly Journal of Economics},
  volume    = {134},
  number    = {4},
  pages     = {2085--2134},
  year      = {2019},
  month     = {11},
  publisher = {Oxford University Press},
  doi       = {10.1093/qje/qjz020}
}

@article{merton1968matthew,
  author    = {Merton, Robert K.},
  title     = {The Matthew Effect in Science},
  journal   = {Science},
  volume    = {159},
  number    = {3810},
  pages     = {56--63},
  year      = {1968},
  publisher = {American Association for the Advancement of Science},
  doi       = {10.1126/science.159.3810.56}
}

@book{cole1973social,
  author    = {Cole, Jonathan R. and Cole, Stephen},
  title     = {Social Stratification in Science},
  publisher = {University of Chicago Press},
  address   = {Chicago},
  year      = {1973}
}

@article{leibowitz2010sigmoid,
  author    = {Leibowitz, Nathaniel and Baum, Barak and Enden, Giora and Karniel, Amir},
  title     = {The exponential learning equation as a function of successful trials results in sigmoid performance},
  journal   = {Journal of Mathematical Psychology},
  volume    = {54},
  number    = {3},
  pages     = {338--340},
  year      = {2010},
  publisher = {Elsevier},
  doi       = {10.1016/j.jmp.2010.01.006}
}

@article{fetta2012peter,
  author    = {Fetta, A.G. and Harper, P.R. and Knight, V.A. and Vieira, I.T. and Williams, J.E.},
  title     = {On the Peter Principle: An agent based investigation into the consequential effects of social networks and behavioural factors},
  journal   = {Physica A: Statistical Mechanics and its Applications},
  volume    = {391},
  number    = {7},
  pages     = {2898--2910},
  year      = {2012},
  publisher = {Elsevier},
  doi       = {10.1016/j.physa.2011.12.053}
}

@article{farias2021peter,
  author    = {Farias, B. and Rap{\^o}so Jr, O. and Penna, T.J.P. and Girardi, D.},
  title     = {The Peter Principle and learning: A safer way to promote workers},
  journal   = {Physica A: Statistical Mechanics and its Applications},
  volume    = {576},
  pages     = {126023},
  year      = {2021},
  publisher = {Elsevier},
  doi       = {10.1016/j.physa.2021.126023}
}

@article{coase1937nature,
  author    = {Coase, Ronald H.},
  title     = {The Nature of the Firm},
  journal   = {Economica},
  volume    = {4},
  number    = {16},
  pages     = {386--405},
  year      = {1937},
  publisher = {Wiley},
  doi       = {10.1111/j.1468-0335.1937.tb00002.x}
}

@incollection{gulick1937notes,
  author    = {Gulick, Luther},
  title     = {Notes on the Theory of Organization},
  booktitle = {Papers on the Science of Administration},
  editor    = {Gulick, Luther and Urwick, Lyndall},
  pages     = {1--45},
  publisher = {Institute of Public Administration},
  address   = {New York},
  year      = {1937}
}

@article{vanfleet1983span,
  author    = {Van Fleet, David D.},
  title     = {Span of Management Research and Issues},
  journal   = {Academy of Management Journal},
  volume    = {26},
  number    = {3},
  pages     = {546--552},
  year      = {1983},
  publisher = {Academy of Management},
  doi       = {10.2307/256261}
}

@article{cotton1986employee,
  author    = {Cotton, John L. and Tuttle, Jeffrey M.},
  title     = {Employee Turnover: A Meta-Analysis and Review with Implications for Research},
  journal   = {Academy of Management Review},
  volume    = {11},
  number    = {1},
  pages     = {55--70},
  year      = {1986},
  publisher = {Academy of Management},
  doi       = {10.5465/amr.1986.4282625}
}

@article{becker1962investment,
  author    = {Becker, Gary S.},
  title     = {Investment in Human Capital: A Theoretical Analysis},
  journal   = {Journal of Political Economy},
  volume    = {70},
  number    = {5, Part 2},
  pages     = {9--49},
  year      = {1962},
  publisher = {The University of Chicago Press},
  doi       = {10.1086/258724}
}

@article{fairburn2001performance,
  title={Performance, Promotion, and the Peter Principle},
  author={Fairburn, James A. and Malcomson, James M.},
  journal={The Review of Economic Studies},
  volume={68},
  number={1},
  pages={45--66},
  year={2001},
  publisher={Oxford University Press},
  doi={10.1111/1467-937X.00159}
}

@article{grabner2013managers,
  title={Managers' Choices of Performance Measures in Promotion Decisions: An Analysis of Alternative Job Assignments},
  author={Grabner, Isabella and Moers, Frank},
  journal={Journal of Accounting Research},
  volume={51},
  number={5},
  pages={1187--1220},
  year={2013},
  publisher={Wiley},
  doi={10.1111/1475-679X.12026}
}

@article{hermelin2007validity,
  title={The Validity of Assessment Centres for the Prediction of Supervisory Performance Ratings: A Meta-Analysis},
  author={Hermelin, Eran and Lievens, Filip and Robertson, Ivan T.},
  journal={International Journal of Selection and Assessment},
  volume={15},
  number={4},
  pages={405--411},
  year={2007},
  publisher={Wiley},
  doi={10.1111/j.1468-2389.2007.00394.x}
}

@article{lacerenza2017leadership,
  title={Leadership Training Design, Delivery, and Implementation: A Meta-Analysis},
  author={Lacerenza, Christina N. and Reyes, Denise L. and Marlow, Shannon L. and Joseph, Dana L. and Salas, Eduardo},
  journal={Journal of Applied Psychology},
  volume={102},
  number={12},
  pages={1686--1718},
  year={2017},
  publisher={American Psychological Association},
  doi={10.1037/apl0000241}
}

\end{document}